\documentclass[a4paper, 11pt, openany]{book_masahito}
\usepackage{book_masahito}

\usepackage{graphicx}
\usepackage{makeidx}
\makeindex
\usepackage[refeq,refpage,english]{nomencl}
\renewcommand{\nomname}{List of Symbols}
\renewcommand{\nompreamble}{\raggedright}
\makenomenclature
\usepackage[fleqn]{amsmath}
\usepackage{amssymb}
\usepackage{ascmac}

\usepackage{bm}
\usepackage{tikz} 
\usepackage{slashed}
\usepackage[utf8]{inputenc}
\usepackage[T1]{fontenc}
\usepackage{url}
\usepackage{newtxtext,newtxmath}
\usepackage{slashbox} 
\usepackage[dvips,colorlinks=true,linkcolor={blue!50!black},citecolor={green!40!black},
  urlcolor={blue!65!black},bookmarksnumbered=true,
  linktoc=page]{hyperref} 
\usepackage{breakurl} 

\graphicspath{{./Figures/}}

\def\beq#1\eeq{\begin{align}#1\end{align}}

\newcommand{\sfP}{\ensuremath{\hat{P}}}
\newcommand{\sfQ}{\ensuremath{\hat{Q}}}
\newcommand{\sfX}{\ensuremath{\hat{X}}}
\newcommand{\sfY}{\ensuremath{\hat{Y}}}
\newcommand{\sfx}{\ensuremath{\hat{y}}}
\newcommand{\sfZ}{\ensuremath{\hat{Z}}}

\newcommand{\Li}{\ensuremath{\textrm{Li}_2}}

\newcommand{\bll}{\blacklozenge}

\newcommand{\bparagraph}[1]{\subsection{#1}}

\newcommand{\qbinom}[2]{\left[\genfrac{}{}{0pt}{}{#1}{#2}\right]}

\newcommand{\CA}{\ensuremath{\mathcal{A}}}
\newcommand{\CD}{\ensuremath{\mathcal{D}}}
\newcommand{\CF}{\ensuremath{\mathcal{F}}}
\newcommand{\bCA}{\ensuremath{\overline{\mathcal{A}}}}

\newcommand{\bC}{\ensuremath{\mathbb{C}}}
\newcommand{\bE}{\ensuremath{\mathbb{E}}}
\newcommand{\bH}{\ensuremath{\mathbb{H}}}
\newcommand{\bP}{\ensuremath{\mathbb{P}}}
\newcommand{\bR}{\ensuremath{\mathbb{R}}}
\newcommand{\bS}{\ensuremath{\mathbb{S}}}
\newcommand{\bT}{\ensuremath{\mathbb{T}}}
\newcommand{\bZ}{\ensuremath{\mathbb{Z}}}

\newcommand{\scA}{\ensuremath{\mathcal{A}}}
\newcommand{\scC}{\ensuremath{\mathcal{C}}}
\newcommand{\scD}{\ensuremath{\mathcal{D}}}
\newcommand{\scF}{\ensuremath{\mathcal{F}}}
\newcommand{\scH}{\ensuremath{\mathcal{H}}}
\newcommand{\scL}{\ensuremath{\mathcal{L}}}
\newcommand{\scM}{\ensuremath{\mathcal{M}}}
\newcommand{\scN}{\ensuremath{\mathcal{N}}}
\newcommand{\scQ}{\ensuremath{\mathcal{Q}}}
\newcommand{\scR}{\ensuremath{\mathcal{R}}}
\newcommand{\scT}{\ensuremath{\mathcal{T}}}
\newcommand{\scV}{\ensuremath{\mathcal{V}}}
\newcommand{\scW}{\ensuremath{\mathcal{W}}}
\newcommand{\scX}{\ensuremath{\mathcal{X}}}

\newcommand{\Teichmuller}{Teichm\"{u}ller }

\def\inv{^{-1}}

\begin{document}
\input{cover_e} 
\frontmatter
\chapter*{Preface to the Japanese Edition}
\phantomsection\addcontentsline{toc}{chapter}{Preface to the Japanese Edition}

Let us start with a question: \textbf{what is quantum field theory (QFT)?}

\bigskip

This book deals with quantum field theories,
in particular quantum field theories with supersymmetry.
However, here in the 21st century, when there are already many excellent books
both on field theory and on supersymmetric theories,
why do we need yet another book now?

When I ask myself this question again,
I recall the time when I started studying high-energy theory.
What my thought at that time was that
one should quickly master the basics of field theory and ``graduate'' from it,
and move on to more ``advanced'' topics, for example in string theory.
It is true that present-day students studying high-energy theory, and in particular string theory,
have many things to learn,
and my thought back then might be justified based on practical necessity.

Of course, it was clear to me that many essential problems
remain in quantum field theory.
Field theory is a gigantic framework, and applying it to various real-world systems
is an important problem in each field; moreover, how to solve the problems of individual field theories
(for example, how to show quark confinement analytically) is
a big problem in itself. 
However, isn't that the case that the basic paradigm of field theory as we know it
already been established by the efforts of our great predecessors,
so that nothing is left to be done about \textbf{the framework of field theory itself}?
Is our job as researchers, born in later generations,
to engage in the game of field theory as ``normal science'',
following the rules established by our predecessors?
This was the question I had at that time.

Needless to say, research on field theory as ``normal science'' is extremely
important, and one can even say that it is the royal road of research on field theory
as part of scientific research.
However, it is also true that in the discipline of physics, great breakthroughs have been made
by returning to fundamental questions,
and it may not be so outlandish to expect that even the huge framework of field theory
is no exception.
It is a fact that the mathematical foundation of field theory has not yet been completed even today,
and with the recent developments of string theory, new perspectives on field theory have begun to emerge.
Reconsidering the formulation of field theory itself anew
may not be entirely useless.
Looking back at history, there are examples where our understanding of field theory
deepened dramatically: for example,
the ideas of renormalization and effective field theory
came to rewrite the textbooks of field theory substantially.

\bigskip

How much have we understood the very essence of quantum field theory? 
What are the exact data specifying a field theory?
Among all possible field theories, how many do we actually know
in the usual sense?
Is there a new formulation of field theory, completely different from
the current formulation?
When we consider the whole theory space of field theories, is there some structure in it?
The questions here are not questions about how to understand a specific field theory
(for example, 4d quantum chromodynamics (QCD)),
but questions concerning field theory itself, such as the formulation and understanding of field theories and their collections.

Of course, these questions have already been considered by many predecessors
(among them, as described later, the idea of renormalization is essential),
and it is also that case that these questions themselves are
often vague unless the problem is set up appropriately.
It is also not at all clear whether they are appropriate questions
which can be answered within our current knowledge.

However, I would like to claim that there is not a complete lack of clues to these questions.
What this book introduces is a way of thinking which may grow into
an important clue in approaching these questions---its
history is old, but its importance has been
recognized anew since the beginning of the 21st century.

In a word, it is to understand the properties of the \textbf{theory space} of supersymmetric gauge theories and the relations among them,
in particular their \textbf{dualities}, by making full use of various \textbf{geometric methods}.
Here let us call such a framework of understanding
\keyword{geometric field theory}{geometric field theory}.
This can be said to be an attempt to translate, so to speak, the whole structure formed by field theories into problems of geometry.

\bigskip

The goal of this book is to explain
this idea of geometric field theory, taking as an example the relation between
\textbf{3d $\mathcal{N}=2$ supersymmetric gauge theories}
and \textbf{the theory of 3-manifolds}.
There, geometry and algebra, such as 2d and 3d hyperbolic geometry, knot theory, and cluster algebras,
appear one after another, depicting the rich world of supersymmetric gauge theories.

Progress in this area has been made only in the last few years,
and no book, including books in English, has so far
tackled this subject head-on.
The purpose of this book is to fill this gap.
That does not mean, however, that the aim is to be eccentric.
Many of the properties of supersymmetric gauge theories treated in this book are
standard ones known since the 1990s.
Rather, one of the claims is that, by reconsidering these standard contents,
\textbf{geometric structures appear naturally}.

The mathematics appearing in this book may not be very familiar to physics students,
and the contents of this book seem to give readers the impression of being
too mathematical.
However, the mathematical structures we discuss are
not artificially created---they are beautiful melodies which can be heard
by listening carefully to the tunes of supersymmetric field theories.

This book is neither a comprehensive textbook nor an introductory book on
supersymmetric field theories,
nor a book in which an authority of the field looks back on the development of the field based on experience.
It is also not a book summarizing contents
established over many years.
Rather, the core of the contents of this book consists of very recent
research results, and in this sense
there is a good chance that the contents of this book will become outdated
sooner or later; I myself hope that this will happen
through future research.
Part of the contents of this book includes the author's arbitrary judgments and biases,
for which I can only ask for the reader's indulgence.

Despite all these shortcomings,
there is one thing this book aims at.
It is to convey to the reader, \textbf{as something happening right now},
the story of encounters and the excitement I have experienced
in trying to understand field theories.
That there exists, even now in the 21st century,
a world in which field theories are vibrantly alive---this is
the modest message I want to convey to myself ten years ago, and to you, who are now holding this book.

\bigskip
\textbf{Princeton, April 2015}

\begin{flushright}%
	\textbf{Masahito Yamazaki}

\end{flushright}

\subsection*{Acknowledgments}

The contents of this book have emerged through my encounters and discussions
with countless people over the past thirty-odd years.
They are too many to name one by one here,
but I would like to take this opportunity to thank all of them.

For the contents directly related to this book, I am especially grateful for discussions with my collaborators, who are also mathematicians,
Yuji Terashima and Kentaro Nagao.\footnote{While
	writing this book, I received the sad news of Kentaro Nagao's passing.
	The short but wonderful time
	in which I could discuss mathematics and physics with him
	is irreplaceable.
	It is truly regrettable that such a wonderful person has been lost.
	I would like to dedicate this book to his memory.
} Yuji Terashima and Kazuya Yonekura gave me comments on the manuscript. Moreover, in relation to the writing of this book, I would like to thank Nima Arkani-Hamed, Benjamin Assel, Francesco Benini, Tudor D. Dimofte, John Estes, Dongmin Gang, Sergei Gukov, Akikazu Hashimoto, Jonathan J. Heckman, Kazuo Hosomichi, Ken Intriligator, Ivan Chi-ho Ip, Rinat Kashaev, Igor Klebanov, Sungjay Lee, Tatsuma Nishioka, Takuya Okuda, Peter Ouyang, Pavel Putrov, Shlomo S. Razamat, Mauricio Romo, Nathan Seiberg, Yuji Tachikawa, J\"{o}rg Teschner, Cumrun Vafa, Herman Verlinde, Noriaki Watanabe, Brian Willett, Edward Witten, and Dan Xie for discussions.

I would like to thank Saiensu-sha, who planned the publication of this book,
and the editors, Kosuke Hirase and, in particular, Ryota Takahashi.
When I received the request to write this book,
I honestly felt some hesitation as to whether it was meaningful for a young researcher
to write such a book in Japanese; however, having been allowed to write as I liked
within the given length
was a precious experience of re-examining my own understanding.
I hope that this modest result will also
be meaningful for the readers.

This book organizes and develops the contents of the talks I have given at various places, in particular
the lectures at the Autumn Meeting of the Physical Society of Japan (2012),
the Research Institute for Mathematical Sciences, Kyoto University (2013), and
National Taiwan University (2013),\footnote{
	One can find the videos of the talks (three hours in total) uploaded to YouTube, starting with the first talk \url{https://www.youtube.com/watch?v=ltCxsYiMFog}. See my YouTube channel \url{https://www.youtube.com/@masahito.yamazaki} for many more videos related to the contents of this book.} as well as
the articles in the magazine Suuri Kagaku \cite{Yamazaki_science_2012,Yamazaki_science_2015}.\footnote{
	The articles in Suuri Kagaku explain the contents of this book in a much shorter length than this book,
	and are especially recommended for busy readers.
}
I would like to thank everyone who gave me feedback on the occasions of these lectures.

I would also like to thank
the Kavli Institute for the Physics and Mathematics of the Universe, the University of Tokyo, and the Institute for Advanced Study, Princeton, which supported my research and writing,
as well as everyone who has supported my scientific research.\footnote{The research on the contents of this book was carried out
	around the time of the Great East Japan Earthquake. Continuing to face the world, and to keep asking myself the meaning of my life, even amid a sense of desperate powerlessness---this inexhaustible feeling was the driving force that pushes me toward research.}

\subsection*{Summary of Contents}

\begin{figure}[t]
	\centering
	\begin{tikzpicture}[inner sep=0]
		\node[anchor=south west] (img) {\includegraphics[scale=0.27]{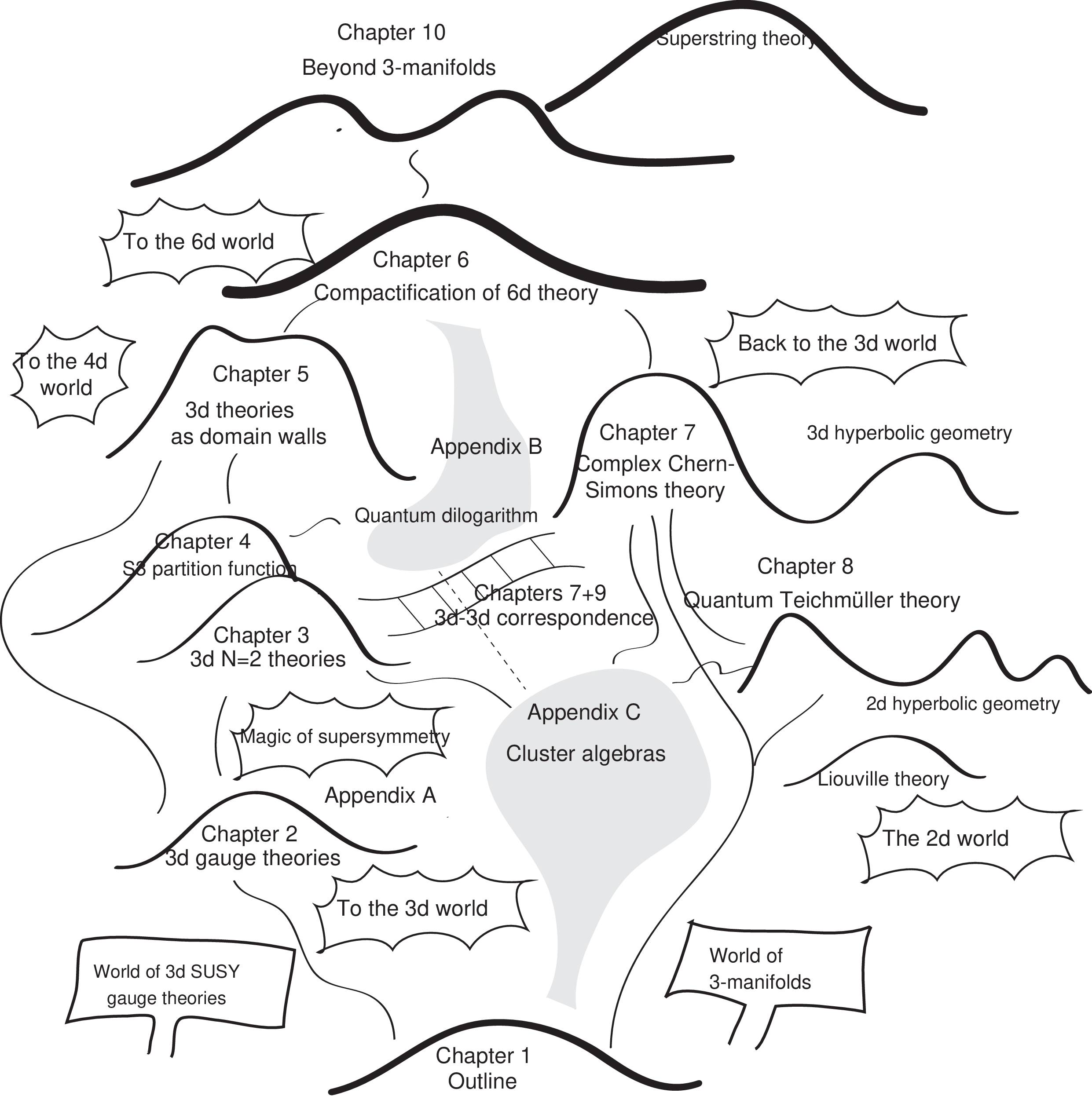}};
		\begin{scope}[shift={(img.south west)},x=0.27pt,y=0.27pt]
			\fill[white] (126,556) rectangle (332,624);
			\node[anchor=north,inner sep=0,align=center] at (229,622) {\tiny Chapter 4\\[1pt]\tiny $S^3$ partition function};
			\fill[white] (448,502) rectangle (714,568);
			\node[anchor=north,inner sep=0,align=center] at (581,566) {\tiny Chapters 7+9\\[1pt]\tiny 3d-3d correspondence};
		\end{scope}
	\end{tikzpicture}
	\caption{A roadmap for traversing the world of this book.
		We move freely back and forth between the world of 3d $\scN=2$ supersymmetric gauge theories and
		the world of 3-manifolds.
		There are no borders in science.}
	\label{fig.flowSGC}
\end{figure}
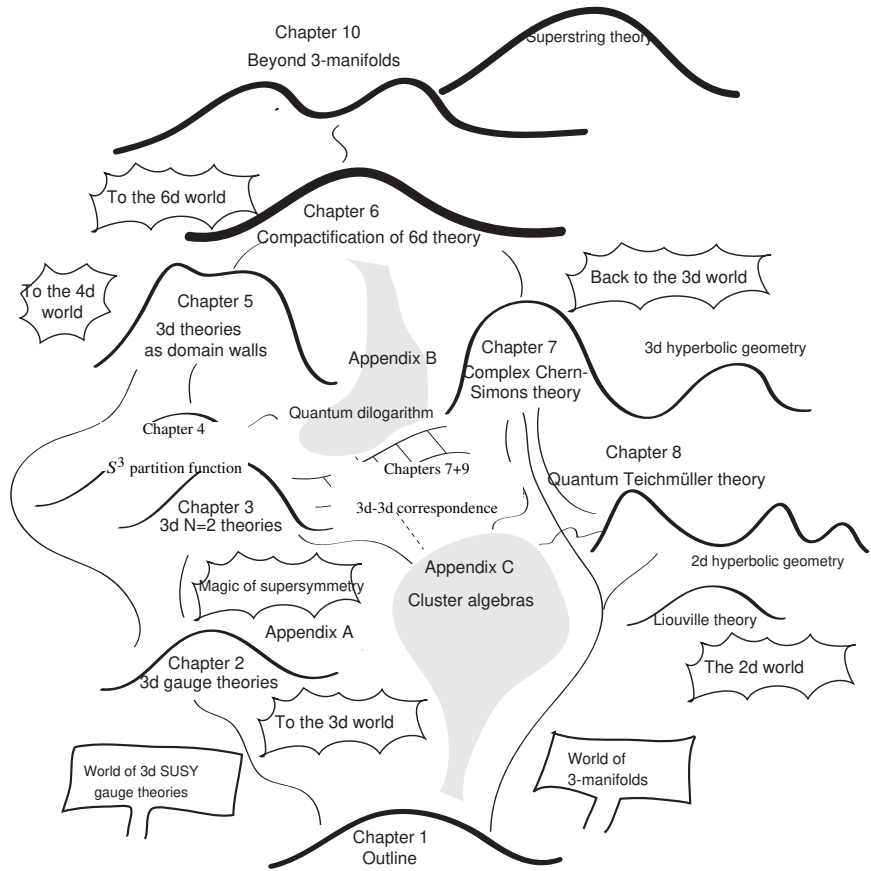

For the convenience of the reader,
let us give an overview of the contents of this book (Fig.~\ref{fig.flowSGC}).

In Chap.~\ref{chap.intro},
before going into specific topics, I would like to explain the basic ideas
in a more general context than in the rest of this book,
and to motivate the later discussions.

After explaining 3d field theories, in particular their ``gluing'', in Chap.~\ref{chap.3dglue},
in the next Chap.~\ref{chap.3dN2} we summarize the basics of theories with 3d $\scN=2$ supersymmetry.
In the next Chap.~\ref{chap.S3}, we reinterpret them in the language of the $S^3$ partition function and
of the quantum dilogarithm function defined in Appendix~\ref{app.dilog}. In Chap.~\ref{chap.wall}, the $S^3$ partition function is
reinterpreted as the expectation value of an operator on a certain finite-dimensional Hilbert space,
and it is clarified that behind this lies the domain wall interpretation of 3d $\scN=2$ theories.

While we have so far exclusively discussed 3d $\scN=2$ gauge theories,
from Chap.~\ref{chap.6d} onward we enter the world of 3-manifolds.
After discussing the compactification of the 6d theory in Chap.~\ref{chap.6d},
Chap.~\ref{chap.complexCS} deals with the complex Chern-Simons theory arising from it, and
3d hyperbolic geometry as its classical solutions.

Moreover, in order to discuss the properties of 3d supersymmetric theories in more detail,
after organizing classical and quantum \Teichmuller theory in Chap.~\ref{chap.Teichmuller}, in the next Chap.~\ref{chap.3mfd} we comment on the 3d $\scN=2$ supersymmetric field theories corresponding to ideal tetrahedral decompositions of 3-manifolds and to braid representations of knots,
and the main text comes to an end in the final Chap.~\ref{chap.conclusion}.
The three Appendices~\ref{chap.SUSYbasic}, \ref{app.dilog}, and \ref{app.clusterapp}
summarize supersymmetric gauge theories, quantum dilogarithm functions, and cluster algebras, respectively.
For the convenience of the reader, an index and a list of symbols are also included.

\subsection*{Prerequisites}
As the main readers of this book, I had in mind
graduate students and motivated undergraduates
interested in elementary particle physics.
I also hope that the book will help researchers in related fields
to learn this area quickly.

In the traditional curriculum on field theory, this book would be positioned
as an advanced course in field theory.
However, the knowledge needed to understand this book is actually not that much,
and I have tried to make the book as self-contained as possible.

A conceptual understanding of the basic ideas of field theory, for example what gauge theories are
and what renormalization is, is desirable,
but technical contents such as the details of perturbative computations with Feynman diagrams are hardly needed.
The basics of 4d $\mathcal{N}=1$ supersymmetry
(e.g.\ the contents of the first seven chapters of the textbook by Wess and Bagger \cite{Wess:1992cp})
are assumed, but are not essential for grasping the overall picture.
For example, knowledge of the superfield formalism is
used in Chap.~\ref{chap.3dN2}, but
only for the simplicity of the explanation, and it is hardly needed in the subsequent chapters.
Therefore, readers who are about to give up at the superfield formalism in Chap.~\ref{chap.3dN2}
can skip it as appropriate and read on.
Moreover, knowledge of string theory should basically not be needed,
except in a few places where it is mentioned superficially.
As for mathematics, apart from basic calculus, complex analysis, differential geometry (e.g.\ differential forms),
and very elementary topology, no highly specialized content is needed.
The necessary mathematical facts, for example on 3d hyperbolic geometry and quantum dilogarithm functions, are
introduced with explanations as they are needed.

Since this book was written mainly with physics readers in mind, physical explanations are emphasized in many places;
however, it is also possible to read it focusing on the more mathematical contents. In that case, one can focus on
the mathematical parts of Chaps.~\ref{chap.complexCS}, \ref{chap.Teichmuller}, and \ref{chap.3mfd} and Appendices~\ref{app.dilog} and \ref{app.clusterapp},
and pick up the earlier chapters depending on interest and need.

The parts in small type and the footnotes in the main text contain somewhat advanced topics and supplementary explanations.
Even if there are parts of them you cannot understand, they do not interfere with understanding
the overall flow, so please keep reading.
The understanding of physics deepens in a spiral, and parts you cannot understand now
may make sense when you come back to them later; this is
one of the pleasures of reading.

\subsection*{Exercises}

To help the reader's understanding, exercises are placed at the end of each chapter.
Many of the exercises are easy if one follows the contents of the main text, but
some deal with somewhat advanced contents or with the contents of supplementary references not directly treated in this book,
so that you may need to think on your own or
consult the literature.
Some of them also do not necessarily have definite answers.
Rather than trying to solve all of them perfectly, I hope you will approach them casually,
as an aid to active learning.
As a rough guide to the difficulty, they are marked for convenience as $[\bll], [\bll\bll]$.
$[\bll]$ indicates problems which can be shown immediately by computation, and
$[\bll\bll]$ somewhat more difficult problems; however, since the classification is arbitrary,
please do not be bothered by the number of $\bll$'s.

\subsection*{Notations and Conventions}

The symbols used in this book are summarized in the list of symbols at the end of the book.

There should be almost no places where the signature of the metric is essential;
in this book, for convenience, I have tried to use
$(-++\cdots +)$, which is convenient for Euclideanization and relatively
common in the supersymmetry literature.
Of course, after Euclideanization the metric becomes
$(++\cdots +)$.

\chapter*{Preface to the English Edition}
\phantomsection\addcontentsline{toc}{chapter}{Preface to the English Edition}

This is an English translation of a book I wrote in Japanese,
published under the title ``Geometries and Structures of Quantum Field Theories,"
by Saiensu-sha in September 2015.

The subject of this book is the celebrated correspondence between
3d $\mathcal{N}=2$ supersymmetric gauge theories and the geometry of 3-manifolds,
which arises when the 6d $(2,0)$ theory is compactified on a 3-manifold---%
what is now known as the 3d-3d correspondence.
I had the good fortune of being involved in the early development of this subject,
and I hope that this book is of help
in conveying the beauty of the subject and my excitement about it.

My aim, however, was not simply to write down such a correspondence 
merely as a dictionary to be memorized.
It was rather to show that the correspondence is an example of
a mathematical structure hidden in the theory space of quantum field theories,
and that this structure, in my opinion,
emerges naturally once we ask what a quantum field theory is in the first place.
In other words, I have tried to write this book in a ``bottom-up'' manner,
as opposed to a ``top-down'' approach starting mechanically from the compactification of the 6d theory.
This is why the book begins not with 3-manifolds,
but with questions about field theory itself,
and why the geometry enters only after we have followed the physics far enough
for it to appear on its own. I have chosen such a presentation in the hope that the
structures discussed in this book will find much wider applicability,
beyond the compactification of 6d theories or supersymmetric theories, and
more generally in the study of the theory space of quantum field theories.

The original plan was to provide an English translation soon after the Japanese version.
A decade has passed instead.
In the meantime the subject has developed considerably,
and I have made no attempt to bring the book up to date:
what the reader finds here is, in content, essentially the book of 2015.
I have taken this opportunity to correct numerous typos and errors
in the original Japanese version, and to make small changes.
The point of view of the book, on the other hand,
seems to me to have only gained in relevance,
and I hope that it still serves as an entry point to the subject.

\subsection*{Acknowledgments}
This work was supported in part by the World Premier
International Research Center Initiative (WPI), MEXT, Japan;
by the JSPS KAKENHI Grant No.~23K25865; by JST, Japan (CREST Grant No.~JPMJCR26XA, Moonshot R\&D Grant No.~JPMJMS256E); and by the IBM-UTokyo-sponsored research.

The author benefited from AI models (such as ChatGPT and Claude) in the preparation of this manuscript. The author used these models to assist with translation and language editing, among other tasks. The author carefully reviewed and edited the content generated by the AI models, and takes full responsibility for the accuracy and integrity of the final manuscript.

\bigskip
\textbf{Tokyo, September 2026}

\begin{flushright}%
	\textbf{Masahito Yamazaki}

\end{flushright}


\tableofcontents

\mainmatter

\chapter{Outline: Geometry of QFTs}
\label{chap.intro}

\begin{abstract}
	What does it mean to understand the \textbf{theory space} of field theories geometrically,
	and why is it necessary?
	In this chapter we explain, step by step,
	some answers to these questions.
	Since the aim is to explain the basic ideas,
	the description is
	somewhat more general than the contents of the subsequent chapters,
	and it should be useful, as an independent introduction in itself,
	for a quick and broader overview
	of the contents of this book.
\end{abstract}

\section{What Is Quantum Field Theory?}

Physics today has been
built upon the unceasing efforts of our predecessors
to understand the natural world.
Among the huge system of physics,
one of the most powerful and elaborate theories is
\keyword{field theory}{field theory}, and in particular
\keyword{quantum field theory}{quantum field theory}. Field theory is
a theory describing a wide range of objects, from the world of
tiny elementary particles to phenomena on cosmological scales,
and it can even be said to be a brilliant asset not only for physics but for
humankind as a whole.

Many readers holding this book will already have some feeling for what field theory is.
Nevertheless, let us dare to ask:
\textbf{what is field theory in the first place?}\footnote{
	Relatedly, there is the long-standing hard problem of whether quantum field theory can be defined
	in a mathematically rigorous manner.
	Such a mathematically rigorous framework is not directly needed for the discussion of this book;
	rather, the focus is on conceptual problems within the realm of physics, namely how to understand and formulate field theory as physics.
	Moreover, in the discussion of this book we mainly
	extract more concrete objects, such as vacuum moduli spaces and
	partition functions, from supersymmetric gauge theories,
	and proceed with the discussion at that level;
	we therefore use only a small part
	(but a very essential part)
	of the rich structure of supersymmetric field theories. Of course, if there were a mathematical formulation of field theory,
	our discussion would likely become much clearer; conversely, we would like to think that
	reconsidering the formulation of field theory physically
	may well contribute to its mathematical formulation.
}

To some readers this question may look strange.
In the preface of this book we wrote that we assume the basic ideas of field theory;
what does it mean to bring it up again now?

Of course, as we will also state just below, it is certainly true that there are already certain
answers to what field theory is,\footnote{Readers who have thought deeply about field theory
	may be able to come up with numerous answers to
	this question: for example,
	approaches to field theory using the S-matrix and its analyticity have a long history, and their value has
	recently been rediscovered in the context of supersymmetric gauge theories.
	However, a comprehensive overview of all approaches to field theory is
	beyond the author's ability, and is not
	necessary for the purposes of this book.} and it is also certain that existing field theory has achieved great success.
The author has no objection at all on this point. However,
what we aim at here is to look at field theory itself afresh from scratch,
and as a starting point for this, the question above will not be useless.

\begin{figure}[t]
	\centering{\includegraphics[scale=0.34]{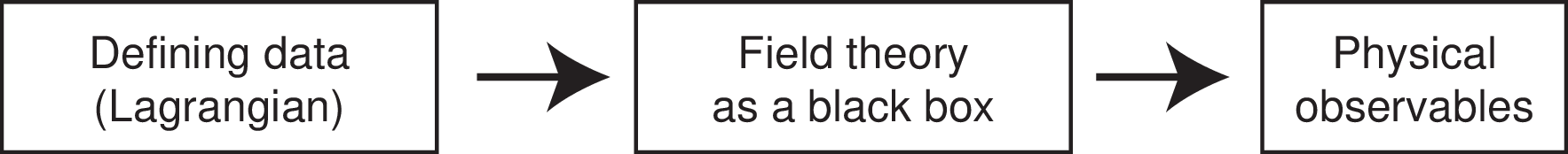}}
	\caption{One schematic picture of field theory: field theory is a black box which takes the Lagrangian as input and physical
		observables as output.}
	\label{fig.scheme}
\end{figure}

In textbooks of field theory, the starting point is in many cases
the \keyword{Lagrangian}{Lagrangian}, a functional of the fields---think,
for example, of the Lagrangian of a free field or the Lagrangian of the $\phi^4$ theory.
Field theory is then nothing but a tool which, starting from this Lagrangian and
making full use of techniques such as Feynman diagrams and renormalization,
extracts physical predictions and physical observables (e.g.\ scattering cross sections).
From this standpoint, field theory is a kind of black box which takes the Lagrangian as
input and physical observables as output, and
learning field theory is nothing but the task of making a
copy of this black box in one's own head (Fig.~\ref{fig.scheme}).

This is a rather crude simplification,\footnote{
	One of the reasons why formulating physical theories in this way is
	crude is the theory-ladenness of observation. Roughly speaking, this refers to the fact that
	a theory is needed in the first place to decide which of the diverse phenomena existing in nature
	should be explained by the theory.
	For example, Kepler tried to associate the planets of the solar system with regular polyhedra;
	in our current understanding of physics, however, the number of planets in the solar system does not need to be
	explained from the fundamental laws of the universe. In this sense, a theory
	specifies not only the input of the black box, but both the input and the output.
	One can think of various similar criticisms, which themselves contain interesting points;
	however, since our purpose is not to dwell on the details of the schematic picture,
	we avoid going deeper into this topic here.\label{foot.observe}} but
it also agrees with the textbook understanding of what physics is.
The natural world is vast and complicated;
in physics, we focus on certain specific properties of natural phenomena,
namely observable physical quantities.
The role of theorists is to create theories describing these physical quantities,
derive the values of the physical quantities from the theories, and
reproduce the observed values of the physical quantities.
In this language, a theory in Fig.~\ref{fig.scheme} is nothing but
a Lagrangian together with its set of
parameters.

\small
In the following we use ``nature'' and ``observable'' loosely.
For example, we do not ask whether something is observable in experiments that
present-day humankind can perform in the universe we currently live in.
What we ask in this book is not such a ``practical'' question, but,
supposing there is a world described by some field theory, and that we could in principle know
all the physical quantities of that world with the highest possible precision, how those physical quantities and
the theory correspond to each other. This is
an assumption often not satisfied in the real world,
but it is sufficient for our purpose of reconsidering field theory itself.

\normalsize

As the reader knows, this paradigm of field theory has achieved great success---the
precise computation of the magnetic moment of the electron in quantum electrodynamics (QED)
and its agreement with experiment
was one of its spectacular examples, and
the standard model of elementary particles, describing the electromagnetic, strong, and weak forces,
can also be described within this paradigm. Similar
successes can be found in large numbers, not only in particle physics but in various fields of physics.

However, when we reconsider field theory itself,
there are in fact many unsatisfactory points in the understanding so far. Let us look at these
one by one in the following.

\section{Renormalization and Effective Field Theory}\label{sec.renormalization}

The arrow in Fig.~\ref{fig.scheme}, namely reading off physical observables given a Lagrangian, is
a problem called the forward problem, which can in principle
be solved (whether this is possible in practice is a separate question). Of course, this is in general a rather non-trivial problem---for
example, quantum chromodynamics (QCD) is a strongly coupled theory at low energies, where non-perturbative
effects are essential, and it is in general difficult to compute physical observables (e.g.\ glueball masses)
analytically, or to verify them experimentally. However, it is not that there is no computational method in principle;
for example, we can put the theory on a lattice and compute with a computer.

In practice, however, the important problem is the inverse.
When we observe nature and obtain the values of physical observables,
what should we do to give the Lagrangian of a field theory?
This is a problem of the class called inverse problems, and
for field theories there is no generally applicable algorithm.
This is not merely a purely theoretical problem; in some contexts it is a very
practical problem. For example, when the LHC (Large Hadron Collider), the accelerator at CERN,
finds some signal,
is there a way to uniquely determine which of the hundreds of models beyond the standard model
is correct \cite{ArkaniHamed:2005px}?

We do not need field theory to notice the difficulty of inverse problems.
When the author learned analytical mechanics, the starting point was also
the Lagrangian.
But how do we obtain the Lagrangian in the first place?
In problems of elementary mechanics the equations of motion (Newton's laws) were already known,
so it was easy to guess the form of the Lagrangian by working backwards from them.
In actual problems, however,
the equations of motion are rarely already known,
and we have to extract the Lagrangian directly
without going through the equations of motion.

In the modern understanding of field theory,
there is a certain answer as to
what the Lagrangian should be.
At its core is the understanding of \keyword{renormalization}{renormalization}
clarified by K.~G.~Wilson \cite{Wilson:1973jj}.
He brought the concept of \textbf{energy scale} into the description of field theory:
the field theories we understand are
\keyword{low-energy effective field theories}{low-energy effective field theory}
of theories at higher energies.

The field theories we currently understand are
theories valid at the energy scales at which we measure physical observables
in the contexts we wish to consider.
At higher energies (the cutoff scale $\Lambda_{\rm UV}$),
the description by the theory in general breaks down,\footnote{
	In field theories called asymptotically free, the field theory description can in principle hold up to
	arbitrary energy scales. However, for example, the standard model of elementary particles
	(to be described shortly) is not asymptotically free (for reference, in contrast,
	an extension of the standard model such as the $SU(5)$ grand unified theory is asymptotically free).
}
and it may be described by a theory going beyond the framework of field theory, such as string theory.
However,
the physics at low energies
does not depend on the details of such theories at high energies,
and can be described by the effective field theory at the energy scale in question. Moreover,
that this effective field theory is naturally described by field theories as we currently understand them
is suggested by the existence of particles, which are irreducible representations of the
\textbf{Poincar\'e symmetry} of spacetime,\footnote{
	In this book we exclusively treat relativistic supersymmetric field theories, and
	do not treat non-relativistic field theories. However, concepts such as dualities of field theories and
	the combining and splitting of field theories by gauging can also be applied to non-relativistic
	field theories, and
	analogues of (part of) the geometric structures developed in this book
	may also exist in non-relativistic theories.
} together with
several assumptions (e.g.\ the locality of the theory).
\small
(This point is emphasized, for example, in the textbook by Weinberg
\cite{Weinberg:1995mt}; however, as he himself admits \cite{Weinberg:1996kw}, his argument has many loopholes. Moreover, field theory itself is also useful for describing non-relativistic phenomena.)
\normalsize

That high-energy physics decouples from low-energy physics is
essential for our understanding of nature.
It is thanks to this renormalization that we can compute
low-energy physics (e.g.\ the trajectory of a bullet)
without knowing
high-energy physics
(e.g.\ the quantum-gravitational behavior of the early universe);
without it, it would surely have been extremely difficult to develop physics.
On the other hand, however, the effects of renormalization
make it difficult to extract high-energy physics from low-energy experiments,
and modern physics stands as a discipline upon this delicate balance.
To put it somewhat jokingly, the existence of physicists and the anthropic principle
require renormalization and the necessity of low-energy effective field theories.

\bigskip

Let us consider field theory from the modern viewpoint of low-energy effective theories.
The data needed to specify a theory
are not that many:
the spacetime dimension, the symmetries of the theory, and the fields transforming in
certain representations under those symmetries.

First, let us fix the symmetries of the theory. Since in this book
we exclusively consider relativistic field theories,
we require symmetry under the \keyword{Poincar\'e group}{Poincare group}
of spacetime.\footnote{In later chapters of this book (e.g.\ Chap.~\ref{chap.S3}) we consider field theories on non-flat spacetimes,
	in which case the Poincar\'e group is replaced by another group depending on the choice of spacetime.
	However, even in those cases, the theory is always first defined on flat spacetime, and its renormalization
	is also mainly considered on flat spacetime, so that the following considerations are unchanged.}

A theory also has internal symmetries commuting with the Poincar\'e group, namely \textbf{local symmetries}
(\keyword{gauge symmetries}{gauge symmetry}) and
\textbf{global symmetries}. Moreover, in this book we exclusively consider theories which further have
\keyword{supersymmetry}{supersymmetry}
(the details of what supersymmetry concretely is are not directly needed in this chapter).

Next, we introduce \textbf{fields} (particles) transforming as irreducible representations under these symmetries.
Irreducible representations of the Poincar\'e group are called,
depending on the representation, scalars (spin $0$), fermions (spin $1/2$),
gauge fields (spin $1$), and so on, and these fields transform in certain representations under the respective
local and global symmetries.

Finally, we write down all the terms of the Lagrangian, a functional of these fields, compatible with all the symmetries.
There are infinitely many such terms,
but, except for a finite number of terms (the renormalizable terms),
the effects of most of the terms are, by dimensional analysis,
suppressed by the \keyword{cutoff scale}{cutoff scale} $\Lambda_{\rm UV}$ of renormalization.
Therefore, at energies (field values) sufficiently smaller than the cutoff scale,
it is in many cases no problem to restrict to the renormalizable Lagrangian.
However, the effects of non-renormalizable terms are only made relatively small by the cutoff,
and it is not that their effects cannot be measured.\footnote{For example, proton decay is represented in the standard model of elementary particles by non-renormalizable terms, whose effects are
	small due to powers of $\Lambda_{\rm UV}$;
	the proton decay experiments at Super-Kamiokande nevertheless give strong experimental constraints on their coefficients.}

\small

In the explanation above, we implicitly assumed that the simple criterion of renormalizability determined by naive dimensional counting agrees with
the more rigorous condition of renormalization determined by a detailed analysis of Feynman diagrams. For why this claim holds,
see the clear exposition by Polchinski \cite{Polchinski:1983gv}.

\normalsize

\bigskip

Since the explanation above was abstract,
let us explain it with examples.
The following two examples are not needed in the rest of this book, but
are nevertheless educational.

Let us consider the following $\phi^4$ theory,
often discussed in textbooks of field theory.
This theory imposes Poincar\'e symmetry in four-dimensional spacetime,
and further imposes the $\mathbb{Z}_2$ global symmetry
$\phi\to -\phi$ on a field $\phi(x)$ which behaves as a scalar under it.
The possible Lagrangian is then determined as
\begin{align}
	\begin{split}
		{\mathcal L}=
		- \frac{1}{2}\partial^{\mu}\, \phi \partial_{\mu} \phi
		 & -\frac{1}{2}m^2 \phi^2-\frac{1}{4!}\lambda \phi^4 \\
		 & -\frac{1}{\Lambda_{\rm UV}^2}\lambda' \phi^6
		-\frac{1}{\Lambda_{\rm UV}^2} \lambda''(\partial^2\phi)(\partial^2
		\phi) + \cdots \ ,
		\label{phi4}
	\end{split}
\end{align}
where the parameters $m$ and $\lambda, \lambda', \lambda''$ have
mass dimensions $1$ and $0$, respectively.
In this formula, the first line consists of the renormalizable terms, and is nothing but the Lagrangian often taken as the starting point in textbooks of field theory.
The second line shows only two of the infinitely many non-renormalizable terms (e.g.\ $\frac{1}{\Lambda_{\rm UV}^{2n}}\phi^{4+2n}$),
whose effects, as can be seen from dimensional analysis, are suppressed by at least $\Lambda_{\rm UV}^2$. These terms
can be neglected at energy scales much smaller than $\Lambda_{\rm UV}$.
As the energy scale goes up,
their effects gradually become non-negligible;
in particular, at scales of order $\Lambda_{\rm UV}$,
the infinitely many non-renormalizable terms give contributions of comparable size,
and the theory requires infinitely many parameters and loses its predictive power.

As a more complicated and realistic example,
let us consider the \keyword{standard model}{standard model} of elementary particles.
In this case, we fix the gauge group to be $SU(3)\times SU(2)\times U(1)_Y$.
Next, we fix the representations of the matter fields (fermions, in the present case quarks, leptons, and neutrinos) under it:
\begin{align}
	\begin{split}
		 & Q_L=(u_L, d_L):\, (\bm{3}, \bm{2})_{\frac{1}{3}} \ , \quad
		u_R:\, (\bm{3}, \bm{1})_{\frac{4}{3}} \ , \quad
		d_R:\, (\bm{3}, \bm{1})_{-\frac{2}{3}} \ ,                     \\
		 & (\nu_L, e_L):\, (\bm{1}, \bm{2})_{-1} \ , \quad
		e_R:\, (\bm{1}, \bm{1})_{-2} \ .
	\end{split}
\end{align}
Here, for example, $Q_L=(\bm{3}, \bm{2})_{\frac{1}{3}}$ means that
the left-handed quark
$Q_L$ is in the $\bm{3}$-representation under $SU(3)$, in the $\bm{2}$-representation under $SU(2)$,
and has charge $\frac{1}{3}$ under $U(1)_Y$.
Adding to these the Higgs particle, a scalar field (and, if necessary, the right-handed neutrino $\nu_R$),
and then writing down all the renormalizable terms of the Lagrangian not violating Lorentz symmetry and gauge invariance,
we obtain the Lagrangian of the standard model.
This Lagrangian has a finite number of parameters
(e.g.\ the gauge coupling constants and the quark masses),
whose values can be determined by comparing with experiments.\footnote{
	Whether the values of the parameters determined in this way are ``natural'' is another
	question. From the standpoint which does not allow fine-tuning of parameters
	(naturalness), unnaturally small values of parameters require
	some reason (e.g.\ related to symmetries or dynamics).
	For example, the mass of the Higgs particle and
	the value of the cosmological constant, which becomes important when coupled to gravity,
	are much smaller than the naturally expected values; these are
	the hard problems known as the hierarchy problem and the cosmological constant problem, respectively.
}

In this way, the low-energy
Lagrangian, and hence the forms of the corresponding physical observables,
can be determined solely from the symmetries and their representations.\footnote{
	However, when the fixed point of the renormalization group (RG) is not a single point but
	a finite-dimensional fixed manifold,
	we can consider marginal deformations of the theory.
	Since the physical observables change depending on the position on the fixed manifold,
	the values of the physical observables are
	not uniquely determined solely from the dimension and the symmetries.
	Supersymmetric gauge theories often have such freedom of marginal
	deformations \cite{Leigh:1995ep}.
} In particular, they do not depend on
the details of the theory at high energies. This is
known as \keyword{universality}{universality}, and plays
an extremely important role, for example, in condensed matter physics.
As an example, the low-energy effective
theory of the (3+1)-dimensional Ising spin system with values in $\mathbb{Z}_2$ is, by symmetry considerations, given by the $\phi^4$ field theory.
In this way, the ultraviolet (UV) theory in the high-energy region
need not even be a continuum field theory;
it can be a lattice theory, or a theory of quantum gravity which is not a field theory.

\bigskip

The concept of renormalization brings the energy scale into the description of theories,
and requires a refinement of the naive picture (Fig.~\ref{fig.scheme}).
When we speak of ``physical observables'', we need to make clear at which energy scale
we consider them.
Correspondingly, the variables of the Lagrangian also need to change.
This is called the \keyword{renormalization group running}{renormalization group running}
of the parameters. Namely, the physics is described by the pair $(L, \vec{c}(\mu))$ of a Lagrangian $L$ and
the parameters $\vec{c}(\mu)$ of the Lagrangian at a certain energy
scale $\mu$.

In this book we mainly consider gauge theories, and study their behavior at low energies. Recall that the gauge field $A_{\mu}$ has the kinetic term
\begin{align}
	L_{\textrm{gauge kin}}=-\frac{1}{2g^2}\int
	\textrm{Tr}\, F_{\mu\nu}F^{\mu\nu} \ ,
\end{align}
where the coefficient $g$ is called the \keyword{gauge coupling constant}{gauge coupling constant}. Here
the field strength is given by $F_{\mu\nu}:=\partial_{\mu}A_{\nu}-\partial_{\nu} A_{\mu}
	-i[A_{\mu}, A_{\nu}]$.
Moreover, for the trace we chose, denoting the generators of the Lie algebra of the gauge symmetry by $T^a$,
the so-called Killing form\footnote{For example, for the $SU(2)$ gauge group, when we take the generators of the Lie algebra to be
	$T^a={\sigma^a}/{2}$ using the Pauli matrices $\sigma^a$,
	the Killing form agrees with the trace in the usual sense.
}
\begin{align}
	\textrm{Tr}(T^a T^b) = \frac{1}{2}\delta_{ab} \ .
	\label{Killing_form}
\end{align} For a simple group $G$, a symmetric $G$-invariant
non-degenerate quadratic form is unique up to the ambiguity of normalization, and it
agrees with the trace here.

If we consider four-dimensional spacetime,
$g$ is dimensionless in mass dimension by dimensional analysis, and its change under the renormalization group
has, in the perturbative expansion in $g$, the form
\begin{align}
	\frac{d}{d\log \mu} g(\mu)= \beta(g)=-\frac{b_0}{16\pi^2} g^3+ b_1 g^5+b_2 g^7+\cdots \ ,
	\label{eq.beta}
\end{align}
where recall that we denote the energy scale by $\mu$.
The function $\beta(g)$ is called the
\keyword{beta function}{beta function}.
The term in $\beta(g)$ proportional to $g^3$ is a one-loop effect, and the sign of its coefficient
$b_0$ is positive for \textbf{asymptotically free} theories.
In this case, even if we start from $g=0$,
the value of $g$ becomes larger as we go to lower energies
due to the effects of renormalization, and eventually, when $g$ becomes of order $1$,
the coupling expansion \eqref{eq.beta} in $g$ loses its usefulness,
and it becomes indispensable to include \textbf{strong coupling} effects.\footnote{
	\label{Banks_Zaks_footnote}
	The flow may also stop by flowing into a fixed point of the renormalization group before the value of the gauge coupling constant $g$ becomes large,
	in which case the analysis is completed within the range of
	\eqref{eq.beta}.
	This happens, for example, when $0<b_0=\epsilon\ll 1$ and
	$b_{1,2, \ldots}=\mathcal{O}(1)$; in this case \eqref{eq.beta}
	has a fixed point at $g=g_*\sim \epsilon^{1/2}$, and the terms from $g^7$ onward in
	\eqref{eq.beta} are
	$\mathcal{O}(\epsilon^{7/2})$ and negligible.
	Fixed points of this kind appear in 4d $SU(N_c)$ QCD with $N_f$ flavors
	when we tune the values of $N_f$ and $N_c$,
	and are called \keyword{Banks-Zaks(-type) fixed points}{Banks-Zaks(-type) fixed point} \cite{Banks:1981nn}. For 4d $\mathcal{N}=1$ theories,
	the end $N_f \lesssim 3N_c$ of the conformal window discussed later corresponds to this.} What this book is concerned with is how to analyze the behavior of field theories
in this strong coupling region.
In particular, in the strong coupling region, a surprising phenomenon called \keyword{duality}{duality}
is found between field theories.

\section{Duality}
\subsection{Why Duality?}

Before explaining duality, let us look back at the discussion so far.

What specifies the Lagrangian of a theory was only the
symmetries and the fields as representations under them. This is an elegant understanding of field theory,
but let us return to our original question:
does this answer our question of
whether there is a way to determine the theory (Lagrangian)
from physical observables obtained by observing nature?

Unfortunately, the answer to this question is no.
Even if it is possible to observe nature,
the observations often contain a lot of unnecessary data (backgrounds and
statistical errors), and
grasping the essence from them and
finding the order behind them, in this case the symmetries,
requires non-trivial insight.
Of course, this is why inverse problems cannot be solved
by algorithms.
If an algorithm deriving physical theories from observations had been established,
physics today would be much more boring than it is now.

However, instead of ending the discussion here, let us go a little further,
since a problem of principle lies hidden here.

Among the defining data of a theory, the spacetime dimension under consideration is in most cases
trivially fixed by the setup of the problem.
Moreover, since the global symmetries of a theory are directly reflected in physical observables,
they should in principle be uniquely determinable
(this is only in principle;
whether it can be carried out in practice is a separate question).
Symmetries can also be spontaneously broken.
However, since theories with and without a symmetry are expected to give
different predictions for physical observables,\footnote{For example, in relativistic field theories, spontaneous breaking of a global symmetry predicts the existence of Nambu-Goldstone modes.}
whether a theory has a symmetry should in principle be decidable
from the observable physical quantities.

What, then, about the other data, the fields appearing in the Lagrangian
(including the gauge fields corresponding to gauge symmetries)?
In fact, whether these data are uniquely determined by natural phenomena
is a rather subtle problem, as described below.
In general, fields other than scalar fields are shy,
and cannot be measured directly.

\bigskip
First, let us consider a spin-$1/2$ fermion $\psi(x)$ (e.g.\ the electron) as a field.
As physical observables we can consider, for example, correlation functions of this field;
the difference from scalar fields is that the correlation function of a single fermion
always vanishes: $\langle \psi(x) \rangle=0$.
This is true as long as Lorentz symmetry is not spontaneously broken.
The non-trivial quantities which can actually be observed are expectation values of Lorentz-invariant quantities
obtained by contracting the fermion indices, for example the quantity called the
fermion condensate,
$\langle \overline{\psi}\psi(x) \rangle$.
In this way, for fields other than scalar fields, the relation between physical observables and fields is
not simple, and care is needed.

This is consistent with the following fact: in 2d spacetime,
a description using bosons and a description using fermions can be equivalent
(this is called \keyword{bosonization}{bosonization} \cite{Coleman:1974bu,Mandelstam:1975hb}).
We do not need its details here,
but very roughly speaking, it can be thought of as identifying the fermion bilinear $\overline{\psi}\psi$ with
a boson $\phi$. As already stated,
this combination is what appears in non-trivial expectation values of fermions,
so it is not hard to imagine that there can be situations where, under this identification,
all the physical observables can be replaced by bosons.

\bigskip

The situation is similar for gauge fields.
Since the gauge field $A_{\mu}$ transforms non-trivially under the Lorentz
group, its own expectation value cannot be measured---non-trivial observables should
always be Lorentz invariant.
In this case, however, there is an even stronger constraint than for fermions:
\textbf{gauge symmetry is by definition
	a redundancy of the description, and observables
	should always be gauge invariant}.
For example, $\langle A_{\mu} A^{\mu}\rangle$ is Lorentz invariant but not gauge invariant,
and hence is not observable.

Of course, being a redundancy of the description does not mean that gauge fields are meaningless.
Gauge fields themselves have non-trivial dynamics, and we can
construct observable physical quantities involving gauge fields. A typical one is the
\keyword{Wilson line}{Wilson line}:
\begin{align}\displaystyle
	W_{\gamma}:=\left\langle \textrm{Tr}\, P \, \exp\left( i \oint_{\gamma} A \right) \right\rangle \ .
	\label{Wilson}
\end{align}
\nomenclature{$W_{\gamma}$}{Wilson line}
Here $\gamma$ is a closed path, and $P$ means that, when we expand the exponential of the (in general non-Abelian) gauge field,
the order of the products is determined along the closed path (path
ordering).
Wilson lines are important physical observables for gauge theories,
and play the role, for example, of an order parameter of confinement.

Since Wilson lines are observables, it is possible, at least in principle, to extract information about the gauge fields
from them. What is subtle here, however, is that
Wilson lines are \textbf{non-local} physical observables.
Here, non-local means that the observable is not determined by the value of the field at a single point of spacetime,
but requires the values of the fields at several points of spacetime, or in general in a certain region;
in the present case, all the values of the gauge field along the closed path $\gamma$ are needed.
In this way, observables of gauge fields
are very different from those of other fields without gauge symmetry, in that they have to be
non-local.

\small

Let us emphasize again that \textbf{gauge symmetry is a redundancy}.
Usually, a symmetry means that
two different physical states have the same properties.
On the other hand, gauge symmetry is not a symmetry in this sense;
it means that there are two states with different gauges which are physically the same.
In fact, as we will see later in concrete examples in the context of supersymmetric gauge theories,
there can even be situations where, on one side of a pair of dual descriptions,
the gauge fields disappear, and
hence the gauge symmetry
disappears.
In this sense it might be more appropriate to call it
\keyword{gauge redundancy}{gauge redundancy} rather than gauge symmetry;
in this book, however, in order not to confuse the reader,
we exclusively use the term gauge symmetry.

In relation to this point,
care is needed in the explanation of the Aharonov-Bohm effect.
This effect has often been explained as
``a clear experimental demonstration that,
in quantum mechanics,
the vector potential and
the gauge field physically exist'',\footnote{The interpretation of this sentence strongly depends on the precise meaning of ``exist''. What is stated below is that there exists a description which does not need gauge fields.}
and the author was deeply
impressed when first hearing this explanation.
However, if we thoroughly adopt the standpoint of regarding gauge fields as a redundancy,
it is possible to remove the gauge field itself
from the description of the Aharonov-Bohm effect
and to describe everything in terms of gauge-invariant quantities \cite{Mandelstam:1962mi}.
In this case, however, the values of the fields
are not determined at a single point of spacetime, but depend on the choice of the path
from infinity to that point.
This can be rephrased as the existence of a non-trivial magnetic flux (the \keyword{Wilson loop}{Wilson loop} defined in \eqref{Wilson}).
From this standpoint,
the Aharonov-Bohm effect should be interpreted
as demonstrating not so much the existence of gauge fields
as the \textbf{non-locality of gauge fields}.\footnote{
	The gauge field itself is described by a local theory,
	but if we try to write everything only in terms of gauge-invariant quantities, the description becomes non-local.
}\footnote{The idea of formulating gauge theories only in terms of gauge-invariant quantities, namely Wilson loops, has a long history (see e.g.\
	Refs.~\cite{Mandelstam:1962mi,Migdal:1984gj}); however,
	directly dealing with its non-locality is at present not easy
	(except in low dimensions).
}

\normalsize

\bigskip

Why, then, do we have to consider such troublesome gauge fields?\footnote{One sometimes sees the explanation that non-Abelian gauge fields are needed
	for field theories to be asymptotically free \cite{Coleman:1973sx}; however,
	as already stated, field theories are in general only low-energy
	effective field theories, so this does not seem essential from today's viewpoint.}

The existence of gauge fields is essential for the traditional formulation of field theory, in particular for its
manifest \textbf{Lorentz symmetry},
\textbf{unitarity}, and the \textbf{locality} of the theory.

An explanation of this can be found in textbooks of field theory;
here let us explain one modern understanding of gauge fields
(this is somewhat off the main topic, but it is an important point).
It is that \textbf{gauge fields are needed to smoothly connect massless spin-1 theories and massive
	spin-1 theories} (the explanation here is close to
Ref.~\cite{BanksBook}).

As follows from the representations of the Poincar\'e group,
a gauge field has two helicities,
$1, -1$, in the massless case, but three, $1, 0, -1$, in the massive case
(this is a consequence of representation theory; intuitively, it means that massless particles
move at the speed of light and cannot be brought to rest by Lorentz transformations).
This is a big difference from scalar fields, whose number of degrees of freedom does not differ
between the massive and massless cases.

However, we have already seen that
energy scales are important
in field theory.
Suppose that there is a massive spin-$1$ particle.
Since there exist, for example, massive W bosons in our world,
this is a ``practical'' concern.
Its mass $m$ has a specific value, and hence determines a specific energy scale.
Now consider the limit where the energy scale $\mu$ we consider is taken to be
much larger than this mass scale,
while $\mu$ is much smaller than the scale $\Lambda_{\rm UV}$ where the effective field theory description
breaks down:
$m \ll \mu \ll \Lambda_{\rm UV}$.
In this case, the mass of the particle can be regarded as almost zero:
$m\simeq 0$.

This limit, however, is rather subtle, since,
as already stated, the numbers of degrees of freedom differ discontinuously between the massive and massless cases.
As one manifestation of this,
note that the propagator of a massive spin-$1$ field (the Proca field)
is given in momentum space by
\begin{align}
	\Delta(p)_{\mu\nu}=-i \frac{\eta_{\mu\nu} +\frac{p_{\mu} p_{\nu} }{m^2}}{p^2+m^2-i \epsilon}
	\label{eq.propagator}\end{align}
(exercise~\ref{ex.propagator}),
which diverges in the limit $m\to 0$.

To solve this problem, we can assume that the spin-$1$ field
is always coupled to a current $J^{\mu}$ satisfying the conservation law $\partial_{\mu} J^{\mu}=0$.
Since the conservation law becomes $p^{\mu} J_{\mu}=0$ after Fourier transformation to momentum space,
the contributions from the terms of the propagator proportional to $p^{\mu}$ vanish,
and the divergence of \eqref{eq.propagator} in the $m\to 0$ limit disappears.

To see more clearly how this has solved the problem of the difference in the degrees of freedom with the massive spin-$1$ field $B_{\mu}$,
let us decompose $B_{\mu}$ as
\begin{align}
	B_{\mu}=A_{\mu}+\partial_{\mu} \theta \ .
	\label{eq.BAtheta}
\end{align}
The degree of freedom $\theta$ then does not contribute to the coupling with the current $J^{\mu}$:
$\int B_{\mu} J^{\mu}=\int A_{\mu} J^{\mu}$.
This was why the divergence of the propagator at $m\simeq 0$ disappeared.
Note here that the decomposition \eqref{eq.BAtheta} has the invariance
\begin{align}
	A_{\mu} \to A_{\mu} +\partial_{\mu} \Lambda \ , \quad
	\theta \to \theta -\Lambda \ .
	\label{eq.Lambda}
\end{align}
This is a redundancy existing in the decomposition \eqref{eq.BAtheta} in the massive spin-$1$ theory; in the massless limit,
the degree of freedom $\theta$ decouples and only $A_{\mu}$ remains,
and the symmetry \eqref{eq.Lambda} appears as a gauge symmetry.

\small
For reference, conversely, we can give a mass to a massless gauge field $A_{\mu}$ by replacing it with
$B_{\mu}$ in \eqref{eq.BAtheta}. This is called the
\keyword{St\"{u}ckelberg mechanism}{Stuckelberg mechanism}.
\normalsize

\bigskip

As we have seen so far,
for fields other than scalar fields,
there was a large gap between
the fields themselves and the actually observable physical quantities.
Moreover, the existence of the gap
is not accidental, but
has appeared necessarily from
fundamental physical requirements,
and gauge fields and gauge symmetries were the prime examples.
Developing this point further, we claim that
\textbf{even if the observable physical quantities in nature are given, the corresponding
	theory (Lagrangian) is not uniquely determined}.
There are several reasons for this.

First, even when a single Lagrangian is given,
different parameter regions of it can represent the same reality.

Next, two (or more) Lagrangians with different fields can describe the same physical observables.

These are collectively called \keyword{duality}{duality}.\footnote{
	As a review discussing duality in a broader sense than this book, Ref.~\cite{Polchinski:2014mva} is useful.}
The basic reason why such ambiguities can exist is that
the fields used in the description of the Lagrangian, and in particular
the gauge fields themselves, are not directly observable.

Finally, it is believed that there exist field theories in reality
which have no Lagrangian
(we will comment later, in Sec.~\ref{subsec.compact_twist}, on an example of a field theory believed to have no Lagrangian,
the \keyword{6d $(2,0)$ theory}{6d (2,0) theory}).
Our scheme (Fig.~\ref{fig.scheme}), which defined field theories using Lagrangians,
is in fact insufficient as a definition of field theory in the first place.
In general, even without a Lagrangian, we can define physical observables using operators and their correlation functions, or even the S-matrix.
Moreover, there are also situations, such as conformal field theories (in particular in two dimensions),
where the symmetries alone impose strong constraints on the physical properties.

In the following, let us look at concrete examples of dualities.
Examples of dualities include the 2d Ising spin system, and
later, in Chap.~\ref{chap.3dglue}, we will
introduce a simple duality transformation in 3d field theories
(\keyword{particle-vortex duality}{particle-vortex duality}).
Below we look at concrete examples of dualities
in three- and four-dimensional field theories with supersymmetry.
In these examples, the constraints from supersymmetry make a precise understanding of the dualities
possible.

\bparagraph{S-Duality of 4d $\mathcal{N}=4$ Theory}\label{subsec.MO}

One of the most famous dualities of 4d supersymmetric gauge theories is
the \keyword{S-duality}{S-duality}, or more generally the
\keyword{$SL(2, \mathbb{Z})$ duality}{SL(2,Z) duality}, of the \keyword{4d $\mathcal{N}=4$ theory}{4d N=4 theory}.

Let us consider the 4d $\mathcal{N}=4$ supersymmetric gauge theory
with gauge group $U(N)$.
Here we do not need a precise understanding of what 4d $\mathcal{N}=4$ supersymmetry is.
It is sufficient to understand that it is a theory containing, in addition to the gauge field, various fields (bosons and fermions, the superpartners of the gauge field),
and that it has so much symmetry that
the Lagrangian is completely determined once we fix the gauge group.

This theory has two real parameters,
the gauge coupling constant $g$ and the theta angle $\theta_{\rm 4d}$.
The Lagrangian then contains the following terms:\footnote{
	In this book we use the standard notation of differential forms.
	For example, for the field strength $F$, which is a 2-form,
	\begin{align*}
		F=\frac{1}{2} F_{\mu\nu}\, dx^{\mu} \wedge dx^{\nu} \ , \quad
		dF=\frac{1}{2} \partial_{\rho} F_{\mu\nu}\, dx^{\rho}\wedge dx^{\mu} \wedge dx^{\nu} \ .
	\end{align*}
	Here $\wedge$ is the wedge product of differential forms, satisfying $dx^{\mu}\wedge dx^{\nu}=-dx^{\nu}\wedge dx^{\mu}$. In particular, in four dimensions,
	\begin{align*}
		dx^{\mu}\wedge dx^{\nu} \wedge dx^{\rho} \wedge dx^{\sigma}=
		\epsilon^{\mu\nu\rho\sigma} dx^1 \wedge dx^2 \wedge dx^3 \wedge dx^4 \ .
	\end{align*}
	$*$ is the Hodge dual:
	\begin{align*}
		*(dx^{\mu} \wedge dx^{\nu}) =\frac{1}{2} \epsilon^{\mu\nu\rho\sigma} dx_{\rho} \wedge dx_{\sigma} \ , \quad
		                                                                                                      (*F)_{\mu\nu}=\frac{1}{2} \epsilon_{\mu\nu\rho\sigma} F^{\rho \sigma} \ .
	\end{align*}
}
\begin{align}
	L_{\textrm{kin} + \theta_{\rm 4d}} & =-
	\int \! d^4 x \, \textrm{Tr}\left[
		                            \frac{1}{2 g^2} F_{\mu\nu} F^{\mu\nu} +\frac{\theta_{\rm 4d}}{32 \pi^2}
		                            \epsilon^{\mu\nu\rho\sigma}F_{\mu\nu}
		                            F_{\rho \sigma}
		                            \right]  \nonumber \\
	                          & =-
	\int \textrm{Tr}\left[
		                            \frac{1}{ g^2} F \wedge *F+\frac{\theta_{\rm 4d}}{8 \pi^2} F \wedge F
		                            \right]
	\ .
\end{align}
When we shift the theta angle $\theta_{\rm 4d}$ by $2\pi$, the change is
\begin{align}
	\begin{split}
		\delta L_{\textrm{kin} + \theta_{\rm 4d}}
		=
		-2\pi   \frac{1}{8\pi^2}   \int  \, \textrm{Tr}
		\left( F\wedge F \right)
		\in 2\pi \bZ
		\ ,
	\end{split}
	\label{delta_L_FF}
\end{align}
where in the last step we used the fact that the integral over a closed manifold of the \keyword{second Chern character}{second Chern character}
\begin{align}
	\textrm{ch}_2(F):=-\frac{1}{8\pi^2} \textrm{Tr}\left( F\wedge F \right)
\end{align}
is an integer (a related discussion will appear again in Sec.~\ref{subsec.CS_term_2}).
Therefore, the path-integral weight $e^{i L_{\textrm{kin} + \theta_{\rm 4d}}}$ is invariant, and $\theta_{\rm 4d}$
is identified with period $2\pi$.

Let us define the \keyword{complexified gauge coupling constant}{complexified gauge coupling constant} $\tau$,
combining the two parameters $g, \theta_{\rm 4d}$, by
\begin{align}
	\tau:=\frac{\theta_{\rm 4d}}{2\pi }+\frac{4\pi }{g^2}i \ .
	\label{complex_gauge}
\end{align}
\nomenclature{$\tau$}{complexified coupling constant}
By definition, $\textrm{Im}(\tau)>0$.
The following fact is then conjectured to hold:

\begin{itembox}[c]{$SL(2, \bZ)$ duality of 4d $\mathcal{N}=4$ $U(N)$ gauge theory}
	The 4d $\mathcal{N}=4$ theory with complexified gauge coupling constant $\tau$
	and the theory with the same Lagrangian, whose
	complexified gauge coupling constant is $SL(2, \mathbb{Z})$-transformed,
	\begin{align}
		\tau'=\frac{a\tau+b}{c\tau+d} \quad (ad-bc=1\, \textrm{and}\, a,b,c,d
		\in \mathbb{Z}) \ ,
	\end{align}
	describe equivalent physics \cite{Montonen:1977sn}.
\end{itembox}

Since the matrices $\{ \pm I_{2\times 2}\}$ act trivially on the complexified gauge coupling constant $\tau$,
the group acting non-trivially on $\tau$ is
$PSL(2, \bZ)=SL(2, \bZ)/\{
	\pm I_{2\times 2}\}$.

This duality is an example where, once the Lagrangian is fixed, an ambiguity remains in the correspondence between its parameters and
the physical observables. Since we are moving on quickly here, we do not describe
the details; roughly speaking, it can be understood as claiming a
symmetry under transformations mixing the electric field $E$ and the magnetic field $B$.

\small

Many checks of this conjecture have been made, and
there are few researchers today who actively doubt its correctness.
However, there is still no rigorous proof of the conjecture, and its verification continues even now.
For example, (1) the agreement of the expectation values of correlation functions of Wilson loops and their S-duals, \keyword['t Hooft loop]{'t Hooft loops}{t Hooft loop} \cite{Gomis:2009ir},
and (2) the agreement of the $S^1\times S^3$ supersymmetric partition functions (superconformal indices) \cite{Romelsberger:2005eg,Kinney:2005ej,Gadde:2009kb,Spiridonov:2011hf}, and of their extensions to $S^1\times S^3/\mathbb{Z}_r$
\cite{Benini:2011nc,Razamat:2013opa},
are relatively recent, and there are many parts of these agreements (in particular for general gauge groups) which
have not been completely proven.
\normalsize

\bigskip
Let us consider the following two elements of $SL(2, \mathbb{Z})$:
\begin{align}
	S:=\left(
	\begin{array}{cc}
		0  & 1 \\
		-1 & 0
	\end{array}
	\right) \ , \quad
	T:=\left(
	\begin{array}{cc}
		1 & 1 \\
		0 & 1
	\end{array}
	\right) \ .
	\label{STgen}
\end{align}
\nomenclature{$S, T$}{generators of $SL(2, \bZ)$}
The $S, T$ defined in this way satisfy the relations
\begin{align}
	S^2=-I_{2\times 2} \ , \quad (ST)^3=I_{2\times 2} \ .
	\label{STrel}
\end{align}
From \eqref{STrel} one can also show
$(TS)^3=1$.

In fact, $S$ and $T$ are generators of $SL(2, \bZ)$, and
any element $\varphi$ of $SL(2, \bZ)$ can be written as
\beq
\varphi=T^{n_1} S T^{n_2} S T^{n_3} \ldots S T^{n_k} \quad (n_i \in \bZ ) \ .
\label{varphiST}
\eeq

\small

This can be proven as follows.\footnote{The following procedure is
	nothing but the continued fraction expansion of the rational number $a/c$.}
Let $c\ne 0$ for the matrix $\left( \begin{array}{cc}a&b \\c&d \end{array}\right)$. Then, by the $T^n$-transformation ($n\in \bZ$)
\begin{align*}
	\left( \begin{array}{cc}1&n \\ 0&1 \end{array}\right).
	\left( \begin{array}{cc}a&b \\c&d \end{array}\right)
	=\left( \begin{array}{cc}a+n c&b+n d \\c&d \end{array}\right) \ ,
\end{align*}
we can make $|a|<|c|$.
After this, by the $S^{-1}$-transformation
\begin{align*}
	\left( \begin{array}{cc}0&-1 \\ 1&0 \end{array}\right).
	\left( \begin{array}{cc}a&b \\c&d \end{array}\right)
	=\left( \begin{array}{cc}-c& -d \\ a & b \end{array}\right) \ ,
\end{align*}
we can exchange $a$ and $c$, so that in the end we have made the value of $|c|$ smaller.
Repeating this a finite number of times, we obtain $c=0$, and then
the $SL(2, \bZ)$ matrix can be written in the form $\pm T^n$ (note that $-I_{2\times 2}=S^2$).

\normalsize

The claim of $SL(2, \bZ)$ duality has thus been reduced to claims about its generators
$S$ and $T$.
The corresponding changes of the complexified gauge coupling constant are
\begin{align}
	S:\, \tau \to -\frac{1}{\tau} \ , \quad
	T:\, \tau \to \tau+1 \ .
\end{align}
Since the $T$-transformation is $\theta_{\rm 4d}\to \theta_{\rm 4d}+2\pi$,
it represents the periodicity of the theta angle. On the other hand,
the $S$-transformation is a much more non-trivial transformation; for example, if we take $\theta_{\rm 4d}=0$ for simplicity,
it is $g\to 4\pi/g$ (i.e.\ $g^2/4\pi \to 4\pi/g^2$). This is
a transformation mapping weak coupling to strong coupling,
and cannot be captured within perturbation theory, which is a weak-coupling expansion.

\small
A similar S-duality is also known for 4d $\scN=2$ theories with
an Abelian gauge group $U(1)^n$,
where the duality group is $Sp(2n, \bZ)$
(for $n=1$, $Sp(2,\bZ)=SL(2, \bZ)$).
This will be used in Sec.~\ref{subsec.duality_wall}.

\normalsize

Moreover, in Chap.~\ref{chap.wall} later we also use
the S-duality for the gauge group $SU(N)$. In this case, the $S$-transformation is known to exchange the gauge group $SU(N)$ with
another gauge group, $SU(N)/\bZ_N$ \cite{Goddard:1976qe}.
Here $\bZ_N$ is the
center of $SU(N)$ (the elements commuting with all the other elements), given by
$\omega \, \textrm{Id}_{N\times N} (\omega^N=1)$; for example, $SU(2)/\bZ_2=SO(3)$.
The difference between $SU(N)$ and $SU(N)/\bZ_N$ is
important for a precise discussion of the $S$-transformation, but in this book we need not worry much about this point.

\bparagraph{4d Seiberg Duality}\label{subsec.Seiberg}

Next, let us look at an example where two theories with different Lagrangians
correspond to the same physics. Many such examples were
found in the 1990s; among them, the most
typical one is \keyword{Seiberg duality}{Seiberg duality}
\cite{Seiberg:1994pq} for 4d $\scN=1$ theories.

In this book we hardly need the claim of Seiberg duality itself;
briefly summarized, the claim is as follows.
Consider 4d $\mathcal{N}=1$ supersymmetric QCD
with gauge group $SU(N_c)$ and $N_f$ flavors.
Namely, we consider
$N_f$ chiral
superfields $Q^{i=1, \ldots, N_f}$ ($\overline{Q}_{i=1, \ldots, N_f}$)
transforming in the fundamental representation $\bm{N_c}$
(anti-fundamental representation $\bm{\overline{N}_c}$) under the gauge group $SU(N_c)$.
This is one theory.

In the other theory, we first take the gauge group $SU(N_f-N_c)$ and $N_f$ flavors:
let us now denote the chiral superfields transforming in the fundamental and anti-fundamental representations of the gauge group by
$q^{i=1, \ldots, N_f}$ ($\overline{q}_{i=1, \ldots, N_f}$).
Moreover, we consider fields (mesons) $M^i{}_j$ transforming trivially under gauge transformations,
and add the superpotential $W=\textrm{Tr}(\overline{q}_i M^{i}{}_{j} q^j)$
constructed from $q, \overline{q}, M$.
This is the other theory.

Let us assume the condition $\frac{3}{2}N_c< N_f< 3 N_c$, called the \keyword{conformal window}{conformal window}.
Then the two theories, namely
4d $\mathcal{N}=1$ supersymmetric $SU(N_c)$ QCD with $N_f$ flavors, and
4d $\mathcal{N}=1$ supersymmetric $SU(N_f-N_c)$ QCD with $N_f$ flavors
with suitable fields (mesons) and superpotential added,
are believed to flow under the renormalization group to the same strongly coupled renormalization group fixed point.
Namely, the two theories
agree at the infrared (IR) fixed point.

\begin{figure}[t]
	\centering\includegraphics[scale=0.35]{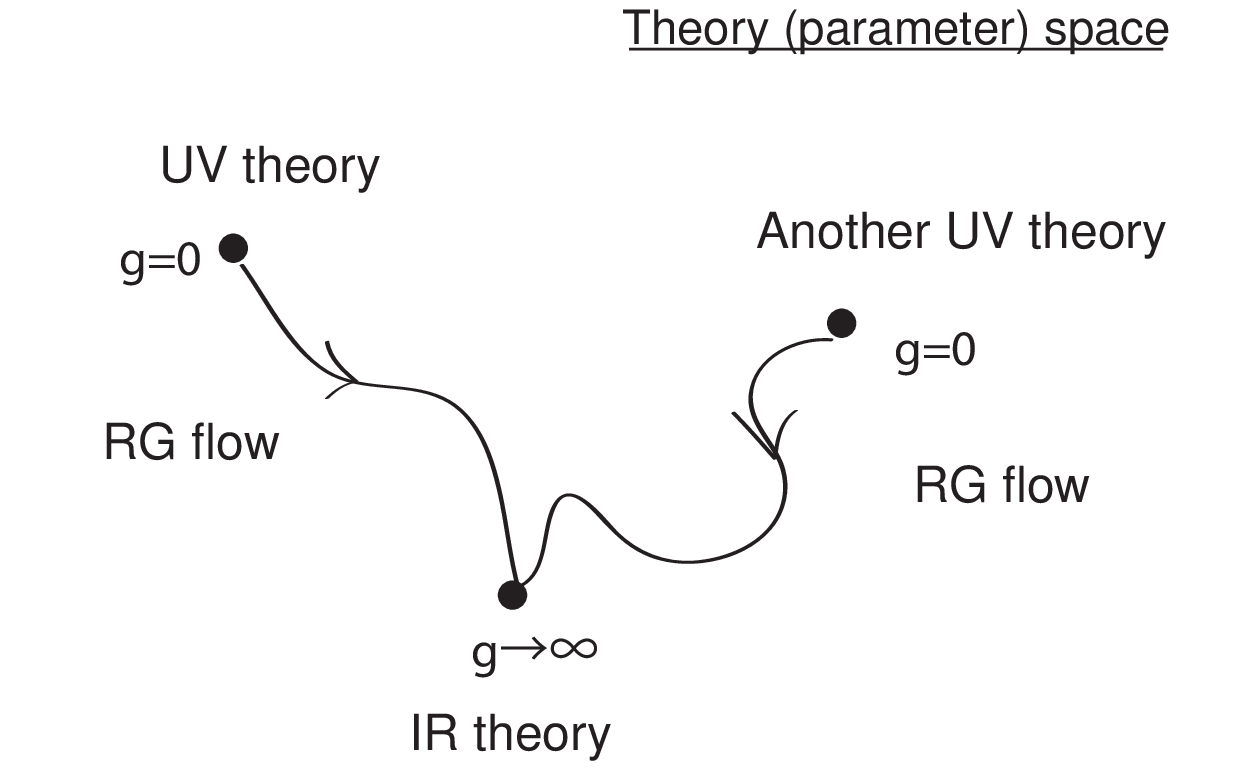}
	\caption{Seiberg duality claims that two UV theories (Lagrangians)
		flow into the same IR fixed point under the renormalization group flow.}
	\label{fig.RGduality}
\end{figure}

Here, rather, the following two points are important. First, this duality is a good example showing that,
given physical phenomena, the notion of the corresponding gauge group is
not uniquely determined, and there is an ambiguity.
The theory with gauge group $SU(N_c)$ describes physics equivalent to that of the theory with gauge group $SU(N_f-N_c)$,
and the gauge groups differ even in their ranks.
The other is that the duality here is not a direct equivalence at the level of the UV Lagrangians,
but \textbf{a duality around the IR fixed point} reached by the renormalization group flow (Fig.~\ref{fig.RGduality}).

\bparagraph{3d Mirror Symmetry}\label{subsec.3d_mirror_intro}

Finally, let us explain an example of duality in 3d supersymmetric theories, which will be important in this book.
It is called \keyword{3d mirror symmetry}{3d mirror symmetry} \cite{Intriligator:1996ex}.
There are several versions of it;
in this book, for example, we discuss
mirror symmetry with 3d $\scN=4$ supersymmetry in Sec.~\ref{subsec.TSUN}.

Here, as an example of mirror symmetry, let us describe
the duality between $\mathcal{N}=2$, $N_f=1$ supersymmetric QED and the $XYZ$ model
(which we call here the
\keyword{2--3 mirror symmetry}{2--3 mirror symmetry}\footnote{
	This is a convenient name in connection with 3-manifolds discussed later,
	but it is not standard terminology, so care is needed when consulting the literature.
}).

\small
Mirror symmetry in 3d supersymmetric field theories
is closely related to mirror symmetry in two dimensions, which is famous in particular in connection with mathematics.
When the gauge group is Abelian, one can show that dimensionally reducing each of the pair of theories of 3d mirror symmetry to two dimensions
gives 2d mirror symmetry
\cite{Aganagic:2001uw}.

\normalsize

One side of the duality is
3d $\mathcal{N}=2$ $N_f=1$ \keyword{supersymmetric QED}{supersymmetric QED} (SQED).
Here supersymmetric QED is a gauge theory with gauge group $U(1)$, and
the number of flavors $N_f$ being $1$ means that
there is one field with charge $+1$ (denoted by $q$) and one with charge $-1$ ($\overline{q}$).

On the other hand, the other theory is called the \keyword{$XYZ$ model}{XYZ model}.
This is not a gauge theory; it consists only of three chiral superfields $X, Y, Z$, and has the
superpotential $W=XYZ$. The names $X, Y, Z$ here are merely conventional,
and have no deep meaning.\footnote{To be sure, the $XYZ$ model here is completely different from the
	spin chain model of the same name. The notation $X, Y, Z$ for the three fields has simply become established.}

The 3d 2--3 mirror symmetry claims
that these two theories are equivalent.

\begin{itembox}[c]{3d 2--3 mirror symmetry}
	3d $\mathcal{N}=2$ supersymmetric QED and
	the 3d $\mathcal{N}=2$ $XYZ$ model are equivalent.\footnote{The equivalence of the two theories is believed to hold
		even when we follow the renormalization group flow back from the IR fixed point to the UV.}
	In particular, the two theories
	flow to the same IR fixed point under the renormalization group.
\end{itembox}

This example shows that a gauge theory can be dual to a non-gauge theory,
and is a good example clearly showing that the gauge group is not uniquely determined for physical phenomena.
Physically, this claims a duality exchanging particles and vortices
in 3d field theories,
and is the 3d version of the duality transformation exchanging electric and magnetic fields in 4d theories (see Chap.~\ref{chap.3dglue}).
Moreover, we will explain evidence for this conjecture in more detail in Chaps.~\ref{chap.3dN2} and \ref{chap.S3} below.
To anticipate, this duality corresponds to
the pentagon identity of the quantum dilogarithm function and to the
\keyword{Pachner move}{Pachner move} in hyperbolic geometry,
and is extremely important for the discussion of this book.

\section{Compactification of 6d $(2,0)$ Theory}

So far we have described several examples of dualities of
supersymmetric field theories.
Many examples of similar dualities are known in the literature, and
listing them in detail would by itself fill this book.
As already stated, one origin of dualities is
the discrepancy between field theories and physical observables,
and the very possibility of dualities is
not surprising. Moreover, one can, for example,
accumulate checks of dualities by computing physical observables.

However, what we would like to ask is how we can understand
the dualities of gauge theories more deeply.
In the examples of dualities discussed so far,
we need to consider parameter regions where the gauge coupling constant is large,
where computational methods based on perturbative expansions such as Feynman diagrams
are not useful.
Can we understand the dualities of gauge theories
more directly and conceptually?

What we would like to introduce here is
the approach of translating dualities into geometric statements.
Namely, we consider geometric objects corresponding to gauge theories,
and reinterpret the dualities of gauge theories as properties on the geometric side.
This is a translation, but it sometimes
maps complicated phenomena into simple statements.

\small
Historically, this approach has been strongly motivated by
string theory; however, as described in this book, the approach itself
does not necessarily require string theory.

\normalsize

Let us explain the approach of translating into geometric problems
in the case of the S-duality of the 4d $\mathcal{N}=4$ theory.
In this case the translation into geometry is based on the following
claim:
\textbf{there exists a certain six-dimensional theory specified by an integer $N(>1)$, and
	when we compactify this theory on a 2d torus with complex structure $\tau$,
	the theory appearing in the remaining four dimensions
	gives the 4d $SU(N)$ gauge theory
	(up to subtleties of the global form of the gauge group).
	Moreover, the complexified gauge coupling constant $\tau$ of the 4d $\mathcal{N}=4$ theory
	is identified with the parameter $\tau$ of the complex structure of the 2d torus}.

The six-dimensional supersymmetric theory mentioned here is
a conformal field theory called the 6d $A_{N-1}$-type $(2,0)$ theory, and its existence is
suggested by considerations from string theory.
We will comment further on the properties of this theory in Chap.~\ref{chap.6d},
but the details are not important here.
What is important is the identification of the 4d complexified gauge coupling constant with the complex structure of the 2d torus.

\small

The \keyword{complex structure}{complex structure} can intuitively be thought of as the freedom to deform a manifold of fixed size.
Alternatively, more mathematically,
it specifies how to introduce complex coordinates $z$ in each local coordinate chart of the manifold.

\normalsize

In the present case, the 2d torus with complex structure $\tau$ can be thought of as obtained from the complex plane $\bC$
under the identification $z\sim z+1 \sim z+\tau $ of its coordinate $z$.
If we fix an orientation of the torus,
we may assume that $\tau$ has $\textrm{Im}(\tau)>0$.

\begin{figure}[t]
	\centering{\includegraphics[scale=0.35]{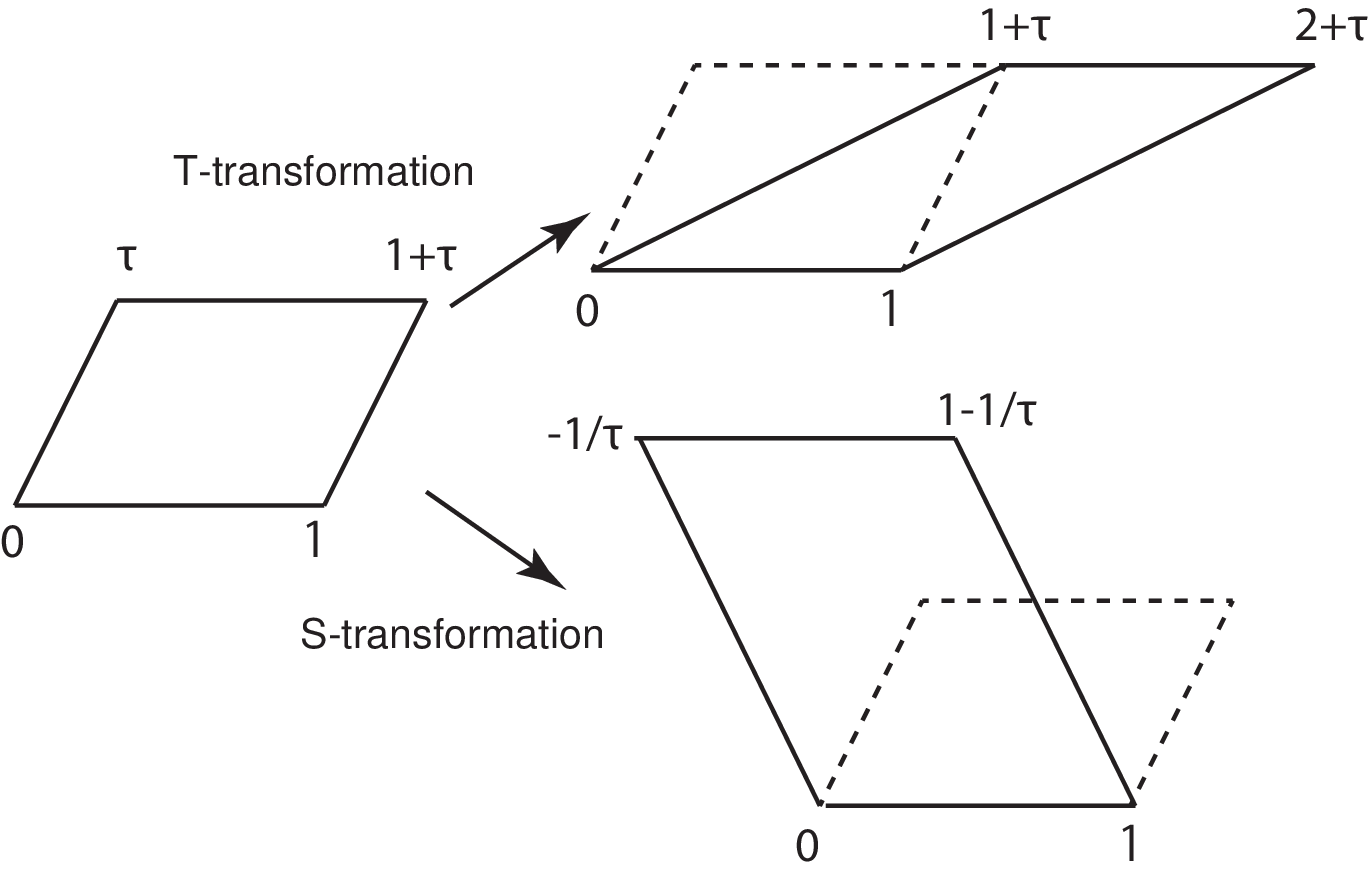}}
	\caption{The complex structure of a 2d torus is given by a complex number $\tau$ satisfying $\textrm{Im}(\tau)>0$,
		and values of $\tau$ identified by $SL(2, \mathbb{Z})$-transformations give the same complex structure.}
	\label{fig.torus}
\end{figure}

The complex structure of a complex torus is thus described by a complex number $\tau$ with $\textrm{Im}(\tau)>0$;
however, not all values of $\tau$ describe different complex structures:
the following $SL(2, \bZ)$-transformations preserve the complex structure (Fig.~\ref{fig.torus}):
\begin{align}
	\tau \to \tau'=\frac{a\tau+b}{c\tau +d} \ , \quad
	\left(
	\begin{array}{cc}
		a & b \\ c&d
	\end{array}
	\right) \in SL(2, \bZ) \ .
	\label{tauSL2Z}
\end{align}
Indeed, when we transform $\tau$ under \eqref{tauSL2Z}, if we correspondingly transform the complex coordinate $z$ as
$z'=z/(c\tau+d)$, we obtain $z'\sim z'+1 \sim z'+\tau'$.
Since this change of variables does not involve the complex conjugate $\bar{z}$ of $z$, it preserves the complex structure.

Here, among the $SL(2, \bZ)$-transformations,
the matrix $-I_{2\times 2}$
acts trivially on $\tau$, so that
dividing by it, what acts non-trivially is $PSL(2, \bZ)=SL(2, \bZ)/ \{
	\pm I_{2\times 2}\}$.
The group $SL(2, \bZ)$ is also the \keyword{mapping class group}{mapping class group} $\textrm{MCG}(T^2)$ of $T^2$, namely
the group of isotopy classes of orientation-preserving diffeomorphisms from $T^2$ to itself,
whose action on $\tau$ factors through $PSL(2, \bZ)$:
\begin{align}
	\textrm{MCG}(T^2)=\frac{\textrm{Diff}^+ (T^2)}{\textrm{Diff}_0(T^2)}
	=SL(2, \mathbb{Z}) \ .
\end{align}

One of the claims is that
this $PSL(2, \mathbb{Z})$-transformation of the 2d torus appearing in the compactification
is also reflected in the 4d $\mathcal{N}=4$ theory obtained as a result of the compactification
(which has $PSL(2, \mathbb{Z})$ symmetry).
In this way,
\textbf{the profound claim of duality of supersymmetric gauge theories
	has been derived from a relatively simple geometric claim}.
In other words,
if we assume the correspondence between gauge theories and geometry,
the non-perturbative properties of the gauge theory (in this case the 4d $\scN=4$ theory) can be derived
without knowing its details.
This is the strength of the geometric approach.

The explanation so far is admittedly somewhat heuristic.
As already stated,
the existence of the six-dimensional field theory has not been shown directly within field theory (without invoking string theory),
and the argument above is often reversed, taking the S-duality of the 4d $\mathcal{N}=4$ theory
as evidence for the existence of the six-dimensional theory.

However, once we accept the existence of the six-dimensional theory,
various consequences follow from it. For example, what happens if we change the manifold on which we compactify
from $T^2$ to more general manifolds?
What kinds of theories appear in the remaining dimensions?
This has been studied since the 1990s \cite{Witten:1997sc}, and
there has been great progress recently, since around 2009 \cite{Gaiotto:2009we}.

For example, let us consider a 2d Riemann surface $\Sigma=\Sigma_{g,h}$ as the manifold,
where $g$ is the genus of the Riemann surface and
$h$ is the number of \keyword{punctures}{puncture}.
A puncture is a hole; as we will discuss later in Chap.~\ref{chap.Teichmuller},
we consider holes of finite size.
Moreover, since the genus is also a kind of hole and this can be confusing,
we use the term ``puncture'' throughout this book.

As described in Chap.~\ref{chap.S3}, as one way of considering supersymmetric field theories
on curved manifolds,
we can consider the \keyword{topological twist}{topological twist} (see Chap.~\ref{chap.6d}).
Depending on the choice of this twist,
it is known that $\mathcal{N}=2$ theories or $\mathcal{N}=1$ theories appear in the remaining four dimensions,
and the case of 4d $\mathcal{N}=2$ theories in particular has been
studied well.
In this case, too, the correspondence between gauge theories and geometry is similar to the case of $T^2$,
and it is known that
\textbf{the mapping class group of the Riemann surface $\Sigma$ can be interpreted as the group formed by the dualities of the 4d $\mathcal{N}=2$ theory}\footnote{For general Riemann surfaces, the mapping class group does not necessarily map strong coupling to weak coupling,
	but maps a strongly coupled theory to another strongly coupled theory. Therefore, this duality is not an S-duality in the usual sense.} \cite{Gaiotto:2009we,Tachikawa:2013kta}.

What we mainly consider in this book is the situation where the six-dimensional theory is
compactified on a 3-manifold $M$.
In the situations considered in this book, $\mathcal{N}=2$
supersymmetry remains in the remaining three dimensions. The purpose of this book is to study
this 3d $\mathcal{N}=2$ theory using the geometry of the 3-manifold $M$, namely the theory of 3-manifolds.
More physically, \keyword{Chern-Simons theory}{Chern-Simons theory} with a complexified gauge group
(Chap.~\ref{chap.3mfd}) appears on the 3-manifold \cite{Terashima:2011qi,Dimofte:2011jd}.

\begin{table}[t]
	\begin{center}
		\caption{By changing the manifold on which we compactify and the way of twisting,
			we can produce, from the 6d theory, field theories in various dimensions with various amounts of supersymmetry.
			Here we consider the case of general manifolds (more precisely, manifolds with general holonomy);
			for more special manifolds we can construct field theories with more supersymmetry. Of the cases in this table, we discuss only two in this book: (1) 4d $\scN=2$ theories obtained by compactification on 2-manifolds, and (2) 3d $\scN=2$ theories obtained by compactification on 3-manifolds. Compactifications and twists are explained in Chap.~\ref{chap.6d}.
		}\label{tab.twist}

		\begin{tabular}{c|c}
			\begin{tabular}{@{}c@{}}manifold for\\ compactification\end{tabular} & \begin{tabular}{@{}c@{}}supersymmetric field theory\\ in the remaining dimensions\end{tabular} \\
			\hline
			\hline
			2-manifold                                                           & 4d $\scN=2$ theory                                                                             \\
			\cline{2-2}
			                                                                     & 4d $\scN=1$ theory                                                                             \\
			\hline
			3-manifold                                                           & 3d $\scN=2$ theory                                                                             \\
			\cline{2-2}
			                                                                     & 3d $\scN=1$ theory                                                                             \\
			\hline
			                                                                     & 2d $\scN=(0, 2)$ theory                                                                        \\
			\cline{2-2}
			4-manifold                                                           & 2d $\scN=(1, 1)$ theory                                                                        \\
			\cline{2-2}
			                                                                     & 2d $\scN=(0, 1)$ theory                                                                        \\
		\end{tabular}
	\end{center}
\end{table}

\small
Historically, compactifications on 2-manifolds were discussed first,
and were later generalized to the case of compactifications on 3-manifolds.
There is also more literature on compactifications on 2-manifolds. Moreover, the geometry of 3-manifolds (see Chap.~\ref{chap.3mfd})
is more complicated than the geometry of 2-manifolds (see Chap.~\ref{chap.Teichmuller}),
and is less familiar to physicists.

In reality, however, the situation is simpler in many respects for compactifications on 3-manifolds.
For example, as we will explain in Chap.~\ref{chap.3mfd}, the correspondence between 3d gauge theories and 3-manifolds
is rather direct,
and there is a correspondence for each step
of their building blocks and of their computations.
On the other hand, the correspondence between 4d gauge theories and 2-manifolds
is more intricate,
and comparing computations on the 4d side and on the 2d side often requires
mathematically rather non-trivial identities.
%

This book aims to explain the correspondence between 3d $\scN=2$ theories and
3d Chern-Simons theory,
but in the process of the explanation we have tried to
make the relations with 4d $\scN=2$ theories and
2-manifolds (and \Teichmuller theory on them)
clear as well.

The discussion in this book is limited to specific compactifications of the 6d theory and
the supersymmetric field theories arising from them.
However, some of the ideas appearing there, for example
the construction of field theories by iterated gauging, dualities of theories in the IR,
field theories on boundaries and domain walls and
their operator interpretations, and geometric structures on the theory space of field theories,
are expected to be applicable to more general field theories
(see Chap.~\ref{chap.conclusion}).

\normalsize
\nextsectionmark{Gauging, Combining and Decomposing Theories}
\section{Gauging, and Combining and Decomposing Gauge Theories}

What we have described so far are dualities of field theories, and how to
understand them.
However, what we would like to discuss in this book is not limited to
dualities between two field theories: in general,
we would like to consider situations where several field theories describe the same physics.
In this book we call this a \keyword{duality web}{duality web}, and each of the several field theories a \keyword{duality frame}{duality frame}.

\small
Note that, as the ``dual'' in the name duality suggests,
duality itself is sometimes used only for the equivalence of two
theories, in which case, for example, a relation among three theories is called
triality. In this book we use duality in a broader sense,
meaning the equivalence of (several) different theories.

\normalsize

Duality in the usual sense discussed so far is
the tip of the iceberg in the richer duality web of field theories.
What plays an essential role here is the idea of
\keyword{gauging}{gauging}.

\bparagraph{Combining and Splitting Theories}

First, let us review what gauging, or its inverse, \keyword{un-gauging}{un-gauging},
is.
Suppose that a theory has a global symmetry. Suppose moreover that
there is a background gauge field $A^{\rm bgd}_{\mu}$ corresponding to the global symmetry.
Here a background field is a field introduced into the Lagrangian as an external field, whose value
is fixed without performing the path integral over it.

As an example of a background gauge field, it is easy to imagine a magnetic field
applied to an experimental apparatus.
The value of the magnetic field can be freely changed by an observer outside the apparatus to see the response,
and is a parameter independent of the dynamics of the theory describing the inside of the apparatus.

More concretely, we assume that the Lagrangian contains the coupling term between the background gauge field and the current $J^{\mu}$ of the global symmetry,\footnote{As we have already seen, this was needed to continuously connect massless and massive spin-$1$ particles.}
\begin{align}
	L_{\textrm{coupling}}=\int \, J^{\mu} A^{\rm bgd}_{\mu} \ .
	\label{JA}
\end{align}

This is similar to the external fields formally introduced in computations of correlation functions in field theory;
because of the Lorentz invariance of the Lagrangian,
$J$ carries a vector index $\mu$. Moreover,
for the invariance of the Lagrangian under the gauge transformation $\delta A^{\rm bgd}_{\mu}=D_{\mu
		}\Lambda$,
$D_{\mu} J^{\mu}=0$ is required.
Namely, $J^{\mu}$ has to be a conserved current.

Now, in this setup, gauging the global symmetry means
adding the kinetic term $-\int \frac{1}{2g^2 } \textrm{Tr}\, F_{\mu\nu}F^{\mu\nu}$ for the background gauge field $A^{\rm bgd}_{\mu}$ to the Lagrangian,
and promoting this global symmetry to a gauge symmetry.
Then $A^{\rm bgd}_{\mu}$, which was an external field, becomes a dynamical field,
and in the path integral we sum over all its configurations.

Since in this book we consider supersymmetric theories, this whole procedure has to be performed
including the superpartners so as to preserve supersymmetry. For example,
the kinetic term of the gauge field has to be accompanied by the kinetic term of the fermions (gauginos).

Conversely, when we have a gauge group and its gauge field $A_{\mu}$, we can define un-gauging:
we remove the kinetic term of the gauge field from the Lagrangian,
and replace the dynamical gauge field by a background gauge field.

This concept of gauging is something we do almost unconsciously when we study field theory.
For example, in quantum chromodynamics (QCD), quarks and anti-quarks transform non-trivially under the $SU(3)$ gauge group;
this can be regarded as gauging the theory in which the quarks are free fermions.
More generally, every gauge theory (with a Lagrangian) can be regarded as
first considering background gauge fields and then gauging them.

\small
In the real world, gauging is not everything; for example, couplings among matter fields
(e.g.\ Yukawa couplings) are also important, and indeed these interactions are also
indispensable in this book. However, since couplings of matter fields often break symmetries,
following the idea of effective field theory that all the terms allowed by symmetries are generated,
the presence or absence of matter interactions can also be reinterpreted as the presence or absence of
global symmetries, and it is sufficient to consider how the symmetries change.

Once the presence or absence of the interaction terms is determined, the strengths of the interactions can be expected to be
fixed to specific values at the IR fixed point (except for the freedom of marginal deformations),
so the discussion so far, focusing on global symmetries and their gauging,
in fact loses almost no generality;
we only need to consider the explicit breaking of global symmetries by interaction terms.

Apart from this, symmetries can be spontaneously broken by the dynamics of the theory.
This is a very important phenomenon, but
here we are concerned with the definition of the theory itself, which is a question prior to such properties of the theory,
so we do not consider it for the moment.

\normalsize

\bigskip

We have explained gauging so far, but it is
something that anyone studying field theory does more or less unconsciously.
What, then, is the benefit of the discussion so far\footnote{
	As an example of a useful application of the concept of gauging, we can mention the argument of
	anomaly cancellation by 't Hooft \cite{'tHooft:1979bh}.
}? Before answering this question, let us continue with a somewhat more abstract discussion.

As a special case of gauging, let us consider the following example.
Suppose that there are two theories $\mathcal{T}_1$ and $\mathcal{T}_2$,
each with a global symmetry $G_1\simeq G_2\simeq G$.
Let us consider gauging its diagonal part $G_{\rm diag}\subset G_1\times G_2$.
By this gauging, the two theories, which originally were independent without interactions,
now interact through the gauged dynamical
gauge field, and
the two theories can now be regarded as a single theory.
Let us denote it by $\mathcal{T}_{1\cup 2}$.

Conversely, given the theory $\mathcal{T}_{1\cup 2}$,
if we un-gauge the gauge group $G$, this theory is
decomposed into the two theories $\mathcal{T}_1$ and $\mathcal{T}_2$.

Here we assumed that the two theories have the same global symmetry; in general, even if the two
global symmetries $G_1$ and $G_2$ are different,
we can perform a similar operation by gauging $H$, as long as they have a common subgroup $H$.

\begin{figure}[t]
	\centering
	\begin{tikzpicture}[inner sep=0]
		\node[anchor=south west] (img) {\includegraphics[scale=0.3]{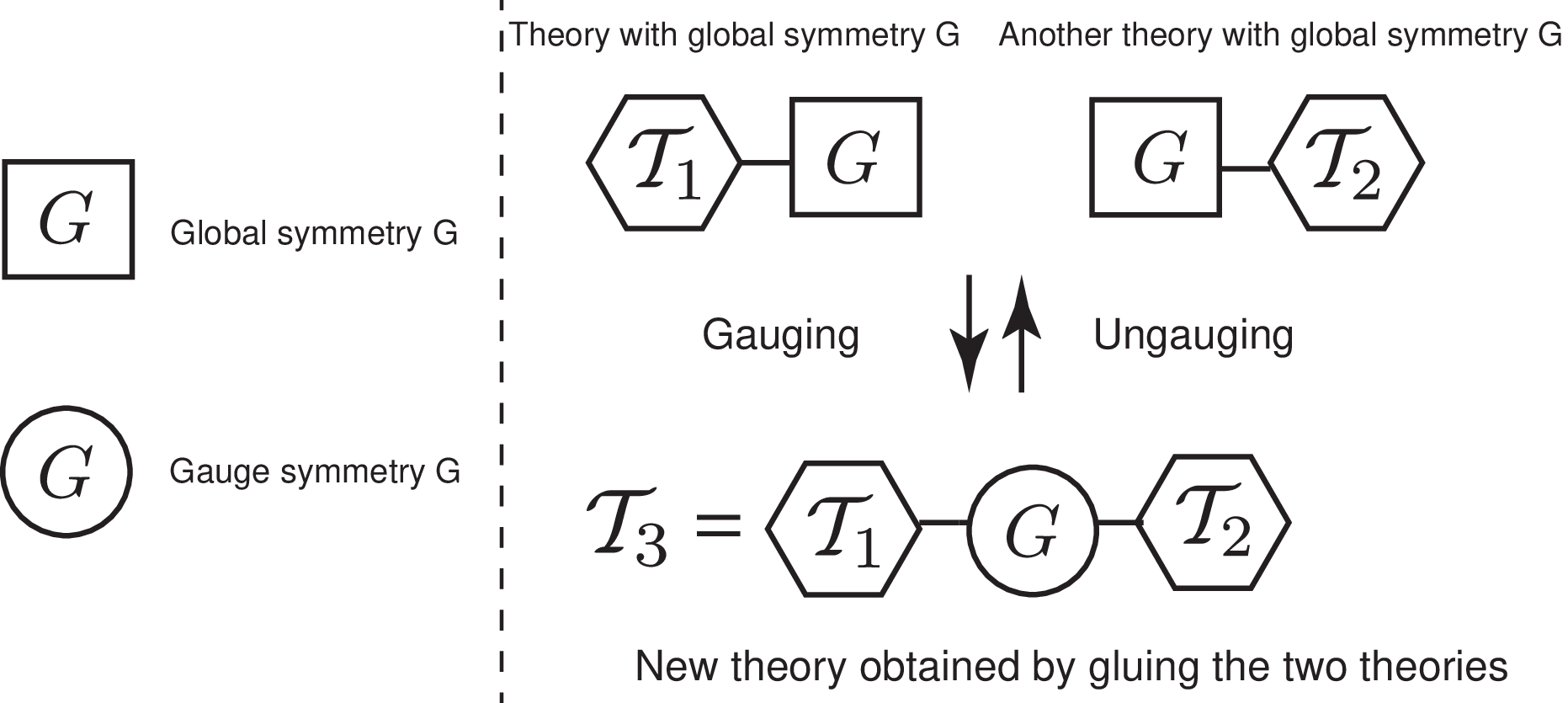}};
		\begin{scope}[shift={(img.south west)},x=0.3pt,y=0.3pt]
			\fill[white] (300,70) rectangle (436,142);
			\node[anchor=east,inner sep=0] at (432,106) {\Large$\mathcal{T}_{1\cup 2}=$};
		\end{scope}
	\end{tikzpicture}
	\caption{When there are two theories $\mathcal{T}_{1}$ and $\mathcal{T}_2$ with global symmetry $G$,
		we can construct a larger theory $\mathcal{T}_{1\cup 2}$ by gauging the diagonal part $G_{\rm diag}\subset G\times G$ of their $G\times G$ symmetry.
		Conversely, by un-gauging the gauge group $G$ of $\mathcal{T}_{1\cup 2}$,
		we can decompose $\mathcal{T}_{1\cup 2}$ into the two theories $\mathcal{T}_{1}$ and $\mathcal{T}_2$.
	}
	\label{fig.gauge}
\end{figure}

In this way, by gauging and un-gauging,
we can combine two theories into one, or conversely
split one theory into two (Fig.~\ref{fig.gauge}).

What is interesting here is that, once we allow the operation of gauging,
we can go beyond the realm of theories described by Lagrangians in the usual sense.

For example, in the 3d $\mathcal{N}=2$ theories discussed in the following chapters,
we discuss the \keyword{topological $U(1)$ symmetry}{topological U(1) symmetry} and its gauging (Chap.~\ref{chap.3dN2}).
This symmetry can be read off indirectly from the Lagrangian, but
\textbf{acts trivially on the fields appearing in the Lagrangian}.
Alternatively, there can also be global symmetries which cannot be (manifestly) read off at the Lagrangian level
(an example is the $SU(N)/\mathbb{Z}_N$ symmetry of the $T[SU(N)]$ theory discussed in Chap.~\ref{chap.wall}). Even if there are such symmetries, they can conceptually be
gauged; however, since the symmetry does not exist in the Lagrangian,
the theories constructed in this way do not in general have Lagrangians.

\bigskip

How, then, is this concept of gauging connected with duality?
First, suppose that there is a pair of dual theories $\mathcal{T}_1$ and $\mathcal{T}_2$.
Suppose moreover that they have a (non-anomalous) global symmetry $G$---since global symmetries should agree between dual theories,
if one of the theories has $G$, the other automatically also has $G$.

Now suppose further that there is another theory $\mathcal{T}_3$, which also has
the global symmetry $G$. We can then combine $\mathcal{T}_1$ and $\mathcal{T}_3$,
or $\mathcal{T}_2$ and $\mathcal{T}_3$, by gauging $G$---since
the starting theories $\mathcal{T}_1$ and $\mathcal{T}_2$ were equivalent,
the theories $\mathcal{T}_{1\cup 3}$ and $\mathcal{T}_{2\cup 3}$ obtained by this combination are also expected to be equivalent.

We have thus obtained a new duality, the duality between $\mathcal{T}_{1\cup 3}$ and
$\mathcal{T}_{2\cup 3}$.

This duality is not so different from the original duality.
However, by combining this idea we can increase the variety of dualities further.

Consider the theory $\mathcal{T}_{1\cup 2\cup 3}$.
Let us assume that the theories $\mathcal{T}_{1\cup 2}$ and $\mathcal{T}_4$ are dual, and that
the theories $\mathcal{T}_{2\cup 3}$ and $\mathcal{T}_5$ are dual. Then
\begin{align}
	\mathcal{T}_{4\cup 3} \simeq
	\mathcal{T}_{1\cup 2} \cup \mathcal{T}_3 \simeq
	\mathcal{T}_{1\cup 2\cup 3}  \simeq
	\mathcal{T}_{1} \cup \mathcal{T}_{2\cup 3} \simeq
	\mathcal{T}_{1 \cup 5} \ ,
\end{align}
so that we find that the theories $\mathcal{T}_{4\cup 3}$ and $\mathcal{T}_{1\cup 5}$ are
equivalent.

In general, if there is a huge theory made by several gaugings,
we can consider a web of dualities among several huge theories
by un-gauging, applying duality transformations to the resulting subtheories, and
gauging again.
Alternatively, we can also break the global symmetries to their subgroups by adding interaction terms to the theories.
By repeating such operations, we can produce sets of several
equivalent theories.

\small
Let us add a remark for careful readers:
to be precise, there is a subtle point in the claims so far.
It is that some of the dualities (for example, the Seiberg duality already discussed)
are dualities at IR fixed points, and do not hold directly at the Lagrangian level.
In general, the operation of gauging introduces new gauge coupling constants, which
also flow, so that the behavior under the renormalization group changes, and we cannot rule out the possibility that
dualities which held before gauging are broken.
However, in the examples of 3d $\scN=2$ theories discussed in this book,
the dualities are strongly suggested by the analysis of $S^3_b$ partition functions and the correspondence with 3-manifolds,
and there is no positive evidence to doubt them.

\normalsize

\bparagraph{Decomposing and Combining Manifolds}

We have so far explained combining and decomposing theories by gauging.
Can we, then, understand this geometrically?

So far we have discussed rather general theories;
to introduce a geometric picture, let us assume that the theories are supersymmetric gauge theories and are given by
compactifications of the six-dimensional theory on manifolds.
Since gauge groups appear when we compactify the six-dimensional theory (e.g.\ on tori or cylinders),
it is not such a far-fetched idea to expect that gauging, and hence the combining and
decomposing of gauge theories, corresponds to decomposing and combining manifolds.

Indeed, developing this claim and summarizing it, we obtain the following:
\textbf{the choice of a duality frame of a supersymmetric gauge theory corresponds to a different way of
	decomposing a manifold. Non-trivial
	dualities of field theories are reinterpreted as the non-uniqueness of decompositions on the geometry side.
	Moreover, properties of field theories, such as their parameters and information about their vacua, are
	reinterpreted as geometric information about the manifolds}.

What we aim for in the following chapters is to flesh out this slogan
in concrete examples.
For this purpose, let us first step into the world of 3d field theories.
Readers who spend every day in four-dimensional spacetime
will encounter a world which, despite being ``low-dimensional'', is surprisingly rich.


\begin{practice}

	\item $[\bll \bll]$ (Bosonization)
	\label{ex.bosonize} Study and summarize the transformation between fermions and bosons in two dimensions (bosonization).

	\item $[\bll]$ (Propagator of the Proca field)
	\label{ex.propagator}
	Derive the propagator \eqref{eq.propagator} of a massive spin-$1$ particle. Hint: since the field $B^{\mu}$
	has $4$ degrees of freedom, in order to obtain a spin-$1$ field with $3$ degrees of freedom (an irreducible representation of the Poincar\'e group)
	we need to impose one Lorentz-invariant condition: $\partial^{\mu}
		B_{\mu}=0$.
	The Lagrangian is a Lorentz-invariant functional constructed from $B^{\mu}$,
	and has to give equations of motion compatible with
	this condition $\partial_{\mu} B^{\mu}=0$ and the Klein-Gordon equation
	$(\Box-m^2) B_{\mu}=0$
	(where $m$ is the mass). From this the Lagrangian is found to be
	\begin{align}
		\begin{split}
			\mathcal{L}_{\rm Proca}=-\frac{1}{4}
			(\partial_{\mu}B_{\nu}-\partial_{\nu} B_{\mu})
			(\partial^{\mu}B^{\nu}-\partial^{\nu} B^{\mu})
			- \frac{m^2}{2} B^{\mu} B_{\mu} \ .
		\end{split}
	\end{align}

	\item $[\bll \bll]$ (Spin-$2$ particles) If we apply exercise~\ref{ex.propagator} and the discussion in the main text to massless and massive spin-$2$ particles, what conditions do we obtain from their continuity?

	\item $[\bll\bll]$  (Formulations of field theory) What formulations of field theory are possible other than the one using Lagrangians described in this chapter?

	\item $[\bll]$  (Examples of dualities of supersymmetric field theories) What dualities of supersymmetric field theories other than those described in this book are known? Look up the literature and
	give two or three examples. Can you use them to create a web of dualities by the operations of gauging and un-gauging?

\end{practice}

\chapter{The Joy of 3d Field Theories}
\label{chap.3dglue}

\begin{abstract}
	In Chap.~\ref{chap.intro}, we introduced a general
	theory of ``gluing QFTs'' via gauging.
	From now on let us specialize to the case where
	the quantum field theories are defined in three spacetime dimensions.
	In three spacetime dimensions, the concept of
	``gluing quantum field theories'' by gauging is
	especially natural, and (as we will see) more general
	than in the case of the more familiar four spacetime dimensions.
	The concept of the $Sp(2n, \mathbb{Z})$-transformation
	introduced in this chapter will play central roles in subsequent chapters.

\end{abstract}

\section{Topological $U(1)$ Symmetry}

\subsection{Why Three Spacetime Dimensions?}

In the rest of this book, most of the time we focus on
quantum field theories in three spacetime dimensions.

This might already discourage some readers from
continuing to read the rest of this book: after all,
our Universe, at least as far as we can see and experience, has four spacetime dimensions,
so why care about three spacetime dimensions, which apparently do not have much to do with our Universe?
This is a completely legitimate question, especially for those
who are not studying formal aspects of quantum field theories and string theories on a daily basis.

We can provide several answers to this question, at different levels of sophistication.
One possible (although technical) answer is that
the tools to analyze supersymmetric gauge theories, which we shall discuss
in Chap.~\ref{chap.S3}, are best developed
in three spacetime dimensions. Another answer is
that three spacetime dimensions are among the most interesting
dimensions obtained from compactifications of the
mysterious six-dimensional theory.
We can also appeal to the more general point that
understanding QFTs in three spacetime dimensions provides a useful toy model for
better understanding QFTs in four spacetime dimensions.

Each of these justifications has some point, of course,
but in this book we would like to emphasize the following two points,
which in my opinion are more fundamental.
Namely, (1) \textbf{there exist Chern-Simons terms in three spacetime dimensions,
		enriching and enlarging the concept of ``gluing QFTs''}, and
	(2) \textbf{the 3d QFTs discussed in this book arise as domain walls (and more generally boundary conditions)
		of four-dimensional QFTs}.

As we shall elaborate below, the two points above
are natural and fundamental problems
for us, who want to study the theoretical problem of
understanding the ``geometry of QFTs''.\footnote{We should also point out that this is
	natural from a more ``practical'' viewpoint as well. For example, in relation to (2),
	let us point out that in real experimental systems, 3d ((2+1)d) systems are quite often realized on the
	boundaries between two materials. For example, if we consider a junction of two different types of
	semiconductors (say GaAs and AlGaAs),
	we observe the quantum Hall effect at the junction, which is described by a
	3d Chern-Simons theory. This is a physical realization of the two ideas (1) and (2).}

In the following we will discuss (1); we will comment on (2) later,
in Chap.~\ref{chap.wall}.

\subsection{Repetition of Gauging}\label{subsec.repeat}

Let us first recall the concept of ``gluing QFTs by gauging'' from
the previous chapter---we consider two QFTs with (non-anomalous) global symmetries $G$,
and gauge the diagonal of the $G\times G$ symmetry.

As commented already in Chap.~\ref{chap.intro},
what is important here is not to perform such a gauging/un-gauging operation
just once; we start from a simple theory,
but the crucial point is that
we \textbf{repeat gauging}, and obtain a web of dualities
and the geometric structures therein.

We encounter one problem here, however.
Suppose that we have a QFT with global symmetry $G$.
We can then gauge that flavor symmetry; however, the
resulting theory no longer has a global symmetry, and
hence no further gauging is possible. This means that we cannot
repeat the process of gauging to obtain more and more complicated theories.

Fortunately, it is not hard to come up with a solution to this problem---there is no need to gauge all the global symmetries;
we can instead choose to gauge only part of the global symmetries, while keeping the rest.
As an example, suppose that we have a QFT with global symmetry $G$.
The trick is to bring in another theory with $G\times G'$ global symmetry,
and gauge the diagonal of the $G\times G$ flavor symmetry.
The resulting theory is a new theory with global symmetry $G'$.
In the next step, we prepare yet another theory with
$G'\times G''$ flavor symmetry, and repeat this operation.

\begin{figure}
	\centering
	\begin{tikzpicture}[inner sep=0]
		\node[anchor=south west] (img) {\includegraphics[scale=0.5]{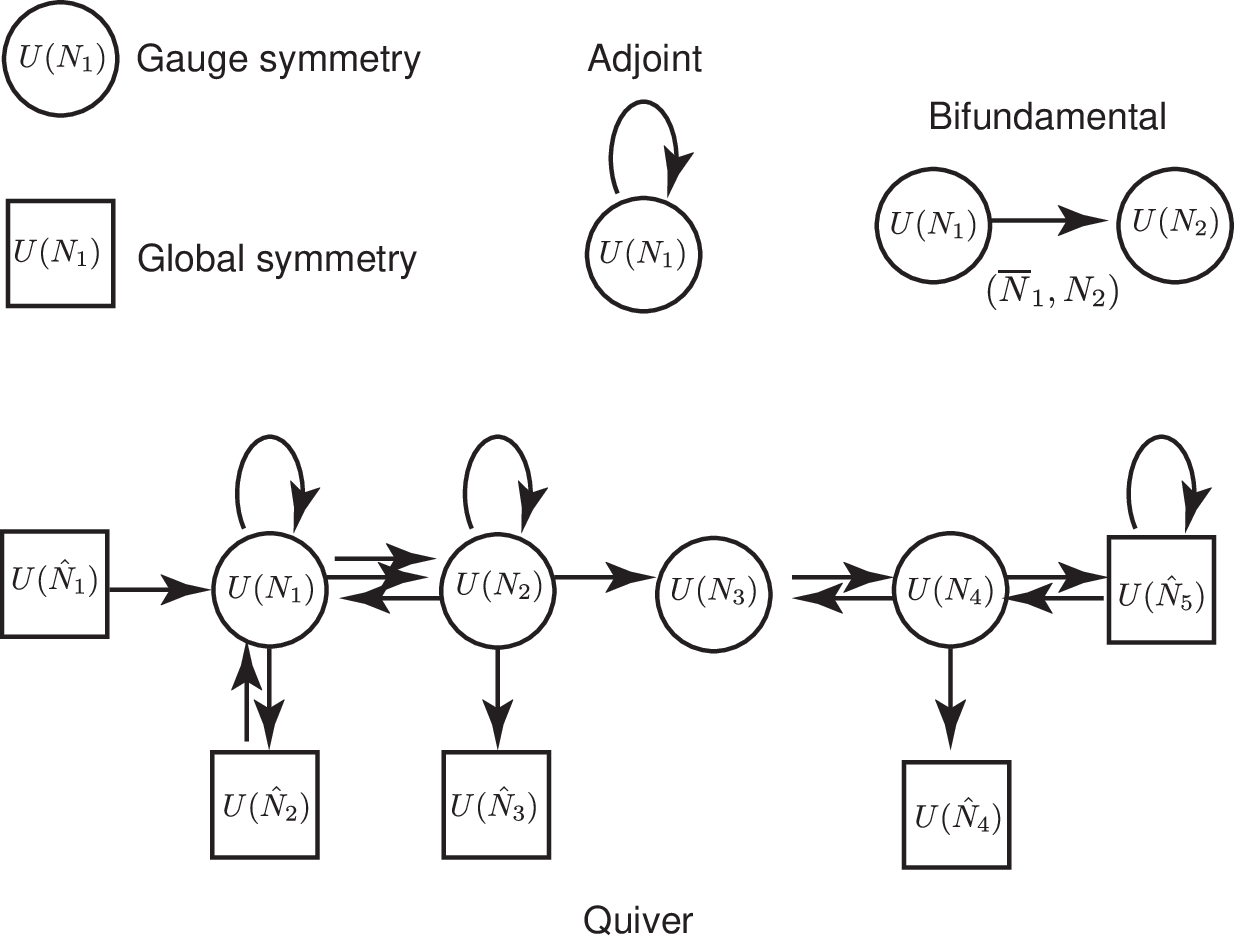}};
		\begin{scope}[shift={(img.south west)},x=0.5pt,y=0.5pt]
			\fill[white] (468,300) rectangle (548,326);
			\node[anchor=center,inner sep=0] at (508,313) {\small$(\bm{N}_1, \overline{\bm{N}}_2)$};
		\end{scope}
	\end{tikzpicture}
	\caption{We represent a gauge symmetry by a circle and
		a global symmetry by a square.
		A matter field is represented by an arrow,
		transforming in the bifundamental representation with
		respect to the symmetries at the starting and ending points of the
		arrow. We can construct complicated
		gauge theories (quiver gauge theories) by combining all these ingredients.
	}
	\label{fig.quiver}
\end{figure}

The simplest example of a theory with
$G\times G'$ global symmetry is a free (i.e.\ non-interacting) field
which transforms, under the $G\times G'$ global symmetry, in the fundamental representation of $G$
and in the anti-fundamental representation (the complex conjugate of the fundamental representation) of $G'$.
Such a representation is also called the
\keyword{bifundamental representation}{bifundamental representation} of $G\times G'$.
In this case, the field theories constructed as above are called
\keyword{quiver gauge theories}{quiver gauge theory} (see Fig.~\ref{fig.quiver}; quiver gauge theories reappear in Sec.~\ref{subsec.TSUN}).

\small
Here a \keyword{quiver}{quiver}\footnote{A quiver is a case for holding arrows; the name likens the arrows of a directed graph, lined up together, to arrows in a quiver.} is simply an oriented graph.
In our context, a vertex means a gauge field (say with gauge group $U(N)$).
This gauge field can either be a dynamical gauge field or a \keyword{background gauge field}{background gauge field}.
We represent this difference graphically by a circle and a
square, as in Fig.~\ref{fig.quiver}.
An arrow represents a matter field transforming in the bifundamental
representation with respect to the gauge groups at its two endpoints.
Namely, if an arrow starts at a vertex with gauge group
$U(N_1)$ and ends at another vertex with gauge group $U(N_2)$,
then we associate with it a bifundamental matter field in the representation $(\bm{N}_1, \overline{\bm{N}_2})$
under $U(N_1 )\times U(N_2)$.

\normalsize

\subsection{Gauging in 3d}

The method discussed so far is applicable in an arbitrary spacetime dimension;
in three dimensions, however, there is another method.
Suppose that we are interested in gauging the background
gauge field $A_{\mu}$.
The global symmetry is then promoted to a
gauge symmetry, and hence it looks as if the global symmetry is lost altogether.
However, it turns out that gauging introduces a new global symmetry,
which we call the topological global symmetry.

To see this, let us define a current
from the field strength $F_{\mu\nu}$ as follows:\footnote{
	For Abelian gauge groups, in the language of vector analysis, \eqref{JstarF} becomes
	$J=\frac{1}{2\pi}*dA=\frac{1}{2\pi}\textrm{rot}\, A$, and \eqref{U1Jconserve} becomes $\textrm{div}\, J =0$. Recall that in three dimensions, when $\textrm{div}\, J=0$, we can (locally) write $J=\textrm{rot}\, A$.
	\label{Jvector}
}
\begin{align}
	J:=\frac{1}{2\pi}* F \quad \textrm{i.e.} \quad J^{\mu}:=\frac{1}{4\pi}\epsilon^{\mu\nu\rho} F_{\nu\rho} \ .
	\label{JstarF}
\end{align}
Note that this is a conserved current by definition:
\begin{align}
	D_{\mu} J^{\mu}=\frac{1}{4\pi}\epsilon^{\mu\nu\rho} D_{\mu} F_{\nu\rho}=0 \ ,
	\label{U1Jconserve}
\end{align}
where the last equation is the Bianchi identity, which you can also verify by direct computation.
In \eqref{JstarF} we have included a factor of $2\pi$ for later convenience.

We can now appeal to the
general argument that
the existence of a conserved current implies the
conservation of its charge $Q_{\rm conserved}:=\int J^0$;
such a charge typically generates a global symmetry of the system.
What is slightly unconventional here is that
the charge here is not the Noether charge associated with a symmetry,
but rather a topological charge, and its conservation is
guaranteed not by the equations of motion
but by a topological reasoning:
note that we did not need any equation of motion
for the proof of the current conservation \eqref{U1Jconserve}.
For this reason, the global symmetry corresponding to
$J^{\mu}$ is often called the
\keyword{topological $U(1)$ symmetry}{topological U(1) symmetry},
or the \keyword{$U(1)_J$ symmetry}{U(1)J symmetry}.

To see this point in more detail,
let us come back to the definition \eqref{JstarF} of the
current for the
topological $U(1)$ symmetry.
Suppose that we consider the electric field created by a point-like electric charge $q$ at the origin.
The electric field is then given by $E^i=F^{i0}=\frac{q x^i}{2\pi r^2}$, where $i,j=1,2$ are indices for the 2d spatial directions.
Note that in two spatial dimensions the strength of the electric field is inversely proportional to the distance.

\begin{figure}
	\centering\includegraphics[scale=0.35]{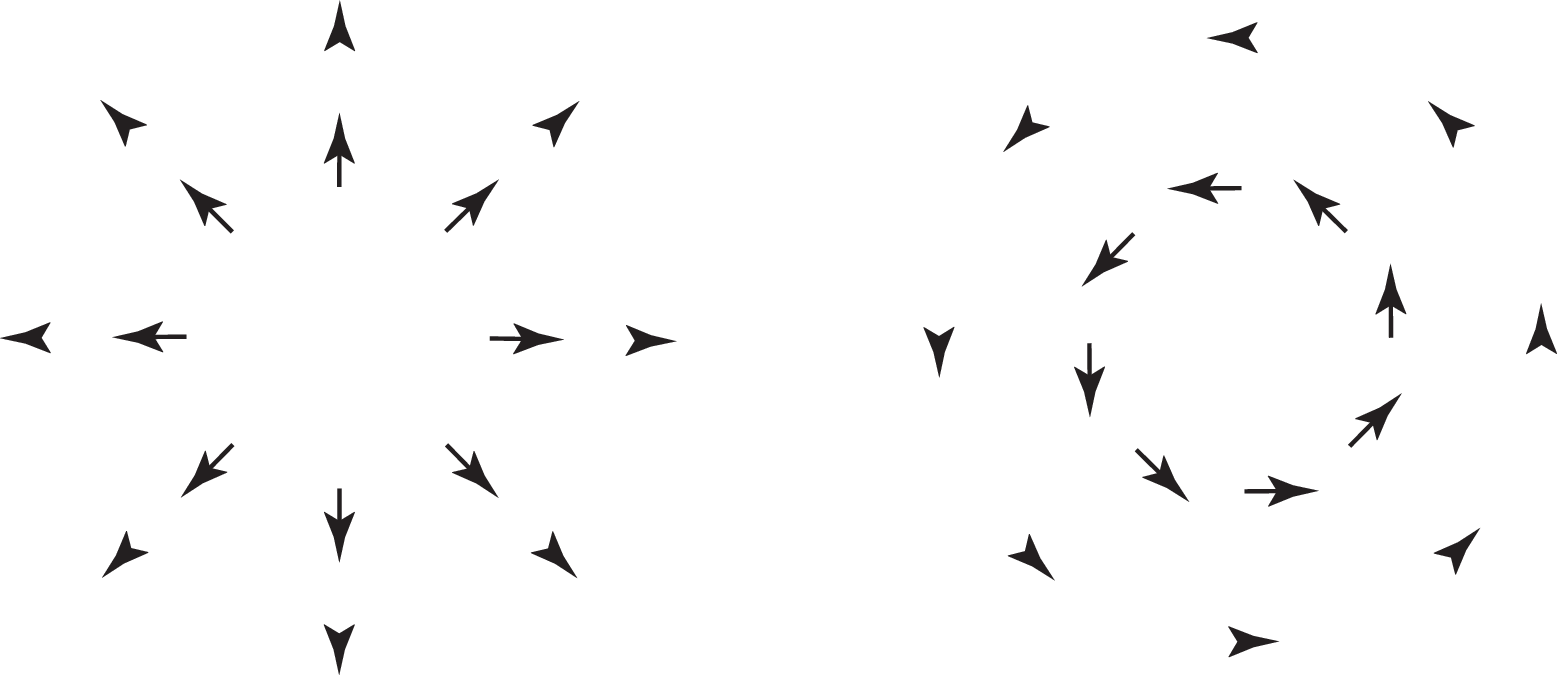}
	\caption{The electric field (left) generates a vortex (right) of the corresponding topological $U(1)$ symmetry.}
	\label{fig.vortex}
\end{figure}

The corresponding current for the
topological $U(1)$ symmetry is then
\begin{align}
	J^i=\frac{1}{2\pi}\epsilon^{ij} \frac{q x_j}{2\pi r^2}  \ .
\end{align}
As shown in Fig.~\ref{fig.vortex},
we have a rotating flow around the origin,
namely a \keyword{vortex}{vortex}.
Notice that our electric field is time-independent, and hence the vortex
extends uniformly along the time direction.
The value of $q$ is an integer
known as the winding number around the origin, and is called the \keyword{vortex charge}{vortex charge}:
\begin{align}
	q=-2\pi\oint dx^i J_i \in \bZ \ .
	\label{vortex_charge}
\end{align}
The winding number is a topological quantity, and
the vortex, which has such a charge, is a soliton.
Summarizing,
\textbf{the electric field of the gauge field creates a vortex of the
	topological $U(1)$ symmetry,
	and the charge with respect to the gauge field is given by the
	charge (winding number) of the vortex.}

What is unusual about the
topological $U(1)$ symmetry is that
none of the fields present in the Lagrangian
(neither the gauge field $A_{\mu}$ nor the bosons/fermions
coupled to the gauge field)
are charged under this symmetry.
However, since the topological $U(1)$ symmetry is a non-trivial symmetry,
it should act non-trivially on something, and we
come to the natural question:
what physical object is charged under the topological $U(1)$ symmetry?

To answer this question, we need the concept of the ``dual'' of the gauge field,
which we now explain.

\section{Duality Transformation}\label{sec.3d_dualize}

To explain the duality transformation of the gauge field,
we start with the simplest possibility, namely
the Lagrangian for a $U(1)$ gauge field:
\begin{align}
	L_{\textrm{Abelian gauge field}}=-\int\! d^3 x\, \frac{1}{4g^2} F_{\mu\nu} F^{\mu\nu}  \ .
\end{align}
Here the field strength $F_{\mu\nu}$ is not an elementary field:
not all the components of $F_{\mu\nu}$ are independent,
since the field strength is represented in terms of the gauge field $A_{\mu}$ as
$F_{\mu\nu}=\partial_{\mu} A_{\nu}-\partial_{\nu} A_{\mu}$.
As we learn in electromagnetism, this implies the Bianchi identity
$\epsilon^{\mu\nu\rho}\partial_{\mu} F_{\nu\rho}=0$.
Conversely, if we have an anti-symmetric field $F_{\mu\nu}$ satisfying the
Bianchi identity, then we can construct the corresponding gauge field
$A_{\mu}$ (of course only up to gauge transformations\footnote{
	See footnote~\ref{Jvector}. In the language of differential forms,
	when we have $dF=0$ for a 2-form $F$, there exists a 1-form $A$
	such that $F=dA$. This is known as the Poincar\'e lemma.
	To be more precise, this lemma does not hold when the 3d spacetime geometry
	has non-trivial topology, but we here consider a flat and in particular simply-connected spacetime
	$\mathbb{R}^{2,1}$ (see Ref.~\cite{Beasley:2014ila} for a recent discussion
	of duality transformations on more general spacetime manifolds).}).
This means that we can treat the field $F_{\mu\nu}$ as an elementary field,
as long as we introduce a Lagrange multiplier $\gamma$ imposing the
Bianchi identity:
\begin{align}
	L=-\int\! d^3 x\, \left[\frac{1}{4g^2} F_{\mu\nu} F^{\mu\nu}
		                  -\frac{1}{2g^2}\gamma \epsilon^{\mu\nu\rho}\partial_{\mu}F_{\nu\rho}\right] \ ,
	\label{Lag1}
\end{align}
where $\gamma$ has mass dimension $1$.
By integrating by parts, we obtain
\begin{align}
	L=-\int\! d^3 x\, \left[\frac{1}{4g^2} F_{\mu\nu} F^{\mu\nu}
		                  +\frac{1}{2g^2}\epsilon^{\mu\nu\rho}\partial_{\mu} \gamma F_{\nu\rho}\right] \ ,
\end{align}
where we assumed that appropriate boundary conditions are imposed at
infinity so that the boundary term vanishes.
This Lagrangian is quadratic in the 2-form $F_{\mu\nu}$,
and hence we can trivially integrate over $F_{\mu\nu}$, to obtain
\begin{align}
	F_{\mu\nu}= -
	\epsilon_{\mu\nu\rho } \partial^{\rho} \gamma
	\ , \quad
	\textrm{i.e.\ }
	\quad
	\partial^{\mu}\gamma=\frac{1}{2}\epsilon^{\mu\nu\rho} F_{\nu\rho} \ .
	\label{Fgamma}
\end{align}
Substituting this result into the original Lagrangian, we obtain the
new Lagrangian
\begin{align}
	L= - \int \! d^3 x\,
	\frac{1}{2g^2} \partial^{\mu} \gamma \partial_{\mu} \gamma \ .
	\label{dualphotonL}
\end{align}

It turns out that the dual scalar
$\gamma$ has a certain periodicity.
To verify this, let us shift the value of $\gamma$ by $g^2$.
The Lagrangian \eqref{Lag1} is then shifted by
\begin{align}
	\delta L=\frac{1}{2}\int d^3 x\,
	\epsilon^{\mu\nu\rho}\partial_{\mu}F_{\nu\rho}
	=\int_{\mathbb{R}^{2,1}} \,
	dF
	=2\pi \int_{S^2_{\infty}} \frac{F}{2\pi}\in 2\pi \mathbb{Z} \ .
\end{align}
Here $S^2_{\infty}$ is the sphere at infinity of $\mathbb{R}^{2,1}$.
Also, the combination $c_1(F):=\frac{F}{2\pi}$ in the final expression
is an example of a so-called characteristic class, known as the first Chern class of a vector bundle over $S^2$,
and in particular its integral over $S^2_{\infty}$ is known to be an integer \cite{MilnorBook}.
This means that, while the Lagrangian itself is shifted,
the value of $e^{iL}$ is kept invariant.
This shows that $\gamma$ has periodicity $g^2$:
\nomenclature{$\gamma$}{dual photon}
\begin{align}
	\gamma \sim \gamma+g^2 \ .
\end{align}

We have therefore seen that
a free $U(1)$ gauge field is equivalent
to a free scalar field $\gamma$ with periodicity $g^2$.
Such a scalar field is called the
\keyword{dual photon}{dual photon}.

The newly obtained Lagrangian is invariant
under the shift of $\gamma$: $\delta\gamma=\textrm{constant}$.
Indeed, the associated Noether current is given by (see exercise~\ref{ex.Noether})
\begin{align}
	J^{\mu}=\frac{1}{2\pi} \partial^{\mu} \gamma\ ,
	\label{dualphotonJ}
\end{align}
which coincides with our previous definition \eqref{JstarF},
thanks to \eqref{Fgamma}.
We have thus learned that
\textbf{the topological $U(1)$ symmetry is nothing but the shift symmetry
	of the dual photon}; it turned out that what is charged under the
topological $U(1)$ symmetry is the dual photon.

In the discussion of the duality transformation so far we discussed only the gauge field; in reality, however,
the gauge field interacts with matter fields (scalars and fermions), so that
we have to perform the duality transformation including them.
For systems of bosons,
this is often discussed in condensed matter physics, where it is called
\keyword{particle-vortex duality}{particle-vortex duality}, or
\keyword{boson-vortex duality}{boson-vortex duality}.
The 3d mirror symmetry described in the next chapter can be said to be
this duality transformation carried out for supersymmetric theories containing
bosons and fermions (see also exercise~\ref{ex.superdual} of the next chapter).
Moreover, even for theories with non-Abelian gauge groups, a similar duality transformation can be carried out
in situations where the gauge group is broken to an Abelian subgroup.
The Coulomb branch of the vacuum moduli space discussed in the next chapter is an example of such a situation.

\bigskip

Now that we have a better understanding of
the topological $U(1)$ symmetry, we can
introduce the associated background gauge field, which we denote by $A^{\rm bgd}_{\mu}$.
Its coupling \eqref{JA} with the current reads (recall
the definition of the current in \eqref{JstarF})
\begin{align}
	\int J_{\mu} A^{\rm bgd \, \mu}
	=\frac{1}{2\pi}\int dA\wedge A^{\rm bgd} \ .
\end{align}
This is an example of the famous
\keyword{Chern-Simons term}{Chern-Simons term}.
To be somewhat more general, given a set of
Abelian gauge fields $(A_i)_{\mu}$,
the Chern-Simons term is given by\footnote{
	For the Chern-Simons term, relativistic invariance follows from the requirement of gauge invariance alone (unlike, e.g., the kinetic term of Yang-Mills fields). This is why quantum Hall systems, which appear in non-relativistic condensed matter systems, are described by relativistic field theories.
}\footnote{
	In quantum Hall systems, when the interactions between Landau levels can be neglected, one sometimes considers
	a corresponding $U(1)$ gauge field for each Landau level, together with \eqref{KCS}.
	In this case the coefficient matrix $k_{i,j}$ is often called the $K$-matrix.
}
\begin{align}
	\begin{split}
		\mathcal{L}_{\rm CS} & =  \sum_{i,j} \frac{k_{i,j}}{4\pi }  A_i\wedge
		d A_j                                                                 \\
		                     & =
		\sum_{i,j}\frac{k_{i,j}}{4\pi}
		\epsilon^{\mu\rho \sigma}(A_i)_{\mu} \partial_{\rho} (A_j)_{\sigma}   \ .
		\label{KCS}
	\end{split}
\end{align}
Here $\{k_{i,j} \}$ is a symmetric matrix, $k_{i,j}=k_{j,i}$,
and a term with $i=j$ ($i\ne j$) is called a diagonal (off-diagonal) Chern-Simons term.\index{Chern-Simons term@Chern-Simons term!off-diagonal@off-diagonal}

Namely, \textbf{the coupling between the
	current of the topological $U(1)_J$ symmetry for the gauge field $A_{\mu}$
	and the background gauge field $A^{\rm bgd}_{\mu}$
	is given precisely by an off-diagonal Chern-Simons term.}

Since the Chern-Simons term contains the gauge field itself (not its field strength)
in the Lagrangian, it is not obvious whether the action is gauge invariant.
As discussed in more detail in Sec.~\ref{subsec.CS_term_2},
the levels $k_{i,j}$ are required to be integers for the path integral to be gauge invariant.

\section{$Sp(2n, \mathbb{Z})$-Action}

Now that we have understood the meaning of the topological $U(1)$ symmetry,
let us return to the problem of combining field theories.
Since a gauged gauge field has
the topological $U(1)_J$ symmetry, which shifts its dual photon,
it is possible to gauge this $U(1)_J$ symmetry.
What happens if we repeat this operation?

In fact, if we repeat this operation twice, it becomes trivial. To see this,
it is convenient to discuss everything in terms of Lagrangians.
As already stated in the previous section, the operation of gauging was represented by an off-diagonal Chern-Simons term:
\begin{align}
	S:\, \mathcal{L}[A] \to
	\mathcal{L}'[B]=\mathcal{L}[A]+ \frac{1}{2\pi }B\wedge
	dA \ .
	\label{SLag}
\end{align}
Here we denoted the background gauge field of the $U(1)_J$ symmetry by $B_{\mu}$. Recall that the background field $A_{\mu}$ of the original global symmetry is gauged after the operation.
Let us also call this operation the $S$-transformation; the reason will become clear shortly.

Repeating this operation twice, by definition we have
\begin{align}
	S^2:\, \mathcal{L}[A] \to
	\mathcal{L}'[C]=\mathcal{L}[A]+ \frac{1}{2\pi }B\wedge
	dA+\frac{1}{2\pi }C\wedge dB  \ .
\end{align}
Here $B$ is a dynamical gauge field, and can be integrated out first. Since $B$ appears linearly in the
Lagrangian, this can be easily integrated, giving the constraint $A+C=0$
(more precisely, $A+C$ is trivial up to gauge degrees of freedom). Therefore $\mathcal{L}'[C]=\mathcal{L}[-A]$, and we find that $S^2$ preserves the theory (up to the sign of $A$).

Since one of our aims was to construct new theories by gauging theories,
this is a disappointing result. However, it is too early to be discouraged. Eq.~\eqref{SLag} was
a transformation adding an off-diagonal Chern-Simons term; then
let us also consider a transformation adding a diagonal Chern-Simons term:
\begin{align}
	T:\, \mathcal{L}[A] \to
	\mathcal{L}'[A]=\mathcal{L}[A]+ \frac{1}{4\pi} A\wedge dA  \ .
	\label{TLag}
\end{align}
What happens, then, if we repeatedly perform the transformations \eqref{SLag} and \eqref{TLag}? In fact, \textbf{the transformations $S, T$ generate $SL(2, \mathbb{Z})$
		transformations, and \eqref{SLag} and \eqref{TLag} correspond to the generators $S, T$ of
		$SL(2, \mathbb{Z})$ (see \eqref{STgen}), respectively}:
namely, the relations \eqref{STrel} hold.

We have already examined $S^2$: the minus sign in $S^2=-1$ is interpreted as reversing the sign of the gauge field (charge conjugation).

The proof of the other relation, $(ST)^3=1$, is similar.
Computing $(ST)^3$ following the definitions, we obtain
\begin{align}
	\begin{split}
		(ST)^3:\, \mathcal{L}[A] \to
		\mathcal{L}'[D]=\mathcal{L}[A] &
		+\frac{1}{4\pi } A\wedge dA
		+ \frac{1}{2\pi }A\wedge dB
		\\
		                               & +\frac{1}{4\pi } B\wedge dB
		+\frac{1}{2\pi }B\wedge dC
		\\
		                               & +\frac{1}{4\pi } C\wedge dC
		+\frac{1}{2\pi }C\wedge dD \ .
	\end{split}
\end{align}
Integrating over the dynamical field $B$ gives $B=-(A+C)$, and
substituting this into the expression above, we obtain
\begin{align}
	\begin{split}
		\mathcal{L}'[D]=\mathcal{L}[A] &
		-\frac{1}{2\pi }A\wedge dC
		+\frac{1}{2\pi }C\wedge dD \ .
	\end{split}
\end{align}
Integrating next over another dynamical field $C$,
we obtain $A=D$, and hence $\mathcal{L}'[D]=\mathcal{L}[A]$. This is
nothing but $(ST)^3=1$. Note that in the discussion so far we did not care about the normalization constants of the partition functions;
by a more precise discussion one can show that
the relations $S^2=-1, (ST)^3=1$ hold including the normalization of the partition functions \cite{Witten:2003ya}.

To summarize,
by pursuing the concepts of the topological symmetries of 3d field theories and their
gauging, we have arrived at the concept of
	\textbf{$SL(2, \bZ)$-transformations acting on 3d gauge theories}.
Let us emphasize here that this $SL(2, \bZ)$-transformation is \textbf{neither a symmetry nor a duality}
of a single theory: an $SL(2, \bZ)$-transformation maps one 3d theory to
another 3d theory, and hence one physical phenomenon to another physical phenomenon.

For us, who are used to studying the properties of a single field theory as in textbooks,
the meaning of such a concept of transformations may not yet be clear.
However, its meaning will become clear in the latter half of this book (e.g.\ in Chaps.~\ref{chap.wall} and \ref{chap.3mfd}).
To anticipate,
this $SL(2, \bZ)$-transformation (and its generalization $Sp(2n,\bZ)$ below) can in fact be interpreted, not as a symmetry or duality of a 3d theory, but as
a duality transformation of another theory (a 4d theory) (Sec.~\ref{subsec.4d_3d_SpZ}).
There we will also discuss how to generalize the $SL(2, \bZ)$-transformations to non-Abelian gauge groups.

\small
The $SL(2, \bZ)$-action on Abelian gauge theories was historically discussed also in the context of the fractional quantum Hall effect \cite{Burgess:2000kj}, and later also with motivations from the AdS/CFT correspondence \cite{Witten:2003ya}.
The case of non-Abelian gauge theories will be discussed (in the context of 3d $\mathcal{N}=2$ theories) in Sec.~\ref{subsec.TSUN}.

\normalsize
The discussion so far can be extended to the case where the Lagrangian has several global $U(1)$ symmetries, $U(1)^n$.
We omit the discussion in this case; the basic idea is the same, only with more kinds of gauge fields,
and the transformations generate the \keyword{modular symplectic group}{modular symplectic group} $Sp(2n, \mathbb{Z})$ (note that $Sp(2, \mathbb{Z})=SL(2, \mathbb{Z})$).

\section{IR Behavior}\label{sec_3d_beta}

The properties of 3d QFTs discussed above in this chapter originate from the fact that
three is an odd number of spacetime dimensions.
We can instead regard ``three'' as a number smaller than four,
and this has rather interesting consequences.

In fact, the effect is rather dramatic for gauge fields:
the behavior under the renormalization group flow is rather different
between three and four spacetime dimensions.

To see this, let us check the mass dimension of the gauge coupling constant $g^2$.
Recall that the mass dimension of the gauge field is the
same as that of the coordinate derivative, namely $1$,
and also that the kinetic term for the gauge field is
given by $-\frac{1}{2g^2}\int \textrm{Tr}F_{\mu\nu}
	F^{\mu\nu}$. This means that the mass dimension of the coupling constant $g$
in $D$ spacetime dimensions is given by
\begin{align}
	[g]=\frac{4-D}{2} \label{g_dim} \ .
\end{align}
This means that the renormalization group equation \eqref{eq.beta}
changes into (note that $\hat g:= g \mu^{(D-4)/2}$ is a dimensionless combination, and hence plays the same role as the coupling constant $g$ in
four spacetime dimensions)
\begin{align}
	\frac{d \hat g}{d\log \mu}=
	\frac{(D-4)}{2} \hat g -\frac{b_0}{16 \pi^2}\hat g^3+\cdots \ .
	\label{eq.beta2}
\end{align}
We therefore expect the following for $D<4$: when we start with weak coupling
$g\sim 0$ in the UV, the value of $g^2$ becomes larger and larger
in the long-wavelength limit.\footnote{
	For $D>4$ the RG flow is reversed and we have
	strong coupling in high-energy regions.
	In this case, since we follow the renormalization group backwards, it is in general not clear whether field theory is sufficient
	for discussing the strong coupling region. However,
	examples are known which suggest that the theory flows into a fixed point of the renormalization group
	before reaching the cutoff scale where the theory breaks down (see Ref.~\cite{Intriligator:1997pq}, and,
	relatedly, the discussion of the 6d $(2,0)$ theory in Sec.~\ref{subsec.compact_twist}).
	\label{foot_inverse_RG}
}
This is very different from the situation in four dimensions, where the low-energy behavior differs greatly depending on the sign of $b_0$ (the one-loop contribution), namely on the choice of the gauge group and
matter fields of the gauge theory.
In particular, in four dimensions a $U(1)$ gauge theory approaches a free theory at low energies, with the value of $g^2$ approaching $0$,
whereas in three dimensions it often happens that, even for $U(1)$ gauge theories,
the value of $g^2$ grows and the theory enters the strong coupling region.\footnote{
	Even for 3d theories, we can consider weakly coupled fixed points, for example by taking the number of flavors to be large.
	This is the 3d analogue of the Banks-Zaks-type fixed points mentioned in footnote~\ref{Banks_Zaks_footnote}.}\footnote{
	As can be seen from this, $D=4$ is a delicate, well-balanced dimension where the mass dimension of the gauge coupling
	is exactly $0$.
	This might be related to the fact that
	(the large part of) the spacetime we live in has four dimensions.
}
\textbf{In this book we often consider 3d gauge theories with
	Abelian gauge groups};
as explained above, even such theories have non-trivial dynamics in their low-energy regions.

\begin{practice}

	\item $[\bll]$ (Noether current for the $U(1)_J$ symmetry) Consider the Lagrangian \eqref{dualphotonL} for the
	dual photon of an Abelian $U(1)$ gauge field. Construct the Noether current for the
	shift symmetry of the dual photon. \label{ex.Noether}

	\item $[\bll]$ (Duality transformation of a free Abelian gauge field) In this chapter, we carried out a duality transformation
	of a free Abelian gauge field in three dimensions, and obtained a free periodic scalar.
	Can we perform a similar duality transformation in
	four spacetime dimensions? What about five dimensions? (cf.\ Sec.~\ref{subsec.4d_3d_SpZ})

	\item $[\bll]$ (Wilson-Fisher fixed point)
	Let us formally take the dimension $D$ to be $D=4-\epsilon$ ($\epsilon$ here is a small positive number)
	in the renormalization group equation \eqref{eq.beta2}.
	Show that there exists a fixed point of the renormalization group equation in the one-loop approximation,
	as long as $b_0$ has the appropriate sign.
	Show in addition that the approximation of neglecting
	two-loop and higher-order contributions is (generically) self-consistent
	when $b_0$ is of order $1$.

\end{practice}

\nextchaptermark{Crash Course on 3d $\mathcal{N}=2$ Theories}
\chapter{Crash Course on 3d Supersymmetric $\mathcal{N}=2$ Theories}
\label{chap.3dN2}

\begin{abstract}
	In this chapter we summarize basic material
	on 3d $\mathcal{N}=2$ supersymmetric field theories.
	Among the many topics covered in this chapter,
	the real mass parameters and the 2--3 mirror symmetry
	will be essential in subsequent chapters.
	We also supersymmetrize the concepts introduced in the previous chapter,
	such as the $Sp(2n, \mathbb{Z})$
	transformations.
\end{abstract}

\section{3d $\mathcal{N}=2$ Supersymmetry}

In the previous chapter we discussed 3d field theories and
their gauging.
In this chapter, we specialize to the case of 3d $\mathcal{N}=2$ supersymmetric
field theories, and see how the discussion of the previous chapter is reflected there.

Let us begin by briefly summarizing the basic material on 3d $\mathcal{N}=2$ supersymmetric field theories,
to the extent needed for the purposes of this book.
The original references are Refs.~\cite{Aharony:1997bx,deBoer:1997kr};
the former in particular contains a review of the basic material and is highly recommended for beginners.
The contents of this chapter are classical results already known in the
1990s; however, we especially highlight the aspects which will be important in subsequent chapters.

The details of the formulation of supersymmetric gauge theories depend very much on the
spacetime dimension as well as on the number of supersymmetries.
This is primarily because of the fermions---some fermions,
such as Weyl fermions and Majorana fermions, do not exist
in an arbitrary spacetime dimension, and moreover
their existence depends on the signature of the
spacetime metric.
This is in sharp contrast with scalar fields,
which exist in any spacetime dimension.

In this book, we consider 3d $\mathcal{N}=2$ supersymmetry.
This has four supercharges, and
can be understood as the
dimensional reduction of
4d $\mathcal{N}=1$ supersymmetry.
This means that a significant part of the theoretical formulation of 3d $\mathcal{N}=2$ supersymmetry is
shared with the more familiar case of 4d $\mathcal{N}=1$ supersymmetry.

In the following, let us briefly summarize the \keyword{superspace}{superspace} formalism of
3d $\mathcal{N}=2$ field theories.
While the superspace formalism is useful for some purposes,
in this book there are very few instances where
we need superpartners explicitly,
and we actually do not need much of the superspace formalism.
We explain the superspace formalism below
mainly because it will be of help to
readers who already have some familiarity with
4d $\mathcal{N}=1$ supersymmetric field theories.
Readers unfamiliar with the superspace formalism
should not dwell too much on it,
and can skip the parts they find unclear as appropriate.
While we will consider a Euclidean signature in the next chapter (discussing supersymmetric localization),
in this chapter we consider the familiar Lorentzian signature of spacetime.

\subsection{Supersymmetry Algebra}
\label{subsec.SUSY_algebra}

In 4d $\mathcal{N}=1$ supersymmetry we consider a
Majorana fermion with four real components.
Correspondingly, we have four supercharges,
which are denoted by
\begin{align*}
	\scQ_{\alpha}, \quad \overline{\scQ}_{\dot{\beta}} \ .
	\quad
	(\alpha=1,2, \quad \dot{\beta}=1,2)
\end{align*}

Here we distinguished the dotted and undotted indices
$\alpha$ and $\dot{\beta}$ in order to indicate that
the 4d Lorentz group is
$SO(3,1)\sim SL(2, \mathbb{R})\times SL(2, \mathbb{R})$\footnote{
	We wrote $\sim$ rather than $\simeq$ here since,
	strictly speaking, we need a Wick rotation: the correct statement is
	$SO(2,2)\simeq SL(2, \mathbb{R})\times SL(2, \mathbb{R})$.
	For $SO(2,2)$ the two spinors
	$(\bm{2}, \bm{1})$ and $(\bm{1}, \bm{2})$ can be taken to be real and are independent,
	whereas for the Wick-rotated $SO(3,1)$ the two spinors become complex and are complex conjugates of each other.
}
and that the two types of indices belong to different representations,
$(\bm{2}, \bm{1})$ and $(\bm{1}, \bm{2})$, respectively.

The situation is different in three dimensions.
Majorana fermions exist in three dimensions,
but they have only two real components, half of those in four dimensions.
A 4d Majorana fermion therefore decomposes into two 3d Majorana fermions.
This explains why the dimensional reduction\footnote{Reducing the number of dimensions by compactification (in this case, $S^1$-compactification) is called \keyword{dimensional reduction}{dimensional reduction}.}
of 4d $\mathcal{N}=1$ supersymmetry gives 3d $\mathcal{N}=2$ supersymmetry.

This can also be explained as follows. The 3d Lorentz group is $SO(2,1)\simeq SL(2,
	\mathbb{R})$, which is a subgroup of the 4d Lorentz group,
embedded diagonally into the two $SL(2, \mathbb{R})$ factors:\footnote{
	After Wick rotation, this embedding becomes $SO(3)\subset SO(4)\simeq
		SO(3)\times SO(3)$. If we take the generators of $SO(4)$ to be
	the generators $J_{\mu\nu}$ $(\mu, \nu=1, \ldots, 4)$ of rotations in the $\mu\nu$-plane
		(more concretely, $(J_{\mu\nu})_{a,b}=i(\delta_{\mu a}\delta_{\nu b}-\delta_{\mu
		b}\delta_{\nu a})$),
		then $SO(3)\times SO(3)$ is
		generated by $\{ J_{12}\pm J_{34}, J_{13} \pm J_{42}, J_{14} \pm J_{23} \}$,
		and we find that $\{J_{12}, J_{23}, J_{31}\}$, corresponding to
		rotations in the $123$-directions, are diagonally embedded into the two $SO(3)$'s.
}
\begin{align}
	SO(2,1) \simeq SL(2, \mathbb{R}) \subset SL(2, \mathbb{R})\times SL(2,
	\mathbb{R})\sim SO(3,1) \ .
\end{align}%
Consequently, there is no distinction between dotted and undotted indices for the supercharges,
and the supercharges are written as
\begin{align*}
	\scQ_{\alpha}, \quad \overline{\scQ}_{\beta} \ ,
	\quad
	(\alpha, \beta=1,2) \ .
\end{align*}
\nomenclature{$\scQ_{\alpha}, \overline{\scQ}_{\beta}$}{supercharges}
\nomenclature{$\alpha, \beta, \ldots$}{spinor indices}

The commutation relations satisfied by the supercharges
are obtained from the dimensional reduction of those
of 4d $\mathcal{N}=1$ supersymmetric theories:
\begin{align}
	\{ \scQ_{\alpha}, \scQ_{\beta} \}=\{ \overline{\scQ}_{\dot{\alpha}},
	\overline{\scQ}_{\dot{\beta}}\}
	=0, \quad \{\scQ_{\alpha}, \overline{\scQ}_{\dot{\beta}} \}=2
		                                                      (\sigma^{\hat{\mu}})_{\alpha \dot{\beta}} P_{\hat{\mu}} \ ,
	\label{4dQQ}
\end{align}
where $\hat{\mu}=0,1,2,3$.
We choose the 3d sigma matrices
$\sigma^{0,1,2}$ to be real:
\begin{align}
	\begin{split}
		 & \sigma^0:=
		\left(
		\begin{array}{cc}
			1 & 0 \\
			0 & 1
		\end{array}
		\right)
		\ ,\quad
		\sigma^1:=
		\left(
		\begin{array}{cc}
			0 & 1 \\
			1 & 0
		\end{array}
		\right)
		\ ,\quad
		\sigma^2:=
		\left(
		\begin{array}{cc}
			1 & 0  \\
			0 & -1
		\end{array}
		\right)
		\ ,                                                    \\
		 & (\sigma^3)_{\alpha \beta}:=i\epsilon_{\alpha \beta}
		=
		\left(
		\begin{array}{cc}
			0  & i \\
			-i & 0
		\end{array}
		\right) \ .
	\end{split}
	\label{sigma_matrix}
\end{align}
Note that this is different from the standard choice of Pauli matrices, where we have replaced
$(\sigma^2, \sigma^3) \leftrightarrow (\sigma^3, -\sigma^2)$.
\nomenclature{$\sigma^{\mu}$}{Pauli matrices}
We then have
\begin{align}
	\{ \scQ_{\alpha}, \scQ_{\beta} \}=\{ \overline{\scQ}_{\alpha},
	\overline{\scQ}_{\beta}\}
	=0 \ , \quad \{\scQ_{\alpha}, \overline{\scQ}_{\beta} \}=2 (\sigma^{\mu})_{\alpha
			                                                   \beta} P_{\mu}
	+i \epsilon_{\alpha \beta} Z  \ ,
	\label{3dQQ}
\end{align}
where $\mu=0,1,2$. Here $Z$ is the so-called
\keyword{central charge}{central charge}, which appears in
theories with extended supersymmetry; it corresponds to the dimensional reduction of the
4d momentum $P^3$.

The commutation relation \eqref{3dQQ}, with $P_{\mu}=-i \partial_{\mu}$,
is realized by the following differential operators:
\begin{align}
	\scQ_{\alpha}=\frac{\partial}{\partial \theta^{\alpha}}-i \sigma^{\mu}_{\alpha \beta}
	\overline{\theta}^{\beta}\partial_{\mu} \ , \quad
	\overline{\scQ}_{\alpha}=\frac{\partial}{\partial \overline{\theta}^{\alpha}}-i \theta^{\beta}
	\sigma^{\mu}_{\beta \alpha} \partial_{\mu} \ .
\end{align}
In order to consider \keyword[superfield]{superfields}{superfield} as irreducible representations,
it is useful to introduce the covariant derivatives
$D_{\alpha}, \overline{D}_{\alpha}$ in superspace:
\begin{align}
	D_{\alpha}=\frac{\partial}{\partial \theta^{\alpha}}+i \sigma^{\mu}_{\alpha \beta}
	\overline{\theta}^{\beta}\partial_{\mu} \ , \quad
	\overline{D}_{\alpha}=-\frac{\partial}{\partial \overline{\theta}^{\alpha}}-i \theta^{\beta}
	\sigma^{\mu}_{\beta \alpha} \partial_{\mu} \ .
	\label{eq.covariantD}
\end{align}
These anti-commute with the supercharges $\scQ_{\alpha}, \overline{\scQ}_{\alpha}$:
\begin{align}
	\{\scQ_{\alpha}, D_{\beta} \} =
	\{\scQ_{\alpha}, \overline{D}_{\beta} \} =
	\{\overline{\scQ}_{\alpha}, D_{\beta} \} =
	\{\overline{\scQ}_{\alpha}, \overline{D}_{\beta} \} =0 \ .
\end{align}

In this book, we choose the positions of spinor indices to be
the same as in 4d (except that there is no distinction between the
indices $\alpha$ and $\dot{\alpha}$).
For example, the indices are raised and lowered by the
epsilon tensors $\epsilon_{\alpha\beta}$ and
$\epsilon^{\alpha\beta}=-\epsilon_{\alpha\beta}$:\footnote{Care is needed when comparing with the
	literature, since in 3d the indices are sometimes raised and lowered by, e.g., $-\epsilon_{\alpha\beta}$.}
\begin{align}
	\theta^{\alpha}=\epsilon^{\alpha\beta} \theta_{\beta} \ , \quad
	\theta_{\alpha}=\epsilon_{\alpha\beta} \theta^{\beta} \ .
	\label{spinor_raise_lower}
\end{align}
We also contract the indices in the same manner as in 4d,
e.g.\
\begin{align}
	\theta \theta=\theta^{\alpha} \theta_{\alpha}\ , \quad
	\theta \sigma^{\mu} \theta=\theta^{\alpha} (\sigma^{\mu})_{\alpha\beta} \theta^{\beta} \ ,
	\quad
	\overline{\theta} \overline{\theta}=\overline{\theta}_{\alpha} \overline{\theta}^{\alpha}\ .
\end{align}

\bparagraph{Superfield Formalism}

A useful way to deal with representations of supersymmetry
is the superfield formalism.
We can obtain 3d $\mathcal{N}=2$ superfields
by the dimensional reduction of 4d $\mathcal{N}=1$
superfields.

A
\keyword{chiral multiplet}{chiral multiplet} $\Phi$ is a superfield satisfying the constraint $\overline{D}_{\alpha} \Phi=0$,
and consists of a complex scalar $\phi$, a complex two-component (3d Dirac) fermion $\psi$ (with its conjugate $\overline{\psi}$), and a complex scalar auxiliary field
$F$:\footnote{Let us check that bosons and fermions have the same number of degrees of freedom.
	Off-shell we have
	\begin{align*}
		\phi: 2\ , \quad  \psi: 2  \times 2 \ ,  \quad F: 2 \ ,
	\end{align*}
	and hence both bosons and fermions have
	$4$ degrees of freedom.
	On-shell we have
	\begin{align*}
		\phi:2\ ,  \quad \psi: 1 \times 2 \ , \quad F:0 \ ,
	\end{align*}
	and hence both have
	$2$ degrees of freedom.
}
\begin{align}
	\Phi:\,\, (\phi, \psi, \overline{\psi}, F)  \ .
\end{align}
Similarly, we can consider an
anti-chiral superfield $\overline{\Phi}$
satisfying the constraint $D_{\alpha} \overline{\Phi}=0$.

The constraint on a \keyword{chiral superfield}{chiral superfield} can easily be solved by using the variable
$y^{\mu}:=x^{\mu}+i \theta \sigma^{\mu} \overline{\theta}$,
which leads to the following expansion of the superfield in the
Grassmann variables $\theta, \overline{\theta}$:
\begin{align}
	\begin{split}
		\Phi(y, \theta) & =\phi(y)+\sqrt{2} \theta \psi(y)+(\theta\theta) F(y)                                 \\
		                & = \phi(x)+\sqrt{2}\theta \psi(x)+(\theta\theta) F(x)
		+i (\theta \sigma^{\mu} \overline{\theta}) \partial_{\mu} \phi(x)                                      \\
		                & \qquad \qquad -\frac{i}{\sqrt{2}} (\theta\theta) \partial_{\mu} \psi(x) \sigma^{\mu}
		\overline{\theta}
		+\frac{1}{4} (\theta\theta)(\overline{\theta}\overline{\theta} )\Box \phi(x) \ ,
	\end{split}
	\label{chiral_expand}
\end{align}
\nomenclature{$\Phi$}{chiral superfield (chiral multiplet)}
where we used identities such as
\begin{align}
	\theta_{\alpha} \theta_{\beta}=\frac{1}{2} \epsilon_{\alpha\beta} (\theta\theta) \ , \quad
	(\theta\sigma^\mu \overline{\theta})(\theta\sigma^\nu \overline{\theta})
	=-\frac{1}{2}(\theta\theta) (\overline{\theta}\overline{\theta}) \eta^{\mu\nu} \ .
\end{align}
These identities can be derived from the fact that $\theta$ is a Grassmann variable
and from the definition of $\sigma^{\mu}$.

\bigskip

Let us next consider a
\keyword{vector multiplet}{vector multiplet}.
This is a multiplet $V$ containing a gauge field
$A_{\mu}$, and is real (i.e.\ $V^{\dagger}=V$).
In addition to the gauge field, the multiplet contains a
real scalar $\sigma$, a gaugino $\lambda,
	\overline{\lambda}$, and a real scalar auxiliary field
$D$:\footnote{Let us again check that bosons and fermions have the same number of degrees of
	freedom. Off-shell we have
	\begin{align*}
		A_{\mu}:3-1=2 \ ,\quad \sigma: 1\ ,  \quad \lambda:2\times 2\ , \quad D:1 \ ,
	\end{align*}
	and hence both have four degrees of freedom.
	On-shell we have
	\begin{align*}
		A_{\mu}: 3-1-1=1 \ , \quad \sigma: 1 \ , \quad\lambda:1\times 2 \ ,  \quad D: 0 \ ,
	\end{align*}
	and hence both have $2$ degrees of freedom.
}
\begin{align}
	V: \, (A_{\mu} ,  \sigma, \lambda, \overline{\lambda}, D ) \ .
\end{align}
\nomenclature{$\sigma$}{vector multiplet scalar}

When we regard the 3d $\mathcal{N}=2$ vector multiplet as the dimensional reduction
of a 4d $\mathcal{N}=1$ vector multiplet, the
scalar field $\sigma$ can be thought of as the component of the 4d gauge field along the compactified direction
($A_3$), and hence is called the
\keyword{vector multiplet scalar}{vector multiplet scalar}.
As we will see later, this scalar parametrizes the Coulomb branch
of 3d $\mathcal{N}=2$ theories, and
is therefore sometimes called the \keyword{Coulomb branch scalar}{Coulomb branch scalar}.
This scalar field has no counterpart in 4d $\mathcal{N}=1$ theories, making
3d $\mathcal{N}=2$ theories richer in comparison,
and it will play central roles in the subsequent discussion of 3d field theories.
In fact, this scalar pairs up with the dual photon $\gamma$---the Nambu-Goldstone mode of the (spontaneously broken) topological $U(1)_J$ symmetry discussed in the previous chapter---into a chiral multiplet (see Sec.~\ref{subsec.monopole}).

The \keyword{vector superfield}{vector superfield} $V$
can be expanded in the Grassmann variables $\theta, \overline{\theta}$.
The vector superfield allows for gauge transformations by a chiral superfield
$\Lambda$:
\begin{align}
	e^V \to e^{-i \overline{\Lambda}} e^V e^{i \Lambda} \ ,
	\label{V_gauge}
\end{align}
and in the so-called Wess-Zumino gauge, which partially fixes this gauge freedom, the
expansion of the superfield reads
\begin{align}
	\begin{split}
		V & =  i\left(\theta \overline{\theta}\right) \sigma(x)
		-\left(\theta \sigma^{\mu} \overline{\theta}\right) A_{\mu}(x)
		+i (\theta\theta)(\overline{\theta} \overline{\lambda}(x))
		\\
		  & \qquad \qquad  -i (\overline{\theta}\overline{\theta}) (\theta \lambda(x))
		+\frac{1}{2 } (\theta\theta)(\overline{\theta}\overline{\theta}) D(x) \ .
	\end{split}
\end{align}
\nomenclature{$V$}{vector superfield (vector multiplet)}
The term $i(\theta\overline{\theta})\sigma(x)$
is specific to 3d, and is the dimensional reduction of the
4d term $\left(\theta \sigma^{3}
	\overline{\theta}\right) A_{3}(x)$.

\section{Lagrangian and Its Parameters}

\bparagraph{Kinetic Term}

In the superfield formalism, the Lagrangian can be written in terms of superfields.

Let us first discuss the kinetic terms.
The kinetic term of a chiral superfield
coupled with charge $q$ to an Abelian vector superfield
is the same as in 4d, and is given by
\begin{align}
	\scL_{\textrm{chiral kin}}=\int\! d^4\theta\,\, \Phi^{\dagger} e^{qV} \Phi \ .
	\label{eq.chiralkin}
\end{align}

Let us next consider the kinetic term for the vector superfield.
In 4d $\mathcal{N}=1$ theories,
we defined the superfield
$W_{\alpha}:=-\frac{1}{4}\overline{D}^2( e^{-V} D_{\alpha} e^{V})$,
which contains the field strength of the gauge field and
is invariant under the gauge transformation \eqref{V_gauge}
by the chiral superfield $\Lambda$.
We then obtain the kinetic terms for the gauge field and its superpartner (the gaugino)
as
\begin{align}
	\mathcal{L}_\textrm{vector kin}= \frac{1}{2g^2}
	\left[
		\int \! d^2  \theta \, \, \textrm{Tr}\, W^{\alpha} W_{\alpha}
		+
		\int \! d^2  \overline{\theta} \, \, \textrm{Tr}\, \overline{W}_{\alpha} \overline{W}^{\alpha}
		\right] \ .
	\label{Lkin1}
\end{align}

The expression \eqref{Lkin1} is still valid for
3d $\mathcal{N}=2$ theories. However,
for 3d $\mathcal{N}=2$ theories
there is another expression:
\begin{align}
	\mathcal{L}_\textrm{vector kin}=- \frac{1}{g^2}
	\int \! d^4  \theta \,\, \textrm{Tr}\, \Sigma^2  \ .
	\label{Lkin2}
\end{align}
We leave it as an exercise to verify that the two expressions
\eqref{Lkin1} and \eqref{Lkin2} give the same action
(exercise~\ref{ex.Lkin}).
Here we defined the superfield
$\Sigma$ by
\begin{align}
	\Sigma:=\frac{i}{2}\epsilon^{\alpha \beta} \overline{D}_{\alpha} \left( e^{-V} D_{\beta} e^{V}\right) \ .
	\label{SigmaV}
\end{align}
\nomenclature{$\Sigma$}{linear multiplet containing the current}
Note that it follows from the definition that $D^2\Sigma=\overline{D}^2 \Sigma=0$ and
$\overline{D}_{\alpha}\Sigma=-iW_{\alpha}$.
We can also verify that $\Sigma$ is invariant under the gauge transformation \eqref{V_gauge}.

The superfield $\Sigma$ is a representation of the supersymmetry algebra called a \keyword{linear multiplet}{linear multiplet},
and its components contain a conserved current.
Indeed, expanding the superfield $\Sigma$ in the Grassmann variables, we obtain (exercise~\ref{ex.Sigma_component})
\begin{align}
	\Sigma= \sigma+ \cdots -i \left(\overline{\theta}\sigma_{\mu} \theta
	\right) \frac{1}{2} \epsilon^{\mu\nu\rho }F_{\nu\rho} -i (\theta \overline{\theta}) D + \cdots \ .
	\label{Sigma_component}
\end{align}
The lowest component of the superfield
$\Sigma$ is the vector multiplet scalar $\sigma$.
Note also that one of its components is (up to the factor $2\pi$)
the current for the topological symmetry,
$\frac{1}{2} \epsilon^{\mu\nu \rho}F_{\nu\rho}=2\pi J^{\mu}$,
with $J^{\mu}$ defined in the previous chapter \eqref{JstarF}.

\bparagraph{Complex Mass}

Let us now go through the parameters of 3d $\scN=2$ theories one by one.
Of course, the gauge coupling constant is one of the parameters.

Next come the masses. In 3d $\mathcal{N}=2$ theories we can consider two types of
masses for chiral multiplets.

In 3d gauge theories we can consider a
\keyword{superpotential}{superpotential} $W(\Phi)$.
This is a holomorphic function of the chiral superfields $\Phi$, and
the corresponding Lagrangian is given by
\begin{align}
	\mathcal{L}_{W}= \int \! d^2\theta \, W(\Phi) + \textrm{(h.c.)} \ .
	\label{superW}
\end{align}
\nomenclature{$W$}{superpotential}
In particular, the coefficient $m_c$ of the quadratic term
$W=\frac{m_c}{2}\Phi^2$ in $W(\Phi)$
gives a mass to the chiral multiplet, and $m_c$ is called
the \keyword{complex mass}{complex mass}
(to distinguish it from the real mass discussed shortly).
The corresponding Lagrangian is
\begin{align}
	\mathcal{L}_{\textrm{complex mass}}=
	-|m_c|^2 |\phi |^2-\frac{m_c}{2} (\psi \psi)-\frac{m_c^*}{2}(\overline{\psi} \overline{\psi})  \ .
\end{align}
Terms of cubic and higher order in the superpotential give interactions (e.g.\ Yukawa couplings);
however, the values of these couplings are not reflected in the $S^3$ partition function
discussed in the next chapter
(except for whether they vanish or not).

What is important here is that the superpotential
$W(\Phi)$ is holomorphic.
\keyword[holomorphy]{Holomorphy}{holomorphy}
is a very important property of 3d $\mathcal{N}=2$ theories.
For example,
just as in the case of 4d $\mathcal{N}=1$ theories \cite{Seiberg:1993vc},
we can prove the non-renormalization theorem for 3d $\mathcal{N}=2$
theories conceptually, without relying on a detailed analysis of
Feynman diagrams.
Holomorphy has also been a crucial tool in the
analysis of the moduli space of
3d $\scN=2$ theories.
Moreover, holomorphy immediately implies that no phase transition occurs.
This is because even if there is a locus in the parameter space where something special happens,
it is a complex submanifold, whose real codimension is therefore at least $2$,
and hence it can always be avoided.

\bparagraph{Real Mass}

In 3d, in addition to complex masses, we can also consider
masses called \keyword{real masses}{real mass}.

Suppose that a chiral superfield $\Phi$ has a non-trivial charge $q$
under a global symmetry (i.e.\ it transforms non-trivially under the symmetry).
Let us denote by $V_{\rm bgd}$
the background vector superfield for the global symmetry.
The real mass $\mu$ for this global symmetry is then obtained
by giving an expectation value
$V_{\rm bgd}=i \mu \theta \overline{\theta}$
to the $\theta \overline{\theta}$-component of $V_{\rm bgd}$
in the kinetic term \eqref{eq.chiralkin} of the chiral multiplet,
and is represented by the following Lagrangian:
\begin{align}
	\mathcal{L}_{\textrm{real mass}} = \int \! d^4 \theta \, \Phi^{\dagger} e^{q V_{\rm
					                                                                        bgd}}
	\Phi
	=-(q\mu)^2 |\phi|^2+ i q \mu \epsilon^{\alpha \beta} \overline{\psi}_{\alpha}
	\psi_{\beta}+\cdots \ .
	\label{Lrealmass}
\end{align}
Note in particular that the fermionic part is $\epsilon^{\alpha \beta}\overline{\psi}_{\alpha}
	\psi_{\beta}$; the same term cannot be written in 4d, since there $\overline{\psi}$ and $\psi$
carry different types of indices.

The real mass will play a crucial role in this book.
Let us therefore add some supplementary comments.

First, as is clear from the definition in the superfield formalism,
the Lagrangian defining the real mass
takes the same form as the
kinetic term for the chiral superfield.
This implies that when we gauge the global symmetry
(i.e.\ promote the background gauge field for the global symmetry into a dynamical gauge field
for a local symmetry),
the former Lagrangian turns into the latter.
In this sense, the real mass $\mu$
and the vector multiplet scalar $\sigma$ can be regarded as essentially the same object.

The global symmetry under consideration has a conserved current,
from which we can construct a conserved charge. Since this conserved charge
commutes with all the generators of the supersymmetry algebra,
it appears as the central charge $Z$ in \eqref{3dQQ}.
In this sense the real mass deforms the supersymmetry algebra.

When we regard the 3d theory as a compactification from 4d, the real mass
can be thought of as specifying the \keyword{holonomy}{holonomy} of the Wilson line of the background gauge field
along the $S^1$-direction.
This again shows that the real mass is incompatible with
4d Lorentz symmetry, and is specific to 3d.
For non-Abelian gauge groups, it is sufficient to turn on the real masses only for the holonomies
in the Cartan subalgebra of the gauge group.
This is because the fundamental group of the
$S^1$ used for the compactification is Abelian,
so that by conjugation with a suitable element of the gauge group
we can always bring the holonomy into the Cartan subalgebra of the gauge group.

Finally, (as is clear from the name) the real mass
$\mu$ is a real parameter (recall that the vector superfield $V$, and hence $\mu$, is real),
and breaks the
holomorphy of 3d $\mathcal{N}=2$ theories.
This means that the holomorphy argument given above for the complex masses does not apply:
a non-trivial jump in physics can happen
on a real codimension-one locus in the parameter space, and the
physics can be discontinuous in the parameter space.
For this reason, the vacuum moduli spaces of 3d $\scN=2$ theories in general change in a rich and interesting manner
as we vary the real masses.

\bparagraph{R-Symmetry}\label{subsec.R_sym}

Among the global symmetries of 3d $\mathcal{N}=2$ theories,
those which do not commute with the supercharges
are called (for historical reasons) \keyword{R-symmetries}{R-symmetry}.
For 3d $\mathcal{N}=2$ supersymmetry, such a symmetry is
$SO(2)_R\simeq U(1)_R$, which rotates the two supersymmetries,
and acts on the Grassmann variables $\theta_{\alpha}, \overline{\theta}_{\alpha}$ as
\begin{align}
	U(1)_R: \theta_{\alpha} \to e^{i \xi}  \theta_{\alpha}\ , \quad
	\overline{\theta}_{\alpha} \to e^{-i \xi}  \overline{\theta}_{\alpha} \ .
\end{align}
With this convention, the superpotential \eqref{superW} has R-charge $2$.

Some care is needed in dealing with such symmetries. For example,
the fields $\phi, \psi, F$ in a chiral multiplet all have the same charge under ordinary global symmetries.
By contrast, for the R-symmetry, if $\phi$ has charge $r$, then the expansion \eqref{chiral_expand} shows that
$\psi$ and $F$ have charges $r-1$ and $r-2$, respectively.
The R-symmetry will play a privileged role when we discuss supersymmetry on curved spaces
in the next chapter.

\section{Chern-Simons Term}\label{subsec.CS_term_1}

In 3d we cannot consider a $\theta$-angle for the gauge field.
Instead, we can write down a Chern-Simons term for the gauge field.
We have already written down the (diagonal) Abelian Chern-Simons term \eqref{KCS}
in the previous chapter. Its version for non-Abelian gauge groups is
given by
\begin{align}
	\mathcal{L}_{\rm CS}= \frac{k}{4\pi} \textrm{Tr}\,\left( A\wedge d A+\frac{2}{3} A\wedge A\wedge A \right)\ .
	\label{LCS}
\end{align}
\nomenclature{$k$}{Chern-Simons level}\index{Chern-Simons level@Chern-Simons level}
Here in \eqref{LCS} the field
$A_{\mu}$ is a gauge field;
it can be either a dynamical gauge field corresponding to a
local symmetry, or a background gauge field corresponding to a
global symmetry.

This Chern-Simons term can be supersymmetrized.\index{Chern-Simons term@Chern-Simons term!supersymmetric@supersymmetric}
For non-Abelian gauge fields, the
resulting expression takes a somewhat complicated form
in the 3d $\mathcal{N}=2$ superfield formalism:
\begin{align}
	\mathcal{L}_\textrm{SUSY CS} & \!=\!  \frac{i k}{4\pi }
	\int \! d^4\theta \int_0^1\!\! dt \, \, \textrm{Tr} \left[ V\, \epsilon^{\alpha\beta} \overline{D}_{\alpha}\! \left( e^{-tV}\! D_{\beta}  e^{tV}\right)\right] \ .
\end{align}
In components, we have
\begin{align}
	\mathcal{L}_\textrm{SUSY CS} & =\frac{k}{4\pi}\textrm{Tr}\,
	\left(
	A\wedge dA+\frac{2}{3} A\wedge A\wedge A
	+\overline{\lambda} \lambda-2 D \sigma
	\right) \ .
	\label{SUSY_CS}
\end{align}

This simplifies for Abelian gauge groups.
When we have multiple Abelian gauge fields $A_i$,
the supersymmetric Chern-Simons term is given by
\begin{align}
	\label{SUSY_CS_Abelian}
	\mathcal{L}_\textrm{SUSY CS} & =\sum_{i,j} \frac{k_{i,j}}{4\pi }
	\int \! d^4\theta \, V_i \Sigma_j  \ ,
\end{align}
where $k_{i,j}$ is a symmetric matrix $k_{i,j}=k_{j,i}$ of Chern-Simons levels, satisfying the quantization condition $k_{i,j}\in \mathbb{Z}$.
This is a natural supersymmetrization of the fact that the Chern-Simons term is $A_i^{\mu} J_{j,\mu}$:
recall that the superfield $V_i$ ($\Sigma_i$) contains $A_i$ ($J_i$) as one of its components.
In components, we have
\begin{align}
	\mathcal{L}_\textrm{SUSY CS} & =\sum_{i,j}\frac{k_{i,j}}{4\pi}\,
	\left(
	A_i\wedge dA_j
	+\overline{\lambda}_i \lambda_j-2 D_i \sigma_j
	\right) \ .
	\label{SUSY_CS_Abelian_components}
\end{align}
Here the term $-2D\sigma$ will play an essential role in
the discussion of localization in the next chapter.

The $\lambda, \overline{\lambda}$ part and the $D\sigma$ part of \eqref{SUSY_CS} are quadratic in the fields,
and can be trivially integrated out as Gaussian integrals; in the end, only \eqref{LCS} for the gauge field
remains non-trivial. This is consistent with the fact that
the Chern-Simons term is topological, i.e.\ independent of the metric of the 3-manifold (see Chap.~\ref{chap.complexCS}),
and that this property persists after supersymmetrization.

This is not the case, however, when the Chern-Simons gauge field couples to matter fields.
In the theories considered below, the gauge fields appearing in the Chern-Simons terms
can have Yang-Mills kinetic terms, and moreover couple through covariant derivatives
to matter fields such as scalars and fermions; such theories are not topological
(for reference, such theories have been studied in detail in recent years,
for example in connection with the physics of M2-branes).

The presence of the Chern-Simons term has dramatic effects on
various properties of the theory, e.g.\ correlation functions and critical exponents.
Indeed, in three dimensions the Chern-Simons term
has a lower mass dimension than the standard Yang-Mills kinetic term.
This means that for a theory with a non-vanishing Chern-Simons term,
the effect of the Chern-Simons term dominates over that of the Yang-Mills kinetic term
in the deep IR region.

\bparagraph{$Sp(2n, \mathbb{Z})$-Action}

Now that the Chern-Simons term has been supersymmetrized,
it is straightforward to supersymmetrize (for Abelian gauge groups) the \keyword{$Sp(2n, \mathbb{Z})$-action}{Sp(2n,Z)-action}
of the previous chapter for 3d $\scN=2$ theories.
Since the Lagrangian for the Chern-Simons term in supersymmetric gauge theories is given by \eqref{SUSY_CS_Abelian},
the supersymmetric versions of the $S$- and $T$-transformations are given by
\begin{align}\begin{split}
		S: & \, \mathcal{L}[V] \to
		\mathcal{L}'[\tilde{V}]:=\mathcal{L}[V]+
		\frac{1}{2\pi } \int d^4\theta \,
		\tilde{V} \Sigma
		\ ,                        \\
		T: & \, \mathcal{L}[V] \to
		\mathcal{L}'[V]:=\mathcal{L}[V]+\frac{1}{4\pi} \int d^4\theta \,
		V \Sigma  \ .
		\label{STLagSUSY}
	\end{split}\end{align}
As in the previous chapter, one can directly verify from these definitions that
$S$ and $T$ satisfy the relations \eqref{STrel} of $SL(2, \mathbb{Z})$.
We will comment on the $SL(2, \bZ)$-action for non-Abelian gauge theories in Sec.~\ref{subsec.TSUN}.

\bparagraph{FI Parameter}

For a $U(1)$ gauge group, we can consider
the corresponding
\keyword{Fayet-Iliopoulos (FI) parameter}{Fayet-Iliopoulos parameter}. The corresponding Lagrangian
can be written, in terms of the background vector multiplet $V_{\rm bgd}=i\zeta \theta \overline{\theta}$,
as
\begin{align}
	\begin{split}
		\mathcal{L}_{\rm FI} & =
		\frac{1}{2\pi }\int \! d^4\theta \, V_{\rm bgd} \Sigma
		= \frac{i\zeta}{2\pi} \int \! d\theta d\overline{\theta} \, \Sigma                                  \\
		                     & = \frac{\zeta}{2\pi} \int \! d^4\theta \, V +\textrm{(total derivative)} \ .
	\end{split}
\end{align}
\nomenclature{$\zeta$}{FI parameter}
This means that
\textbf{the FI parameter is nothing but the real mass parameter for the
	$U(1)_J$ global symmetry, which shifts the dual photon
	of the $U(1)$ gauge field}.

The normalization of the FI parameter here is not the standard one in four dimensions;
however, it turns out to be convenient in 3d, for example in the discussion of
3d mirror symmetry.

\bparagraph{Parity Anomaly}\label{subsec.parity_anomaly}

In the discussion so far, the level of the Chern-Simons term has been a parameter,
and it seems that we can simply set it to zero. However,
there are cases where $k\ne 0$ is required
for the consistency of the theory. Moreover,
even if we start with a theory with $k=0$,
a Chern-Simons term can be generated
as we go to low energies. This is the
quantum generation of the Chern-Simons term in three dimensions, namely the
\keyword{parity anomaly}{parity anomaly}
\cite{Redlich:1983dv,Redlich:1983kn}
(see also the exposition in Ref.~\cite{Dunne:1998qy}).

Let us consider a (two-component) Dirac fermion $\psi$
with mass $\mu$, and let us assume that the fermion
couples to a background gauge field $A_{\nu}$:
$\scL=-\overline{\psi}\left(\slashed{\partial}-i\slashed A+\mu\right) \psi$,
written here in the mostly-plus signature used throughout this book,
so that the Dirac equation gives the mass shell $p^2=-\mu^2$.
Let us consider the limit of large $\mu$.
Naively, the degrees of freedom of $\psi$ then decouple
from the low-energy physics,
and it seems that nothing is left behind from $\psi$.

It turns out, however, that this conclusion is incorrect.
In three dimensions the fermion mass term breaks parity (exercise~\ref{mass_parity}),
and correspondingly the remaining theory acquires, as a one-loop correction,
a Chern-Simons term with level
\begin{align}
	k_{\rm eff}=\frac{\textrm{sgn}(\mu)}{2} \ .
	\label{parity_anomaly}
\end{align}
Note that the Chern-Simons term \eqref{LCS} breaks parity, since its definition
contains $\epsilon^{\nu\rho\sigma}$.
This is the parity anomaly.
Since the level of the Chern-Simons term has to be an integer,
for the consistency of the theory with respect to the gauge symmetry
we need to add, when necessary, a Chern-Simons term with a half-integer level
to the Lagrangian from the outset.

More generally, when the fermion has charge $q$ under the background gauge field,
the level of the effective Chern-Simons term obtained from its real mass $\mu$ is
\begin{align}
	k_{\rm eff}=q^2 \frac{\textrm{sgn}(\mu)}{2} \ .
	\label{parity_anomaly_2}
\end{align}
When there are multiple global symmetries,
off-diagonal Chern-Simons terms are also generated in general (see \eqref{keff}
in the exercises).

\small

The derivation of the one-loop correction to the Chern-Simons term (the parity anomaly)
by an explicit evaluation of Feynman diagrams will be discussed in
exercise~\ref{parity_derivation}. Here let us briefly comment on
an alternative derivation.

In the path-integral formalism, the fermion integral gives the
determinant of the Dirac operator, $\textrm{det}\, i \sigma^{\mu} (\partial_{\mu} + A_{\mu})$.
The potential problem is that this determinant might change under gauge transformations.

To see what kind of change could arise,
let us consider the 4d gamma matrices $\gamma_{\mu}:=\left(
	\begin{array}{cc} \sigma_{\mu} & 0             \\
	             0                 & -\sigma_{\mu}
	\end{array}
	\right)$. These contain two copies of the 3d ones,
and hence our 3d determinant $\textrm{det}\, i \sigma^{\mu} (\partial_{\mu} + A_{\mu})$ can be identified with the
square root of the 4d determinant,
$\textrm{det}^{\frac{1}{2}} i \gamma^{\mu} (\partial_{\mu} + A_{\mu})$.
The advantage of going to four dimensions is that there exists $\gamma^{5}$, which
anti-commutes with all the gamma matrices $\gamma_{0, 1, 2,3}$; hence each eigenvalue
always pairs up with another eigenvalue of the
same absolute value and the opposite sign. We can then define the square root of the determinant
by collecting the positive eigenvalues only.

The subtlety is that, while the determinant itself is invariant under gauge transformations,
the square root can change its sign. To analyze this, consider
a one-parameter family of gauge fields continuously connecting
the two 3d gauge field configurations related by the gauge transformation.
Regarding the direction of this deformation (i.e.\ the parameter) as the time direction,
we obtain a 4d gauge field.
Whether or not the sign of the square root changes can then be translated into
the problem of determining the parity (even or odd) of the number of zero eigenvalues
of the 4d Dirac operator.
By the index theorem, this is related to the 4d instanton number (mathematically, $\pi_3(G)$)
(see exercise~\ref{ex.instanton}(b) of Chap.~\ref{chap.complexCS}). This anomaly can be canceled by a 3d Chern-Simons term
(see Sec.~\ref{subsec.CS_term_2}).

This derivation of the parity anomaly is essentially the same
as the derivation of the $SU(2)$ anomaly in four dimensions \cite{Witten:1982fp}, except for the difference in spacetime dimensions.
A similar argument applies in 11 dimensions \cite{Witten:1996md}.

\normalsize

\section{Vacuum Moduli Space}

We next turn to the
vacuum moduli space of 3d $\mathcal{N}=2$ theories.
In general, one of the
characteristic features of supersymmetric field theories
is that the lowest-energy states are not isolated,
but rather form a non-trivial moduli space.
This moduli space is called
the
\keyword{vacuum moduli space}{vacuum moduli space}.
For 3d $\mathcal{N}=2$ theories,
the moduli space is a K\"{a}hler manifold (in general with singularities),\footnote{
	We do not need the details of the mathematical definition of K\"{a}hler manifolds in this book.
	What is important here is the weaker property that it is a complex manifold.
} and is also an interesting object mathematically. This is one manifestation of the
holomorphy of 3d $\mathcal{N}=2$ theories.

In order to identify the vacua
of the classical theory, we simply need to write down the potential from the Lagrangian
and set it to zero.
The full form of the potential can be obtained by expanding the Lagrangian discussed so far.
The important term here is the following contribution to the potential, which arises from the expansion
of the kinetic term for the chiral multiplet \eqref{eq.chiralkin}:
\begin{align}
	\int\! d^3 x\,  (q^a_i \sigma_a) ^2 |\phi_i|^2 \ ,
	\label{sigma2phi2}
\end{align}
where we assumed that the chiral multiplet $\phi_i$ has charge $q^a_i$
under the $U(1)_a$ symmetry in the Cartan of the gauge group.
Recall that the vector multiplet scalar $\sigma$ becomes a real mass parameter when
the vector multiplet is ungauged; the term \eqref{sigma2phi2} then corresponds to the mass term \eqref{Lrealmass}.

Since the term \eqref{sigma2phi2} is always non-negative,
minimization of the energy
requires either $\sigma_a$ or $\phi_i$ to vanish.
This means that the vacuum moduli space is divided into three types of
\keyword{branches}{branch}: the \keyword{Coulomb branch}{Coulomb branch},
the \keyword{Higgs branch}{Higgs branch}, and the \keyword{mixed branch}{mixed branch} in between:
\begin{align}
	\begin{split}
		 & \textrm{Coulomb branch:} \,  \sigma_a\ne 0 \ , \quad \textrm{all}\, \phi^i=0 \ ,                \\
		 & \textrm{mixed branch:}\, \textrm{some}\, \sigma_a\ne 0\ , \quad\textrm{some}\,  \phi^i\ne 0 \ , \\
		 & \textrm{Higgs branch:} \, \textrm{all}\, \sigma_a=0\ , \quad \phi^i\ne 0 \ .                    \\
	\end{split}
\end{align}
The names originate from the fact that
in the Coulomb branch the Abelian gauge group remains unbroken
and the Coulomb interaction of its gauge fields survives,
whereas in the Higgs branch the gauge symmetry is completely broken (up to discrete parts)
by the expectation values of the scalar fields.

\small

In this book we also consider theories with (effective) Chern-Simons terms.
In this case all the matter fields are massive, and only the effect of the
Chern-Simons term for the gauge field (which is also a mass term for the gauge field)
remains at low energies, giving rise to a gapped vacuum.
This is sometimes called the \keyword{topological branch}{topological branch}.

\normalsize

The moduli spaces of 3d $\mathcal{N}=2$ theories have a very rich and complicated structure.
First, each of the three types of branches above can have multiple components,
which are connected with each other in intricate ways.
Moreover, once we take the parameters (complex masses, real masses, FI parameters) into account,
the moduli space is deformed depending on their values: for example,
when we turn on a complex mass the chiral superfield becomes massive,
and the corresponding Higgs branch disappears. The same happens when
the gauge field becomes massive due to a Chern-Simons term.

Furthermore, while the discussion above was classical, in practice we need to take
quantum corrections into account. In particular,
for $\mathcal{N}=2$ theories with non-Abelian gauge groups,
the Coulomb branch can be lifted by quantum corrections \cite{Affleck:1982as};
moreover, the Coulomb branch and the Higgs branch can be smoothly connected
into a single branch \cite{Aharony:1997bx} (see also Ref.~\cite{Hashimoto:2014nwa} for a recent discussion).

We will discuss one of the simplest (but still rather non-trivial) examples of such quantum corrections below,
but before that, let us introduce the notion of the monopole operator as a preparation.

\bparagraph{Monopole Operator}\label{subsec.monopole}

The discussion so far is incomplete in one respect.
We mentioned before that the vector multiplet scalar $\sigma$
is the coordinate of the Coulomb branch.
However, due to the constraints from holomorphy (as follows from 3d $\mathcal{N}=2$ supersymmetry),
the Coulomb branch is a K\"{a}hler manifold, and in particular a complex manifold.
This means that its coordinate should be complex.
If $\sigma$ is the real part, what is the imaginary part?

In the Coulomb branch the gauge group is broken to its Abelian subgroup,
and we have a field $\sigma$ for each $U(1)$ factor.
This means that the imaginary part should also
be a scalar field determined from the $U(1)$ gauge field.
We have already encountered such a scalar field in the previous chapter:
the dual photon $\gamma$ of the $U(1)$ gauge field.
The duality transformation in the previous chapter dealt only with the gauge field,
and transformed it into the dual photon.
We can supersymmetrize this duality transformation
by using the superfield formalism:
a 3d $\mathcal{N}=2$ vector multiplet of a supersymmetric Abelian gauge theory
is transformed into an $\mathcal{N}=2$
chiral multiplet containing $\gamma$ as one of its components (exercise~\ref{ex.superdual}).

Since the dual photon
$\gamma$ has periodicity $g^2$,
it is natural to remove the periodicity by exponentiating the combination $\sigma+i\gamma$.
The resulting operator is the so-called
\keyword{monopole operator}{monopole operator}\index{monopole operator@monopole operator!Vpm@$\mathcal{V}_{\pm}$}:\footnote{
	In the literature the monopole operator is often denoted by $V$; we nevertheless
	denote it by $\mathcal{V}$, to distinguish it from the vector superfield $V$.}
\begin{align}
	\mathcal{V}\sim \exp\left( \frac{2\pi }{g^2}(i\gamma+\sigma)\right) \ .
	\label{monopoleop}
\end{align}
\nomenclature{$\mathcal{V}$}{monopole operator}
We have thus found that \textbf{the monopole operator parametrizes the Coulomb branch}.
Note, however, that this expression is valid for small $g^2$; in the IR
it receives quantum corrections, as we will see in an example later
(see Fig.~\ref{fig.quantumCoulomb}).

As is clear from the definition, the monopole operator $\scV$
is charged under the $U(1)_J$ symmetry (the shift $\delta\left(\frac{\gamma}{g^2} \right)=\frac{\vartheta}{2\pi}$ of $\gamma$),
and transforms by a phase:
\begin{align}
	U(1)_J: \, \scV\to e^{i\vartheta } \scV \ .
	\label{U1J}
\end{align}

When we insert a monopole operator at a point,
we can realize the situation of Fig.~\ref{fig.vortex}
with the roles of the electric field and the $U(1)_J$ current exchanged. In Fig.~\ref{fig.vortex},
a particle charged under the original gauge field creates an electric field around it,
and hence a vortex for the $U(1)_J$ symmetry;
now, a monopole operator
carrying $U(1)_J$ charge creates a vortex for the gauge field around it
(there is, however, also a difference:
corresponding to the fact that in 3d the dual of a gauge field is a scalar field,
in Fig.~\ref{fig.vortex} we considered a particle extended in the time direction,
whereas the monopole operator here is inserted at a point in spacetime, i.e.\ is not extended in
the time direction).
By appropriately supersymmetrizing this setup,
we can obtain a definition of the monopole operator (which is valid also at low energies);
see e.g.\ Ref.~\cite{Intriligator:2013lca}.

The definition of the monopole operator requires a field which is not directly contained in the
original Lagrangian.
Such an operator is called a \keyword{disorder operator}{disorder operator}.
The discussion in Chap.~\ref{chap.3dglue} suggests the existence of a duality transformation
mapping this operator into an operator written in terms of the ordinary fields in the Lagrangian.
Let us see below a concrete example of such a duality.

\section{2--3 Mirror Symmetry}

To conclude this chapter, we
discuss an example of the vacuum moduli space and the quantum corrections to it.
In particular, let us verify the 2--3 mirror symmetry
mentioned in Sec.~\ref{subsec.3d_mirror_intro}
at the level of the vacuum moduli space.

Here we consider 3d $\mathcal{N}=2$ supersymmetric QED (SQED).
Let us begin with a classical analysis.
In the Higgs branch the scalar fields $q, \overline{q}$ have VEVs;
however, these VEVs themselves are not gauge invariant, since the fields transform non-trivially
under the gauge symmetry. The
gauge-invariant quantity is the VEV of their combination,
the meson $M:=q \overline{q}$.
This is a single complex number, whose phase rotates
under the axial rotation $U(1)_A$ (under which both
$q$ and $\overline{q}$ have charge $1$).

Let us next consider the Coulomb branch.
As explained already, the classical Coulomb branch is parametrized by the
monopole operator $\scV$ \eqref{monopoleop},
and hence, since $\gamma$ is periodic, it is given by a cylinder $\mathbb{R}\times S^1_{\gamma}$---$\sigma$ parametrizes the
$\mathbb{R}$-direction, and $\gamma$ the $S^1_{\gamma}$-direction.
In particular, the $U(1)_J$-rotation \eqref{U1J} is the rotation along $S^1_{\gamma}$.

The vacuum moduli space should be the union of these two branches.
There is one problem here, however.
The two branches should intersect at the single point $\sigma=\phi=0$;
however, this cannot be the case, since
the $U(1)_J$ symmetry acts trivially on the Higgs branch,
while it acts on the Coulomb branch without a fixed point.
This is a contradiction.

\begin{figure}[t]
	\centering{\includegraphics[scale=0.35]{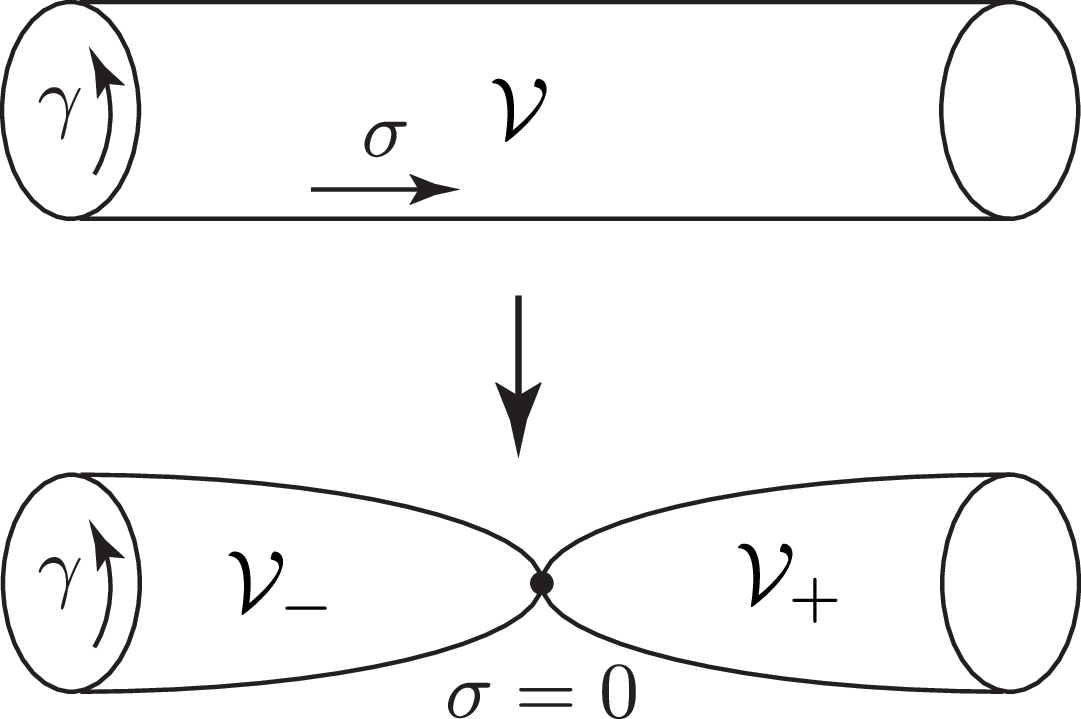}}
	\caption{The classical Coulomb branch $\mathbb{R}^1\times S^1$
		splits into two branches by quantum corrections, which are connected
		at the point $\sigma=0$.}
	\label{fig.quantumCoulomb}
\end{figure}

This contradiction can be resolved by taking quantum corrections
(i.e.\ finite-$g$ effects) into account.
Since the gauge coupling constant $g$ appears explicitly in the
definition of the monopole operator $\scV$ \eqref{monopoleop},
it is natural to expect that it is the Coulomb branch which receives quantum corrections.
Suppose that the quantum corrections split the Coulomb branch into two branches
at the point $\sigma=0$.
The point $\sigma=0$ is then a fixed point of the $U(1)_J$-action,
and the Coulomb branch can be connected with the Higgs branch without running into
a contradiction (Fig.~\ref{fig.quantumCoulomb}).
The one-loop quantum correction to the metric on the Coulomb branch can be computed,
and it supports this picture.

The Coulomb branch is then described by two monopole operators
\begin{align}
	\scV_+\sim e^{\frac{2\pi}{g^2}(\sigma+i \gamma)} \ , \quad
	\scV_{-}\sim e^{-\frac{2\pi}{g^2}(\sigma+i \gamma)} \ ,
\end{align}
where we wrote $\sim$ because of the quantum corrections.

\begin{figure}[t]
	\centering{\includegraphics[scale=0.35]{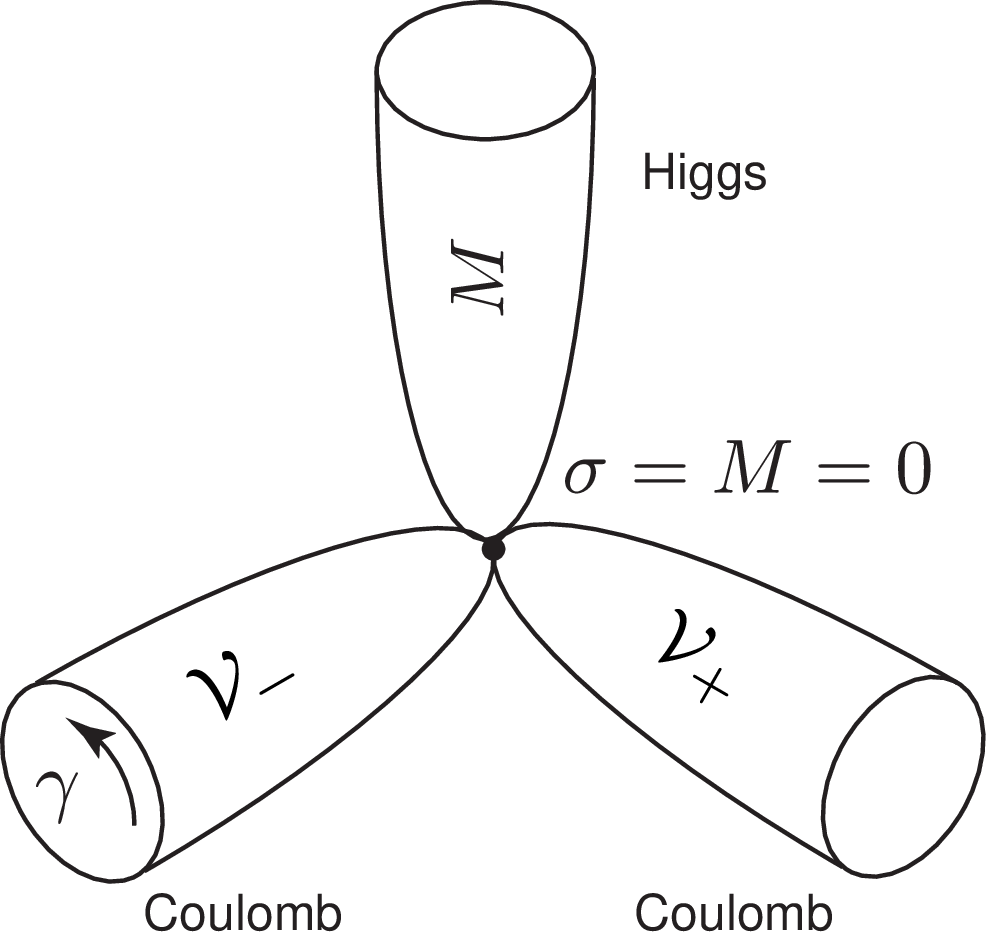}}
	\caption{The vacuum moduli space of 3d $\mathcal{N}=2$ SQED, with quantum corrections taken into account.
		Three branches, two Coulomb branches and one Higgs branch,
		meet at a single point, where we have conformal symmetry.}
	\label{fig.quantumCoulombHiggs}
\end{figure}

Combining the quantum-corrected Coulomb branch with the Higgs branch,
we obtain Fig.~\ref{fig.quantumCoulombHiggs}.
The interesting point to note from the figure is that the
three branches look symmetric.
In fact, the metric on the Coulomb branch, after the one-loop quantum correction is taken into account,
is given by \cite{deBoer:1997kr}\footnote{For Abelian gauge groups
	the one-loop correction is claimed to be exact, even including non-perturbative effects \cite{deBoer:1997kr}.}
\begin{align}
	ds^2=\frac{1}{4} \left(\frac{1}{g^2}+\frac{1}{|\sigma|} \right) d\sigma^2
	+
	\left(\frac{1}{g^2}+\frac{1}{|\sigma|} \right)^{-1} d\gamma^2  \ .
\end{align}
Taking the low-energy limit $g\to \infty$
and performing an appropriate change of variables,
we can verify that this coincides with the metric on the Higgs branch, namely the
flat metric on $\bC$.

This strongly supports the duality with the $XYZ$ model
introduced in Chap.~\ref{chap.intro}.
Recall that the $XYZ$ model has the superpotential
$W=XYZ$ for three chiral superfields $X, Y, Z$, leading to the scalar potential
\begin{align}
	V=\left|\frac{\partial W}{\partial X}\right|^2
	+\left|\frac{\partial W}{\partial Y}\right|^2
	+\left|\frac{\partial W}{\partial Z}\right|^2
	=|YZ|^2+|ZX|^2+|XY|^2 \ .
\end{align}
The vacuum moduli space therefore has three branches:
\begin{align}
	\begin{split}
		 & \left\{ X\ne 0, \quad Y=Z=0 \right\}, \qquad
		\left\{ Y\ne 0, \quad Z=X=0  \right\},          \\
		 & \left\{ Z\ne 0, \quad X=Y=0 \right\}  \ .
	\end{split}
\end{align}
These can be identified with the three branches
of the vacuum moduli space of $N_f=1$ SQED,
with the identification
\begin{align}
	M\leftrightarrow X\ , \quad
	\scV_{+}\leftrightarrow Y \ , \quad
	\scV_{-} \leftrightarrow Z \ .
\end{align}

\small

It is often said that 3d mirror symmetry exchanges
the Coulomb branch and the Higgs branch.
This is correct for 3d $\mathcal{N}=4$ theories,
but not for the 3d $\mathcal{N}=2$ theories
discussed here. Indeed, in the 2--3 mirror symmetry discussed here,
the $XYZ$ model has only Higgs branches (the theory simply has no gauge field),
one of which corresponds to the Higgs branch of 3d $N_f=1$ SQED.
More generally, in $\mathcal{N}=2$ supersymmetric gauge theories with
non-Abelian gauge groups,
we cannot even clearly distinguish between the Coulomb
branch and the Higgs branch, once quantum corrections are taken into account.

\normalsize

\bigskip

Let us summarize the correspondence so far.
The fields and symmetries of
3d $\mathcal{N}=2$ SQED
are summarized in Table~\ref{tab.SQED}.
Most of the entries of this table follow from the discussion so far.

The exceptions are the $U(1)_A$ and $U(1)_R$ charges of
the monopole operators $\scV_{\pm }$.
These arise from the parity anomaly explained above.
When we give masses to the quarks $q, \overline{q}$
and consider the Coulomb branch,
the parity anomaly \eqref{parity_anomaly_2} implies the existence of the
Chern-Simons term
\begin{align}
	\frac{k^{\rm eff}_{A/R, g} }{2\pi} \int A_{U(1)_{A/R}}\wedge dA_{U(1)_{\rm gauge}}
	= k^{\rm eff}_{A/R, g}  \int d^3x\, A_{\mu, U(1)_{A/R}} J^{\mu}_{U(1)_J} \ .
\end{align}
This shifts the $U(1)_{A/R}$ current $J^{\mu}_{A/R}$ by $k^{\rm eff}_{A/R, g} J^{\mu}_{U(1)_J}$,
since the current appears in the Lagrangian as
\begin{align}
	\int A_{\mu, U(1)_{A/R}} J^{\mu}_{U(1)_{A/R}}
\end{align}
(see \eqref{JA}).
This means that the $U(1)_A$ and $U(1)_R$ symmetries
mix with the $U(1)_J$ symmetry,
which determines the $U(1)_A$ and $U(1)_R$ charges of $\scV_{\pm }$\footnote{Be careful about the
	normalization of the $U(1)_J$ generator when reading off the charges; here the generator is normalized to have period $2\pi$.}
(exercise~\ref{ex.monopole_charge}).

\begin{table}[t]
	\centering
	\caption{The fields of $N_f=1$ SQED and their symmetry charges.}
	\begin{tabular}{c|c|c|c|c|c}
		                   & $q$ & $\overline{q}$ & $M$  & $\scV_+$ & $\scV_-$ \\
		\hline
		\hline
		$U(1)_{\rm gauge}$ & $1$ & $-1$           & $0$  & $0$      & $0$      \\
		\hline
		$U(1)_A$           & $1$ & $1$            & $2$  & $-1$     & $-1$     \\
		\hline
		$U(1)_J$           & $0$ & $0$            & $0$  & $1$      & $-1$     \\
		\hline
		$U(1)_R$           & $0$ & $0$            & $0$  & $1$      & $1$      \\
	\end{tabular}
	\label{tab.SQED}
\end{table}

Keeping only the gauge-invariant quantities $M, \scV_+, \scV_-$
in Table~\ref{tab.SQED} and writing down their charges, we obtain Table~\ref{tab.charge2}.
One can easily recognize these
as nothing but the global symmetries of the $XYZ$ model
preserving the superpotential
$W=XYZ=M \scV_+ \scV_-$.

\begin{table}[t]
	\centering
	\caption{The fields of the $XYZ$ model and their symmetry charges.}
	\begin{tabular}{c|c|c|c}
		         & $M=X$ & $\scV_+=Y$ & $\scV_-=Z$ \\
		\hline
		\hline
		$U(1)_A$ & $2$   & $-1$       & $-1$       \\
		\hline
		$U(1)_J$ & $0$   & $1$        & $-1$       \\
		\hline
		$U(1)_R$ & $0$   & $1$        & $1$        \\
	\end{tabular}
	\label{tab.charge2}
\end{table}

Once the global symmetries are identified,
we also obtain a natural identification of the corresponding real mass parameters (FI parameters).
Note that SQED has one extra parameter, namely the gauge coupling constant $g$.
However, this parameter disappears ($g\to \infty$) when we flow to the IR fixed point under the
RG flow; hence the parameter counting matches.

The mass parameters are relevant deformations of this fixed point,
and change the vacuum moduli space.
For example, when we turn on the real mass parameter for the $U(1)_A$
symmetry, the meson $M$
becomes massive and the Higgs branch is lifted.
Similarly, the two Coulomb branches are lifted when we turn on
the real mass parameter for the $U(1)_J$ symmetry.

In this section we discussed a check of the
2--3 mirror symmetry in terms of the
vacuum moduli space.
Other non-trivial checks of the duality include the
match of the Witten indices (exercise~\ref{Witten_index}),
as well as the match of the parity anomaly, which is a $\bZ_2$-anomaly
(exercise~\ref{parity_anomaly_matching}).
In the next chapter, we will present yet another check
using the $S^3$ partition function---the $S^3$ partition function
is a powerful tool for analyzing 3d $\scN=2$ theories, and
will be an indispensable companion in our journey through the rest of this book.

\begin{practice}

	\item
	$[\bll \bll]$ (For supersymmetry aficionados)
	Take your favorite textbook on 4d $\mathcal{N}=1$ global supersymmetry,
	for example the textbook by Wess and Bagger \cite{Wess:1992cp},
	and rewrite its description of 4d $\scN=1$ supersymmetric gauge theories (such as the superspace formalism)
	for the case of 3d $\scN=2$ theories.
	The discussion of this chapter will be of some help.
	Which parts change qualitatively? Are there new ingredients
	specific to 3d $\scN=2$ theories, which have no counterparts in 4d $\scN=1$ theories?

	\item
	$[\bll]$ (Superfield in components) Expand the superfield into components, and verify \eqref{Sigma_component}.
	\label{ex.Sigma_component}

	\item
	$[\bll]$ (3d $\mathcal{N}=4$ theories) Let us consider 3d $\scN=4$
	theories, which have twice as much supersymmetry as the theories in the main text.
	Such a theory (with global supersymmetry) has $\scN=4$ vector multiplets
	containing gauge fields, and $\scN=4$ hypermultiplets
	containing only scalars and fermions.
	Any 3d $\scN=4$ theory can be regarded as a 3d $\scN=2$ theory.
	Work out the decomposition of each of the 3d $\scN=4$ multiplets
	into 3d $\scN=2$ multiplets.
	\label{N4ex}


	\item
	$[\bll]$ (Two expressions for the gauge kinetic term)
	Verify the equivalence of the two expressions \eqref{Lkin1} and \eqref{Lkin2} for the
	kinetic term of an Abelian gauge field. Hint: use \eqref{eq.covariantD},
	and verify that the difference between the two Lagrangian densities is a total derivative.
	\label{ex.Lkin}

	\item
	$[\bll \bll]$ (Duality transformation in supersymmetric theories) In Sec.~\ref{sec.3d_dualize}
	we have seen that in three dimensions a $U(1)$ gauge field is dual to a
	periodic compact scalar. Supersymmetrize this fact (to 3d $\scN=2$ supersymmetry),
	and show that a 3d $U(1)$ vector multiplet is dual to a
	chiral multiplet.

	\small

	A similar argument works for 2d $\scN=(2,2)$ theories
	(this time mapping a scalar field into another scalar field),
	and this is one of the crucial steps in the
	derivation of mirror symmetry by Hori and Vafa \cite{Hori:2000kt}.

	\normalsize
	\label{ex.superdual}

	\item
	$[\bll]$ (Mass term and parity) Verify that in three dimensions the Dirac mass term
	breaks parity. This is consistent with the fact that the Chern-Simons term generated in the IR
	by the parity anomaly breaks parity.
	Next, can we preserve parity in special circumstances
	when we have several Dirac fermions?
	Also, compare this with the case of four dimensions.
	Hint: in 3d theories, a parity transformation
	acts on the spatial coordinates $x^{1,2}$ as $x^1\to x^1, x^2\to -x^2$,
	and not as $x^1\to -x^1, x^2\to -x^2$; the latter is nothing but a
	rotation by angle $\pi$ in the $x^{1,2}$-plane.

	\label{mass_parity}

	\item\label{Witten_index}
	$[\bll]$ (Match of Witten indices)

	Let us define the
	\keyword{Witten index}{Witten index} by
	\begin{align}
		I:=\textrm{Tr}(-1)^F \ .
	\end{align}

	When a $U(1)$ gauge theory with a level-$k$ Chern-Simons term
	has fields $\Phi_i$ with charges $q_i$,
	the Witten index of the theory is given by \cite{Intriligator:2013lca}
	\begin{align}
		I=|k|+\frac{1}{2} \sum_i q_i^2  \ .
	\end{align}
	Use this formula to show that the Witten indices of
	3d $N_f=1$ SQED and the $XYZ$ model match.

	Hint: in the computation of the Witten index of the $XYZ$ model,
	we need to set all the real mass parameters to non-zero values.


	\item
	$[\bll]$  (Parity anomaly from an explicit one-loop computation)

	For the parity anomaly, we commented in the main text on the
	more abstract derivation using the index theorem.
	Here let us instead verify it more directly,
	by explicit one-loop Feynman diagram computations.
	\label{parity_derivation}

	Let us here consider a $U(1)$ background gauge field.
	Since the Chern-Simons term is quadratic in the
	gauge field, its effect should appear in the
	photon propagator.

	\begin{enumerate}
		\item
		      Convince yourself that the contribution from the Feynman diagram of Fig.~\ref{fig.polarization},
		      where a fermion loop runs inside the photon propagator,
		      is given by\footnote{
			      Here and below we suppress the overall factors which are common to all the terms,
			      namely the minus sign for the closed fermion loop and the powers of the coupling constant;
			      they do not affect the discussion of the parity anomaly,
			      and we fix the overall sign at the end by matching with \eqref{parity_anomaly}.
		      }
		      \begin{align}
			      \begin{split}
				       & \int \frac{d^3 p}{(2\pi)^3} \textrm{Tr}\left(
				      \gamma^{\nu} \frac{1}{i(\slashed{p} + \slashed{q})+\mu}
				      \gamma^{\sigma} \frac{1}{i\slashed{p} +\mu}
				      \right)                                                  \\
				       & \qquad =\int \frac{d^3 p}{(2\pi)^3} \textrm{Tr}\left(
				      \gamma^{\nu} \frac{-i(\slashed{p} + \slashed{q})+\mu}{(p+q)^2+\mu^2}
				      \gamma^{\sigma} \frac{-i\slashed{p} +\mu}{p^2 +\mu^2}
				      \right) \ .
			      \end{split}
		      \end{align}

		      \begin{figure}[t]
		      \centering
		      \begin{tikzpicture}[inner sep=0]
			      \node[anchor=south west] (img) {\includegraphics[scale=0.25]{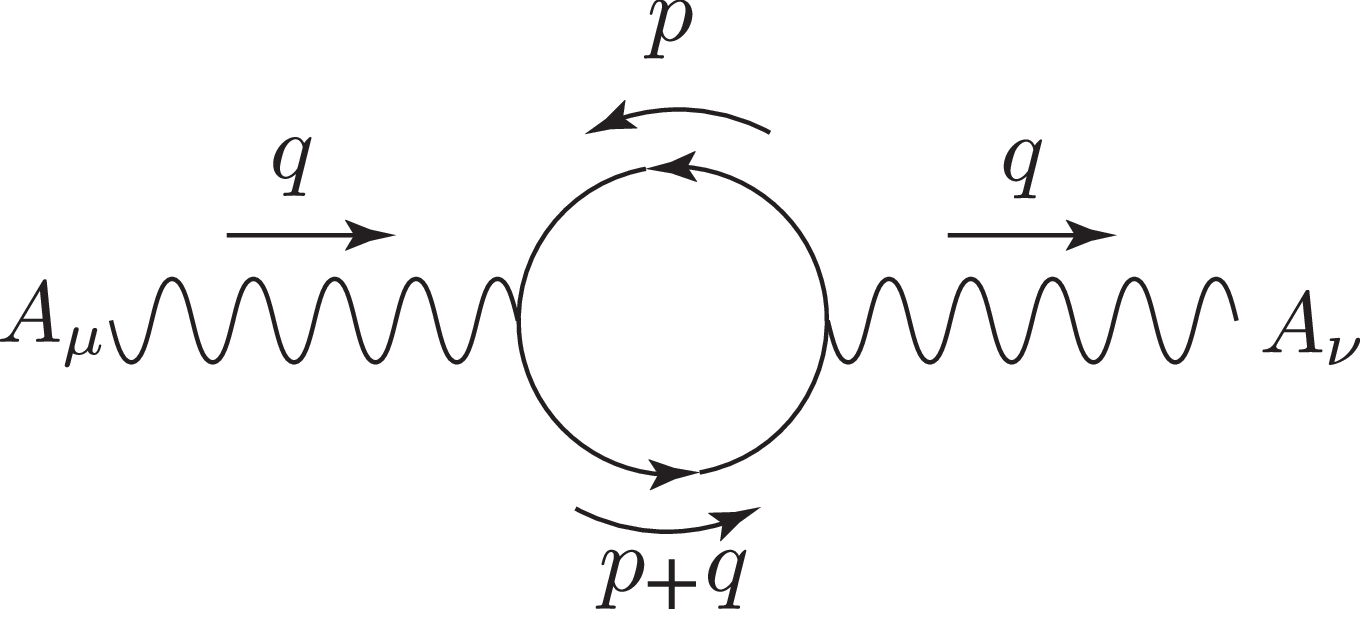}};
			      \begin{scope}[shift={(img.south west)},x=0.25pt,y=0.25pt]
				      \fill[white] (604,126) rectangle (654,170);
				      \node[anchor=west,inner sep=0] at (606,148) {\scriptsize$A_{\sigma}$};
				      \fill[white] (0,126) rectangle (48,170);
				      \node[anchor=east,inner sep=0] at (46,148) {\scriptsize$A_{\nu}$};
			      \end{scope}
		      \end{tikzpicture}
			      \caption{A Feynman diagram
				      for a one-loop correction to the gauge field propagator.
				      This is a standard Feynman diagram discussed in many QFT textbooks;
				      however, in three dimensions, unlike in four dimensions, it contains a parity-breaking term,
				      giving rise to the parity anomaly.
			      }
			      \label{fig.polarization}
		      \end{figure}
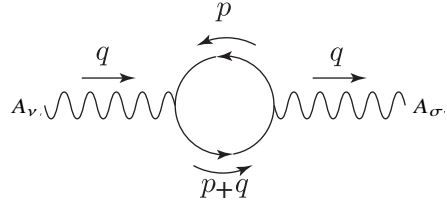

		\item
		      Let us consider the trace of the gamma matrices.
		      Since we are here interested only in the parity-breaking
		      Chern-Simons term, we need to keep only the term which arises from
		      the trace of three gamma matrices:
		      \begin{align}
			      \textrm{Tr}[\gamma^{\nu} \gamma^{\rho} \gamma^{\sigma} ]
			      =-2 \epsilon^{\nu\rho\sigma} \ .\footnote{
				      Here we take $(\gamma^0, \gamma^1, \gamma^2)=(i\sigma^3, \sigma^1, \sigma^2)$, which satisfies
				      $\{\gamma^{\nu}, \gamma^{\rho}\}=2\eta^{\nu\rho}$ in the mostly-plus signature,
				      together with $\epsilon^{012}=1$.
				      Note that the trace is real here, unlike in the mostly-minus signature;
				      the opposite choice of $\gamma^0$ flips the sign of the induced level.
			      }
			      \label{epsilon}
		      \end{align}
		      Note that this term does not exist in four dimensions.
		      Extracting the contribution of \eqref{epsilon}, we obtain
		      \begin{align}
			      A_{\nu}\left( 2i\mu\, \epsilon^{\nu\rho\sigma} q_{\rho}
			      \int \frac{d^3p}{(2 \pi)^3}
			      \frac{1}{\left[(p+q)^2+\mu^2\right] (p^2 +\mu^2) }
			      \right) A_{\sigma} \ .
			      \label{parity_integral}
		      \end{align}

		\item
		      The integral \eqref{parity_integral} can be evaluated by introducing Feynman parameters,
		      as in standard Feynman diagram computations.
		      Use (if needed) the formulas
		      \begin{align}
			       & \frac{1}{AB}=\int_0^1 dx \frac{1}{  \left( x A+(1-x)B \right)^2   } \ ,               \\
			       & \int \frac{d^3 l}{(2\pi)^3}\frac{1}{(l^2+\Delta)^2}=\frac{1}{8\pi \sqrt{\Delta}} \ ,  \\
			       & \int_0^1 dx \frac{1}{\sqrt{\mu^2+x(1-x) q^2}} =\frac{2}{q} \arcsin\left(
			      \frac{q}{\sqrt{4\mu^2+q^2}} \right)\ ,
		      \end{align}
		      to show that the Feynman diagram gives
		      \begin{align}
			       & \frac{i\mu}{2\pi }\, \epsilon^{\nu\rho\sigma} A_{\nu} q_{\rho} A_{\sigma} \,
			      \frac{1}{q}\arcsin\left( \frac{q}{\sqrt{4\mu^2+q^2}} \right) \nonumber \\
			       & \qquad
			      \overset{|\mu|\to\infty}{\longrightarrow}
			      \frac{i\,\textrm{sgn}(\mu)}{4\pi }
			      \epsilon^{\nu\rho\sigma} A_{\nu} q_{\rho} A_{\sigma} \ .
		      \end{align}
		      After Fourier transformation ($q_{\rho}\to -i\partial_{\rho}$),
		      and including the factor $\frac{1}{2}$ of the quadratic effective action, this gives
		      $\frac{1}{4\pi}\frac{\textrm{sgn}(\mu)}{2} \epsilon^{\nu\rho\sigma} A_{\nu} \partial_{\rho} A_{\sigma}$,
		      which is nothing but the Chern-Simons term with level \eqref{parity_anomaly}.

		\item
		      Generalize the argument above to semisimple non-Abelian gauge fields:
		      first repeat the computation above with non-Abelian color factors, and then
		      compute the one-loop correction to the three-point vertex of the gauge field
		      in the same manner, to find the cubic part $A\wedge A\wedge A$ of the Chern-Simons term
		      with the correct coefficient.

		\item
		      As another generalization, consider multiple $U(1)$ global symmetries $U(1)_i$, labeled by $i, j, \ldots$,
		      and suppose that we have many fermions, where each fermion labeled by $f$
		      has charges $(q_f)_i$ with respect to $U(1)_i$. Let us turn on the real mass $\mu_i$ for each $U(1)_i$.
		      Show that the one-loop induced Chern-Simons term has both diagonal and off-diagonal components,
		      and is given by
		      \begin{align}
			      k^{\rm eff}_{ij}:=k_{ij}+\frac{1}{2} \sum_{f: \textrm{fermion}} (q_f)_i (q_f)_j
			      \textrm{sgn}\left[\sum_l (q_f)_l \mu_l \right] \ .
			      \label{keff}
		      \end{align}

		\item
		      The argument above is only at one loop.
		      However, since the Chern-Simons term is topological and its level is quantized,
		      we expect that higher-loop contributions vanish.
		      See \cite{Coleman:1985zi} for a direct Feynman-diagram
		      proof of this fact for Abelian gauge groups.

	\end{enumerate}

	\item\label{parity_anomaly_matching}
	$[\bll]$ (Matching of the parity anomaly)

	Whether or not we can consistently gauge the Abelian global symmetries $i, j, \ldots$ is decided by the
	presence or absence of the parity anomaly,
	namely by whether the level $k^{\rm eff}_{ij}$ of the
	effective Chern-Simons term (as given in \eqref{keff})
	is an integer or a half-integer:
	\begin{align}
		k_{ij}+\frac{1}{2} \sum_{f: \textrm{fermion}} (q_f)_i (q_f)_j
		\in \mathbb{Z} \ ,
		\label{keff_2}
	\end{align}
	where we dropped the sign factor in \eqref{keff} since we are interested only in whether the result is an integer or a half-integer.

	Suppose that we have two UV theories flowing into the same IR fixed point under the
	RG flow.
	Then whether or not we can consistently gauge the global symmetries,
	namely whether $k^{\rm eff}_{ij}$ is an integer or a half-integer,
	should match between the two theories.
	This is called the \keyword{parity anomaly matching}{parity anomaly matching}.
	Unlike the anomalies in four dimensions, the anomaly here in three dimensions is a $\bZ_2$-anomaly,
	and hence the resulting constraint is not very strong;
	it is nevertheless a useful check of dualities.

	Verify the parity anomaly matching for
	the 2--3 mirror pair. More concretely, show that
	\begin{align}
		k^{\rm eff}_{RR}\in \bZ+\frac{1}{2} \ , \quad  k^{\rm eff}_{RA}, k^{\rm eff}_{RJ}, k^{\rm eff}_{AJ}, k^{\rm eff}_{AA}, k^{\rm eff}_{JJ}\in \bZ \ .
	\end{align}

	Hint: note that the summation over fermions in \eqref{keff} is taken over all the fermions
	in the Lagrangian; in particular, the vector superfield contains the gaugino. Note also that the R-charge of a
	chiral superfield and that of the fermion inside the multiplet differ by $1$ (see Sec.~\ref{subsec.R_sym}).

	\item
	$[\bll]$ (Charges of monopole operators)
	\label{ex.monopole_charge}

	Complete the argument of the main text, and show that the
	charges of $\mathcal{V}_{\pm}$ are as given in Table~\ref{tab.SQED}.

	\item
	$[\bll\bll]$ ($\mathcal{N}=2$ $U(1)$ theory with $N_f$ flavors) In this chapter
	we discussed the quantum-corrected vacuum moduli space of the $U(1)$ gauge theory with $N_f=1$ flavor.
	What about the case of $N_f=2$, or more general $N_f$?
	What happens to the vacuum moduli space when we turn on real mass parameters?\footnote{In Chap.~\ref{chap.wall}
		we will discuss a similar theory, a 3d $U(1)$ theory with $N_f=2$, albeit with $\scN=4$ supersymmetry.}

\end{practice}

\chapter{$S^3$ Partition Function as a Tool}\label{chap.S3}

\begin{abstract}
	We discuss aspects of rigid supersymmetric gauge
	theories on curved spaces and their supersymmetric localization,
	focusing on the $S^3_b$ partition function
	of 3d $\mathcal{N}=2$ theories.
	One of the goals of this chapter is to provide a dictionary between
	supersymmetric gauge theories and
	mathematical
	expressions involving quantum dilogarithm functions,
	so that we can move freely between the two.
\end{abstract}

\section{Partition Functions as Functors}\label{subsec.functor}

As we have seen in Chap.~\ref{chap.3dN2},
3d $\mathcal{N}=2$ theories have a number of interesting properties.
While this is an attractive feature of the theory, at the same time
its complexity can be a burden for beginners.
Some readers might already have run out of breath in the previous chapter.
Can we replace the contents of the previous chapter, or at least their main parts,
by a simpler discussion, while still keeping their essence?

As we have described in Chap.~\ref{chap.intro},
a quantum field theory can be regarded as giving a procedure to
compute physical observables.
From this viewpoint, if we wish to prove, for example, that
two QFTs are dual, we should list all the possible observables of the two theories,
compute the corresponding observables in each theory, and show that they coincide.

Of course, it is a tremendous amount of work to carry this out, and it is not realistic;
it is extremely rare to find a QFT all of whose physical observables
can be computed exactly.
If that is the case, how about defining, for a given theory, a quantity which is easy to compute,
and using it to represent the physics?\footnote{Similar ideas have been used in various different branches of physics and mathematics, for example, in the context of knot invariants in knot theory.}

\begin{figure}[t]
	\centering
	\begin{tikzpicture}[inner sep=0]
		\node[anchor=south west] (img) {\includegraphics[scale=0.33]{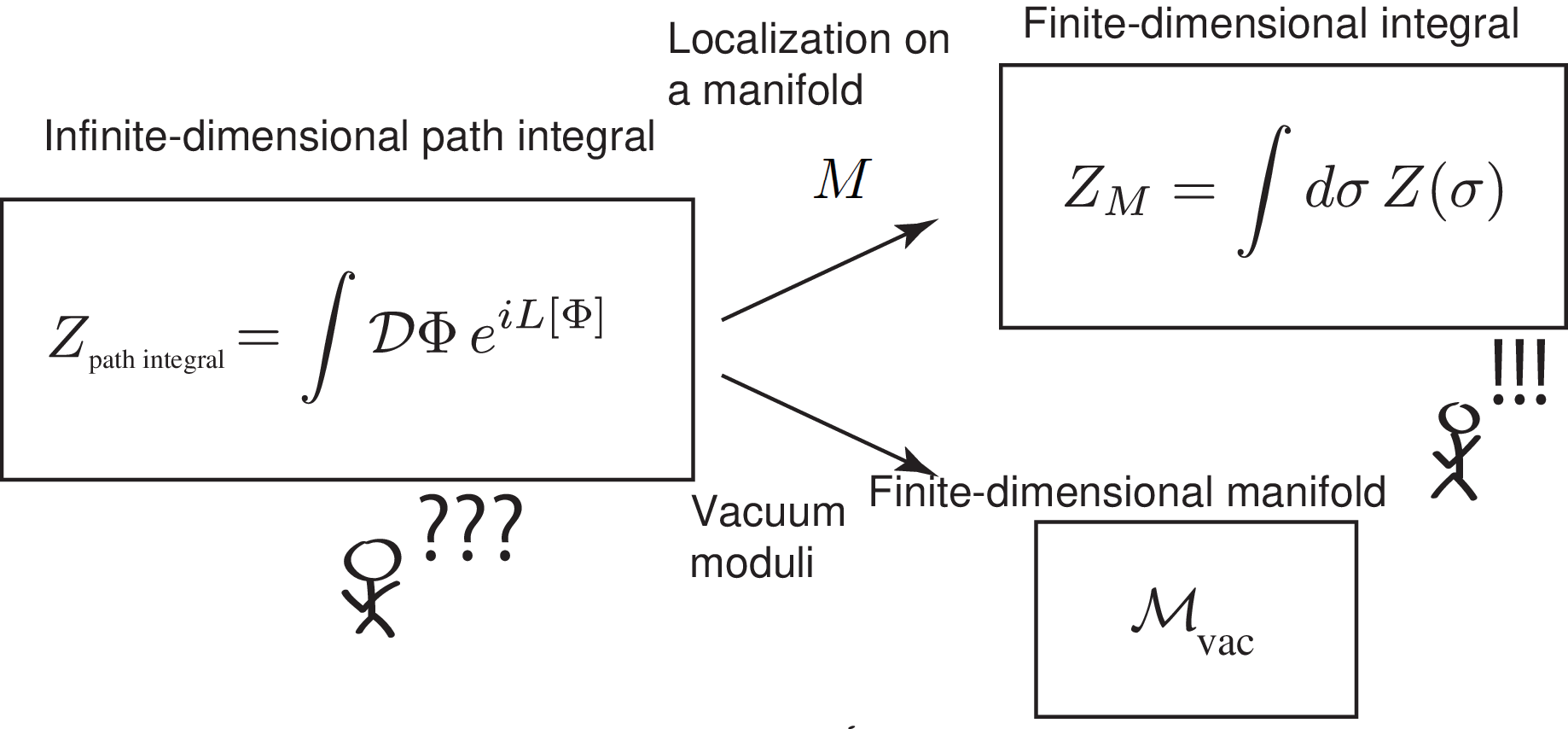}};
		\begin{scope}[shift={(img.south west)},x=0.33pt,y=0.33pt]
			\fill[white] (14,166) rectangle (392,262);
			\node[anchor=west,inner sep=0] at (22,214) {\small$Z_{\textrm{path integral}}=\int \mathcal{D}\Phi \, e^{iL[\Phi]}$};
		\end{scope}
	\end{tikzpicture}
	\caption{We replace a complicated QFT by a partition function
		exactly computable by \keyword{supersymmetric localization}{supersymmetric localization}.
		After this process
		the data of a QFT (which is defined by an infinite-dimensional path integral)
		is turned into a concrete identity involving finite-dimensional integrals and sums.}
	\label{fig.Z}
\end{figure}

Of course, it would be too much to expect that
just two or three quantities contain all the information
of all the physical observables: such a quantity is a kind of ``tinted glasses'',
and there are things which become invisible through this filter.
However, if we consider quantities with good properties,
can we not efficiently extract important information about the theory?
If we could do this, dualities of QFTs
would be replaced by concrete mathematical identities stating that
two QFTs have the same partition function (Fig.~\ref{fig.Z}).\footnote{More mathematically,
	the partition function can be regarded as an (incomplete) \textbf{``functor''} from the
	\textbf{\keyword{category}{category} formed by QFTs}
	to the category of more concrete numbers or functions
	(we used here the terminology of category theory; readers unfamiliar with it can safely skip this.
	The definitions can easily be found by searching).
	Since there is no rigorous mathematical formulation of QFTs underlying this statement,
	this understanding is not at all mathematically rigorous; however, this idea is
	sometimes useful as a framework of thinking.
	Indeed, by considering the category formed by QFTs, and in particular the notion of morphisms there,
	we are naturally led to the notion of domain walls in the next chapter.}
We have already used the same idea: in the discussion of the 2--3 mirror symmetry in the previous chapter,
we extracted from two 3d $\mathcal{N}=2$ theories the K\"{a}hler manifolds called
the vacuum moduli spaces, and discussed their agreement.

\subsection{Why $S^3$ Partition Function?}

In this chapter, as an example of such a good physical quantity,
we consider the \keyword{$S^3$ partition function}{S3 partition function}
for the 3d $\mathcal{N}=2$ gauge theories discussed in the previous chapter.

The $S^3$ partition function, as the name suggests,
is the partition function of the theory obtained by Euclideanizing a 3d theory
and placing it on the 3d sphere $S^3$.
Divergences of the path integral include IR divergences and UV divergences.
Since $S^3$ is a compact manifold with a finite size (radius), there are no IR divergences.
Therefore, by appropriately regularizing the UV divergences, we can obtain a finite quantity.
This is the partition function we wish to consider.

\small
Care is needed for the dependence on the regularization of UV divergences.
Indeed, in standard textbooks on quantum field theory, the partition function is used as a quantity
for normalizing correlation functions, or as a quantity whose derivatives (when external fields are added)
give correlation functions, and the value of the partition function itself is never discussed.
This is because, when we add counterterms to the Lagrangian,
the value of the partition function changes depending on the values of their coefficients.

The situation is different in three spacetime dimensions
(and more generally in odd spacetime dimensions).
Let us expand the $S^3$ partition function in the radius $r$ of $S^3$:
\begin{align}
	\log Z=c_3 r^3 +\cdots + c_0 r^0+ \mathcal{O}\left(\frac{1}{r}\right) \ .
\end{align}
Here, the coefficients of the divergent parts depend on the choice of regularization:
for example, the value of $c_3$ depends on the value of $c$ if we consider
the cosmological constant term
\begin{align}
	c \int d^3 x \sqrt{g}
\end{align}
as a counterterm.
However, the constant term $c_0$ does not have such an ambiguity.
Namely, in three spacetime dimensions, the value of $c_0$ cannot be changed by
counterterms invariant under diffeomorphisms:\footnote{
	In truth this is not correct: there exists a gravitational version of the Chern-Simons term
	(the \keyword{gravitational Chern-Simons term}{gravitational Chern-Simons term}\index{Chern-Simons term@Chern-Simons term!gravitational@gravitational}) discussed in Chap.~\ref{chap.complexCS},
	and unless we take its effect properly into account, we cannot determine the precise value of the partition function
	including its normalization \cite{Closset:2012vp}. This is an analogue of the
	\keyword{framing anomaly}{framing anomaly} in Chern-Simons theory without matter fields
	(the dependence of the value of the partition function on the choice of
	trivialization of the normal bundle of the manifold).
}
for example, if we try to use the curvature $R$, then since $R$ is
a second derivative of the metric, it has mass dimension $2$,
and the dimensions do not match in odd dimensions. Therefore, we can understand that
the ``UV-regularized'' $S^3$ partition function actually refers to $c_0$.

\normalsize

The partition function is a priori defined by an infinite-dimensional
path integral. However, for 3d $\scN=2$ theories,
thanks to \keyword{localization}{localization} explained below,
the path integral reduces to a finite-dimensional integral,
and the integral expression can be written down
\textbf{mechanically} from the Lagrangian.\footnote{At least when the theory has a Lagrangian description.}
This is the reason why the $S^3$ partition function is so powerful.

\small
The $S^3$ partition function has been studied extensively
since 2009 \cite{Kapustin:2009kz,Jafferis:2010un,Hama:2010av}.
It has been used for checks of various dualities of 3d theories
and in the derivation of the
$N^{3/2}$ scaling law for M2-branes \cite{Drukker:2010nc,Herzog:2010hf};
moreover, its relation to
\keyword{entanglement entropy}{entanglement entropy} is known~\cite{Casini:2011kv},
and it has been shown to be a quantity which decreases under the RG flow in 3d field theories (the so-called \keyword{$F$-theorem}{F-theorem}) \cite{Myers:2010tj,Jafferis:2011zi,Casini:2012ei}.

Finite-dimensional integral expressions of partition functions by localization (or other methods)
have also been obtained in other dimensions.
Representative examples include,
in four dimensions,
$S^1\times S^3$ for $\scN\ge 1$ (the 4d $\scN\ge 1$ superconformal index)
\cite{Romelsberger:2005eg,Kinney:2005ej} and
$S^4$ for $\scN\ge 2$ \cite{Pestun:2007rz},
and in $(1+1)$ dimensions $S^2$ \cite{Benini:2012ui,Doroud:2012xw}, the 2d disk $D^2$ \cite{Honda:2013uca,Sugishita:2013jca,Hori:2013ika} and $T^2$ (the elliptic genus) \cite{Gadde:2013dda,Benini:2013xpa}.
We can also find relations between different partition functions.
For example, by taking limits of the 4d $S^3/\bZ_n\times S^1$ partition function,
we can derive the 3d $S^1\times S^2$ and $S^3$ partition functions, and moreover the 2d $S^2$ partition function
(see Ref.~\cite{Yamazaki:2013fva} and references therein).

\normalsize

\section{Comments on Supersymmetric Localization}

A detailed discussion of the computation of the
$S^3$ partition function by supersymmetric localization could in itself fill a whole book,
and hence we do not go into the details of localization in this book (see e.g.\ the reviews \cite{Marino:2011nm,Cremonesi:2014dva,Hosomichi:2014hja} for interested readers.
The original reference \cite{Kapustin:2009kz} also contains careful details of the computation,
and is educational).
In the following we take a pragmatic approach:
we ask the readers to become familiar with the $S^3$ partition function
as users of the results of localization, regarded as a tool.

Having said that, rather than regarding the partition function as a complete black box,
it is also important to
have some feeling for its inner workings.
We therefore very briefly comment on the derivation of the
partition function below (readers in a hurry can skip directly to the
integral expression of the partition function in Sec.~\ref{sec.S3_integral}).
In the following we denote by $r$ the radius (the overall scale)
of $S^3$.

\bparagraph{Supersymmetry on $S^3$}

To discuss the $S^3$ partition function, we need to put the theory on
$S^3$, a curved space.
Since an ordinary Lagrangian is given on a flat spacetime $\mathbb{R}^{2,1}$ with Lorentzian signature,
we first Euclideanize the theory.\footnote{
	This analytic continuation is rather subtle for fermions, since
	the existence of a particular type of fermion depends on the signature of the metric.
	For example, 3d Majorana fermions exist in Lorentzian signature,
	but not in Euclidean signature.
	Therefore, when we analytically continue to Euclidean signature,
	we need to formally double the number of Majorana fermions,
	making $\psi$ and $\overline{\psi}$ independent.
} Next we need to take into account
the effect of the curvature of $S^3$.
Here we would like to keep at least one of the
four supercharges $\scQ, \overline{\scQ}$.
In the following we denote (one of) the remaining supercharge(s) by $\scQ$.

To consider supersymmetry on a curved space, we need an appropriate spinor,
and whether such a spinor exists depends on the manifold under consideration.
Fortunately, while there are no \keyword{Killing spinors}{Killing spinor} in the usual sense on $S^3$,
their relatives (e.g.\ \keyword{conformal Killing spinors}{conformal Killing spinor})
exist, and by using them we can write down the supersymmetry transformations
on $S^3$.\footnote{
	By a conformal transformation of the Killing spinor equation $D_{\mu}\epsilon=0$ in flat spacetime
	to $S^3_{b=1}$, we obtain the conformal Killing spinor equation
	\begin{align}
		D_{\mu} \epsilon=\pm  \frac{i}{2r}\gamma_{\mu} \epsilon \ ,
	\end{align}
	which has two solutions (for a fixed sign $\pm$).
	By using these two supercharges, we can write down the supersymmetry transformations on $S^3_{b=1}$.
}
In this case, terms whose coefficients are powers of the inverse radius $1/r$
appear in the Lagrangian and in the supersymmetry transformations. Of course, all these terms vanish
when we go back to flat spacetime by taking $r\to\infty$.

\small
The supersymmetry transformations can be constructed order by order in powers of $1/r$,
or, more systematically, by starting from supergravity
and giving expectation values to the auxiliary fields in the supergravity multiplet \cite{Festuccia:2011ws,Closset:2012ru}.
We will not discuss these details in this book.
\normalsize

Since the partition function depends on the choice of the metric on $S^3$ and
of the auxiliary fields of supergravity, a priori it could depend on infinitely many data.
It has been shown, however, that in reality it depends only on a very limited part of the data
(a transversely holomorphic foliation) \cite{Closset:2013vra},
and for $S^3$ this freedom of deformation can be represented by
a one-parameter deformation of the standard metric (the round metric) \cite{Hama:2011ea}:
\begin{align}
	S^3_b:= \left\{ (x_1, \!\cdots \! ,  x_4)\! \in\! \mathbb{R}^4 \,\big|\,\, b^2
	(|x_1|^2+|x_2|^2)\! + \! b^{-2}(|x_3|^2+|x_4|^2) \! = \! r^2
	\right\} \ .
	\label{S3bmetric}
\end{align}
\nomenclature{$b$}{deformation parameter of $S^3$}
This metric has $SO(4)$ isometry for $b=1$, but
for $b\ne 1$ the isometry $SO(4)$ is broken to $SO(2)\times SO(2)$.
In this book we exclusively use \eqref{S3bmetric}; however,
the metric giving the same one-parameter deformation of the partition function
is far from unique. For example, it has been shown in
Ref.~\cite{Imamura:2011wg} that the same answer is obtained
from a metric preserving $SO(3)\times SO(2)$ symmetry and the associated deformation
of the supersymmetry transformations.\footnote{
	When we use $S^3_b$ described by \eqref{S3bmetric}, the parameter $Q:=b+b^{-1}$ appearing later
	can only take values $Q\ge 2$, whereas when we use the $S^3$ of Ref.~\cite{Imamura:2011wg}
	we can consider more general values of $Q$.
}
Since the dependence on the radius $r$ can be recovered by dimensional analysis,
in the following we set $r=1$ for simplicity.

\begin{figure}[t]
	\centering\includegraphics[scale=0.37]{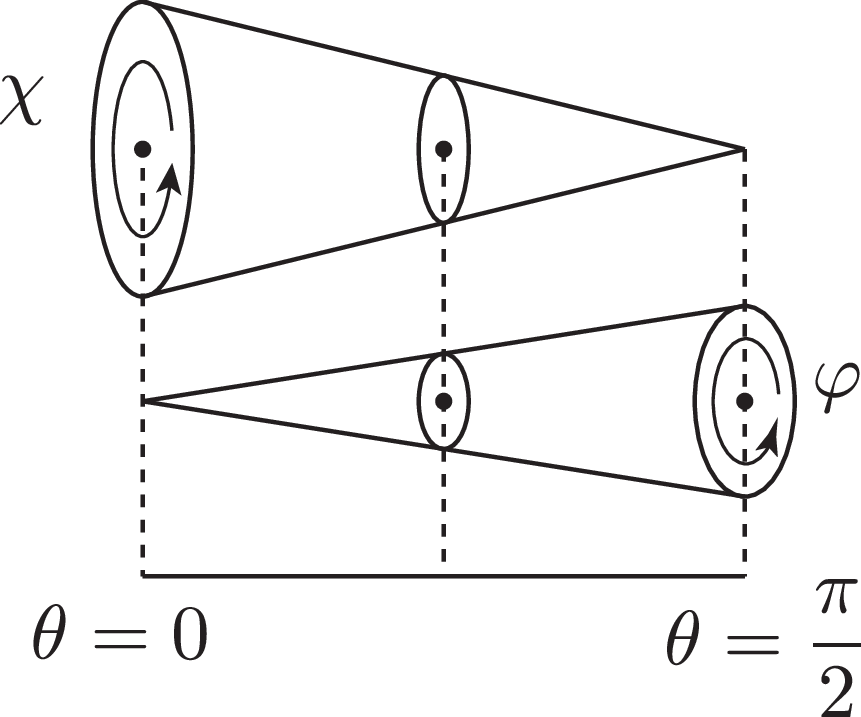}
	\caption{The geometry $S^3_b$ is a space in which $U(1)^2$, parametrized by $\varphi, \chi$,
		is fibered over the interval $\left[0, \frac{\pi}{2}\right]$ parametrized by $\theta$.
		The value of $b$ changes the
		relative size of the two $U(1)$ fibers.}
	\label{fig.S3fibration}
\end{figure}

For concrete computations it is useful to choose coordinates.
In the parametrization
\begin{align}
	 & (x_1, x_2, x_3, x_4)\!=\!\left(\frac{1}{b} \sin \theta \sin \varphi, \frac{1}{b} \sin
	\theta \cos \varphi, b \cos \theta \sin \chi, b \cos \theta \cos
	\chi\right) \ , \label{S3coordinate_new} \\
	 & 0\le \theta\le \frac{\pi}{2} \ , \quad
	0\le \varphi \le 2\pi \ , \quad
	0 \le \chi \le 2\pi \ , \nonumber
\end{align}
the natural metric induced from $\bR^4$ is
\begin{align}
	ds^2
	=(b^{-2} \cos^2\theta+b^2 \sin^2 \theta) d\theta^2
	+b^{-2} \sin^2 \theta d\varphi^2+b^2 \cos^2 \theta d\chi^2 \ .
	\label{S3metric}
\end{align}
This is a space in which $U(1)^2$, parametrized by $\varphi, \chi$,
is fibered over the interval $\left[0, \frac{\pi}{2}\right]$ parametrized by $\theta$.
At the two end points of the interval, one of the two
$U(1)$ fibers shrinks (Fig.~\ref{fig.S3fibration}).

\bparagraph{Supersymmetric Localization}

Suppose now that we have a Lagrangian on the curved manifold $S^3$,
preserving a supercharge $\scQ$.
As the next step, we use this $\scQ$ to perform the
\keyword{localization}{localization}
of the path integral.

The partition function in the path-integral formulation is given by
\begin{align}
	Z=\int\! \mathcal{D} \Phi \,\, e^{-L[\Phi]}
\end{align}
(since we have already Euclideanized the theory,
there is no factor of $i$ in the exponent).
Let us deform this partition function by a $\scQ$-exact term $\scQ V$ as
\begin{align}
	Z(t)=\int\! \mathcal{D} \Phi \,\, e^{-L[\Phi]-t\scQ  V} \ .
\end{align}
Let us moreover assume that $\scQ^2 V=0$ holds.
This is automatic when $\scQ^2=0$, but this need not necessarily be the case.
In practice, $\scQ^2$ is in most cases either zero or a translation along a certain direction.
Let us then consider the derivative of the
deformed partition function with respect to $t$:
\begin{align}
	\frac{dZ(t)}{dt}=-\int \! \mathcal{D}\Phi\,\, \left( \scQ  V\right)  e^{-
			                                                                     L-t \scQ V }
	=-\int \! \mathcal{D}\Phi\,\,  \scQ  \left( V e^{- L-t  \scQ V} \right) \ .
\end{align}
Here we used the invariance of the Lagrangian under the supercharge $\scQ$ ($\scQ L=0$),
as well as $\scQ^2 V=0$.
This means that $Z(t)$ is independent of $t$, and
we may compute the original partition function $Z(t=0)$
in the $t\to \infty$ limit of $Z(t)$.\footnote{
	This is an explanation often found in the literature; to be precise, however, we need to assume two things:
	(1) the deformed action $L(t):=L+t\scQ V$ has a non-degenerate kinetic term, and (2) no new saddle points come in from infinity
	in the field configuration space. Care is needed as to whether these are satisfied \cite{Witten:1992xu}.
	In the case of the $S^3$ partition function, however, this issue turns out not to be important.
}

In the limit $t\to\infty$, the term linear in $t$ in the exponent first has to vanish.
Namely, the condition
\begin{align}
	\scQ V=0
\end{align}
has to be satisfied.
Such a field configuration is called a saddle point,
and we parametrize it by (a set of) field variables $\sigma(x)$.
Our partition function then receives a contribution from
the value of the partition function at the saddle point.
We denote this by
$Z_{\rm classical}(\sigma)$.

Once we choose a saddle point, the $\mathcal{O}(t^1)$ terms in the exponent vanish.
However, the $\mathcal{O}(t^0)$ terms still remain.
We can therefore consider fluctuations of the fields around the saddle point solution.
When we expand around the saddle point, the first-order variation vanishes by the definition of the saddle point.
Hence what appears next is the second-order variation, which is Gaussian in the fluctuations of the fields,
and can therefore be integrated. Let us call this contribution
$Z_{\textrm{1-loop}}(\sigma)$.
Since in the present case we may take the limit $t\to \infty$,
the saddle point approximation (WKB approximation) becomes exact.

The partition function is obtained by
combining the two contributions above
and integrating over the saddle points:
\begin{align}
	Z=\int_{\textrm{saddle}} \scD \sigma\,\, Z_{\textrm{classical}}(\sigma)
	Z_{\textrm{1-loop}}(\sigma) \ .
	\label{eq.1-loop1}
\end{align}

So far the argument has been rather general, and holds for any $V$.
For a general choice of $V$, the saddle point solutions still allow complicated field configurations $\sigma(x)$.
However, with a clever choice of $V$, it can happen that only saddle point solutions
with trivial dependence on spacetime are allowed. In this case, the path integral \eqref{eq.1-loop1}
reduces to a finite-dimensional integral over constant $\sigma$:
\begin{align}
	Z=\int_{\textrm{saddle}} d \sigma\,\, Z_{\textrm{classical}}(\sigma)
	Z_{\textrm{1-loop}}(\sigma)
	\label{eq.1-loop2}
\end{align}
(namely, $\scD \sigma$ is replaced by $d\sigma$).
In this case the $S^3$ partition function is a finite-dimensional integral; this allows us to discuss it
in a mathematically rigorous manner, and moreover to make use of the various techniques of
\keyword{matrix models}{matrix model} as $0$-dimensional field theories.
Numerical methods also become applicable.

\small

For the $S^3_b$ (and $S^1\times S^2$) partition functions of the 3d $\scN=2$ theories we consider,
choices of $V$ are known for which the result reduces to a finite-dimensional integral.
The choice of $V$ is not unique; for example,
for a fermion $\psi$ (a gaugino in a vector multiplet or a fermion in a chiral multiplet)
we can consider
\begin{align}
	V=\textrm{Tr}\left( (\scQ\psi)^{\dagger} \psi +(\scQ\overline{\psi})^{\dagger} \overline{\psi}  \right) \ .
\end{align}
In this case
\begin{align}
	\scQ V=\textrm{Tr}\left( (\scQ\psi)^{\dagger} (\scQ\psi)+  (\scQ\overline{\psi})^{\dagger} (\scQ\overline{\psi})\right) \ ,
\end{align}
which is expected to be positive definite, as can be checked explicitly.
Depending on the choice of $V$, we may obtain the same expression, or an expression which looks different.

Although we do not discuss this in this chapter,
let us add that localization itself is useful even when the integral does not reduce to a finite-dimensional one.
For example, there are situations where the path integral of a field theory in a certain dimension
reduces to a path integral in lower dimensions;
the localization of 5d $\scN=2$ theories on 3-manifolds introduced in Chap.~\ref{chap.6d}
is a good example. As another example,
it is known that correlation functions of ($1/8$-Bogomol'nyi-Prasad-Sommerfield (BPS)) Wilson lines (inserted on $S^2$)
in 4d $\scN=2$ theories on $S^4$ reduce to 2d Yang-Mills theory \cite{Giombi:2009ds}.

\normalsize

\section{Integral Expression for $S^3_b$ Partition Function}\label{sec.S3_integral}

Let us now summarize the concrete answer for the $S^3_b$ partition function
after localization.

\bparagraph{Saddle Point}

The saddle point equations are
\begin{align}
	\begin{split}
		 & F_{\mu\nu}=0\ , \quad D=\frac{\sigma}{\sqrt{b^2 \sin^2\theta+b^{-2}\cos^2\theta}}  \ , \\
		 & \sigma=\textrm{constant} \ , \quad \phi=F=0
	\end{split}
\end{align}
(note that fermions do not have VEVs, since we impose 3d Lorentz symmetry).
For $b=1$, this simplifies to $D=\sigma=\textrm{constant}$.

What is important here is that the values of all the fields at the saddle point are constant,
independent of the coordinates of $S^3$.
The infinite-dimensional path integral therefore reduces to a finite-dimensional integral.
Since the fundamental group of $S^3$ is trivial,
it follows from $F_{\mu\nu}=0$ that the gauge field is
trivial (up to gauge degrees of freedom).\footnote{
	The conclusion changes when we consider the manifold $S^3/\mathbb{Z}_n$, the quotient of $S^3$ by a discrete group,
	since its fundamental group is non-trivial \cite{Gang:2009wy,Benini:2011nc,Alday:2012au}.
} Moreover, the fields $\phi, F$ of the chiral superfield both vanish, so that their contribution is trivial;
the only fields with non-trivial values are the scalar field $\sigma$ of the vector superfield
and the scalar auxiliary field $D$.

\bparagraph{Classical Part}

Since the only scalar fields with non-trivial values at the saddle point are
$D$ and $\sigma$,
the classical contribution comes only from
the terms in the Lagrangian containing $D$ and $\sigma$:
these are the Chern-Simons terms and the FI parameters.

First, the contribution from the term $-2D\sigma$ in the Chern-Simons term \eqref{SUSY_CS} is
\begin{align}
	\begin{split}
		\mathcal{L}_{\textrm{SUSY CS}}
		 & \supset \frac{k}{4\pi} \int_{S^3} (-2D \sigma)                                      \\
		 & =
		\frac{k}{4\pi} \int_{S^3} \frac{-2\sigma^2}{\sqrt{b^2\sin^2\theta+b^{-2}\cos^2\theta}} \\
		 & = -k\pi \sigma^2 \ ,
	\end{split}
\end{align}
where in the last step we performed the integral using the metric \eqref{S3metric}. The contribution to the
partition function is therefore
\begin{align}
	Z_{\textrm{CS}}=\exp\left(-i k\pi \sigma^2 \right) \ .
	\label{ZCS_diagonal}
\end{align}
More generally, when we consider a Chern-Simons theory with
off-diagonal components, we obtain
\begin{align}
	Z_{\textrm{CS}}=\exp\left( -i\pi \sum_{i,j} k_{i,j} \sigma_i \sigma_j\right) \ .
	\label{ZCS_offdiagonal}
\end{align}

The effect of the FI parameter can be obtained similarly. Indeed,
as we have seen in the previous chapter, introducing an FI parameter
is the same as including an off-diagonal Chern-Simons term of level $1$
between the gauge symmetry and the
topological $U(1)_J$ symmetry, whose scalar field takes the value $\zeta$.
Substituting $k_{i,j}=k_{j,i}=1$ into \eqref{ZCS_offdiagonal}, we therefore obtain
\begin{align}
	Z_{\textrm{FI}}=\exp\left(-2\pi i \zeta\sigma  \right) \ .
	\label{FI_S3}
\end{align}

\bparagraph{One-Loop Contribution}

The one-loop contribution is given by the product of the contributions from each of the
chiral multiplets and vector multiplets.

First, the one-loop part for a vector multiplet is
\begin{align}
	Z_{\textrm{vector multiplet}}=\frac{1}{|W|} \prod_{\substack{\alpha: \textrm{positive roots}\\ \textrm{of the gauge group}}} \sinh \pi b \alpha(\sigma)  \sinh \pi b^{-1} \alpha(\sigma)   \ ,
\end{align}
where we used the language of roots of the gauge group.
Here $|W|$ is the number of elements of the \keyword{Weyl group}{Weyl group} of the gauge group.
More concretely, when we consider a $U(N)$ gauge theory
and represent its Cartan subalgebra by $\sigma_1, \ldots, \sigma_N$,
the positive roots are given by $\{ \sigma_i-\sigma_j \}_{1\le i<j \le N}$, and
\begin{align}
	Z_{\textrm{vector multiplet}}=\frac{1}{N!}  \prod_{1\le i <j \le N } \sinh \pi b (\sigma_i-\sigma_j) \sinh \pi b^{-1} (\sigma_i-\sigma_j) \ .
\end{align}
Here we used the fact that the Weyl group of $U(N)$ is the symmetric group $\mathfrak{S}_N$ of degree $N$,
and hence $|\mathfrak{S}_N|=N!$.
When we specialize to $b=1$, we obtain
\begin{align}
	Z_{\textrm{vector multiplet}}\Big|_{b=1}=\frac{1}{N!}  \prod_{1\le i <j \le N } \frac{1}{4}
	(e^{2\pi \sigma_i}-e^{2\pi \sigma_j}) (e^{-2\pi \sigma_j}-e^{-2\pi \sigma_i})
	\ .
\end{align}
This is the integration measure (Haar measure) of the $U(N)$ gauge group.

The simplest case is when the gauge group is Abelian. In this case,
the \keyword{one-loop determinant}{one-loop determinant} is trivial, irrespective of the value of $b$:
\begin{align}
	Z_{\textrm{Abelian vector multiplet}}=1 \ .
\end{align}

\bigskip

Let us next consider a chiral multiplet.
Consider a chiral multiplet with charge $q_i$ under global or local symmetries,
and suppose that its \keyword{R-charge}{R-charge} is $r$. The contribution of this chiral multiplet
to the one-loop determinant is then
\begin{align}
	Z_{\textrm{chiral multiplet}}=
	\prod_{w: \textrm{weights of the representation}}\, s_b\left(\frac{iQ}{2}(1-r)-w(\sigma)\right)
	\ .
	\label{chiralZ}
\end{align}
Here $s_b(z)$ is the \textbf{quantum dilogarithm function}
defined in Appendix~\ref{app.dilog}, and
its argument is given by the complexified $\sigma$:
\begin{align}
	w(\overline{\sigma}):=i\frac{Q}{2} (1-r)-w(\sigma)=-\sigma_R-w(\sigma) \ ,
	\quad \sigma_R:=-i\frac{Q}{2} (1-r) \ .
	\label{bar_sigma}
\end{align}
\nomenclature{$\overline{\sigma}$}{complexified $\sigma$}
Here $\sigma_R$ can be regarded as the expectation value of the scalar field associated with the
background vector multiplet for the R-symmetry;
its value is required to be non-zero in order to preserve part of the supercharges on the curved space
(this is a kind of topological twist discussed in Sec.~\ref{subsec.twist}).
Finally, we defined $Q$ by
\begin{align}
	Q:=b+b^{-1} \ .
	\label{Q_def}
\end{align}\nomenclature{$Q$}{deformation parameter of $S^3$ ($=b+b^{-1}$)}

\bparagraph{Supplementary Comments}

Let us here add several supplementary remarks.

\begin{Kenumerate}

	\item One of the most important
	properties of the $S^3_b$ partition function is that \textbf{the partition function
		does not depend on the gauge coupling constant}.
	This is because the kinetic term of the gauge field is itself $\scQ$-exact.\footnote{
		This does not hold in general. For example, the partition function of 4d $\mathcal{N}=2$ theories
		on $S^4$ \cite{Pestun:2007rz} has a non-trivial dependence on the gauge coupling constant.
	}

	The gauge coupling constant has a small value in the Lagrangian theory in the UV,
	but its value grows under the RG flow and becomes large near the IR fixed point
	(recall the discussion in Sec.~\ref{sec_3d_beta}).
	Therefore, if we compute the partition function of the UV theory,
	its value at the same time gives the value of the partition function
	at the IR fixed point of the theory.
	Since the strongly coupled theory at the IR fixed point is in many cases difficult to
	describe directly from a Lagrangian, this fact is
	very convenient for practical computations, and is also convenient for checking
	dualities which hold only at the IR fixed point.

	\item

	The $S^3$ partition function was written as an integral over the variable $\sigma$
	taking values in the Cartan subalgebra of the gauge group.
	Here $\sigma$ is the Cartan part of the vector multiplet scalar,
	which is also the variable parametrizing the Coulomb branch;
	for this reason the integral of the partition function is sometimes described as
	an integral over the Coulomb branch.

	Note, however, that what is usually called the Coulomb branch is the vacuum moduli space of the theory
	on the flat spacetime $\mathbb{R}^{2,1}$, which is very different from
	the vacuum moduli space of the theory placed on the curved spacetime $S^3$.
	Indeed, on $S^3$, for example, scalar fields acquire masses due to the curvature of $S^3$
	(roughly $1/r$, where $r$ is the radius of $S^3$), and the Higgs branch is lifted.
	Moreover, when we introduce an FI parameter, for example,
	the Coulomb branch in flat spacetime is lifted, whereas the integral over $\sigma$
	in the partition function still runs over all real numbers.

	\item

	So far we have considered only the $S^3$ partition function;
	we can also consider expectation values of physical observables.
	As long as the observables are preserved by the supercharge $\scQ$,
	the localization argument goes through.

	As an example, we can consider a gauge field (it can be a gauge field for either a gauge symmetry or a flavor symmetry)
	and its Wilson loop in a representation $R$.
	The definition of the Wilson loop was given in \eqref{Wilson};
	here we also combine it with the Coulomb branch scalar $\sigma$ in order to preserve supersymmetry:
	\begin{align}
		\langle W \rangle=\textrm{Tr}_R\, \exp\left( \int_{S^1}  \! ds \,\,\left[
			                                                                   i A_{\mu} \dot{x}^{\mu} +\sigma |\dot{x}|
			                                                                   \right] \right) \ .
		\label{Wilson_line_S3}
	\end{align}
	Here $S^1$ is the $S^1$ fiber located at either of the two end points $\theta=0, \frac{\pi}{2}$
	of the base interval, when we regard $S^3$ as a $T^2$-fibration
	(Fig.~\ref{fig.S3fibration}).\footnote{When the value of $b^2$ is rational,
		we can consider more general Wilson loops \cite{Tanaka:2012nr}.}
	Namely, it is the $S^1$ parametrized by $\chi$ ($\varphi$) at $\theta=0$ ($\theta=\frac{\pi}{2}$).
	Since $A_{\mu}=0$ and $\sigma=\textrm{constant}$ at the saddle point, the contribution to the partition function is
	\begin{align}
		Z_{\textrm{Wilson loop}}=\textrm{Tr}_R\, \exp\left(2\pi b^{\pm 1}  \sigma \right) \ ,
	\end{align}
	where for the sign $\pm$ we choose
	plus (minus) for $\theta=0$ ($\theta=\frac{\pi}{2}$).
	\small

	That \eqref{Wilson_line_S3} preserves part of the supersymmetry can be checked
	by directly acting with the supersymmetry transformations.
	The supersymmetry transformations of $A_{\mu}, \sigma$ on $S^3$ are the same as those in flat spacetime
	(except that we have Euclideanized):\footnote{We can also consider the other
		supersymmetry transformation, generated by $\epsilon$:
		\begin{align}
			\delta A_{\mu}=- \frac{i}{2} \overline{\lambda} \sigma_{\mu} \epsilon \ , \quad
			\delta \sigma= - \frac{1}{2} \overline{\lambda} \epsilon\ ;
		\end{align}
		the conclusion is the same.
	}
	\begin{align}
		\delta A_{\mu}=\frac{i}{2} \overline{\epsilon} \sigma_{\mu} \lambda \ , \quad
		\delta \sigma= - \frac{1}{2} \overline{\epsilon} \lambda\ .
	\end{align}
	This is the dimensional reduction of the corresponding 4d $\mathcal{N}=1$ expression
	\begin{align}
		\delta A_{\hat{\mu}}=\frac{i}{2} \overline{\epsilon} \sigma_{\hat{\mu}} \lambda \ , \quad
		\hat{\mu}=0,1,2,3
	\end{align}
	(recall that $(\sigma_{3})_{\alpha \beta}=i\epsilon_{\alpha\beta}$
		\eqref{sigma_matrix}, and that spinor indices are raised and lowered by $\epsilon^{\alpha\beta}$ \eqref{spinor_raise_lower}). We therefore have
		\begin{align}
			\delta \left( i A_{\mu} \dot{x}^{\mu} +\sigma |\dot{x}| \right)
			=-\frac{1}{2}\, |\dot{x}| \, \overline{\epsilon} (P+1) \lambda \ ,
		\end{align}
		where $P:=\frac{\sigma^{\mu} \dot{x}_{\mu} }{ |\dot{x}| }$ squares to the identity, $P^2=1$ (so that $\frac{1\pm P}{2}$ are projection operators), when
	$\frac{\dot{x}_{\mu}}{{|\dot{x}|}}$ is constant. Therefore,
		among the two supersymmetry spinors $\overline{\epsilon}$ remaining on $S^3$,
		the half satisfying $ \overline{\epsilon} (P+1) =0$ is preserved in the presence of the Wilson line.

		\normalsize

		We can also compute the expectation values of
		\keyword{vortex loops}{vortex loop}, which are dual to Wilson loops.
		Vortex loops act on the $S^3_b$ partition function as difference operators
		\cite{Kapustin:2012iw,Drukker:2012sr}; they do not commute with Wilson loops,
		and together they form an algebra (the algebra of supersymmetric loop operators).

		\item
		By changing the topology of the manifold $S^3$, we can define other partition functions;
		for example, the $S^2\times S^1$ partition function \cite{Kim:2009wb,Imamura:2011su}
		and the partition function on the \keyword{lens space}{lens space} $S^3/\mathbb{Z}_p$ have been considered.
	These partition functions have natural decompositions,
	in which sums rather than integrals appear.
	We will comment on this decomposition later, for example in Sec.~\ref{sec.6d_revisited}.

\end{Kenumerate}


\section{Translation into Partition Functions}

Since the partition function is a convenient tool,
the best thing is to use it.
Let us examine concretely how the contents discussed in the previous chapter
are translated into the language of the $S^3$ partition function.

\bparagraph{Complex Mass}

First, when we have chiral superfields with a complex mass,
what is their $S^3_b$ partition function?
This question is easily answered using physical considerations.
First, recall that the $S^3_b$ partition function does not depend on the gauge coupling constant, and hence
on the RG scale. Next, once this scale becomes sufficiently smaller than the mass of the chiral superfields,
the chiral superfields can be regarded as having infinite mass,
and we can first perform the path integral over them and remove them
(this is often expressed as ``integrate out'').
Their $S^3_b$ partition function should therefore be trivial.

Let us check this explicitly.
Let us represent the complex mass by the superpotential
$W=m_c XY$,
where $X$ and $Y$ are chiral superfields; in particular, if we choose
$X=Y$, this is the familiar mass term.

The contribution from the chiral superfields $X, Y$ to the one-loop determinant is,
according to \eqref{chiralZ},
\begin{align}
	Z_{XY}=
	s_b\left(
	\frac{iQ}{2}(1-r_X)-q_X \mu
	\right)
	s_b\left(
	\frac{iQ}{2}(1-r_Y)-q_Y \mu
	\right) \ .
	\label{eq.XY}
\end{align}
Here, since the superpotential has to have R-charge $2$ and flavor charge $0$,
we have
\begin{align}
	r_X+r_Y=2 \ , \quad q_X+q_Y=0 \ ,
\end{align}
so that the sum of the arguments of the two quantum dilogarithm functions in \eqref{eq.XY} is $0$,
and the relation \eqref{sbinv} implies $Z_{XY}=1$, as expected.

\bparagraph{Parity Anomaly}

Let us next see
what happens in the $S^3_b$ partition function
when we make the real mass $\mu$ large.
When we take $\mu\to \infty$ in the partition function \eqref{chiralZ},
the asymptotic form of the function $s_b(x)$
(\eqref{sb_asymp} in the appendix) depends on the sign of the charge $q$,
and we obtain (ignoring for now the overall constant factor)
\begin{align}
	s_b\left( \overline{\sigma} \right) \xrightarrow[\overline{\sigma}\to \infty]{} e^{-\frac{i \pi}{2}
			                                                                                \overline{\sigma}^2} \ .
\end{align}
This can be interpreted as representing the quantum correction (parity anomaly) \eqref{parity_anomaly}
to the Chern-Simons term with level $k_{\rm eff}=\frac{1}{2}$.
In other words, even without knowing anything about the derivation of the parity anomaly
as in exercise~\ref{parity_derivation}, the same fact has been derived from mathematical manipulations alone.

\small
Since $\overline{\sigma}$ includes the effects not only of flavor symmetries but also of the R-symmetry
(see \eqref{bar_sigma} and the explanation there), note that the parity anomaly here
is not restricted to Chern-Simons terms for flavor symmetries, but includes
diagonal Chern-Simons terms for the R-symmetry, as well as
off-diagonal Chern-Simons terms between flavor symmetries and the R-symmetry
(this is also reflected in the discussion around \eqref{Delta_level} in Chap.~\ref{chap.3mfd} later).

\normalsize

\bparagraph{Mirror Symmetry}

As already stated, dualities of field theories imply identities of partition functions.
Let us write down the concrete identity in the case of
the mirror symmetry discussed in the previous chapter.

The fields appearing in the two theories and their charges under the global symmetries
are summarized in Table~\ref{tab.SQED} and Table~\ref{tab.charge2} of the previous chapter.
First, for $N_f=1$ SQED, note that two quantum dilogarithm functions appear from the two fields $q, \overline{q}$,
and that the monopole operators $\scV_{\pm}$ do not contribute to the $S^3$ partition function,
since they are not fields appearing in the Lagrangian.
Denoting the real masses for $U(1)_J$ and $U(1)_A$ by $\zeta$ and $\mu$, respectively,
the partition function is
\begin{align}
	\begin{split}
		Z_{\textrm{SQED}} & =\int\! d\sigma\,  e^{-2\pi i \zeta \sigma} s_b\left(\frac{iQ}{2}+\sigma-\mu\right)
		s_b\left(\frac{iQ}{2}-\sigma-\mu\right) \ .
	\end{split}
\end{align}
On the other hand, in the XYZ model there are three chiral multiplets, and correspondingly
three quantum dilogarithms appear; since this is not a gauge theory, no integral is needed:
\begin{align}
	\begin{split}
		Z_{\textrm{XYZ}} & =
		s_b\left(\frac{iQ}{2}-2\mu\right) s_b\left(\mu+\zeta\right) s_b\left(\mu-\zeta\right) \ .
	\end{split}
\end{align}
In this case, the mathematical identity derived from the duality is
\begin{align}
	Z_{\textrm{SQED}}=Z_{\textrm{XYZ}} \ .
	\label{ZQED_ZXYZ}
\end{align}
This formula can be proven by setting
$-s=r=\frac{iQ}{2}-\mu$ in the \keyword{pentagon identity}{pentagon identity} of the quantum dilogarithm (\eqref{Ramanujan} in the appendix).

\small
To be precise, the two sides of \eqref{ZQED_ZXYZ} may differ by
a constant factor (independent of $\mu, \zeta$) (see the comment on the gravitational Chern-Simons term at the end of Sec.~\ref{subsec.functor}). In this book we do not discuss constant factors of partition functions (independent of the real masses and the levels of the Chern-Simons terms).
\normalsize

\subsection{Dimensional Reduction}\label{sec.3d_dim_red}

So far we have obtained a partition function depending on one parameter $b$.
When we obtain a quantity with such a parameter,
it is in general important to consider what happens at special values or in limits of the parameter.

As such a limit, in the rest of this chapter
let us consider the limit $b\to 0$ of the $S^3_b$ partition function.
As shown in \eqref{sblimit} of Appendix~\ref{app.dilog},
mathematically this is the limit where
the quantum dilogarithm function reduces to the classical dilogarithm function.
Since the quantum dilogarithm function diverges as $b\to 0$,
the integrand of the $S^3_b$ partition function, which is an integral of their products,
also diverges.
Namely, we find that there exists a function $\mathcal{W}(\vec{\sigma})$ such that
\begin{align}
	Z_{S^3_b} \xrightarrow[b\to 0]{} \int d\vec{\sigma} \, \exp\left[
		                                                           \frac{1}{2\pi i b^2} \mathcal{W}(2\pi b \vec{\sigma}) +\mathcal{O}(b^0)\right] \ ,
	\label{Zdiverge}
\end{align}
where we kept the combination $2\pi b \sigma$ finite as $b\to 0$.
Since we have already removed the IR and UV divergences of the partition function,
why does a divergence appear again?

\begin{figure}[t]
	\centering\includegraphics[scale=0.3]{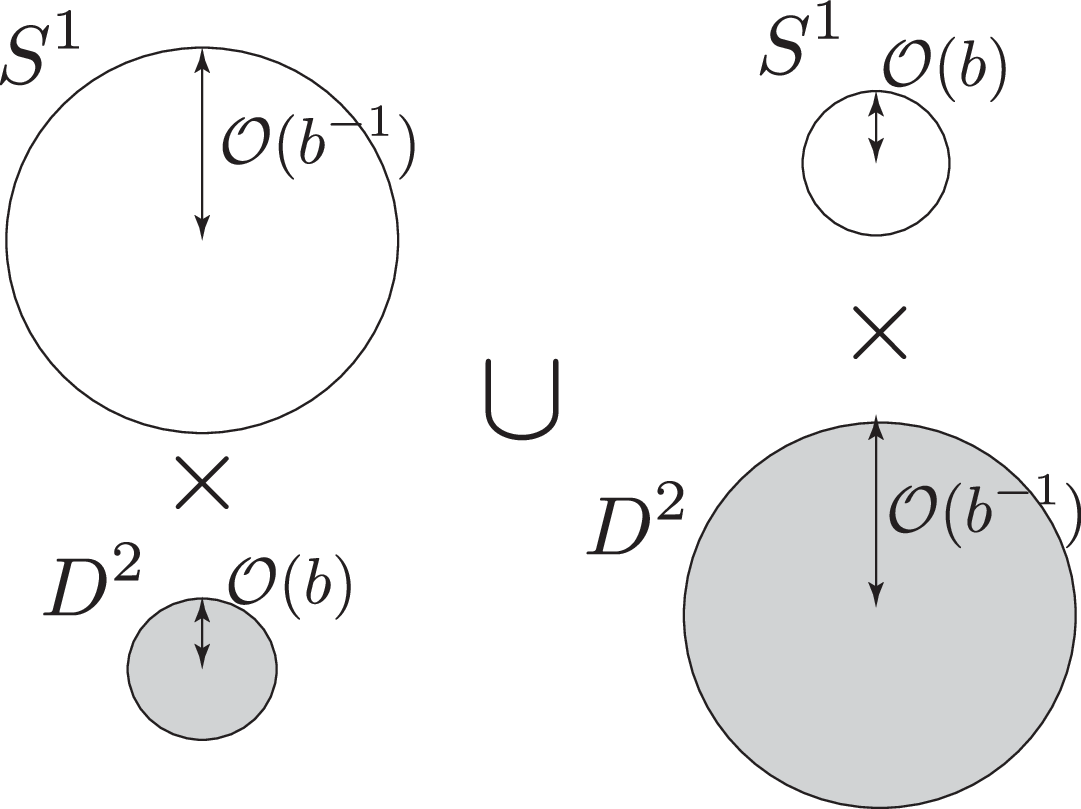}
	\caption{$S^3$ is obtained by gluing two solid tori $S^1\times D^2$
		along their boundaries $T^2$ by an $S$-transformation. In the limit $b\to 0$,
		one of the solid tori becomes $S^1_b\times D^2_{b^{-1}}\to S^1_b\times \bR^2$,
		and the 3d field theory is dimensionally reduced to two dimensions.}
	\label{fig.S3Heegaard}
\end{figure}

This divergence has a physical meaning.
To see this, recall from the fibration structure of $S^3_b$ (Fig.~\ref{fig.S3fibration}) that
the parts $\theta\in [0, \frac{\pi}{4}]$ and $\theta\in [\frac{\pi}{4}, \frac{\pi}{2}]$ are
each given by a solid torus $S^1\times D^2$,
and that $S^3_b$ is obtained by gluing their two boundaries $T^2$
by an $S$-transformation (Fig.~\ref{fig.S3Heegaard}).
This is the \keyword{Heegaard decomposition}{Heegaard decomposition} of $S^3$ (see Chap.~\ref{chap.3mfd}
for more general 3-manifolds).

Now, when we take $b$ to be small,
the two solid tori scale as
$S^1_b\times D^2_{b^{-1}}$ and
$S^1_{b^{-1}}\times D^2_{b}$.
In particular, the area of $D^2_{b^{-1}}$ scales as $b^{-2}$ and diverges,
which is reflected in the power of $b$ in \eqref{Zdiverge}. Namely, formally we can write
\begin{align}
	S^3_b \rightarrow \mathbb{R}^2\times S^1_b \ .
	\label{S3_b_limit}
\end{align}

We thus find that
\textbf{in the classical limit $b\to 0$ of the quantum dilogarithm, $S^3_b$ approaches $\mathbb{R}^2\times S^1_b$,
	where the radius of $S^1$ is given by $b$}.
Namely, \textbf{the classical limit of the quantum dilogarithm is the $S^1$-compactification
	of 3d $\mathcal{N}=2$ theories to two dimensions}.

The $S^1$-compactification of 3d $\mathcal{N}=2$ theories gives 2d
$\mathcal{N}=(2,2)$ theories.
Here, since in 2d the rotation group is $SO(2)$ and its irreducible representations are
one-dimensional (labeled by the helicity), the 3d supercharges $\scQ_{\alpha}, \overline{\scQ}_{\alpha}$
become $\scQ_{+}, \scQ_{-}, \overline{\scQ}_{+}, \overline{\scQ}_{-}$;
$(2,2)$ means that there are two supercharges each with $+$/$-$ helicity.
The Lagrangian of this theory is obtained by the dimensional reduction of
the Lagrangian of the 3d $\mathcal{N}=2$ theory.
For 2d $\mathcal{N}=(2,2)$ theories,
see for example Ref.~\cite{MirrorBook}.

Note that in \eqref{Zdiverge} we kept $2\pi b \sigma$ finite,
rather than $\sigma$, when taking the limit $b\to 0$.
This is because what we should consider in the 2d limit is not the parameter $\sigma$ of the 3d theory,
but the parameter of the 2d theory obtained by dimensional reduction.
Denoting the radius of the $S^1$ by $R$ (as already explained, $R=b$ here),
the natural variable of the 2d theory is
\begin{align}
	\sigma_{\rm 2d}:=2\pi R\, \sigma=2\pi b\, \sigma \ ,
	\label{sigma_2d}
\end{align}\nomenclature{$\sigma_{\rm 2d}$}{2d variable obtained from $\sigma$ by dimensional reduction ($=2\pi b \sigma$)}
and \eqref{Zdiverge} is the statement that we keep $\sigma_{\rm 2d}$ finite;
the function $\mathcal{W}(2\pi b\vec{\sigma})$ in \eqref{Zdiverge} is $\mathcal{W}(\vec{\sigma}_{\rm 2d})$.

The function $\mathcal{W}(\vec{\sigma})$ depends only on the vector multiplet scalars,
and is called the \keyword{effective twisted superpotential}{effective twisted superpotential}.
In a 2d $\mathcal{N}=(2,2)$ theory, if all the matter fields are
massive, then at energy scales below their mass scale
we can first integrate out the matter fields,
and obtain a theory with only the gauge fields and their superpartners.
The theory thus obtained has a
\textbf{twisted superpotential}, which is a term in the Lagrangian allowed
in 2d $\mathcal{N}=(2,2)$ theories:
\begin{align}
	\int d\theta_{+} d\overline{\theta}_{-}\,\,  \scW(\Sigma_{\rm 2d}) +
	\int d\theta_{-} d\overline{\theta}_{+}\,\,  \overline{\scW}(\overline{\Sigma}_{\rm 2d}) \ .
\end{align}
\nomenclature{$\scW(\Sigma)$}{twisted superpotential}
Here $\Sigma_{\rm 2d}$ is the superfield defined in two dimensions:
\begin{align}
	\Sigma_{\rm 2d}:=\overline{D}_{+} \left( e^{-V}D_{-} e^V \right) \ ,
	\quad
	\overline{\Sigma}_{\rm 2d}:=D_{+} \left( e^{V}\overline{D}_{-} e^{-V} \right) \ ,
	\label{Sigma_2d}
\end{align}
so that $\overline{\Sigma}_{\rm 2d}$ is the complex conjugate of $\Sigma_{\rm 2d}$
(for an Abelian gauge group they reduce to $\overline{D}_{+}D_{-}V$ and $D_{+}\overline{D}_{-}V$).
This is related, up to normalization, to the dimensional reduction of the linear multiplet $\Sigma$ of \eqref{SigmaV}
as $\Sigma\leftrightarrow \Sigma_{\rm 2d}+\overline{\Sigma}_{\rm 2d}$,
and the lowest component of $\Sigma_{\rm 2d}$ is
a complex scalar, whose real part is (up to normalization) the variable $\sigma_{\rm 2d}$ of \eqref{sigma_2d}.
In 3d language, this scalar is
the combination $\sigma+i\gamma$ appearing in the definition of the monopole operator \eqref{monopoleop};
in particular, its imaginary part has period $g^2$ (exercise~\ref{2d_Sigma}).

The explicit form of the twisted superpotential can be computed from \eqref{Zdiverge},
and is given by the sum of a quadratic polynomial and classical dilogarithm functions:
\begin{align}
	\scW(\vec{\sigma})=\textrm{(quadratic polynomial in $\sigma$)} + \sum \textrm{$\mathrm{Li}_2$} \ .
	\label{scW_sum}
\end{align}
The meaning of the quadratic part is simple: it originates from the quadratic terms already present
in 3d ($S^3_b$) (see \eqref{ZCS_diagonal} and \eqref{ZCS_offdiagonal}),
namely it is the dimensional reduction of the Chern-Simons terms.

What about the classical dilogarithm part, which is the limit of the quantum dilogarithm function (see \eqref{sblimit})?
The clue here is that the linear part in the $S^3$ partition function
represented the FI parameter (see \eqref{FI_S3}). The FI parameter remains
an FI parameter $\zeta_{\rm 2d}$ after dimensional reduction;
however, in 2d $\mathcal{N}=(2,2)$ theories it is complexified by the theta angle of the gauge field\footnote{In 3d, the FI parameter could be understood as an off-diagonal Chern-Simons term for the $U(1)_J$ global symmetry. The off-diagonal Chern-Simons term $\int A\wedge F$ gives $\int F$ after dimensional reduction.}
\begin{align}
	\mathcal{L}_{\textrm{theta angle}}= \frac{\theta_{\rm 2d}}{2\pi} \, F_{01} \ .
\end{align} Let us denote this by $t$:
\begin{align}
	t:=\zeta_{\rm 2d}-i \theta_{\rm 2d} \ .
\end{align}

The reason we have discussed the FI parameter is that,
when the twisted superpotential contains classical dilogarithm functions,
their effect can be understood as changing the value of the FI parameter $t$:
\begin{align}
	t_{\textrm{eff FI}}(\sigma_{\rm 2d}):= \frac{\partial \scW(\sigma_{\rm 2d})}{\partial \sigma_{\rm 2d}} \ .
	\label{t_eff}
\end{align}
Substituting $\scW\sim \textrm{Li}_2(e^{\sigma_{\rm 2d}})$ here (see \eqref{sblimit}; whether we have $e^{\sigma_{\rm 2d}}$ or $e^{-\sigma_{\rm 2d}}$ depends on the sign of the charge),
we obtain, up to an overall sign and a term linear in $\sigma_{\rm 2d}$,
\begin{align}
	\begin{split}
		t_{\textrm{eff FI}}(\sigma_{\rm 2d}) & \sim \log \sinh\left(\frac{\sigma_{\rm 2d}}{2}\right)
		\sim \sum_{n\in \mathbb{Z}} \log \left(\sigma_{\rm 2d}+2\pi i n \right) \\
		                                     & =\sum_{n\in \mathbb{Z}} \log \left[2\pi R \left(\sigma+\frac{i n}{R} \right)\right] \ ,
	\end{split}
	\label{t_eff_sum}
\end{align}
where we used the infinite product representation of the hyperbolic sine function
\begin{align}
	\sinh(z)= z\prod_{n=1}^{\infty}
	\left(
	1+\frac{z^2}{\pi^2 n^2}
	\right) \ .
\end{align}
The last expression \eqref{t_eff_sum} can be understood from the $S^1$-compactification of the 3d theory: the sum over integers $n$
represents the sum over \keyword[Kaluza-Klein (KK) modes]{KK modes}{Kaluza-Klein modes} (Fourier modes along the $S^1$-direction) with masses $n/R$,
and the term $\log (\sigma_{\rm 2d})$ for $n=0$ represents the one-loop correction to the FI parameter
when the matter fields are integrated out.\footnote{By the non-renormalization theorem for $\mathcal{N}=(2,2)$ theories,
	this result is correct to all orders in perturbation theory. Moreover, since there are no non-perturbative effects when the gauge group is Abelian, this result is expected to be exact \cite{deBoer:1997kr}.}
We have thus understood the structure of \eqref{scW_sum}.

Once the twisted superpotential is obtained,
we can determine the vacua of the 2d $\mathcal{N}=(2,2)$ theory:
\begin{align}
	\exp\left(
	\frac{\partial \scW(\vec{\sigma}_{\rm 2d})}{\partial \vec{\sigma}_{\rm 2d}}
	\right)=1  \ .
	\label{2dvac}
\end{align}
In the language of the effective FI parameter \eqref{t_eff}, this can be interpreted as representing
the periodicity of its imaginary part, the theta angle $\theta_{\rm 2d}$.\footnote{More physically, the theta angle in 2d induces a constant electric field in the vacuum,
	but part of its effect is canceled by pair creation, and a periodicity appears \cite{Coleman:1976uz}. As clearly explained in Ref.~\cite{Coleman:1976uz}, this is special to two dimensions.} The vacua of the 2d $\mathcal{N}=(2,2)$ theory thus obtained (a deformation of a conformal field theory by mass parameters\footnote{This mass is the 2d version of the real mass in 3d, and is called the \keyword{twisted mass}{twisted mass}.}) have no moduli, and form a set of isolated points.

\normalsize

By taking limits of the 3d $S^3$ partition function,
we can derive various properties of 2d twisted superpotentials.
We leave it as an exercise to carry this out for each of the properties and dualities
discussed in the previous chapter.

For example, one can show that the pentagon identity of the quantum dilogarithm function,
which represents the 2--3 mirror symmetry discussed in the previous chapter,
reduces in the limit $b\to 0$ to the pentagon identity of the classical dilogarithm function (see Appendix~\ref{app.dilog}).

\bigskip

As we have seen so far, the $S^3$ partition function is
a quantity which is relatively easy to handle, and contains a considerable amount of information
about 3d $\scN=2$ theories. In particular, the $S^3$ partition function is written in terms of quantum dilogarithm functions,
and \textbf{relations among quantum dilogarithm functions can be reinterpreted as statements about 3d $\scN=2$ theories}.
In fact, if we pursue this method thoroughly, a considerable part of the 3d-3d correspondence
can be inferred solely from known mathematical identities of quantum dilogarithm functions (see Chap.~\ref{chap.3mfd}).
Confident readers are encouraged to think about what can be done using only the contents so far and Appendix~\ref{app.dilog}, before reading on to the next chapter (exercise~\ref{ex.sb_translate}). In this book, however, we take a somewhat different route,
and return once again to the concept of gauging described in Chap.~\ref{chap.intro}.
In the process, it will gradually become clear what kind of physical and mathematical structures
lie behind the identities of dilogarithm functions.

\begin{practice}

	\item $[\bll]$ ($S^3_b$ partition function of 3d $\mathcal{N}=4$ theories)
	Find the one-loop determinants in 3d $\mathcal{N}=4$ theories.
	In particular, verify that
	\begin{align}
		Z_\textrm{$\scN=4$ vector multiplet}=Z_\textrm{$\scN=2$ vector multiplet} \ ,
	\end{align}
	and that, for $b=1$,
	\begin{align}
		Z_\textrm{$\scN=4$ hypermultiplet}=\frac{1}{2\cosh \pi \sigma} \ .
	\end{align}
	Here the R-charge of the hypermultiplet is fixed to $r=\frac{1}{2}$
	by the $\mathcal{N}=4$ R-symmetry.
	Hint: for the latter, regard the hypermultiplet as a pair of
	$\mathcal{N}=2$ chiral multiplets with weights $\pm\sigma$ and R-charges $r=\frac{1}{2}$,
	apply \eqref{chiralZ} to each of them, and combine the results
	using \eqref{sbinv} and \eqref{recursion}.
	The former means that the partition function of the $\mathcal{N}=2$ multiplet
	inside the $\mathcal{N}=4$ vector multiplet is trivial. Recalling that the R-charge of this $\mathcal{N}=2$ multiplet is $1$ (which is related to the fact that the R-charge of the gauge field has to be $0$), this reduces to the discussion around \eqref{eq.XY}. For the multiplets of 3d $\mathcal{N}=4$ supersymmetry, see exercise~\ref{N4ex} and Appendix~\ref{chap.SUSYbasic}.

	\item $[\bll]$ (Example of a duality and an identity of $S^3_b$ partition functions)
	Verify mathematically that the identity
	\begin{align}
		\int d\sigma \frac{e^{-2\pi i \zeta \sigma}}{2\cosh \pi \sigma}
		=\frac{1}{2\cosh \pi \zeta}
	\end{align}
	holds. Also verify that this formula represents a duality between two 3d
	$\mathcal{N}=4$ supersymmetric gauge theories at the level of the $S^3_{b=1}$ partition function.

	\item $[\bll \bll]$ (More general Abelian mirror symmetry) Verify at the level of the $S^3_{b=1}$ partition function that the following two 3d $\mathcal{N}=2$ theories are dual (this duality is due to Ref.~\cite{Dorey:1999rb}).

	In one theory, the gauge group is $U(1)^r$, and there are $N$ chiral multiplets $q_i$ with charges $R_i^a$ ($i=1, \cdots, N, \, a=1, \cdots, r$). In this theory we can consider Chern-Simons terms $\kappa_{ab}$, $r$ FI parameters $\zeta_a$, and $N-r$ real masses $m_i$. Here $m_i$ themselves are $N$ parameters, but $r$ of them can be absorbed by gauge transformations, so that only $N-r$ of them are non-trivial parameters.

	In the other theory, we replace $r$ by $N-r$. Namely, the gauge group is $U(1)^{N-r}$, and there are $N$ chiral multiplets $\overline{q}_i$ with charges $S_i^u$ ($i=1, \cdots, N, \, u=1, \cdots N-r$). In this theory we can consider Chern-Simons terms $\overline{\kappa}_{uv}$, $N-r$ FI parameters $\overline{\zeta}_u$, and $r$ real masses $\overline{m}_i$.

	For these two theories to be dual, the condition
	\begin{align}
		\sum_{i=1}^N R_i^a S_i^u=0
	\end{align}
	is necessary. Moreover, the parameters need to be appropriately identified, for example $\zeta_a$ with $\overline{m}_i$, and
	$m_i$ with $\overline{\zeta}_u$.

	\item $[\bll]$ ($S^3_{b=1}$ partition function of ABJM theory)
	Let us consider the following theory, called the 3d ABJM (Aharony-Bergman-Jafferis-Maldacena)
	theory \cite{Aharony:2008ug}. In the language of 3d $\mathcal{N}=2$ theories,
	the gauge group is $U(N)_k\times U(N)_{-k}$, where the two gauge groups have levels $k$ and $-k$.
	Let us consider chiral superfields $A_1, A_2$ transforming as $(N, \overline{N})$ under this gauge group,
	and fields $B_1, B_2$ transforming as $(\overline{N}, N)$.
	With an appropriate choice of the superpotential, this theory
	in fact has 3d $\mathcal{N}=6$ supersymmetry.
	Find the form of the corresponding $S^3_{b=1}$ partition function, and show that it is given by
	\begin{align}
		\begin{split}
			 & Z_{\rm ABJM} =
			\frac{1}{(N!)^2}
			\int \prod_{j=1}^N \frac{d\sigma_j}{2\pi}
			\int \prod_{j=1}^N \frac{d\rho_j}{2\pi}
			\\
			 & \quad
			\frac{\prod_{1\le j<k \le N}  \left(2 \sinh \pi (\sigma_j-\sigma_k) \right)^2
				\left(2 \sinh \pi (\rho_j-\rho_k) \right)^2
			}{\prod_{1\le j, k \le N} \left(2 \cosh \pi (\sigma_j-\rho_k) \right)^2 }
			e^{-i\pi k \left( \sum_{j=1}^N  \sigma_j^2-\sum_{j=1}^N \rho_j^2 \right)} \ ,
		\end{split}
	\end{align}
	where we set all the real masses for the global symmetries to zero.
	Next, perform the integral for $k=1$ and $N=1,2$ (for example, by using the residue theorem).
	Hint: see the paper \cite{Okuyama:2011su}. The author has a personal attachment to this problem, having also
	considered the case of more general $N$ in Ref.~\cite{Putrov:2012zi}. See e.g.\ the review \cite{Hatsuda:2015gca}
	for recent developments.

	\item $[\bll]$  (Numerical evaluation of quantum dilogarithm functions)
	For some of the quantum dilogarithm identities in Appendix~\ref{app.dilog},
	verify numerically (by choosing suitable values of the variables) that
	the identities, e.g.\ the pentagon identity, hold.
	In general it is not easy to prove whether an identity of $s_b(x)$ holds,
	but such numerical checks are useful in practice.

	Note that the $S^3$ partition function is in general written as an integral of quantum dilogarithm functions;
	when $b=1$, the quantum dilogarithm function can be written in terms of the classical dilogarithm
	function $\textrm{Li}_2(x)$ (see \eqref{b1}). This is
	a convenient limit; in practice, however, the quantum dilogarithm function at $b=1$
	has a rapidly oscillating phase, so that numerical computations at $b=1$
	require some ingenuity.\footnote{See \cite{Gang:2018huc} for a related discussion.}

	\item $[\bll]$  (Dimensional reduction of Chern-Simons terms)
	Consider the Chern-Simons terms of 3d $\mathcal{N}=2$ theories and their dimensional reduction.
	Find the resulting twisted superpotential of the 2d $\mathcal{N}=(2,2)$ theory.

	\item $[\bll]$ ($SL(2, \mathbb{Z})$-transformations in 2d)
	The $SL(2, \mathbb{Z})$ of 3d theories induces, after dimensional reduction,
	2d $SL(2, \mathbb{Z})$-transformations.
	How does this act on the twisted superpotential?
	Verify that this action is consistent with the result obtained from the $b\to 0$ limit of the $S^3$ partition function.

	\item $[\bll]$ (2d twisted chiral multiplet)\label{2d_Sigma}
	When we define a 2d multiplet by \eqref{Sigma_2d},
	show that it is identified with the dimensional reduction of the 3d linear multiplet $\Sigma$ in \eqref{SigmaV}
	as $\Sigma\leftrightarrow \Sigma_{\rm 2d}+\overline{\Sigma}_{\rm 2d}$.
	Also expand it in components, and show that the lowest component of $\Sigma_{\rm 2d}$ is,
	in 3d language, the combination $\sigma+i\gamma$ appearing in the definition of the monopole operator \eqref{monopoleop}.
	Using the periodicity of the dual photon $\gamma$ shown in Chap.~\ref{chap.3dglue},
	check also that its imaginary part has period $g^2$.

	\item $[\bll\bll]$ (Translating between quantum dilogarithm functions and 3d $\mathcal{N}=2$ theories)
	For each of the properties of the quantum dilogarithm functions
	discussed in Appendix~\ref{app.dilog} (and references therein),
	come up with an interpretation in the language of 3d $\mathcal{N}=2$ theories.

	\label{ex.sb_translate}

\end{practice}


\chapter{Physics of Domain Walls}\label{chap.wall}

\begin{abstract}
	In this chapter, we will see that the $S^3_b$ partition function
	discussed in the previous chapter can be reinterpreted as a wave function
	of a 1d quantum mechanics.
	We will next explain that the wave function interpretation exists
	because our 3d $\mathcal{N}=2$ theories
	arise on the boundary (domain wall) of 4d $\mathcal{N}=2$ theories.
	In this setup, the $Sp(2n, \bZ)$-transformation of the 3d theory is identified with the
	duality transformation of the 4d theory,
	and when we supersymmetrize the whole setup,
	the $S^3_b$ partition function is identified with an operator
	acting on the $S^4_b$ partition function.
\end{abstract}

\section{$SL(2, \mathbb{Z})$-Action Revisited}\label{subsec.wave_function}

In the previous chapter we introduced the $S^3_b$ partition function as a tool to study
3d $\scN=2$ theories. One crucial ingredient in the discussion of
Chaps.~\ref{chap.3dglue} and~\ref{chap.3dN2} was the $SL(2, \mathbb{Z})$-action.
It is therefore natural to ask what the $SL(2, \mathbb{Z})$-action looks like
when viewed through the filter of the $S^3_b$ partition function.

Following the notation of Chap.~\ref{chap.3dglue},
we consider a 3d $\mathcal{N}=2$ theory coupled to the
background gauge field $A^{\rm bgd}$ for a $U(1)$ global symmetry.
Let us denote by $\sigma$ the Coulomb branch scalar of the vector multiplet
$V_{\rm bgd}$ containing the gauge field $A^{\rm bgd}$.
We also denote the $S^3_b$ partition function by $Z(\sigma)$.

The action of $T^k$ is to add a level-$k$ Chern-Simons term for $A^{\rm bgd}$,
which translates into (recall \eqref{ZCS_diagonal})
\begin{align}
	T^k:\, Z(\sigma) \to Z(\sigma)\, e^{-ik \pi \sigma^2 } \ .
	\label{ZT}
\end{align}
The action of $S$
was to add an off-diagonal Chern-Simons term.
Denoting by $\sigma'$ the Coulomb branch scalar
inside the vector multiplet containing the newly introduced background gauge field,
we obtain (recall \eqref{ZCS_offdiagonal})
\begin{align}
	S:\, Z(\sigma) \to Z'(\sigma')=\int d\sigma \,\,  e^{-2 i\pi \sigma \sigma'}
	Z(\sigma) \ .
	\label{ZS}
\end{align}
This means that \textbf{the $S$-transformation is nothing but a
	Fourier transformation}.

From these concrete expressions, we can easily check that the
actions of $S$ and $T$ on the $S^3_b$ partition function
defined above satisfy the $SL(2, \mathbb{Z})$ relations \eqref{STrel}.
This is in close parallel with the
discussion of Chap.~\ref{chap.3dglue} at the Lagrangian level.
For example, $S^2=C$ follows immediately from the fact that $S$ is a Fourier transformation,
and hence squares to the identity (up to sign). Working out the action of $S^2$,
\begin{align}
	S^2:\, Z(\sigma) \to \int d\sigma' e^{-2i\pi \sigma' \sigma''}
	\int d\sigma \,  e^{-2i\pi \sigma \sigma'}
	Z(\sigma) \ .
\end{align}
Integrating out $\sigma'$, we obtain
\begin{align*}
	\int d\sigma \,  \delta(\sigma+\sigma'')
	Z(\sigma) =Z(-\sigma'')\ ,
\end{align*}
demonstrating $S^2=C$.
That the charge conjugation $C$ reverses the sign of
$\sigma$ can be understood
from the fact that
$\sigma$ is in the same multiplet as the gauge field (and hence follows the same transformation
rule), that the gauge field in this normalization contains the charge
(the covariant derivative is $D_{\mu}=\partial_{\mu}+i A_{\mu}+\cdots $),
and that the charge flips its sign under charge conjugation.

We can similarly demonstrate $(ST)^3=1$ (exercise~\ref{ex.ST}).


\bigskip

Since $S$ and $T$ generate $SL(2, \bZ)$, we already know the action of an arbitrary element of $SL(2, \bZ)$.
Let us work out the details.
Consider for example the matrix
\begin{align}
	M=T^{k}S T^lS=
	\left(
	\begin{array}{cc}
		kl-1 & -k \\
		l    & -1
	\end{array}
	\right) \ .
\end{align}
The partition function $Z(\sigma)$ then changes as
\begin{align}
	Z_{\rm new}(\sigma'')=\int d\sigma' \, e^{-2i \pi \sigma'' \sigma'}
	e^{-il \pi \sigma'^2} \! \int d\sigma \, e^{-2i \pi \sigma' \sigma }
	e^{-ik \pi \sigma^2} Z(\sigma) \ ,
	\label{eq.TSTS}
\end{align}
where a product of $SL(2, \bZ)$ matrices is to be read from
left to right: in the example above, first $T^k$ and then $S$.
The expression \eqref{eq.TSTS} is Gaussian (quadratic) with respect to $\sigma'$, which can be integrated out,
and we obtain (apart from an overall constant factor, see exercise~\ref{ex.canonical})
\begin{align}
	Z_{\rm new}(\sigma'')
	=\int d\sigma\, e^{2 i\pi\left(
			                \frac{1}{2l} \sigma''^2
			                + \frac{1}{l}\sigma'' \sigma +\frac{1}{2}\left(\frac{1}{l}-k \right) \sigma^2
			                \right)
		                }
	\, Z(\sigma) \ .
\end{align}
The result is an integral operator whose integral kernel
is given by the exponential of a quadratic expression.

We can generalize this computation.
Consider a general $SL(2, \bZ)$ matrix
$M=
	\left(
	\begin{array}{cc}
		a & b \\
		c & d
	\end{array}
	\right)
$,
and assume that $c\ne 0$.
We do not lose generality by this assumption, since
when the $SL(2, \bZ)$ matrix
has $c=0$, $M$ can be written as
$T^k$, whose action on the $S^3_b$ partition function we have already identified.
The action of $M$ is then identified to be
\begin{align}
	Z_{\rm new}(\sigma')
	=\int d\sigma \, e^{2\pi i W_{M}(\sigma', \sigma)}\, Z(\sigma) \ ,
	\label{W1}
\end{align}
with the quadratic function $W_{M}(\sigma', \sigma)$:
\nomenclature{$W_M(\sigma', \sigma)$}{generating function of the $SL(2, \bZ)$-transformation $M$}
\begin{align}
	W_{M}(\sigma', \sigma)
	:=-\frac{1}{2}\frac{d}{c}\sigma'^2+\frac{1}{c} \sigma' \sigma
	-\frac{1}{2}\frac{a}{c} \sigma^2 \ .
	\label{W2}
\end{align}

We can verify that this gives the correct answer for our
previous example $T^k S T^lS$.
We can also check that the composition of the operators found above
is compatible with the
product of two $SL(2, \bZ)$ matrices.
Namely, given two $SL(2, \bZ)$ matrices $M_1, M_2$,
we obtain (see exercise~\ref{ex.Wcompose})
\begin{align}
	\begin{split}
		 & \int d\sigma'\, e^{2\pi i W_{M_2}(\sigma'', \sigma')} \,
		\int d\sigma\, e^{2\pi i W_{M_1}(\sigma', \sigma)}\, Z(\sigma) \\
		 & \qquad \qquad=
		\int d\sigma\, e^{2\pi i W_{M_1 M_2}(\sigma'', \sigma)}\, Z(\sigma) \ .
		\label{Wcompose}
	\end{split}
\end{align}
We can then prove \eqref{W1} and \eqref{W2}
by induction on the number of $S$'s when
an element of $SL(2, \bZ)$ is expressed as a product of
$S$ and $T$.

\bigskip
What did we learn from this computation?
It turns out that
the function $W_M(\sigma', \sigma)$ appearing in the integral kernel
is the generating function of a \keyword{canonical transformation}{canonical transformation},
something we are very familiar with in analytical mechanics.
In fact, defining the canonical momenta $p, p'$ from \eqref{W2} by
\begin{align}
	\begin{split}
		-p': & =\frac{\partial W_M(\sigma', \sigma)}{\partial \sigma'}
		=-\frac{d}{c}\sigma'+\frac{1}{c}\sigma \ ,                     \\
		p :  & =\frac{\partial W_M(\sigma', \sigma)}{\partial \sigma}
		=-\frac{a}{c}\sigma+\frac{1}{c}\sigma' \ ,
	\end{split}
	\label{pp'}
\end{align}
we obtain
\begin{align}
	\left(\sigma' \,\, p' \right)
	=
	\left( \sigma \,\, p \right)
	\left(\begin{array}{cc} a & b \\ c&d \end{array}\right) \ .
	\label{eq.spsp}
\end{align}
This is precisely the canonical transformation from the
pair of canonically conjugate variables $(\sigma,p)$
to another pair $(\sigma',p')$.

We can go further and show
that the integral transformation \eqref{W1}
is the quantum-mechanical version of the
canonical transformation.
Denoting $-2\pi i F(\sigma):=\log Z(\sigma)$,
\eqref{W1} reads
\begin{align}
	e^{-2\pi i F_{\rm new}(\sigma')}=\int d\sigma \, e^{2 \pi i
			                                                 \left(W_M(\sigma',\sigma)-F(\sigma)\right)}  \ .
	\label{W3}
\end{align}
In the saddle point approximation we obtain
\begin{align}
	p:=\frac{\partial F(\sigma)}{\partial \sigma}=\frac{\partial W_M(\sigma',\sigma)}{\partial \sigma} \ ,
\end{align}
and we are back to our previous definition \eqref{pp'}.
In this saddle point approximation,
\eqref{W3} is the Legendre transformation, familiar from thermodynamics.

More directly, we can understand the transformation
\eqref{W1} as a transformation of the wave function of a quantum-mechanical system.
The position operator $\hat{\sigma}$
and its canonical conjugate $\hat{p}$ satisfy the canonical commutation relations
\begin{align}
	[\hat{p}, \hat{\sigma}]=i \hbar \ , \quad
	                                    [\hat{\sigma}, \hat{\sigma}]=[\hat{p}, \hat{p}]=0 \ ,
\end{align}
and they act on the wave function $Z(\sigma)$ as
\begin{align}
	\hat{\sigma} Z(\sigma)=\sigma Z(\sigma) \ , \quad
	\hat{p}\, Z(\sigma)=\frac{i}{2\pi } \frac{\partial}{\partial \sigma} Z(\sigma) \ .
\end{align}
The other pair of canonically conjugate variables, $\hat{\sigma}^{\prime}$ and $\hat{p}^{\prime}$,
act on the wave function $Z_{\rm new}(\sigma')$ in a similar manner
(we have taken our ``Planck constant'' to be $\hbar=\frac{1}{2\pi}$).\footnote{
	The sign of the commutation relation is opposite to the more familiar $[\hat{x}, \hat{p}]=i\hbar$.
	This reflects the sign in the definition $-2\pi i F:=\log Z$: since $Z=e^{iS/\hbar}$ with $S=-F$,
	the momentum $p=\partial F/\partial \sigma$ defined above is minus the usual momentum $\partial S/\partial \sigma$.
	With this choice $\hat{p}$ reduces to $p$ in the saddle point approximation,
	$\hat{p}\, e^{-2\pi i F}=\frac{\partial F}{\partial \sigma} e^{-2\pi i F}$.
	The same orientation appears in quantum \Teichmuller theory, where the coordinate $\hat{Z}$ and the momentum $\hat{Z}''$ of \eqref{Z_comm_here} satisfy $[\hat{Z}, \hat{Z}'']=-2i\hbar$.}
These facts are consistent with \eqref{eq.spsp} and its operator version
under \eqref{W1}. For example, we have
\begin{align}
	\begin{split}
		 & \left( a\hat{\sigma} + c \hat{p} \right) Z_{\rm new}(\sigma')                               \\
		 & \quad
		=\int \! d\sigma \, \left(a\sigma+c \frac{\partial W_M(\sigma', \sigma)}{\partial
			                                    \sigma} \right) e^{2\pi i W_M(\sigma', \sigma)} Z(\sigma) \\
		 & \quad
		=\int \! d\sigma \, \left(a\sigma+cp \right) e^{2\pi i W_M(\sigma',
				                                             \sigma)} Z(\sigma)                             \\
		 & \quad
		=\int \! d\sigma \, \sigma' e^{2\pi i W_M(\sigma', \sigma)} Z(\sigma)
		=\hat{\sigma}' Z_{\rm new}(\sigma') \ .
	\end{split}
\end{align}
As an example, for $M=S$ the corresponding canonical transformation
exchanges position and momentum,
and \eqref{W1} reduces to the Fourier transformation, as we have seen before.

We have therefore come to the following conclusion:
\textbf{our $S^3_b$ partition function is the wave function of a state in a certain Hilbert space},
and the transformation \eqref{W1} is nothing but the
canonical transformation of the wave function.
Namely, we have
\begin{align}
	Z(\sigma)=\langle \sigma | Z\rangle \ , \quad |Z\rangle \in \scH_{S^3} \ ,
	\label{Z_sigma_overlap}
\end{align}
where we hereafter denote the Hilbert space by $\scH_{S^3}$.
In this viewpoint, the transformation rule \eqref{W1} of the partition function
is written as
\begin{align}
	\int d\sigma\, \langle \sigma' |\sigma \rangle  \langle \sigma | Z\rangle
	=\langle  \sigma' |Z \rangle \ , \quad
	\langle \sigma' | \sigma\rangle:=e^{2\pi i W_M(\sigma', \sigma)}  \ ,
\end{align}
which is the
\textbf{completeness condition for the states $|\sigma\rangle$}, familiar from
quantum mechanics:
\begin{align}
	\int d\sigma\, |\sigma \rangle \langle \sigma|=1  \ .
\end{align}
Namely, \textbf{
	the choice of the states $|\sigma \rangle$
	depends on the choice of the polarization in quantum mechanics,
	and the change of the polarization under the $Sp(2n, \bZ)$
	canonical transformation induces the
	corresponding $Sp(2n, \mathbb{Z})$-transformation
	of the 3d $\mathcal{N}=2$ theory}; here we used the fact that $Sp(2n, \mathbb{R})$ is the group of canonical transformations in quantum mechanics.

Here are some supplementary remarks.

\begin{Kenumerate}

	\item

	\textbf{We can identify the dimensional-reduction limit $b\to 0$
		of the $S^3_b$ partition function
		in Sec.~\ref{sec.3d_dim_red}
		with the semiclassical limit in this section.}
	In fact, the semiclassical limit is the saddle point approximation
	in quantum mechanics, which is exactly what we considered in
	Sec.~\ref{sec.3d_dim_red} (recall \eqref{Zdiverge}).

	\item Mathematically, we can regard \eqref{W1}
	as giving a projective unitary representation\footnote{
		In quantum mechanics a state is specified by a ray in the Hilbert space; namely,
		the state is unchanged when we multiply the wave function by a constant.
		Constant multiples are therefore ignored also in the representation,
		and the representation becomes a projective representation.
		In order to lift the projective representation to a representation in the ordinary sense,
		we need to consider the \keyword{metaplectic group}{metaplectic group}
		$Mp(2, \bR)$, which is a double cover of $Sp(2,\bR)$ \cite{ShaleLinear,WeilMetaplectic,LerayLagrangian}.
	}
	of the group of canonical transformations
	$Sp(2, \bR)$ by unitary operators on an infinite-dimensional Hilbert space.\footnote{
		This representation is faithful.
	}
	Indeed, if we denote by $\widehat{M}$ the unitary operator associated with the matrix $M$,
	then it follows from \eqref{Wcompose} that
	\begin{align}
		\widehat{M}_1\circ \widehat{M}_2
		= \widehat{M_1\cdot M_2} \ ,
		\label{Mproduct}
	\end{align}
	where $\circ$ is the product of unitary operators, and
	$\cdot$ is the product of matrices.\footnote{
		The representation of the metaplectic group as unitary operators
		appears naturally, even before quantum mechanics,
		in its forerunner, optics:
		Fresnel optics is mathematically nothing but
		a representation of the metaplectic group $Mp(4, \bR)$.
		This example is very educational physically
		(see Ref.~\cite[Chapter 1]{GuilleminSternberg}).
	}

	\item

	In standard textbook problems in quantum mechanics,
	there are in most cases natural choices of positions and momenta (for example,
	tied to how we do measurements),
	and it is not really necessary or natural to
	consider a general $SL(2, \mathbb{R})$-transformation.

	However, this is a special situation in the general discussion of quantization,
	and indeed in the quantum-mechanical systems discussed later in this book
	there are no such canonical choices.
	This means that in quantization we need to specify which variables we
	regard as positions, and which as their conjugate momenta.
	This choice is known as a \keyword{polarization}{polarization},
	and our $Sp(2n, \bR)$-transformation is a canonical transformation
	changing the polarization.

	\item

	In 3d $\scN=2$ theories we considered the $Sp(2n, \bZ)$-subgroup of the
	$Sp(2n, \bR)$-transformations.
	This originated from the quantization of the levels of the Chern-Simons terms.
	In the viewpoint of 4d $\mathcal{N}=2$ theories discussed in the next section,
	this has a different origin, namely the Dirac quantization condition of electric and magnetic charges.

\end{Kenumerate}

\section{3d Theories as Domain Walls}

\bparagraph{Domain Wall Operator}
We have seen that the $SL(2, \mathbb{Z})$-equivalence class
$[\scT]$ of a 3d $\mathcal{N}=2$ theory $\scT$
can be identified with a state $|\mathcal{T}\rangle$ in a Hilbert space.

We can push this understanding a little further, and regard the
equivalence class $[\scT]$ as an operator $\hat{\scT}$
on the Hilbert space.

\begin{figure}[t]
	\centering\includegraphics[scale=0.3]{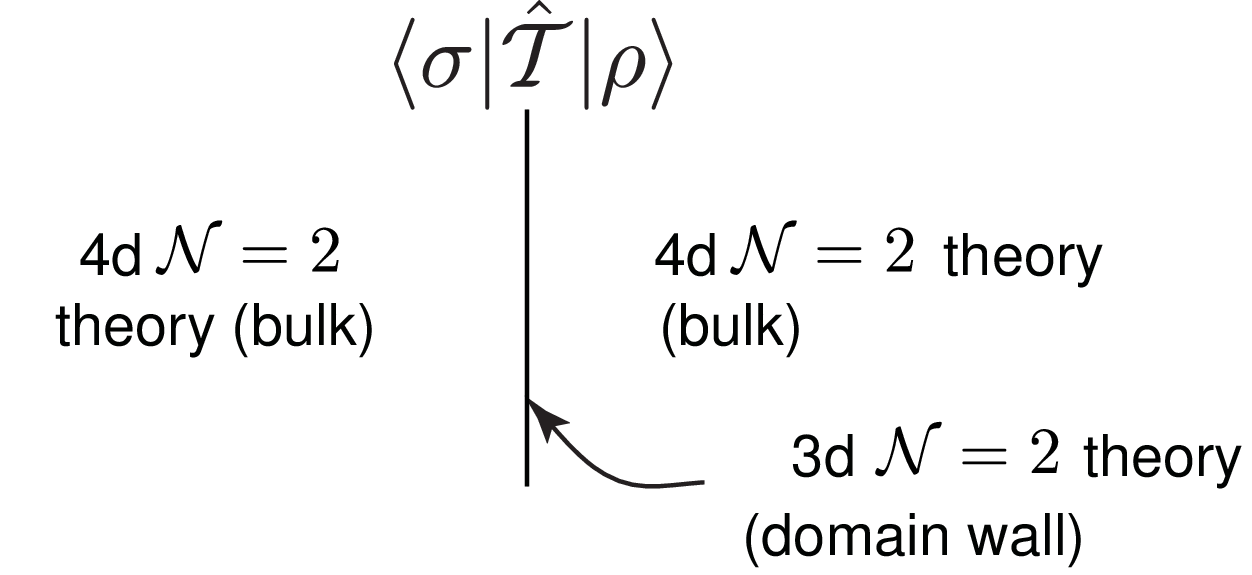}
	\caption{A 3d $\mathcal{N}=2$ theory $\mathcal{T}$ lives on a domain wall between two 4d $\mathcal{N}=2$ theories. The operator $\hat{\mathcal{T}}$ then acts on the two states $|\sigma\rangle, |\rho\rangle$ corresponding to the bulk theories, and its expectation value gives the $S^3_b$ partition function of the 3d theory.
	}
	\label{fig.overlap}
\end{figure}

Suppose that the 3d $\mathcal{N}=2$ theory has another $U(1)$ global symmetry,
in addition to the $U(1)$ global symmetry we have been discussing.
Denoting by $\rho$ the Coulomb branch scalar of the vector multiplet for this
new $U(1)$ symmetry, \eqref{Z_sigma_overlap} should now depend on two variables $\sigma, \rho$,
and hence could be written both as
$\langle \sigma | - \rangle$ and as $\langle - | \rho \rangle$.
It is therefore natural to expect the following expression:
\begin{align}
	Z(\sigma, \rho)=\langle \sigma | \hat{\mathcal{T}} | \rho \rangle \ ,
	\label{Z_sigma_overlap_2}
\end{align}
where $\hat{\mathcal{T}}$ is an operator which should be determined from the theory
$\mathcal{T}$ (see Fig.~\ref{fig.overlap}).

To summarize,
\textbf{
	the $S^3_b$ partition function of a 3d $\mathcal{N}=2$ theory $\mathcal{T}$
	(with $U(1)\times U(1)$ global symmetry)
	is given by the expectation value of an operator
	$\hat{\mathcal{T}} \in \mathrm{Aut}(\mathcal{H}_{S^3})$
	acting on the Hilbert space $\mathcal{H}_{S^3}$ of a certain quantum-mechanical system.
}

Motivated by identities of the quantum dilogarithm function,
we have now arrived at the standpoint of regarding
a 3d gauge theory $\mathcal{T}$ as an operator $\hat{\mathcal{T}}$ on a certain Hilbert space.
We can also apply this idea to concrete 3d $\scN=2$ theories.
For example, the 3d $\mathcal{N}=2$ mirror symmetry discussed in Chap.~\ref{chap.S3}
was represented by the pentagon relation of the quantum dilogarithm function;
the latter is known to follow from its operator version,
and hence the duality is naturally lifted to an identity of operators (we will comment on this later in Sec.~\ref{subsec.23mirror}).

The considerations so far have been purely mathematical, and
many questions remain unanswered.
First of all, what is this ``certain Hilbert space'', and how does it arise?
What does it mean for a 3d QFT to be a state (an operator) in a Hilbert space?
The latter in fact sounds far-fetched:
in the textbook treatment of QFT we fix a QFT, construct a Hilbert space,
and then operators act on the Hilbert space, whereas here
the QFT \textbf{itself} plays the role of an operator,
something we never hear of in textbooks.

It turns out that there are natural answers to these questions.
First, a partition function, which is a number, can be promoted to an operator
on a Hilbert space by going to one dimension higher, namely to
4 dimensions.\footnote{More mathematically,
	going to a higher dimension is a version of \keyword{categorification}{categorification}.}
This is because we can regard the newly added direction as a time direction,
and canonically quantize the 4d theory, to obtain a Hilbert space on the remaining 3d directions.

Once we obtain a Hilbert space,
an operator on it is a map from the Hilbert space to (another) Hilbert space.
Since the Hilbert space $\scH$ corresponds to a 4d QFT and the operator $\hat{\scT}$
to a 3d QFT $\scT$, we arrive at the following:
\textbf{we should regard a 3d QFT as a \keyword{domain wall}{domain wall}
	sandwiched between two 4d QFTs}.\footnote{A domain wall has codimension $1$.
	In general, we can consider defects of various codimensions other than domain walls: for example,
	the vortices discussed in Chap.~\ref{chap.3dglue} have codimension $2$.
	Among these defects, domain walls are special in that, since they have codimension $1$,
	we can consider different theories on their two sides.
}

To put it another way, if we are interested in the relation between two four-dimensional field theories,
then we should consider a domain wall between the two.
This is intuitively a very natural idea---if we want to
know the relation between two materials in the lab,
why not combine the two materials?\footnote{However, the degrees of freedom on the domain wall
	are not uniquely determined even if we specify the 4d theories on its two sides;
	there is a rather huge freedom.}

In this setup, the 4d theory, as contrasted with the 3d domain wall,
is often called the \textbf{bulk}.

Since the boundary 3d theory has $\mathcal{N}=2$ supersymmetry (which was crucial for the
supersymmetric localization of the $S^3_b$ partition function),
it is natural to expect the bulk 4d theory also to have supersymmetry.
From the viewpoint of the bulk, the presence of the boundary
breaks part of the 4d Lorentz symmetry, and hence
part of the 4d supersymmetry (the commutation relation \eqref{4dQQ}
says that we obtain a translation generator from the anti-commutator of supersymmetry generators).
This suggests that the bulk theory should have even more supersymmetry,
and the minimal such choice is 4d $\mathcal{N}=2$ supersymmetry.

\bparagraph{Duality Domain Walls}\label{subsec.duality_wall}

Let us consider domain walls in general.
Recall that a domain wall is a wall of codimension $1$.
Given two 4d QFTs $\mathcal{T}_{1}, \mathcal{T}_2$,
we can put them on the regions $x^3<0$ and $x^3>0$ of
$\mathbb{R}^{3,1}$, respectively, where $x^3$ is one of the spatial coordinates of $\mathbb{R}^{3,1}$.
Suppose that we have degrees of freedom localized at the interface
$x^3=0$. Let us denote the theory on this 3d interface by $\mathcal{T}_{1,2}$.

Since a domain wall interpolates between
two 4d QFTs, it is natural to regard it as
a map (morphism) between the two QFTs, namely
\begin{align}
	\hat{\mathcal{T}}_{1, 2} \in \mathrm{Hom}(\mathcal{T}_1, \mathcal{T}_2) \ .
\end{align}
Here the asymmetry between the two theories $\mathcal{T}_1$
and $\mathcal{T}_2$ arises from the orientation.
When we exchange the positions of the two with respect to $x^3<0$ and $x^3>0$,
we obtain the inverse element of $\hat{\mathcal{T}}_{1,2}$:
\begin{align}
	\hat{\mathcal{T}}_{2, 1}=\hat{\mathcal{T}}_{1,2}^{-1} \in \mathrm{Hom}(\mathcal{T}_2, \mathcal{T}_1) \ .
	\label{T_inverse}
\end{align}

The domain wall interpretation also naturally gives a product structure
for the morphisms.
Given two domain walls
\begin{align}
	\hat{\mathcal{T}}_{1, 2} \in \mathrm{Hom}(\mathcal{T}_1, \mathcal{T}_2) \ , \quad
	\hat{\mathcal{T}}_{2, 3} \in \mathrm{Hom}(\mathcal{T}_2, \mathcal{T}_3)  \ ,
\end{align}
we can gradually decrease the volume of the region of the theory $\mathcal{T}_2$,
so that the two domain walls coincide.
The resulting domain wall, a ``bound state''
of the two domain walls, interpolates between the two theories
$\mathcal{T}_1$ and $\mathcal{T}_3$, and hence
defines an element of $\mathrm{Hom}(\mathcal{T}_1, \mathcal{T}_3)$.
We use this as the definition of $\hat{\mathcal{T}}_{1, 2}\circ \hat{\mathcal{T}}_{2, 3}$, the product of the two domain walls:\footnote{
	However, the limit where the two domain walls collide
	is a singular limit, and care is needed regarding its regularization.
	In this book we consider supersymmetric gauge theories, and moreover discuss their localized partition functions,
	where the IR and UV divergences have already been regularized;
	we therefore expect this issue to be avoided.
}
\begin{align}
	\hat{\mathcal{T}}_{1,2} \circ \hat{\mathcal{T}}_{2,3} \in
	\mathrm{Hom}(\mathcal{T}_{1}, \mathcal{T}_{3}) \ .
	\label{T_product}
\end{align}

In this way, domain walls naturally have a product structure, and together they form
an algebra.
In more mathematical terms, we have obtained a ``category of QFTs'':
the objects of this category are individual QFTs,
and the morphisms between two QFTs are domain walls
between the two QFTs.

\bigskip

Let us come back to our problem.
In the discussion so far, a 3d domain wall was given as a morphism between
two 4d gauge theories; in our problem, however, we were dealing with maps
from a Hilbert space to itself.
What we should consider is therefore a domain wall between the same 4d gauge theory.
Moreover, since the domain wall has to be specified by an element of $Sp(2n, \bZ)$,
there should be a counterpart in the 4d theory.
As stated in Sec.~\ref{subsec.MO}, for 4d $\scN=2$ theories with an
Abelian gauge group $U(1)^n$, the duality group is precisely $Sp(2n, \bZ)$.

\begin{figure}[t]
	\centering\includegraphics[scale=0.3]{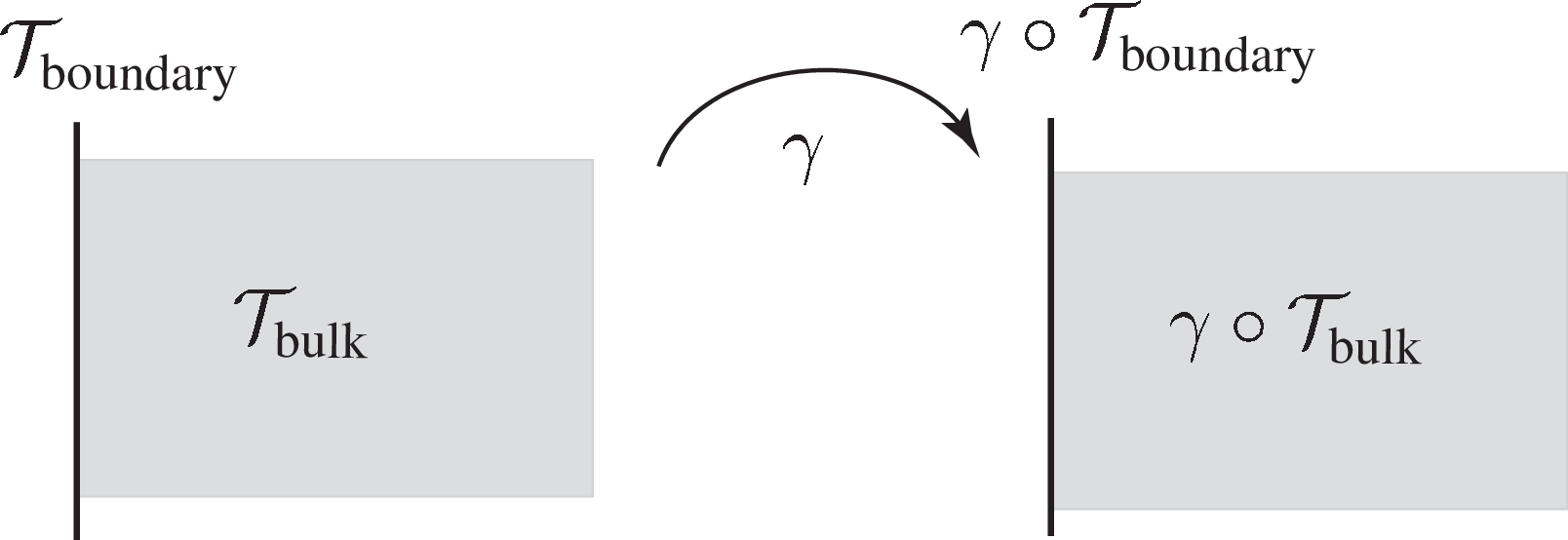}
	\caption{Suppose that a 3d $\mathcal{N}=2$ theory lives on the boundary of a 4d $\mathcal{N}=2$ theory (the bulk). The action of a duality transformation $\gamma$ in the bulk then induces, on the boundary, a transformation which maps
		one theory (boundary condition) to another theory (boundary condition).
	}
	\label{fig.duality_4d3d}
\end{figure}

In general, we expect that
\textbf{the $Sp(2n, \mathbb{Z})$-transformations as dualities of 4d gauge theories are identified with
	the $Sp(2n, \mathbb{Z})$-transformations on 3d field theories}
(Fig.~\ref{fig.duality_4d3d}).
In the present case, we therefore consider a domain wall on whose two sides we have
a theory and the theory transformed from it by the
\keyword{duality group}{duality group}.
This is called a \keyword{duality domain wall}{duality domain wall}.

When a 3d theory appears as the boundary of a 4d theory, the dynamical gauge fields in 4d look like external fields from the viewpoint of the 3d theory, and hence look like background gauge fields for global symmetries: \textbf{the gauge symmetries of the 4d gauge theory are identified with
	global symmetries of the 3d field theory}.
Since the 4d theory now exists on both sides, if the 4d gauge group is $G$, the 3d theory has to have
a $G\times G$ symmetry.\footnote{More precisely,
	this statement is in practice somewhat too strong. As stated in Sec.~\ref{subsec.MO}, the S-dual of the 4d $\mathcal{N}=4$ theory with gauge group $G=SU(N)$ is not the theory itself but the one with $G^{\vee}=SU(N)/\mathbb{Z}_N$; hence, if we consider the duality domain wall corresponding to the S-duality of the 4d theory, the global symmetry of the 3d field theory is $G\times G^{\vee}$.}

\bparagraph{4d and 3d $Sp(2n, \mathbb{Z})$ Actions}\label{subsec.4d_3d_SpZ}

We can verify the relation between 4d and 3d
more directly.
Suppose that we have a 3d theory with a
$U(1)$ global symmetry,
and consider the $Sp(2, \bZ)=SL(2,\bZ)$-action on the theory.

It is not hard to translate the
$T$-transformation into the 4d gauge theory. By integrating by parts, we obtain
\begin{align}
	\frac{k}{4\pi }
	\int_{x^3=0}\!  A\wedge dA
	=\frac{k}{4\pi } \int \theta(x^3)\, F\wedge F \ ,
	\label{T_step}
\end{align}
where $\theta(x^3)$ is the step function
\begin{align}
	\theta(x^3)=\begin{cases}
		            1 & x^3>0 \\
		            0 & x^3<0
	            \end{cases} \ .
\end{align}
This simply means that the value of the $\theta_{\rm 4d}$-angle
jumps by $2\pi k$ (by $2\pi$ for the $T$-transformation, $k=1$).\footnote{The jump of the theta angle is discussed, for example,
	in the context of topological insulators.
	The theta angle is restricted to $0$ or $\pi$ when we impose the $\bZ_2$ time-reversal symmetry,
	and between the 4d regions where its value is $0$ and $\pi$ we obtain
	a 3d Chern-Simons term with half-integer level.
	This is reflected in the half-integer quantization of the Hall conductivity.
	In our $T$-transformation, the value of the theta angle jumps by $2\pi$, and hence
	the Chern-Simons level is quantized to be an integer.
} We can supersymmetrize this argument.
Since the boundary breaks the 4d Lorentz invariance and hence
part of the 4d supersymmetry, the bulk theory should have 4d $\mathcal{N}=2$ supersymmetry.
We thus come to the conclusion that
the $T$-transformation of a
3d $\mathcal{N}=2$ theory
can be identified with the
$T$-transformation ($\tau\to \tau+1$)
acting on the complexified gauge coupling \eqref{complex_gauge}
of the 4d $\mathcal{N}=2$ Abelian gauge theory.

Applying a similar manipulation to the $S$-transformation, we obtain
\begin{align}
	\frac{1}{2\pi }
	\int_{x^3=0}\! B\wedge dA
	=\frac{1}{2\pi } \int \theta(x^3)\, F_B\wedge F_A \ ,
\end{align}
which corresponds to adding $\int F_B \wedge F_A$ in the 4d gauge theory.
To see the meaning of this term,
let us add this term to the Lagrangian of a $U(1)$ gauge theory:\footnote{This expression is confusing: the subscripts $A, B$ label the types of gauge fields,
	while $A, F$ themselves denote the gauge fields and their field strengths.}
\begin{align}
	\frac{1}{g^2} F_A \wedge * F_A + F_B\wedge F_A
	=\frac{1}{g^2} F_A \wedge * F_A + A_B\wedge dF_A+ \textrm{(total derivative)} \ .
	\label{4d_dualize}
\end{align}
Here we set the 4d theta angle $\theta_{\rm 4d}$ to $0$ for simplicity.
On the right-hand side, $A_B$ plays the role of a Lagrange multiplier
imposing the Bianchi identity $dF_A=0$. Hence,
if we regard $A_B$ as a dynamical field, $F_A$ can be regarded as an independent field in itself,
and can be integrated out first. Since the Lagrangian is quadratic in $F_A$, the integral is easy,
and as a result we are left with
\begin{align}
	g^2 F_B\wedge * F_B \ .
\end{align}
This is the same as the original Lagrangian up to the exchange of $A$ and $B$, except
that the value of the gauge coupling constant is inverted, $g\to g^{-1}$.
This is nothing but the S-duality in the sense of 4d gauge theories.
By supersymmetrizing this S-duality as well, we infer that the $S$-transformation of 3d $\mathcal{N}=2$ theories
corresponds to the $S$-transformation of 4d $\mathcal{N}=2$ theories.\footnote{Note that the argument here is the 4d version of the duality transformation discussed in Sec.~\ref{sec.3d_dualize}, which maps a 3d $U(1)$ gauge field to a
	periodic scalar field; the intermediate manipulations are exactly the same,
	except for the differences in the indices of the differential forms.}

We have therefore established the precise relation between 4d and 3d $SL(2, \mathbb{Z})$-transformations.
Note that their physical meanings are rather different between 4d and 3d:
in 4d they are dualities of a single theory, whereas in 3d they are transformations from one theory to another.
This strongly supports our interpretation that our 3d theory should be
regarded as a domain wall in a 4d field theory.

Historically, the $SL(2, \mathbb{Z})$-transformations on 3d field theories
were often motivated by the $SL(2,\bZ)$-duality of 4d theories, and in this sense this explanation
goes partly against the historical order;
note, however, that, as we have seen in this book so far, the $SL(2, \bZ)$-transformations of 3d gauge theories
arise naturally when we pursue the concept of gauging, without any reference to four dimensions.
\subsection{$S^3_b$ Partition Function and $S^4_b$ Partition Function}\label{subsec.S3inS4}

We have been describing the idea that our 3d theory is a
domain wall of a four-dimensional theory.
So far we have considered the case where the 3d theories live
on the flat spacetime $\bR^{2,1}$.
However, what we discussed in Chap.~\ref{chap.S3} was the $S^3_b$ partition function,
and hence, in order to make the discussion there consistent with that of this chapter,
the four-dimensional geometry should correspondingly be a four-manifold containing $S^3_b$.
This four-manifold should also preserve the $U(1)\times U(1)$ symmetry preserved by $S^3_b$.
Moreover, it has to preserve at least
the supercharge used in the localization of the
$S^3_b$ partition function.

One natural candidate for a four-manifold satisfying these criteria
is $S^4_b$, where the parameter $b$ describes the deformation
\begin{align}
	S^4_b: \quad b^2(x_1^2+x_2^2)+b^{-2}(x_3^2+x_4^2)+x_5^2=r^2 \ .
	\label{S4b}
\end{align}
What is important here is that
localization computations on this manifold have already been done
for 4d $\mathcal{N}=2$ theories \cite{Hama:2012bg}
(for $b=1$ this is the metric of the ordinary $S^4$; for localization in this case, see
Ref.~\cite{Pestun:2007rz}).
For our discussion below, however,
we do not need any details of \eqref{S4b} or of the localization there.

\begin{figure}[t]
	\centering\includegraphics[scale=0.3]{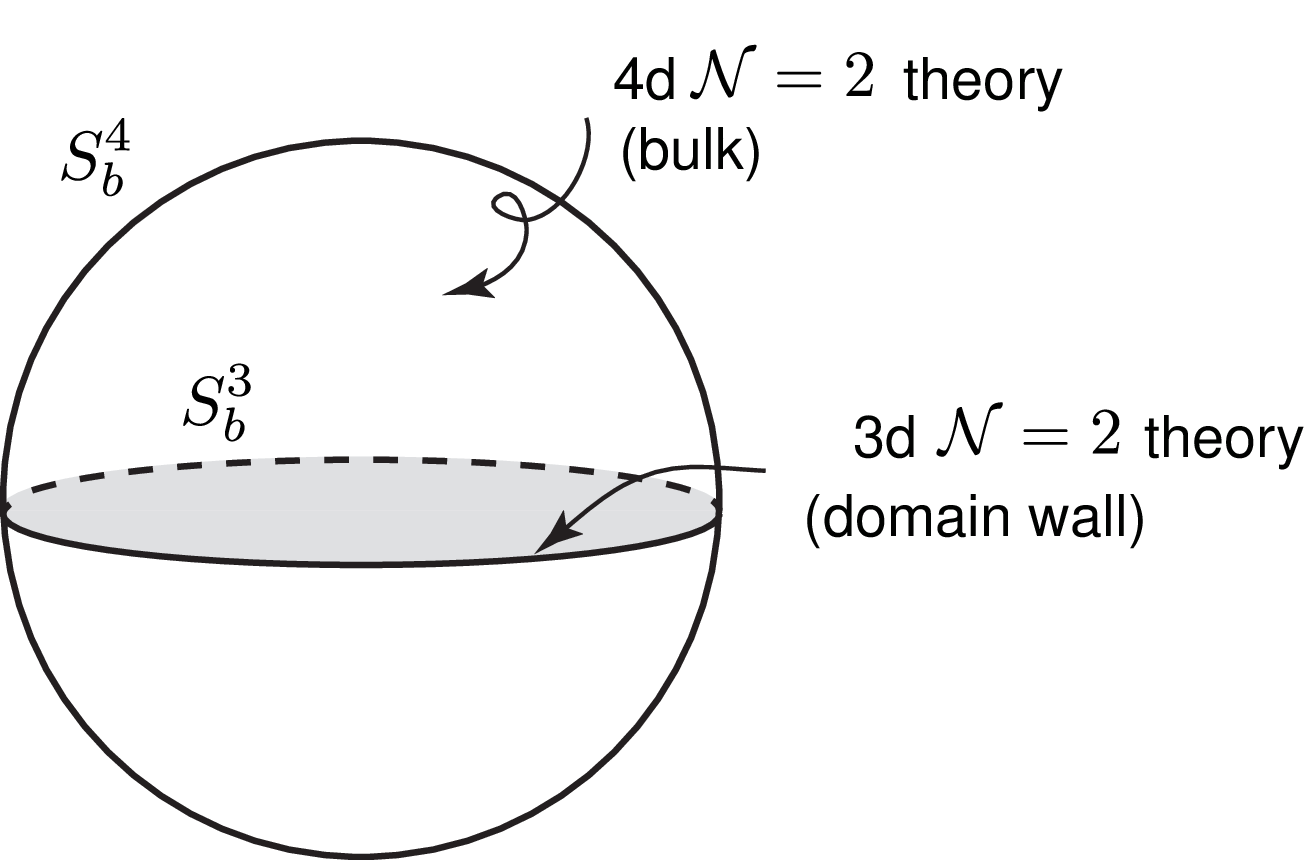}
	\caption{So far in this chapter we have considered situations where a 3d theory is realized as a domain wall
		of a 4d theory. Here, by placing the 4d theory on $S^4_b$ and the 3d theory on $S^3_b$, which is part of $S^4_b$,
		we can apply exact computations by supersymmetric localization to
		the combined 4d-3d system.
	}
	\label{S3inS4}
\end{figure}

Rather, what is important for us is that this $S^4_b$ contains
$S^3_b$ (see \eqref{S3bmetric}) as the great sphere $x_5=0$,
and that we can therefore consider a combined system consisting of
a 4d $\scN=2$ theory on $S^4_b$ and
a 3d $\scN=2$ duality domain wall placed on its equator $S^3_b$ (Fig.~\ref{S3inS4}).

Let us comment on the $S^4_b$ partition function in more detail.
The $S^4_b$ partition function, after supersymmetric localization, takes the form
\beq
Z_{\rm 4d}(S^4_b)=\int \! da \,\, \nu(a)\,  \overline{Z^{\rm Nek}_{a,m}(q)} Z^{\rm Nek}_{a,m}(q) \ .
\label{Z_4d}
\eeq
\nomenclature{$Z^{\rm Nek}_{a,m}(q)$}{Nekrasov partition function}
\nomenclature{$a$}{4d Coulomb branch parameter}
Let us explain the meaning of the variables. First, $a$ is the 4d Coulomb branch parameter, the counterpart of the 3d Coulomb branch parameter $\sigma$ (we will discuss the Coulomb branch of 4d $\mathcal{N}=2$ theories in more detail in Chap.~\ref{chap.6d}).
$q:=e^{2\pi i \tau}$ is (the set of) the complexified coupling constant(s) in 4d,
and $m$ denotes the (complex) mass parameters of the theory.
$\nu(a)$ represents the classical and one-loop contributions in localization,
and $Z^{\rm Nek}$ is the instanton partition function introduced and computed by Nekrasov (the \keyword{Nekrasov instanton partition function}{Nekrasov instanton partition function}) \cite{Nekrasov:2002qd} (more precisely, the parameters $\epsilon_1,\epsilon_2$ of the so-called $\Omega$-background are chosen such that
$b^2=\epsilon_1/\epsilon_2$).
In this book we do not need any details of the Nekrasov partition function.
Intuitively, it is sufficient to think of the Nekrasov partition function $Z^{\rm Nek}$ (and its complex conjugate $\overline{Z^{\rm Nek}}$) as the contribution from the northern (southern) hemisphere of $S^4$.

What happens, then, when we place a 3d domain wall on the equator?
The partition function in this case contains, in addition to the previous expression \eqref{Z_4d}, the contribution from the 3d theory \cite{Drukker:2010jp}:
\beq
Z_{\rm 4d+3d}(S^4_b \supset S^3_b)=\int da da' \,\nu(a) \nu(a')\, \overline{Z^{\rm Nek}_{a,m}} \, Z_{\rm 3d}(a,a'; m) \, Z^{\rm Nek}_{a',m} \ ,
\label{Z4d3d}
\eeq
where $Z_{\rm 3d}(a,a'; m)$ is the $S^3_b$ partition function of the theory on the domain wall,
and $a, a'$ are real masses (including FI parameters) of the 3d theory.
Note that this expression has the same form as \eqref{Z_sigma_overlap_2}.
The more precise meaning of this analogy will become clearer in Sec.~\ref{sec.3d_as_AGT}.

\nextsectionmark{Supplement: $T[SU(N)]$ Theory}
\section[{Supplement: Extension to Non-Abelian Gauge Groups: $T[SU(N)]$ Theory}]{Supplement: Extension to Non-Abelian\\ Gauge Groups: $T[SU(N)]$ Theory} \label{subsec.TSUN}

\subsection{$T[SU(N)]$ Theory}

The theories we have discussed so far are described by
3d Chern-Simons terms and their supersymmetrizations,
and are relatively simple as 3d theories.
The situation is different, however, when the bulk 4d theory has a non-Abelian gauge group $G$.
Recall that in Chap.~\ref{chap.3dglue}
the $Sp(2n, \bZ)$-transformations were defined only for Abelian gauge groups---this is
because it was difficult to carry out the duality transformation for non-Abelian gauge groups.\footnote{For example, for non-Abelian gauge theories the field strength $F=dA+ A\wedge A$ is non-linear, so that \eqref{4d_dualize} is no longer correct.}

In the discussion so far, the key point was the existence of dualities in 4d $\scN=2$ theories.
As a typical example, let us consider the 4d $\scN=4$ theory with gauge group $SU(N)$,
discussed in Chap.~\ref{chap.intro}.
Its duality was the $SL(2, \bZ)$-duality (if we ignore the difference between $SU(N)$ and $SU(N)/\bZ_N$).

Since $SL(2, \bZ)$ is generated by the $S$- and $T$-transformations \eqref{STgen},
it suffices to consider the domain walls corresponding to these two.
For the $T$-transformation, we can use exactly the same argument as for Abelian gauge theories (see \eqref{T_step}),
and we immediately find that a Chern-Simons term for the gauge group $G$ appears.

What about the $S$-transformation, then?
In this case the situation is more complicated.
The previous argument using the duality transformation
does not apply to non-Abelian gauge groups,
and we need a different argument.

\begin{figure}[t]
	\centering\includegraphics[scale=0.45]{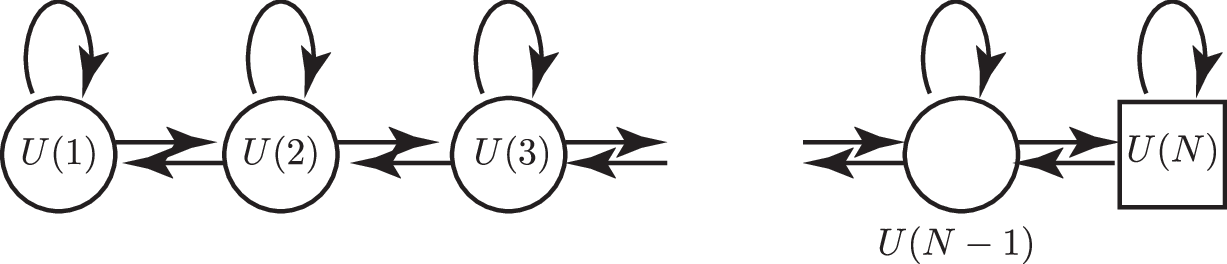}
	\caption{The quiver of the $T[SU(N)]$ theory. This theory has 3d $\mathcal{N}=4$ supersymmetry;
		this quiver is drawn by regarding it as a theory with $\mathcal{N}=2$ supersymmetry.
		Namely, the vertices and the arrows represent $\mathcal{N}=2$ vector multiplets and chiral multiplets, respectively.
		The distinction between square and round vertices is the same as in Fig.~\ref{fig.quiver}.}
	\label{fig.TSUN}
\end{figure}

As already stated in Chap.~\ref{chap.intro}, the $S$-transformation of the 4d
$\mathcal{N}=4$ theory is replaced by a simple operation
from the viewpoint of the 6d theory, namely M5-branes.
If we follow the string dualities starting from M-theory to type IIB string theory,
the $S$-transformation there is easily understood as an exchange of branes
in type IIB string theory. Using this understanding,
we can construct the $S$-transformation (for 3d theories) in the case of non-Abelian gauge groups.
Stating only the result \cite{Gaiotto:2008ak}, it is as follows: the
3d $\scN=4$ theory described in Fig.~\ref{fig.TSUN} has $SU(N)\times SU(N)/\bZ_N$ global symmetry,
and the claim is that the $S$-transformation is the operation of gauging the diagonal part of
one of its $SU(N)$ symmetries and the $SU(N)$ global symmetry of the original theory.
Since in the present case the bulk has 4d $\mathcal{N}=4$ supersymmetry,
the theory on the domain wall correspondingly has half the number of supersymmetries,
namely 3d $\mathcal{N}=4$ supersymmetry.

One strong piece of evidence for this claim will be given later when we discuss the $S^3_b$ partition function of this theory;
here let us look in more detail at the simplest case, \keyword[{$T[SU(2)]$ theory}]{$T[SU(2)]$}{T[SU(2)] theory}.
This theory is an example discussed in Ref.~\cite{Intriligator:1996ex}, and is known to be mapped to itself
under 3d mirror symmetry.
This theory has 3d $\scN=4$ supersymmetry,
but here let us describe it in the language of the 3d $\scN=2$ supersymmetric theories discussed in Chap.~\ref{chap.3dN2} (see Appendix~\ref{chap.SUSYbasic}).

\small

As we will see shortly, we in fact consider the deformation of the 3d $\mathcal{N}=4$ theory to an $\mathcal{N}=2$ theory by a real mass (the \keyword{$\mathcal{N}=2^*$ theory}{N=2* theory}). For simplicity, in this book we call all of these the \keyword[{$T[SU(N)]$ theory}]{$T[SU(N)]$}{T[SU(N)] theory} theory.

\normalsize

First, the gauge group is $U(1)$, and there is a corresponding vector superfield $V$.
Moreover, there are two electrons $Q_1, Q_2$ with charge $+1$ and two positrons $\overline{Q}_1, \overline{Q}_2$ with charge $-1$, and in addition a neutral chiral superfield $\Phi$. Note that $V$ and $\Phi$ form an $\mathcal{N}=4$ vector multiplet $(V, \Phi)$, and that $Q_i$ and $\overline{Q}_i$ form $\mathcal{N}=4$ hypermultiplets $(Q_i, \overline{Q}_i)$.

The fields in the theory and their charges are summarized in Table~\ref{tab.TSU2}.
This time we show only the fields appearing directly in the Lagrangian, and do not show the monopole operators in the table.
Moreover, among the $U(1)$ global symmetries, we show only those which do not
cause a parity anomaly at low energies when the corresponding real masses are turned on.

This theory has an $SU(2)$ global symmetry exchanging the flavors, and
the topological $U(1)_J$ global symmetry.
In fact, this $U(1)_J$ symmetry is enhanced to $SO(3)_J=(SU(2)/\bZ_2)_J$ in the quantum theory.
This $SO(3)_J$ symmetry is a quantum symmetry which does not exist directly in the Lagrangian,
but it can be confirmed indirectly using monopole operators.
Since there are thus two global symmetries, by identifying their background gauge fields with the bulk gauge fields,
we can couple the theory to two 4d theories (the $SU(2)$ gauge theory and its $S$-dual, namely the $SO(3)=SU(2)/\bZ_2$ gauge theory).
According to the claim stated above,
the $S$-transformation is the transformation which gauges one of the $SU(2)\times SO(3)_J$ symmetries of $T[SU(2)]$.

\begin{table}[t]
	\centering
	\caption{Fundamental fields in the Lagrangian of the $T[SU(2)]$ theory,
		and their charges under gauge/global symmetries.}
	\begin{tabular}{c|c|c|c|c|c}
		                   & $Q_1$         & $Q_2$         & $\overline{Q}_1$ & $\overline{Q}_2$ & $\Phi$ \\
		\hline
		\hline
		$U(1)_{\rm gauge}$ & $1$           & $1$           & $-1$             & $-1$             & $0$    \\
		\hline
		$U(1)_A$           & $1$           & $-1$          & $-1$             & $+1$             & $ 0 $  \\
		\hline
		$U(1)_J$           & $0$           & $0$           & $0$              & $0$              & $0 $   \\
		\hline
		$U(1)_m$           & $+1$          & $+1$          & $+1$             & $+1$             & $-2 $  \\
		\hline
		$U(1)_R$           & $\frac{1}{2}$ & $\frac{1}{2}$ & $\frac{1}{2}$    & $\frac{1}{2}$    & $1$    \\
	\end{tabular}
	\label{tab.TSU2}
\end{table}


For the computation of the $S^3_b$ partition function we use only 3d $\mathcal{N}=2$ supersymmetry,
but the $T[SU(2)]$ theory has $\mathcal{N}=4$ supersymmetry.
Among the symmetries in Table~\ref{tab.TSU2}, the $U(1)_m$ symmetry is the symmetry which, when its real mass is non-zero,
breaks the 3d $\mathcal{N}=4$ R-symmetry $\textrm{Spin}(4)=SU(2)_N\times SU(2)_R$
to its $\mathcal{N}=2$ R-symmetry (an $SO(2)$ subgroup).
Therefore, when we turn on the real mass $m$ for the $U(1)_m$ symmetry,
the supersymmetry is broken to $\mathcal{N}=2$.
This parameter has a known 4d counterpart,
namely the mass deforming the 4d $\mathcal{N}=4$ theory into the $\scN=2^*$ theory
(we will mention the 4d $\scN=2^*$ theory in the next chapter in connection with the once-punctured torus). This deformation is known to still
have $SL(2,\bZ)$-duality.

\begin{figure}[t]
	\centering{\includegraphics[scale=0.32]{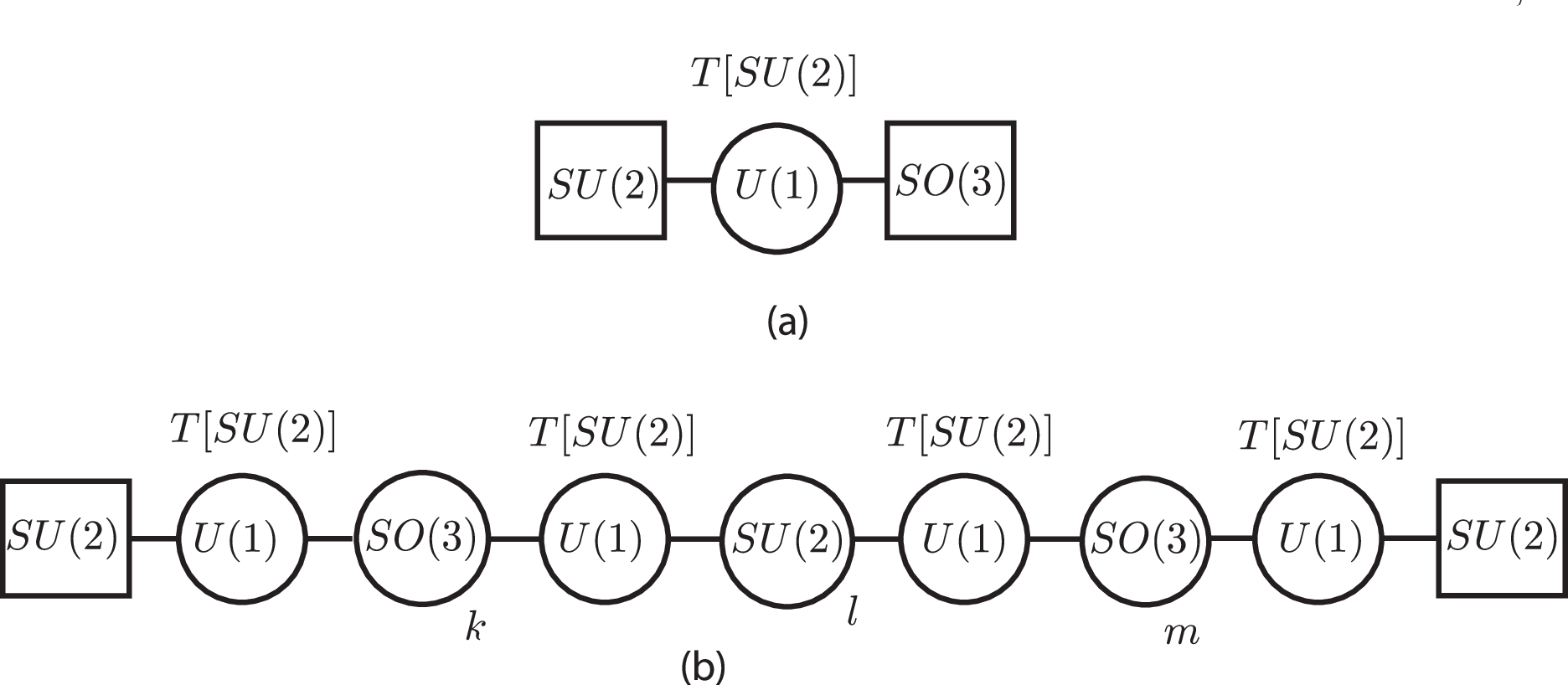}}
	\caption{Duality domain walls corresponding to the $SL(2, \bZ)$-duality of the 4d $\scN=4$ theory. Squares represent global symmetries,
		and circles gauge symmetries. (a) The $S$-transformation corresponds to the $T[SU(2)]$ theory with $SU(2) \times SO(3)_J$ global symmetry. (b) An example of a theory corresponding to a general element of $SL(2, \bZ)$ (a product \eqref{varphiST} of $S$ and $T$) obtained by combining them. Here $k, l,\cdots$ next to the circles mean that we include Chern-Simons terms with levels $k, l, \cdots$ for the corresponding gauge groups.}
	\label{STfigure}
\end{figure}

For the theories corresponding to more general elements of $SL(2,\bZ)$, we can repeat the operation of gauging $T[SU(2)]$ while adding Chern-Simons terms (Fig.~\ref{STfigure}). In this case,
we gauge both of the $SU(2)\times SO(3)_J$ global symmetries of the $T[SU(2)]$ theory;
note in particular that this includes the operation of gauging the $SO(3)_J$ symmetry, which does not exist in the classical Lagrangian (except for its Cartan subgroup $U(1)_J$).

\subsection{Partition Function of $T[SU(N)]$ Theory}

Let us now return to the partition function discussed in Sec.~\ref{subsec.S3inS4}.
While \eqref{Z4d3d} was a general expression, let us here consider in particular the case where the
theory on the 3d domain wall is the $T[SU(2)]$ theory.
Following the rules discussed in Chap.~\ref{chap.S3} (see Table~\ref{tab.TSU2}),
we can write down the $S^3_b$ partition function of the mass-deformed $T[SU(2)]$ theory:
\begin{align}
	\begin{split}
		 & Z_{T[SU(2)]}(\mu,\zeta;m)                                                                                                                          \\
		 & \, =s_b(2m)\int\! d\sigma \,\frac{s_{b}(\mu+\sigma-m+\frac{iQ}{4}) s_{b}(\mu-\sigma-m+\frac{iQ}{4})}  {
			                               s_{b}(\mu+\sigma+m-\frac{iQ}{4}) s_{b}(\mu-\sigma+m-\frac{iQ}{4}) }\, e^{-2\pi i \zeta \sigma} \ .
		\label{Skernelgauge}
	\end{split}
\end{align}
Here we denoted the FI parameter by $\zeta$, and the real mass for $U(1)_A$ ($U(1)_m$) by $\mu$ ($m$),
with the charges read off from Table~\ref{tab.TSU2}; the factor $s_b(2m)$ comes from the adjoint chiral multiplet $\Phi$.
To save space, we used \eqref{sbinv} to move some of the $s_b(x)$ to
the denominator. For the fact that this expression can be regarded as the integral kernel of a kind of Fourier transformation, see Ref.~\cite{Galakhov:2013jma}. Since the expression \eqref{Skernelgauge} is complicated,
let us simplify it here by specializing the parameters. Imposing the condition $m=0$, under which the supersymmetry is enhanced to $\scN=4$, and the condition $b=1$, under which the symmetry of $S^3$ becomes $SO(4)$,
\eqref{Skernelgauge} becomes
\begin{align}
	\begin{split}
		 & Z_{T[SU(2)]}(\mu,\zeta;0)                                                                                         \\
		 & \, =\int\! d\sigma \,\frac{s_{b=1}(\mu+\sigma+\frac{i}{2}) s_{b=1}(\mu-\sigma+\frac{i}{2})}  {
			                        s_{b=1}(\mu+\sigma-\frac{i}{2}) s_{b=1}(\mu-\sigma-\frac{i}{2}) }\, e^{-2\pi i \zeta \sigma} \\
		 & \, =\frac{1}{4}\int\! d\sigma \,\frac{e^{-2\pi i \zeta \sigma}}{\cosh\pi(\mu+\sigma) \cosh\pi(\mu-\sigma)} \ .
		\label{Skernelgauge_b1}
	\end{split}
\end{align}
By closing the integration contour, this evaluates to
\begin{align}
	\begin{split}
		Z_{T[SU(2)]}(\mu,\zeta;0)
		 & = \frac{1}{2}\frac{e^{2\pi i \mu \zeta }-e^{-2\pi i \mu \zeta }}{2\sinh(2\pi \mu)\sinh(\pi \zeta)} \\
		 & =  \frac{i \sin (2\pi \mu \zeta )}{2\sinh(2\pi \mu)\sinh(\pi \zeta)} \ .
		\label{Skernelgauge_b2}
	\end{split}
\end{align}
By a similar computation, for the more general 3d $\scN=4$ theory $T[SU(N)]$ we obtain, for $b=1$ (up to an overall phase factor) \cite{Nishioka:2011dq},
\begin{align}
	Z(\vec{m}, \vec{\zeta})=\frac{1}{N!}\frac{\sum_{\sigma\in
			                                    \mathfrak{S}_N}{(-1)^{\sigma} e^{2\pi i \sum_{i} m_{\sigma(i)}\zeta_i}}}{\Delta(\vec{m}) \Delta(\vec{\zeta})} \ ,
	\label{NTY_formula}
\end{align}
where $\vec{m}$ and $\vec{\zeta}$ are the mass and the FI parameters of the $T[SU(N)]$ theory
(for $N=2$ they are $\vec{m}=(\mu, -\mu)$ and $\vec{\zeta}=(\frac{\zeta}{2}, -\frac{\zeta}{2})$, which reproduces \eqref{Skernelgauge_b2}),
	the sum is over the permutations $\sigma$ of $N$ elements, and
\begin{align}
	\Delta(\vec{x}):=\prod_{i<j} 2\sinh \pi (x_i-x_j) \ .
\end{align}
This is a natural generalization of the integral kernel of the Fourier transformation to
$N$ interchangeable variables.\footnote{The reference \cite{Nishioka:2011dq}
	also proposed a generalization of \eqref{NTY_formula}. For a systematic proof, see Ref.~\cite{Assel:2014awa}.
}

The analysis of more general cases is now similar. For example,
for $\varphi=ST^k S$ the partition function reads
\beq
\int \!d\sigma \, Z_{T[SU(2)]}(\mu,\sigma;m)\, e^{-i \pi k \sigma^2}\, Z_{T[SU(2)]}(\sigma, \mu';m) \ .
\label{Zeg}
\eeq
If we define the integral kernels $S, T$ by
\begin{align}
	\begin{split}
		 & S_{(\sigma,\sigma';m)}:=Z_{T[SU(2)]}(\sigma,\sigma';m) \ ,            \\
		 & T_{(\sigma,\sigma')}:=e^{-i \pi \sigma^2} \delta(\sigma-\sigma') \ ,
	\end{split}
	\label{Tkernelgauge}
\end{align}
then this reads
\beq
\int \! d\sigma' d\sigma'' \, S_{(\sigma,\sigma';m)} T^k_{(\sigma',\sigma'')} S_{(\sigma'',\sigma''';m)}=(ST^kS)_{(\sigma,\sigma''';m)} \ .
\eeq
Written this way, it is straightforward to generalize this to a general
element of $SL(2, \bZ)$, and the result is again
written as a combination of the integral kernels $S, T$.

In fact, exactly the same expressions as the integral kernels \eqref{Tkernelgauge} appear
in Liouville theory and in \Teichmuller theory.
This is not a coincidence, and already suggests a relation with the contents of Chap.~\ref{chap.Teichmuller}.
The reason for this will become clear in the next chapter (in particular in Sec.~\ref{sec.3d_as_AGT}).


\begin{practice}

	\item $[\bll]$ ($SL(2, \bZ)$-transformation of the $S^3_b$ partition function)

	Use the $S$- and $T$-transformations \eqref{ZS}, \eqref{ZT} of the $S^3_b$ partition function
	to directly verify $(ST)^3=1$ for the $S^3_b$ partition function.
	\label{ex.ST}

	\item $[\bll]$ (Composition of canonical transformations)
	Show \eqref{Wcompose} from the definitions
	\eqref{W1} and \eqref{W2}.

	\label{ex.Wcompose}

	\item $[\bll]$ (Determination of constant factors)
	In \eqref{W1} and \eqref{W2}
	we neglected constant factors (see e.g.\ \eqref{eq.TSTS}).
	Write down the complete $SL(2, \bZ)$-transformation rule for the $S^3_b$ partition function, including the constant factors. What is the constant factor for a general
	$SL(2, \mathbb{Z})$ matrix?
	\label{ex.canonical}

	\item $[\bll]$ ($Sp(2n, \bR)$-transformations in quantum mechanics)

	Let us consider a free particle described by the Hamiltonian
	\begin{align}
		\hat{H}=\frac{\sfP^2}{2m} \ ,
	\end{align}
	where the momentum (its conjugate coordinate) is denoted by $\sfP$ ($\sfQ$).
	Let us study the energy eigenstates of this free particle,
	\begin{align}
		\hat{H} |\Psi\rangle =E |\Psi \rangle \ .
	\end{align}
	Denoting the eigenstates of $\sfQ$ by $|Q\rangle$,
	we can explicitly find the wave function $\Psi(Q)=\langle Q|\Psi\rangle$.

	How, then, do the Hamiltonian and the wave function transform under
	canonical transformations ($SL(2, \bR)$-transformations) of $\sfP, \sfQ$? (Note that $\sfP$ and $\sfQ$ do not
	commute when we consider the Hamiltonian.)

	\item $[\bll]$ ($S^3_b$ partition function of the $T[SU(2)]$ theory)

	Verify \eqref{Skernelgauge}, \eqref{Skernelgauge_b1}, and \eqref{Skernelgauge_b2}. Hint: close the integration contour and appeal to the residue theorem.

	\item $[\bll\bll]$ (3d mirror symmetry of the $T[SU(2)]$ theory)\label{TSU2_self_mirror}
	Verify that the partition function \eqref{Skernelgauge} satisfies
	\beq
	Z_{T[SU(2)]}(\mu,\zeta;m)=
	Z_{T[SU(2)]}\left(\frac{\zeta}{2},2\mu;-m\right) \ .
	\eeq
	This is a manifestation of the fact that the $T[SU(2)]$
	theory is self-mirror (up to the change of sign of $m$).
	The factors of $2$ here reflect the normalization of the FI parameter:
	with the conventions of Chap.~\ref{chap.S3} used here, the mass $\mu$ of the
	$SU(2)$ flavor symmetry and the FI parameter $\zeta$ are exchanged as $\mu\leftrightarrow\frac{\zeta}{2}$.

	If this is difficult, evaluate the integral $Z_{T[SU(2)]}(\mu,\zeta;m)$ in the limit $b\to 0$
	by the saddle point approximation, and verify that the result has the symmetry above.

	\item $[\bll]$ (Junction of domain walls)
	In the main text of this chapter, we defined a product
	structure from the coincident limit of parallel domain walls.
	What kind of algebraic structure do we obtain, then,
	if we consider junctions of several such domain walls (Fig.~\ref{fig.junction})?

	\begin{figure}[t]
		\centering\includegraphics[scale=0.3]{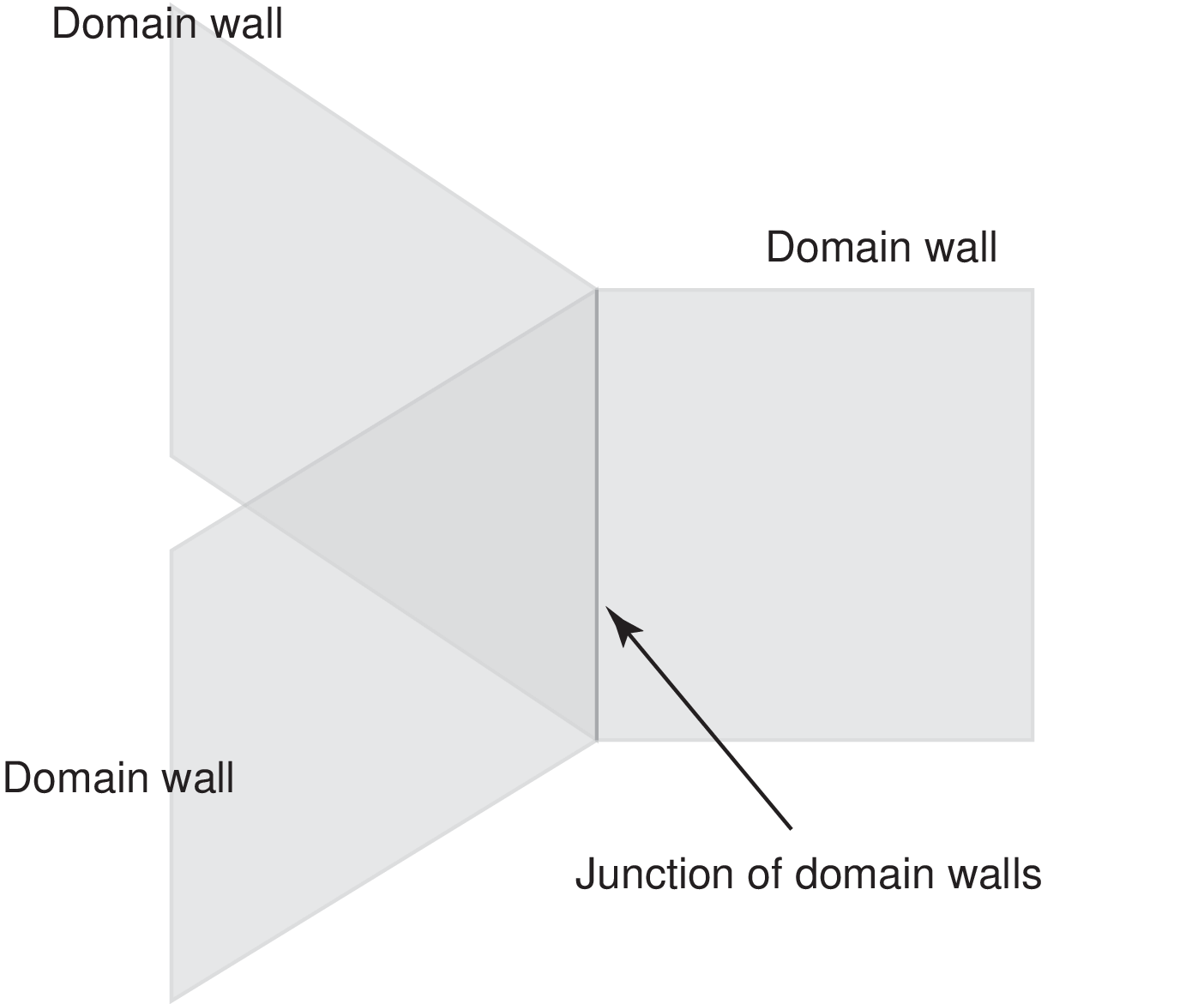}
		\caption{A junction of domain walls.
			A domain wall has codimension $1$ in the bulk, and
			a domain wall junction has codimension $2$.}
		\label{fig.junction}
	\end{figure}

\end{practice}

\chapter{Compactification of 6d $(2,0)$ Theory}\label{chap.6d}

\begin{abstract}
	We explain the compactification of the 6d $(2,0)$ theory on a 3-manifold.
	We shall find that we obtain a 3d complex Chern-Simons theory on
	the 3-manifold.
\end{abstract}

\section{Emergence of Topological Quantum Field Theory}

\subsection{6d Theory Revisited}

In the previous chapter, our 3d field theory was regarded as a domain wall of a four-dimensional field theory.
We also computed the $S^3_b$ partition function,
and saw that (at the level of the $S^3_b$ partition function) the $Sp(2n, \mathbb{Z})$-equivalence class $[\scT]$ of a 3d theory $\mathcal{T}$ can be regarded as an operator
$\hat{\scT}$ acting on the Hilbert space $\mathcal{H}_{S^3}$ determined by a four-dimensional QFT.
This Hilbert space can be interpreted as the Hilbert space on the spatial $S^3$,
where we have the four-dimensional theory on $S^3\times \bR$
with $\bR$ being the time direction:\footnote{The ``four-dimensional
	Hilbert space'' here is a much smaller Hilbert space than that of the four-dimensional theory in the standard sense. We are discussing the $S^4_b$ partition function
	preserving supersymmetry, and hence we do not consider those states
	which do not contribute to the $S^4_b$ partition function.}
\begin{align}
	\textrm{Four-dimensional theory}: S^3\times \bR \leadsto \mathcal{H}_{S^3}  \ .
	\label{4d_S3R}
\end{align}

In the previous chapter we considered the 4d $\scN=4$ theory and its mass deformation (the 4d $\scN=2^*$ theory). We can ask whether we can generalize the discussion to a general
4d $\scN=2$ theory.

Since the S-duality of 4d $\scN=2$ theories was crucial in the discussion before,
the key is to discuss 4d $\scN=2$ theories whose dualities are manifest.

One fundamental idea was already presented in Chap.~\ref{chap.intro}, where
the 4d theory was obtained by compactifying the
6d $(2,0)$ theory on a 2-manifold $\Sigma_{g,h}$ (geometrization of QFT!).
Combining this with \eqref{4d_S3R}, we should consider
\begin{align}
	\textrm{6d $(2,0)$ theory}: S^3\times \bR \times \Sigma \ .
	\label{6d_compactify}
\end{align}

\small
When $\Sigma=\Sigma_{1,0}=\bT^2$, we have the four-dimensional $\mathcal{N}=4$
theory, where the mapping class group $SL(2, \bZ)$ of $\bT^2$ was identified with the
$SL(2, \bZ)$ duality group (acting on $\tau$ through $PSL(2, \bZ)$).
When we add a puncture and consider $\Sigma_{1,1}$, this should correspond to
a natural deformation of the 4d $\scN=4$ theory preserving $\mathcal{N}=2$ supersymmetry.
When we regard the 4d $\scN=4$ theory as a theory with $\mathcal{N}=2$ supersymmetry,
the $\scN=4$ vector multiplet decomposes into an $\scN=2$ vector multiplet and
an $\scN=2$ chiral multiplet (taking values in the adjoint representation),
and we can give a mass to this $\scN=2$ chiral multiplet (this theory is often called the 4d $\scN=2^*$ theory). The value of this mass can be identified with
the holonomy around the hole of the puncture (the ``size'' of the hole).
Even when we add a puncture to the torus $T^2$ to obtain $\Sigma_{1,1}$, its
mapping class group remains $SL(2, \bZ)$, and this is
naturally identified with the S-duality of the 4d $\scN=2^*$ theory (the S-duality is preserved under the mass deformation).

\normalsize

When we consider a more general Riemann surface $\Sigma_{g,h}$,
we obtain a certain
4d $\mathcal{N}=2$ theory. In Chap.~\ref{chap.intro} this theory was denoted by $\mathcal{T}[\Sigma]$.
We then expect that
\textbf{the duality group of the 4d $\scN=2$ theory is again given by the
	mapping class group of the Riemann surface}.
That this expectation actually works was proposed in Ref.~\cite{Gaiotto:2009we},
and it has subsequently passed many non-trivial checks. In this book
most of those details are not really needed.

\subsection{Translation into TQFT}\label{subsec.TQFT}

In the rest of this chapter we discuss such a compactification in detail.
Before coming to the details, however,
let us first see whether we can say anything just from the existence of the six-dimensional theory
and the choice of the compactification \eqref{6d_compactify}.

\begin{figure}[t]
	\centering\includegraphics[scale=0.23]{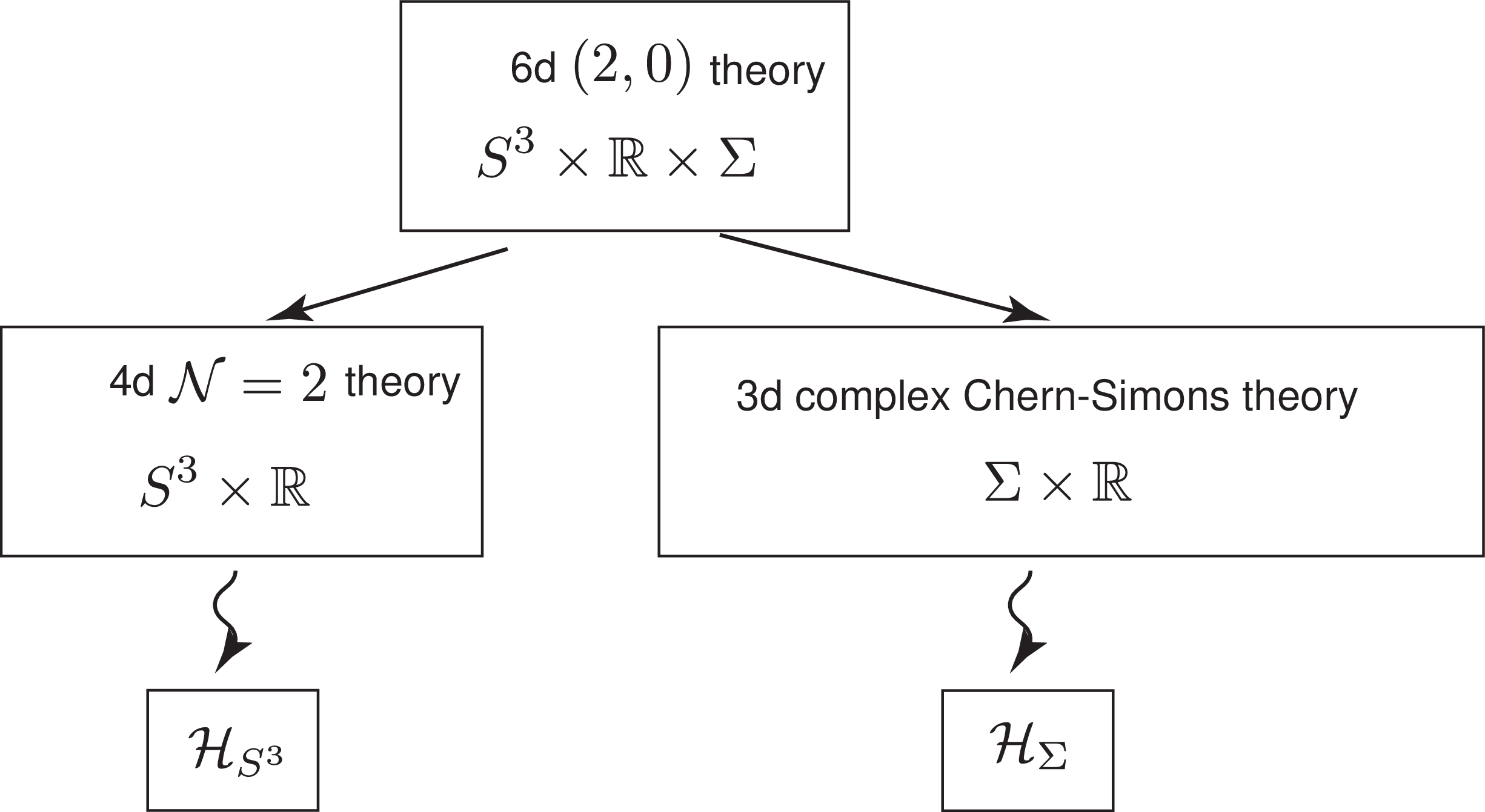}
	\caption{When we consider the 6d theory on $S^3\times \bR\times \Sigma$,
		we can change the order of the compactification.
		When we compactify on $\Sigma$ first,
		we obtain a 4d $\mathcal{N}=2$ theory on $S^3\times \bR$ and its Hilbert space $\mathcal{H}_{S^3}$,
		whereas when we compactify on $S^3$ first, we obtain
		a 3d complex Chern-Simons theory on $\bR\times \Sigma$ and its Hilbert space $\mathcal{H}_{\Sigma}$.}
	\label{fig.Fubini}
\end{figure}

In compactifications of QFTs we often invoke ``Fubini's theorem in QFT'',
where we exchange the order of compactifications.\footnote{When we have different scales of compactification, we compactify according to their scales, with smaller scales compactified first.
	An exchange of the order of compactifications amounts to a change of the relative scales of the different compactification directions, and a careful analysis is needed of its effects on physics.}
Above we first compactified on $\Sigma$;
if we instead first compactify on $S^3$, a 3d theory appears on $\bR\times \Sigma$.
When we regard $\bR$ as time in this theory and do canonical quantization,
we obtain a Hilbert space on $\Sigma$ (Fig.~\ref{fig.Fubini}).
This is the setup we encountered in the previous chapter.

Let us now translate the contents of the previous chapter into the language of
3d field theories.

In the previous chapter, we came to the viewpoint that 3d theories are domain walls of 4d theories.
For the translation, we need to count the number of remaining directions.

\begin{figure}[t]
	\centering\includegraphics[scale=0.25]{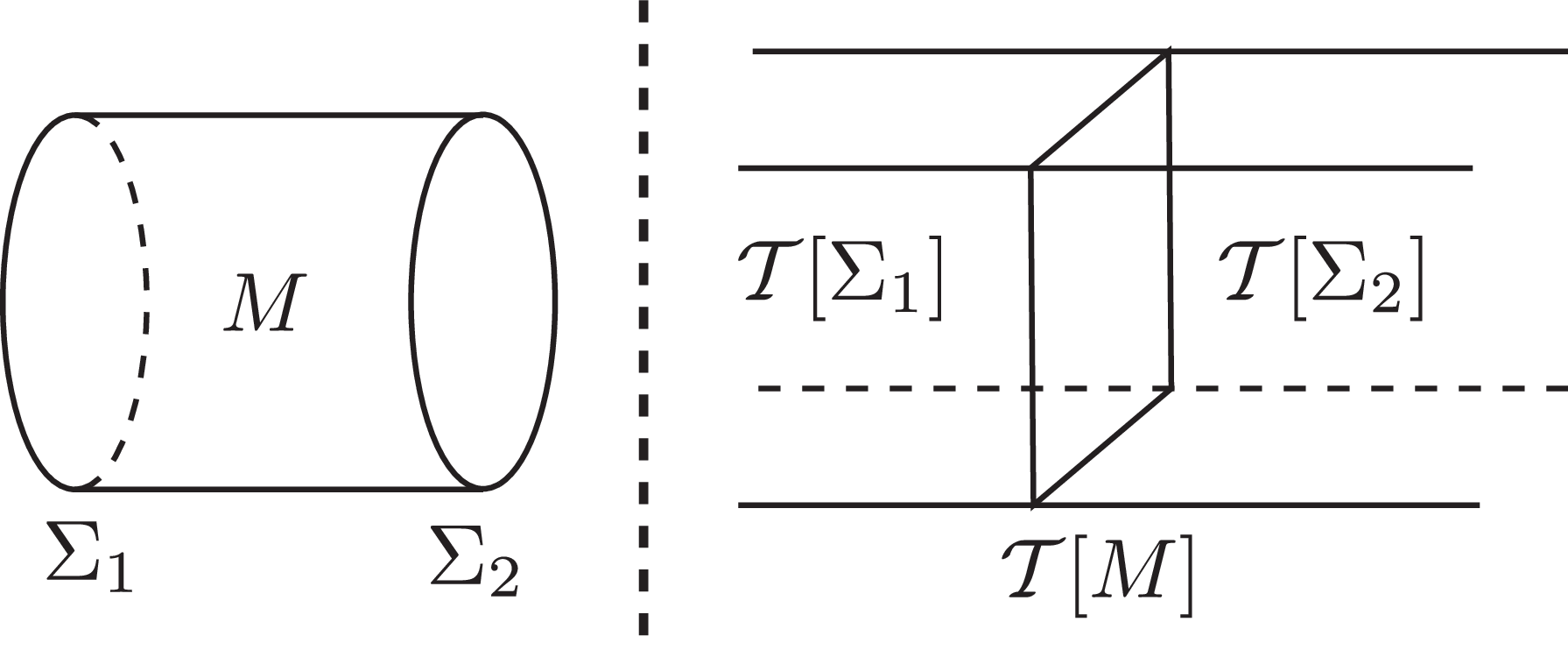}
	\caption{A 3d $\mathcal{N}=2$ domain wall between 4d $\mathcal{N}=2$ theories
		is translated into a cobordism between 2d manifolds
		in the transverse directions of 6d.}
	\label{fig.betweenboundary}
\end{figure}

Since 4d $\mathcal{N}=2$ theories are associated with 2d Riemann surfaces,
the domain wall theories with 3d $\mathcal{N}=2$ supersymmetry
should be associated with 3-manifolds connecting two 2d Riemann surfaces.
Namely, the 3-manifold satisfies $\partial M=(-\Sigma_1)\cup \Sigma_2$,
where the two Riemann surfaces are denoted by $\Sigma_1, \Sigma_2$,
and the orientation reversal of $\Sigma_1$ is denoted by $-\Sigma_1$.
Such a manifold $M$ is called a \keyword{cobordism}{cobordism} between
$\Sigma_1$ and $\Sigma_2$ (Fig.~\ref{fig.betweenboundary}).

\begin{figure}[t]
	\centering\includegraphics[scale=0.25]{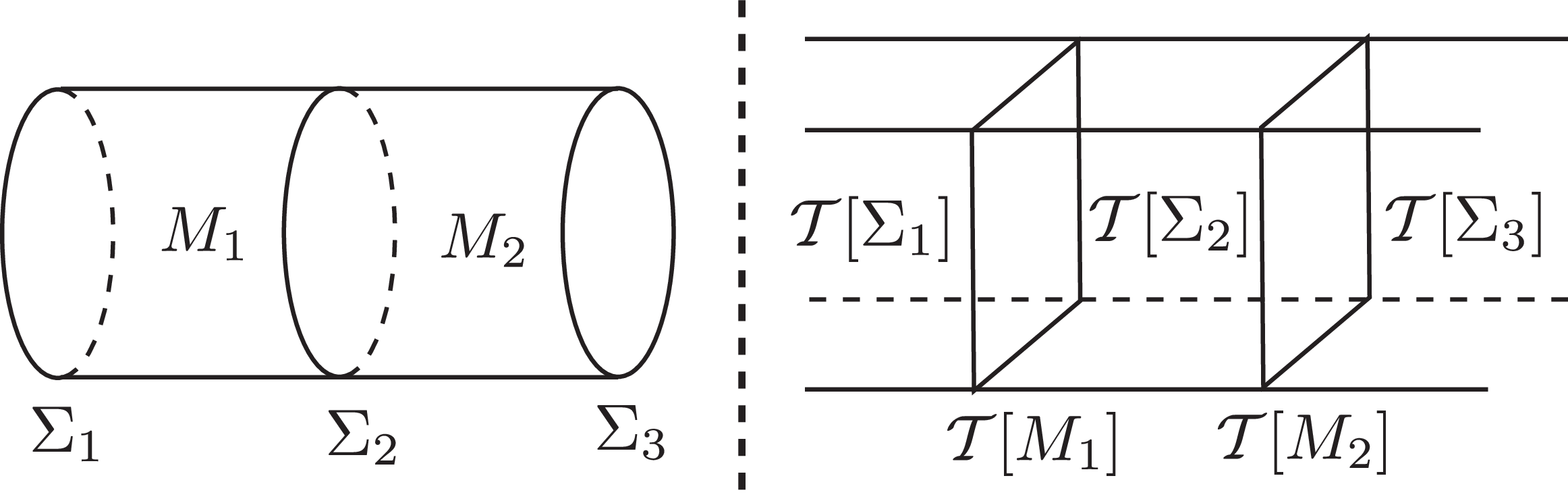}
	\caption{The product of two 3d $\mathcal{N}=2$ domain walls is
		translated into the product of two cobordisms in the transverse directions in six dimensions.}
	\label{fig.translate2}
\end{figure}

\begin{figure}[t]
	\centering\includegraphics[scale=0.25]{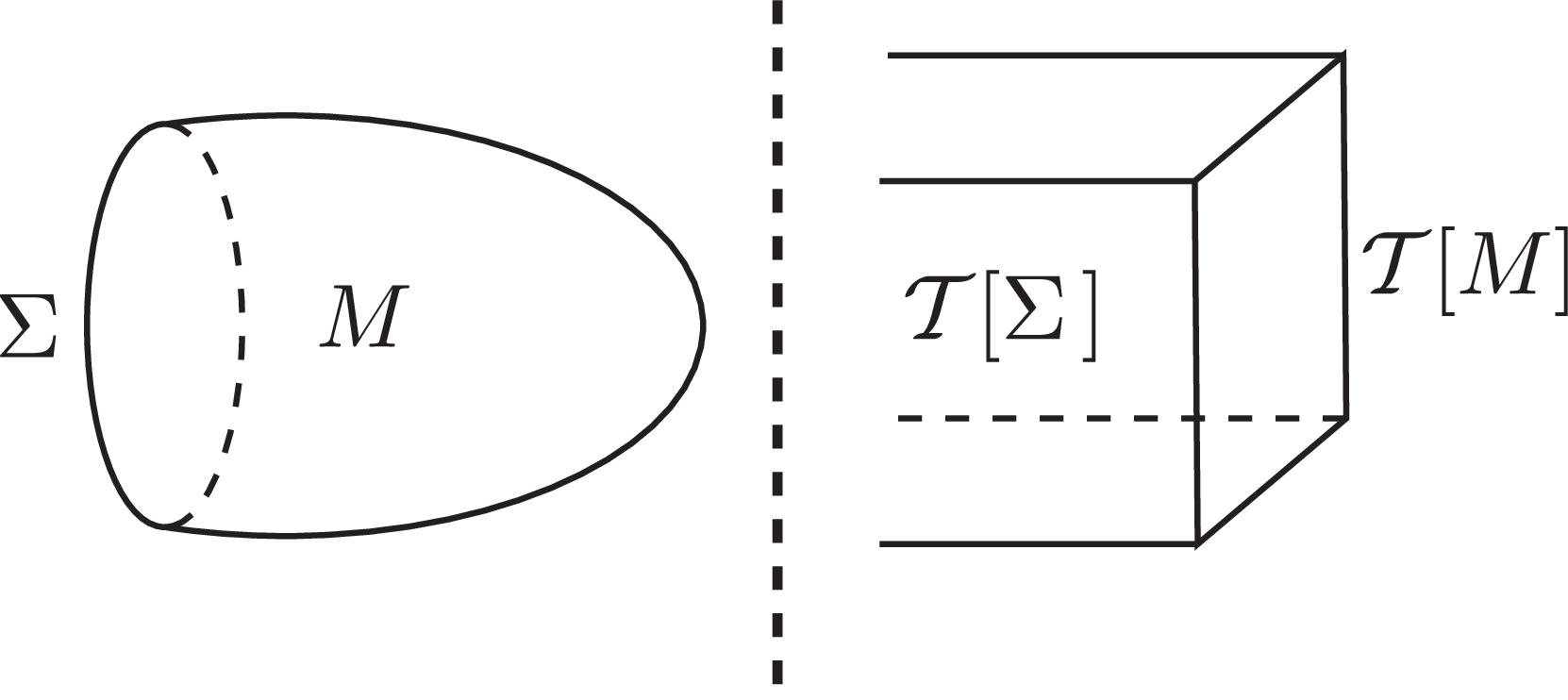}
	\caption{That a 3d $\mathcal{N}=2$ gauge theory is a boundary theory of a four-dimensional $\scN=2$ theory is translated, in the transverse directions in six dimensions,
		into the statement that the 3-manifold has the 2-manifold as its boundary.}
	\label{fig.withbounary}
\end{figure}

Let us rewrite the domain wall axioms \eqref{T_inverse}, \eqref{T_product}
in this language.
First, for a cobordism $M$ between $\Sigma_1$ and $\Sigma_2$, we obtain a map
\begin{align}
	\hat{M} \in \mathrm{Hom}(\mathcal{H}_{\Sigma_1}, \mathcal{H}_{\Sigma_2}) \ .
	\label{axiom1}
\end{align}
Next, if we have two cobordisms
$\partial M_1=(-\Sigma_1)\cup \Sigma_2$ and $\partial M_2=(-\Sigma_2)\cup \Sigma_3$,
then we have yet another cobordism by gluing $M_1$ and $M_2$ along $\Sigma_2$:
$\partial (M_1 \cup_{\Sigma_2} M_2)=(-\Sigma_1)\cup \Sigma_3$.
We then have the associated morphisms (see Fig.~\ref{fig.translate2}; product read from left to right, as in \eqref{Mproduct}):
\begin{align}
	\widehat{M_1 \cup M_2}=
	\hat{M_1}\cdot \hat{M_2} \in \mathrm{Hom}(\mathcal{H}_{\Sigma_1}, \mathcal{H}_{\Sigma_3}) \ .
	\label{axiom2}
\end{align}
When $\Sigma_2$ is the empty set $\emptyset$, the associated Hilbert space $\mathcal{H}_{\emptyset}$
is given by $\bC$, and the morphism associated with the
cobordism $\partial M=-\Sigma_1$ is given by
\begin{align}
	\hat{M} \in \mathrm{Hom}(\mathcal{H}_{\Sigma_1}, \bC)
	\label{axiom3}
\end{align}
(Fig.~\ref{fig.withbounary}).

Finally, given a cobordism $M$ between $\Sigma_1$ and $\Sigma_2$,
if we consider the cobordism $-M$ with the opposite orientation, $\partial(-M)=(-\Sigma_2)\cup\Sigma_1$, we have
\begin{align}
	\widehat{(-M)} \in \mathrm{Hom}(\mathcal{H}_{\Sigma_2}, \mathcal{H}_{\Sigma_1}) \ ,
	\label{axiom4}
\end{align}
and $\mathcal{H}_{-\Sigma}$ is the dual $\mathcal{H}_{\Sigma}^*$ of $\mathcal{H}_{\Sigma}$ (hereafter, following the notation of physics, we denote the dual of a state $| -\rangle\in \mathcal{H}_{\Sigma}$ by
$\langle -|\in \mathcal{H}_{\Sigma}^*$).

Eqs.~\eqref{axiom1}, \eqref{axiom2}, \eqref{axiom3} and \eqref{axiom4} are nothing but what is known as the
\keyword{Atiyah-Segal axioms}{Atiyah-Segal axioms} \cite{AtiyahTopological,SegalBook}
for \keyword{topological quantum field theories}{topological quantum field theory} (TQFTs).
Here, a topological quantum field theory is a special field theory whose properties do not depend on
the metric of the manifold on which the theory is defined.

From the considerations so far, it is natural to expect that
the theory appearing on the 3-manifold is
a certain topological quantum field theory.
Indeed, as we will see shortly, this theory turns out to be one of the most
typical topological quantum field theories, namely 3d Chern-Simons theory.

In the axioms of topological quantum field theory,
the cobordism $M$ with $\partial M=(-\Sigma_1)\cup \Sigma_2$ corresponds to an operator $\hat{M}$.
To consider its expectation value, we choose elements
$|\rho \rangle\in \mathcal{H}_{\Sigma_1}$ and $|\sigma \rangle \in\mathcal{H}_{\Sigma_2}$,
so that
\beq
Z_{\rm CS}[M]=\langle \sigma | \hat{M}  | \rho  \rangle \ ,
\eeq
as in \eqref{Z_sigma_overlap_2}.
Note that there is no non-trivial time evolution here, since we now consider a
topological quantum field theory whose Hamiltonian is trivial.

\small

In particular, what happens when $M=\Sigma\times I$ ($I$ is the interval $[0,1]$) and we identify the two ends of $I$ to obtain $S^1$? In this case the two states $|\rho \rangle$ and $|\sigma \rangle$ are identified,
and we take the trace over the states. Moreover, instead of the simple direct product, if we consider the
\keyword{mapping torus}{mapping torus} twisted by an element $\varphi$ of the
mapping class group of $\Sigma$,
\begin{align}
	(\Sigma\times S^1)_{\varphi} = (\Sigma\times [0,1]) / (x,0) \simeq (\varphi(x), 1) \ ,
	\label{mapping_torus}
\end{align}
then we can change (the topology of) the 3-manifold depending on the choice of $\varphi$.
In this case, $\varphi$ is promoted to an operator $\hat{\varphi}$ on the Hilbert space of the
topological quantum field theory, and its trace gives the
partition function on $(\Sigma\times S^1)_{\varphi}$:\footnote{However, this quantity diverges, for example, when $\hat{\varphi}=\textrm{id}$.}
\beq
Z_{\rm CS}[(\Sigma\times S^1)_{\varphi}]
=\textrm{Tr}_{\mathcal{H}_{\Sigma}}\, \hat{\varphi} \ .
\label{tracevarphi}
\eeq
We will comment on mapping tori again in Sec.~\ref{subsec.map_torus}.

\normalsize

\section{Compactification and Topological Twist}\label{subsec.compact_twist}

\bparagraph{6d $(2,0)$ Theory}

While the discussion so far suggested that we obtain a TQFT
on a 3-manifold, we have not yet identified precisely which TQFT it is.
To answer this question, let us first quickly summarize the properties of the six-dimensional theory.

The \keyword{6d $(2,0)$ theory}{6d (2,0) theory} is a
six-dimensional CFT.
As discussed in footnote~\ref{foot_inverse_RG} in Sec.~\ref{sec_3d_beta},
gauge theories are non-renormalizable in five and more spacetime dimensions, and
the directions of the RG flows are opposite to the usual ones,
so that it is not easy to discuss the existence of fixed points
within the framework of effective field theories.
The existence of the 6d $(2,0)$ theory was originally suggested by string theory;
however, no direct proof of its existence from field theory has been given,
and at least a Lagrangian in the usual sense (one which manifestly preserves both Lorentz symmetry and non-Abelian gauge invariance) is not known at the time of writing of this book,
nor is there any guarantee that it exists.

The field content of the theory is
strongly constrained by 6d $(2,0)$ supersymmetry:
its $(2,0)$ multiplet consists of a
self-dual totally anti-symmetric 2-form $B_{\mu\nu}$ (satisfying $dB=* dB$),
two Weyl fermions $\lambda_1, \lambda_2$, and five scalars $\phi_1, \cdots, \phi_5$.
Note that theories in six dimensions are in general chiral, and the numbers of supersymmetries can be different for right-movers and left-movers; we can consider e.g.\ $\mathcal{N}=(2,0), (1,1), (1,0)$
theories,\footnote{
	That we write $\mathcal{N}=(2,0)$ rather than $\mathcal{N}=(0,2)$ is the standard convention
	in the literature.
} where $(\mathcal{N}_L, \mathcal{N}_R)$ means that we have
$\mathcal{N}_L$ ($\mathcal{N}_R$) supersymmetries for the left-movers (right-movers).

\small
Just to be sure, let us check that the numbers of degrees of freedom of bosons and fermions
are the same. Since in this book we do not go into the question of which
auxiliary fields appear in the supersymmetry multiplets,
we should count the degrees of freedom on-shell.

First, when counting the degrees of freedom of the 2-form $B_{\mu\nu}$,
the time direction cancels against a spatial direction, so that we only need to consider the remaining
four spatial directions (this is similar to counting the on-shell degrees of freedom of a 4d gauge field as $4-2=2$).
Since $B_{\mu\nu}$ has totally anti-symmetric indices, and moreover the number of degrees of freedom is halved by the constraint $dB=* dB$,
we obtain $\frac{4 (4-1)}{2}\times \frac{1}{2}=3$.
Together with the five scalar fields, the bosonic degrees of freedom are $8$.
On the other hand, there are $16$ fermionic components, which are halved on-shell to give
$16/2=8$. The degrees of freedom therefore match.

\normalsize

The five scalar fields $\phi_1, \ldots, \phi_5$ transform in the $\bm{5}$-representation
under the $SO(5)_R\simeq Sp(4)_R$ R-symmetry.

\bigskip

What is important for us is the following fact:
6d $(2,0)$ theories are known to have an ADE classification,
and in this book we consider the one called the $A_N$-type.
\textbf{When we compactify the 6d $A_{N}$-type $(2,0)$ theory on $S^1$ with radius $R$,
	we obtain the 5d $\scN=2$, $SU(N+1)$ pure Yang-Mills theory}:\footnote{A Yang-Mills theory without matter fields is called a \keyword{pure Yang-Mills theory}{pure Yang-Mills theory}.}\textbf{
	\begin{align}
		\begin{split}
			 & \textrm{6d $A_{N}$-type $(2,0)$ theory} : \mathbb{R}^5\times S^1 \\
			 & \longrightarrow  \,
			\textrm{5d $\mathcal{N}=2$ $SU(N+1)$ pure Yang-Mills theory} : \mathbb{R}^5  \ .
		\end{split}
		\label{5d_compact_2}
	\end{align}
	Moreover,
	the gauge coupling constant of the 5d $\scN=2$ theory is then given by\footnote{
		Note that the $R$-dependence of \eqref{g5R} is inverse to that of the usual formula for the dimensional reduction of gauge theories. For the meaning of this, see exercise~\ref{ex.6d_Sdual}.
	}
	\begin{align}
		g_5^2= 2\pi R \ .
		\label{g5R}
	\end{align}}\nomenclature{$g_5$}{gauge coupling constant of the 5d $\mathcal{N}=2$ theory}
Since $g_5^2$ and $R$ are proportional, at weak coupling the radius $R$ is small, and the spacetime is of course five-dimensional;
at strong coupling, however, the radius $R$ becomes large, and the dimension of spacetime increases to six.
In other words, the claim is that when we start from the 5d $\scN=2$ theory and follow its RG flow
backwards, there exists a UV fixed point in the strong coupling limit $g_5\to \infty$,
which is the 6d $(2,0)$ theory (see also footnote~\ref{foot_inverse_RG} in Sec.~\ref{sec_3d_beta}).
Note that, from the previous formula \eqref{g_dim}, $g_5^2$ has mass dimension $-1$,
which agrees with the dimension of the radius $R$.

\small

The 6d $(2,0)$ theories were originally obtained by considering type IIB string theory on ADE-type singularities \cite{Witten:1995zh},
and correspondingly they have an ADE classification. The gauge group of the 5d theory is then also an ADE-type gauge group (more concretely, $A_N=SU(N+1), D_N=SO(2N), E_6, E_7, E_8$).
For the $A_N$-type with large $N$, the corresponding gravity solutions of M-theory
in the sense of the AdS/CFT correspondence are known, and they provide indirect evidence for the existence of the 6d theory.
Moreover, while in the usual dimensional reduction we discard the Kaluza-Klein modes
along the compactified direction, in the present case
the Kaluza-Klein modes of the 6d theory along the $S^1$-direction
can be understood as instantons of the 5d theory, and the 6d degrees of freedom seem to be already
contained in 5d.

The 5d $\scN=2$ theory and the 6d $(2,0)$ theory are the theories appearing on the branes called
D4-branes and M5-branes in 10d type IIA string theory and 11d M-theory, respectively.
Therefore, this lift from 5d to 6d is also a lift from
10d type IIA string theory to 11d M-theory.

\normalsize

In order to be convinced of \eqref{5d_compact_2},
let us examine what happens to the 2-form $B_{\mu\nu}$
when we compactify the 6d $(2,0)$ theory on $S^1$.
Under the $S^1$-compactification, the compactified direction (let us call it the $5$-direction)
is no longer a spacetime index, and hence from $B_{\mu \nu} \quad (\mu, \nu=0, 1, \ldots, 5)$ we obtain
\begin{align}
	A_{\hat{\mu}}=B_{\hat{\mu} 5} \ , \quad B_{\hat{\mu} \hat{\nu}} \ , \quad \hat{\mu}, \hat{\nu}=0, 1, \cdots, 4 \ .
\end{align}
However, since we had the self-duality condition $dB=*_6 dB$ in 6d,
we in fact only need to consider $A_{\hat{\mu}=0, \cdots, 4}$ as independent degrees of freedom, which can be regarded as the gauge field of the 5d theory.

Note also that the 5d $\mathcal{N}=2$ theory has
the same R-symmetry, $SO(5)_R\simeq Sp(4)_R$, as the 6d $\mathcal{N}=(2,0)$ theory.\footnote{
	This will become important later in the discussion of the topological twist.
}

If we compactify the 5d theory on $S^1$, we come back to the claim already stated in Chap.~\ref{chap.intro}, namely that the 4d $\scN=4$ theory
follows from the $T^2$-compactification of the 6d $(2,0)$ theory (see exercise~\ref{ex.6d_Sdual}).

\bparagraph{Topological Twist}

We next consider the six-dimensional theory on
3-manifolds and 2-manifolds.

We have already considered supersymmetry on the curved space $S^3$
in Chap.~\ref{chap.S3}.
There, since appropriate spinors exist on $S^3$, we could use them to
write down the supersymmetry transformations.
However, this method has one drawback: when we consider more general manifolds, such
nice spinors do not exist.
In this case another method is known, namely the
\keyword{topological twist}{topological twist}\footnote{However, the supersymmetry in Chap.~\ref{chap.S3} is also a kind of topological twist, in the sense that we turn on a flux for the R-symmetry, so that there is not such a big difference between the two.} (in the following we sometimes simply call it the twist).

While we do not discuss the topological twist in detail (see e.g.\ Ref.~\cite{MirrorBook}),
one way to explain it is that we identify part of the Lorentz symmetry of spacetime
with part of the R-symmetry of the supersymmetry algebra; more practically, we turn on a flux
which induces such an identification.
Since the supercharges transform under the Lorentz symmetry and at the same time under the R-symmetry,
when the two symmetries are identified we can expect a component $\scQ$ which behaves as a scalar
(while still obeying anti-commuting statistics). If this is the case, the scalar $\scQ$ can be defined without any problem on any manifold,
and hence the supersymmetry algebra can be defined.

In the twisted theory, the energy-momentum tensor becomes
$\scQ$-exact, so that the theory becomes a metric-independent \keyword[topological theory]{topological}{topological theory}
theory. This is why the procedure above is called the topological twist.
In contrast with the twisted theory, the original
metric-dependent theory is sometimes called the \keyword[physical theory]{physical}{physical theory} theory.

\small
In the situation we wish to consider now, we compactify the 6d theory on a manifold and perform the topological twist
only along those directions, while for the remaining directions we do not twist and keep the physical theory.
This situation is often called a
\textbf{partial topological twist}.
\normalsize

\bigskip

There is one problem with this discussion. As already stated, the 6d theory is
still a theory with many mysteries; in particular, a Lagrangian (manifestly preserving Lorentz symmetry and gauge invariance) is not known, so that we cannot actually carry out the twist.

However, the 5d
$\scN=2$ theory obtained by the $S^1$-compactification of the 6d theory is already known (see \eqref{5d_compact_2}), and hence, if the 6d theory is on a space containing $S^1$,
we can first compactify along that direction and then consider the topological twist.
Namely, in
\begin{align}
	\textrm{6d $(2,0)$ theory} : \mathbb{R}^4\times \Sigma \longrightarrow
	 &
	\,
	\textrm{4d $\mathcal{N}=2$ theory} : \mathbb{R}^4
	\label{6d4d}
\end{align}
we consider compactifying one of the directions of $\mathbb{R}^4$ on $S^1$:
\begin{align}
	\begin{split}
		 & \textrm{6d $(2,0)$ $A_N$-type theory} : \mathbb{R}^3\times S^1 \times \Sigma \\
		 & \longrightarrow  \,
		\textrm{5d $\mathcal{N}=2$, $SU(N+1)$ pure Yang-Mills theory} : \mathbb{R}^3 \times \Sigma \ .
	\end{split}
	\label{5d_compact}
\end{align}
The twist of the 5d theory on $\Sigma$ can then be carried out.
Moreover, if we change the order of compactifications and compactify on $\Sigma$ before $S^1$,
the 4d theory $\mathcal{T}[\Sigma]$ is $S^1$-compactified to a 3d theory:
\begin{align}
	\textrm{4d $\mathcal{N}=2$ theory} : \mathbb{R}^3\times S^1 \longrightarrow
	\textrm{3d $\mathcal{N}=4$ theory} : \mathbb{R}^3 \ .
	\label{4d3d}
\end{align}
Here, since the number of supersymmetries does not change between 4d and 3d,
we have $\mathcal{N}=4$ supersymmetry in 3d (see Appendix~\ref{chap.SUSYbasic}).

Since the discussion so far has been abstract, let us make it more concrete.
First, let us consider the 5d $\scN=2$ theory on a Riemann surface $\Sigma$. The
Lorentz symmetry of $\Sigma$ is then the rotation group $SO(2)$ rotating the local frames of its tangent space.
We identify this with a subgroup of the $SO(5)$ R-symmetry;
the remaining R-symmetry is then $SO(2)\times SO(3)\simeq U(1)\times \left(SU(2)/\bZ_2\right)$.
Since the R-symmetry in 4d with $\mathcal{N}$ supersymmetries is
$U(\mathcal{N})$ (Appendix~\ref{chap.SUSYbasic}),
this suggests that
the 4d theory $\mathcal{T}[\Sigma]$ has $\scN=2$ supersymmetry.
Note that the 4d theory is compactified on $\bR^3\times S^1$, as in \eqref{4d3d}.

\bigskip

A similar discussion can be carried out for a 3-manifold $M$.
In this case, for the corresponding 3d $\mathcal{N}=2$ theory we consider
the specific space $S^3_b$;
recall that in the discussion of Chap.~\ref{chap.S3} (see \eqref{S3_b_limit}), $S^1_b$ appeared naturally in the $b\to 0$ limit of $S^3_b$.
Therefore, if we regard the direction of this $S^1$ as the 6d direction,
we obtain\footnote{We will come back to this limit in Chap.~\ref{chap.complexCS}.}
\begin{align}
	\begin{split}
		 & \textrm{6d $(2,0)$ theory}: S^3_b \times M_3  \overset{b\to 0}{\longrightarrow}       \\
		 & \qquad \qquad \textrm{5d $\mathcal{N}=2$ pure Yang-Mills theory}: \bR^2\times M_3 \ ,
	\end{split}
\end{align} where the radius of $S^1$ is given by $b$. Namely,
\textbf{the deformation parameter $b$ of $S^3$ in 3d $\scN=2$ theories was identified with
	the deformation parameter $b$ of the quantum dilogarithm,
	and these are related to the gauge coupling constant of the
	5d $\scN=2$ supersymmetric Yang-Mills theory by
	\begin{align}
		2\pi R=2\pi b=g_5^2 \ .
		\label{g5d}
	\end{align}
}The ``quantum'' in the quantum dilogarithm meant the lift of the 5d theory to the 6d theory, namely,
in the language of string theory, the lift of D4-branes to M5-branes, and
the lift of type IIA string theory to M-theory.

When we twist the 5d $\mathcal{N}=2$ theory on a 3-manifold,
the rotational symmetry of the tangent space of the 3-manifold is $SO(3)$,
and in this case the remaining symmetry is $SO(2)$.
Since the R-symmetry in 3d with $\mathcal{N}$ supersymmetries is
$SO(\mathcal{N})$ (Appendix~\ref{chap.SUSYbasic}), this
means that the 3d theory $\mathcal{T}[M]$ has $\scN=2$ supersymmetry.

Note that, as the counterpart of \eqref{4d3d}, the 3d theory $\mathcal{T}[M]$ is $S^1$-compactified:
\begin{align}
	\textrm{3d $\mathcal{N}=2$ theory} : S^3_b\overset{b\to 0}{\sim} \mathbb{R}^2\times S^1_b \longrightarrow
	\textrm{2d $\mathcal{N}=(2,2)$ theory} : \mathbb{R}^2 \ .
	\label{3d2d}
\end{align}

\small
The theories $\mathcal{T}[\Sigma], \mathcal{T}[M]$ defined in this way could depend on geometric structures of $\Sigma$ and $M$, such as
the metric. However, as we have emphasized so far, we are interested only in the
low-energy behavior of the theories, and not in the information about the sizes of
$\Sigma$ and $M$; hence we do not distinguish between a metric $g_{\mu \nu}$ and its conformal transformation $e^{\phi} g_{\mu\nu}$.
Namely, the theories depend only on the conformal class of the metric.
From this we find, for example, that for a 2d surface $\Sigma$
we only need to consider the complex structure of $\Sigma$ (the moduli of deformations of its shape) (see Sec.~\ref{subsec.Teich}).\footnote{This is a non-trivial statement, and is not necessarily true in general, for example for compactifications of 6d $(1,0)$ theories. For compactifications on 3-manifolds as in the present case, we can combine
	the fact that the 6d theory has scale invariance (so that it is invariant when we simultaneously rescale the 3-manifold and the remaining three dimensions)
	with the fact that, if we consider e.g.\ the $S^3$ partition function in the remaining three directions,
	it does not depend on the energy scale.}
This freedom of deforming the complex structure is identified with the freedom of deforming the theory $\mathcal{T}[\Sigma]$
while preserving $\mathcal{N}=2$ supersymmetry and conformal symmetry. For example, when $\Sigma=T^2$,
its complex structure is the modulus $\tau$ of the torus discussed around \eqref{tauSL2Z},
which was identified with the
complexified gauge coupling constant \eqref{complex_gauge}.

\normalsize

\section{Identification of the Theory on 3-Manifolds}

So far we have determined the number of supercharges of the 3d theory from the discussion of the twist;
however, the original question still remains: what is the theory appearing on the 3-manifold?
We have already obtained a hint that it is a topological quantum field theory,
but we would like to understand more precisely what it is.

For this purpose, we need a somewhat more detailed discussion
of the topological twist.
In the following we give the answer from the viewpoints of the topological twist and of
direct localization.
We first discuss the case of 3-manifolds in Sec.~\ref{subsec.twist}
and the case of 2-manifolds in Sec.~\ref{subsec.Hitchin},
and then summarize the relation with the Alday-Gaiotto-Tachikawa (AGT) correspondence
in Sec.~\ref{sec.3d_as_AGT}.

Regarding the appearance of 3d complex Chern-Simons theory,
in addition to the explanation in this chapter, there is also an interpretation by topological string theory \cite{Cecotti:2011iy},
and in Chap.~\ref{chap.3mfd} we will also give an explanation using ideal tetrahedral decompositions of 3-manifolds and properties of quantum dilogarithm functions.
The contents of these explanations are not exactly the same, but
complement each other.

\bparagraph{Twist and BPS Condition: 3d}\label{subsec.twist}

In the 5d $\mathcal{N}=2$ theory there are five
scalar fields $\phi_1, \ldots, \phi_5$,
which transform in the $\bm{5}$-representation under the $SO(5)$ R-symmetry.
Recall that under the action of the twist described above,
the $SO(5)$ symmetry decomposes into $SO(3)\times SO(2)$, and
$SO(3)$ was mixed with the $SO(3)$ of the tangent bundle along the 3-manifold $M$.
Correspondingly, the scalar fields also decompose into
a $1$-form $\phi_{\mu}=\phi_1, \phi_2, \phi_3$ on $M$
and scalar fields $\phi_4, \phi_5$ on $M$.

In the compactification on the 3-manifold $M$, the three scalar fields $\phi_{\mu}$
combine with the gauge field $A_{\mu}$ along $M$ into the complex combination
$\mathcal{A}=A+i\phi$.
If we examine the conditions for field configurations to preserve supersymmetry, namely the
BPS equations of the twisted 5d $\mathcal{N}=2$ theory,
the resulting equation is
\begin{align}
	\mathcal{F}=d\mathcal{A}+\mathcal{A}\wedge \mathcal{A}=0 \ .
	\label{complex_flat}
\end{align}
This is the equation for a \keyword{flat connection}{flat connection} of the complexified gauge group
$G_{\mathbb{C}}$.
The moduli space obtained by dividing the solution space of this equation by gauge transformations is the
\keyword{moduli space of flat connections}{moduli space of flat connections}.

\small

The BPS equations for the remaining scalar fields $\phi_{4,5}$
require that they commute with the complex gauge field $\mathcal{A}$.
When the flat $G_{\mathbb{C}}$-connection is irreducible,
only the trivial element commutes with $\mathcal{A}$,
so that we can forget about $\phi_{4,5}$.
What we mainly discuss in this book are
$SL(2,\mathbb{C})$ flat connections with a geometric interpretation, and they are irreducible.

\normalsize

The theory appearing in the 3d directions is expected to have equations of motion giving
\eqref{complex_flat}. Such a
theory is nothing but the Chern-Simons theory for the complex connection $\mathcal{A}$:
\begin{align}
	\mathcal{L}_\textrm{complex CS} \sim \textrm{Tr}\, \left(\scA\wedge d\scA +\frac{2}{3}\scA \wedge \scA \wedge \scA \right) \ .
\end{align}
We have thus reached the conclusion that \textbf{the theory on the 3-manifold $M$ is the Chern-Simons theory with the complex gauge group $G_{\mathbb{C}}$}. This theory will be discussed in more detail
in the next chapter. To summarize,
\textbf{when we consider a 3d $\scN=2$ theory on $\bR^2\times S^1$, its vacua
	correspond to flat $G_{\bC}$-connections on the 3-manifold $M$}.

In the discussion so far we identified the theory on the 3-manifold $M$
from the twist and the BPS conditions;
we can also prove more directly that 3d Chern-Simons theory appears on $M$:
we place the 5d $\scN=2$ theory on $S^2\times M$, directly carry out localization, and
show that 3d complex Chern-Simons theory appears on $M$.
This computation was carried out in Refs.~\cite{Lee:2013ida,Cordova:2013cea}
(see also Refs.~\cite{Yagi:2013fda,Luo:2014sva}):
\begin{align}
	Z_{\textrm{5d $\scN=2$ theory}}[S^2\times M]=Z_{\textrm{complex Chern-Simons}}[M] \ .
	\label{5d3d_derive}
\end{align}
The details of this formula, for example the correspondence of the parameters, will be explained in more detail in
Chap.~\ref{chap.complexCS}.
This result can be said to be an unquestionable proof that the theory on the 3-manifold $M$
is complex Chern-Simons theory. Interestingly,
the result above applies to a general gauge group $G$. Moreover,
since the 3d theory has to include all the flat connections in the path integral,
it becomes clear that what we should consider is not, for example, 3d gravity, but in fact
3d complex Chern-Simons theory (on this point, see
Sec.~\ref{subsec.CS_classical}).
Furthermore, in the process of deriving \eqref{5d3d_derive},
one can confirm that the gauge group actually becomes $G_{\bC}$ rather than $G$, due to the effect of gauge fixing.
For details, see Refs.~\cite{Lee:2013ida,Cordova:2013cea}.\footnote{
	As a difference between the two references, Ref.~\cite{Cordova:2013cea} first constructs an off-shell formalism of the 5d $\mathcal{N}=2$ theory and then applies it to $S^2\times M_3$ (obtained from the 6d geometry $S^3_b\times M_3$ by reduction along the Hopf fiber). On the other hand, Ref.~\cite{Lee:2013ida} places the 5d $\mathcal{N}=2$ theory on $S^2\times M_3$ in the on-shell formalism, then lifts it off-shell, and applies it to the 6d geometry $(S^1\times S^2)\times M_3$. For computations by localization, off-shell supersymmetry transformations are ultimately needed.
}

\subsection{Twist and BPS Condition: 2d}\label{subsec.Hitchin}

A discussion similar to the one so far can be repeated in the 2d case.
Since the starting point is the same 5d
BPS equations as before, it is natural to expect similar equations.
Previously we mixed the $SO(5)_R$ symmetry with the $SO(3)$ acting on the tangent bundle of $M$, so that an $SO(2)$ subgroup remained; this time we mix it with the $SO(2)$ acting on the tangent bundle of $\Sigma$, and then an
$SO(3)\times SO(2)$ subgroup remains (this roughly corresponds to exchanging $SO(3)$ and $SO(2)$ in the previous discussion).
Two of the scalar fields are converted into $1$-forms on the 2d surface $\Sigma$,
which complexify the gauge field along $\Sigma$;
if we define the complex connection $\mathcal{A}_{\mu=1,2}=A_{\mu}+i\phi_{\mu}$, the
BPS equations again become the equation for flat $G_{\bC}$-connections:
\begin{align}
	\mathcal{F}=d\mathcal{A}+\mathcal{A}\wedge \mathcal{A}=0 \ .
	\label{flat_F_2d}
\end{align}
It is natural that flat $G_{\mathbb{C}}$-connections appear both for 2-manifolds and 3-manifolds,
since the theory on a 2-manifold should be the restriction of the theory on a 3-manifold
to its 2d boundary.

When expanded into real and imaginary parts, \eqref{flat_F_2d} becomes
\begin{align}
	F-\phi\wedge \phi=0\ , \quad
	D_A \phi=0 \ .
	\label{flat_F_2d_2}
\end{align}
These two equations, supplemented by the equation
\begin{align}
	D_A \star \phi=0 \ ,
	\label{Hitchin_third}
\end{align}
are called the
\keyword{Hitchin equations}{Hitchin equation} \cite{HitchinSelfDuality}
(the third equation also follows from the BPS equations; see exercise~\ref{ex.BPS_ASD}).
The third equation can be regarded as a gauge-fixing condition for the complexified gauge transformations:
each orbit of a (reductive) flat $G_{\bC}$-connection under $G_{\bC}$ gauge transformations contains
a solution of \eqref{Hitchin_third}, which is unique up to $G$ gauge transformations \cite{MR965220,MR887285}.
Namely, the solution space of the Hitchin equations (the \keyword{Hitchin moduli}{Hitchin moduli}) gives
the space of flat $G_{\bC}$-connections.\footnote{The BPS equations for the remaining three scalar fields $\phi_{3,4,5}$
	require that they commute with the complex gauge field $\mathcal{A}$.
	These scalar fields are not important for understanding the Coulomb branch of the
	4d $\mathcal{N}=2$ theories discussed in this book, but
	are indispensable, more generally, for studying the branch structure of the vacuum moduli space \cite{Xie:2014pua}.}

\small
To explain this a little more precisely: the moduli space of solutions of the Hitchin equations
is a manifold called a
\keyword{hyperK\"{a}hler manifold}{hyperKahler manifold}, which has a triplet of complex structures $I, J, K$.
The Hitchin moduli gives the space of
flat $G_{\bC}$-connections when we choose a
special one as the complex structure.\footnote{The space of flat $G_{\bC}$-connections has
	the symplectic form \eqref{symplectic_flat}. In the usual convention, this symplectic form is
	holomorphic with respect to the complex structure $J$.}
\normalsize

As in the case of the 3-manifold $M$,
when we consider the 4d $\scN=2$ theory on $\bR^3\times S^1$
(see \eqref{4d3d}), we expect the Hitchin moduli
to appear. This is indeed the case, and the claim is that
\textbf{when we consider the 4d $\mathcal{N}=2$ theory $\mathcal{T}[\Sigma]$ on $\bR^3\times S^1$,
	its Coulomb branch moduli space is given by
	the Hitchin moduli of $\Sigma$}.
This fact is useful in Chap.~\ref{chap.Teichmuller}
for understanding the geometric considerations there physically.

\bigskip

Let us here add an explanation of the Coulomb branch of 4d $\mathcal{N}=2$ theories
(see also Appendix~\ref{chap.SUSYbasic}).
The supermultiplet of 4d $\mathcal{N}=2$ supersymmetry containing the
gauge field $A_{\mu}$ is called the
\keyword{vector multiplet}{vector multiplet},
and contains a complex scalar field $\phi$
transforming in the adjoint representation of the gauge group.

In the language of 4d $\mathcal{N}=1$ supersymmetry,
the $\mathcal{N}=2$ vector multiplet splits into
an $\mathcal{N}=1$ chiral multiplet $(\phi, \psi)$ and
an $\mathcal{N}=1$ vector multiplet $(A_{\mu}, \lambda)$.
Since the gauge field $A_{\mu}$ transforms in the adjoint representation of the gauge group, its superpartner,
the scalar field $\phi$, also transforms in the adjoint representation, as expected.

In the vacua of 4d $\mathcal{N}=2$ theories,
we can consider situations where the complex scalar field $\phi$ has an expectation value.
Since $\phi$ takes values in the adjoint representation of the gauge group,
when it has an expectation value
the gauge group is broken to its Abelian part (Cartan subalgebra) $U(1)^{r_G}$,
where $r_G$ is the rank of the gauge group $G$.
Note that the adjoint representation transforms trivially under the Abelian subgroup of the gauge group.

The vacuum moduli space determined in this way is called the
\keyword{Coulomb branch}{Coulomb branch}\footnote{We imagine that it is called a branch because several ``branches'' grow out of a particular point (for example, a point with conformal invariance).} of
4d $\mathcal{N}=2$ theories.

\small

This $\phi$ is analogous to the
Coulomb branch scalar $\sigma$ in the vector multiplet of 3d $\mathcal{N}=2$ theories (Chap.~\ref{chap.3dN2}):
in both cases there is a scalar field taking values in the adjoint representation in the same supersymmetry multiplet as the gauge field,
and this scalar field has an expectation value.
However, compared with the case of Chap.~\ref{chap.3dN2}, the number of supersymmetries is now doubled.

Note that the Coulomb branch here is not the Coulomb branch in the
usual sense, namely the one considered on $\mathbb{R}^4$,
but the Coulomb branch considered on $\bR^3\times S^1$. Under the $S^1$-compactification,
the component of the gauge field along $S^1$ becomes a scalar field, and its complexification by the dual photon (see Sec.~\ref{subsec.monopole}) gives new coordinates of the Coulomb branch,
so that the dimension of the Coulomb branch after the $S^1$-compactification is
twice the dimension of the Coulomb branch before the compactification.
Moreover, this suggests that the Hitchin moduli has
torus fibers over a base space,
which is indeed true as a property of the Hitchin moduli.

\normalsize

\subsection{Supplement: Relation with the AGT Correspondence}\label{sec.3d_as_AGT}

Before concluding this chapter,
let us comment on the relation with the
\keyword{AGT correspondence}{AGT correspondence} \cite{Alday:2009aq}.
The purpose is to explain that the 3d-3d correspondence is
a natural generalization of the AGT correspondence.
If we assume the AGT correspondence,
this can also be regarded as giving an alternative derivation of the 3d-3d correspondence.

Let us first explain the AGT correspondence (due to limited space we cannot explain everything in this section,
so the reader should focus on grasping the overall structure).
The AGT correspondence claims a correspondence between
the $S^4$ partition function of 4d $\scN=2$ theories and \keyword{Liouville theory}{Liouville theory}:
\begin{align}
	\begin{split}
		 & \textrm{$S^4$ partition function of the 4d $\scN=2$ theory $\mathcal{T}[\Sigma]$}                 \\
		 & \qquad\qquad \qquad \longleftrightarrow \textrm{Liouville theory on the 2d surface $\Sigma$}  \ .
	\end{split}
\end{align}

Liouville theory is a huge topic in itself, and since it is hardly mentioned in this book except in this section,
we do not need its details. What we need here is the following:
first, it is a conformal field theory whose Lagrangian is given by\footnote{Incidentally, the parameter $b$ gives its
	central charge by $c_L:=1+6 (b+b^{-1})^2$.}
\beq
S=\int d^2 z \left(\partial \phi \bar{\partial} \phi + \pi \mu e^{2b\phi} \right) \ .
\label{LiouvilleAction}
\eeq
Another point important for our purposes is the
conformal block decomposition of correlation functions.
In general, it is known that in conformal field theories the correlation functions of operators
decompose into a part holomorphic in the complex structure of the Riemann surface and
an anti-holomorphic part.
In particular, if we consider correlation functions of the operators $V_{\alpha}:=e^{2\alpha
			\phi}$,
we obtain a decomposition of the following form:
\beq
\Big\langle \prod_i V_{E_i}  \Big\rangle_{\Sigma}=\int d\alpha \,
\nu(\alpha) \,  \overline{\scF_{\alpha,E}({\tau})}\scF_{\alpha,E}(\tau) \ .
\label{Liouvillecorrelator}
\eeq
\nomenclature{$\scF_{\alpha,E}({\tau})$}{conformal block}
Here $\scF_{\alpha,E}$ is called the
\keyword{conformal block}{conformal block}, and
$\alpha$ and $E$ are each
sets of variables;\footnote{$\alpha$ are the internal momenta, and $E$ the external momenta.}
we integrate over $\alpha$, while $E$ is fixed.
Moreover, $\tau$ ($\bar{\tau}$) are the parameters of the complex structure of $\Sigma$ (and their complex conjugates).
$\nu(\alpha)$ is a certain integration measure, whose details are not particularly needed here.

This has exactly the same form of decomposition as the $S^4$ partition function in \eqref{Z_4d}.
The AGT correspondence claims that the two decompositions are identical.
Namely,
\beq
Z_{\rm 4d}[S^4]=\Big\langle \prod_i V_{E_i} \Big\rangle_{\Sigma} \ ,
\eeq
and
\begin{align}
	\begin{split}
		 & \textrm{Nekrasov partition function $Z^{\rm Nek}_{a,m}(\tau)$ of the 4d $\scN=2$ theory $\mathcal{T}[\Sigma]$} \\
		 & \qquad \longleftrightarrow \textrm{conformal block $\mathcal{F}_{\alpha,E}(\tau)$ of Liouville theory}         \\
		 & \qquad\qquad\quad \textrm{on the 2d surface $\Sigma$} \ .
	\end{split}
	\label{ZNek_F}
\end{align}
The identification of the parameters is\footnote{
	In the standard notation $\alpha=\frac{b+b^{-1}}{2}+a$; since in this book we do not use the coordinate $\alpha$ explicitly,
	it is sufficient here to understand only that $\alpha$ and $a$ correspond to each other.
}
\beq
\alpha=a, \quad E=m \ .
\eeq
Moreover, the integration measure $\nu(\alpha)$ in \eqref{Liouvillecorrelator} agrees with the measure $\nu(a)$ of the $S^4$ partition function in \eqref{Z_4d}.

\bigskip
The discussion so far was about $S^4_b$; what happens, then, to the corresponding 2d theory
if we further consider a domain wall on $S^3_b$ as in the previous chapter (Fig.~\ref{S3inS4})?
Moreover, since the 3d $\mathcal{N}=2$ theory on the domain wall thus obtained (or rather its equivalence class under $Sp(2n, \bZ)$-transformations)
could itself be regarded as an operator,
is there a corresponding operator in Liouville theory as well?

To understand this, recall that the domain wall was associated with an element of
the duality group of the 4d $\mathcal{N}=2$ theory. Recalling that the duality group of 4d $\mathcal{N}=2$ theories was identified with
the mapping class group of the 2d Riemann surface,
we find that what we should consider is an element $\varphi\in \mathrm{MCG}(\Sigma)$ of the mapping class group.
Moreover, we find that what the operator version $\hat{\varphi}$ of $\varphi$ should act on is the counterpart of the Nekrasov partition function,
namely the conformal block.
Namely, first,
\begin{align}
	\hat{\scT} \longleftrightarrow \hat{\varphi} \ ,
\end{align}
\nomenclature{$\varphi$}{element of the mapping class group of a Riemann surface}
and they act on the wave functions
\eqref{ZNek_F}
\begin{align}
	Z^{\rm Nek}_{a,m}(\tau)=\langle a,m |Z^{\rm Nek}(\tau) \rangle
	\longleftrightarrow \scF_{\alpha, E}(\tau)=\langle  \alpha, E |\scF(\tau) \rangle  \ ,
\end{align}
respectively. Here $| a,m \rangle=|\alpha,E\rangle$ and
$|Z^{\rm Nek}(\tau) \rangle=|\scF(\tau) \rangle $ are two different kinds of states;
for example, the wave function $Z^{\rm Nek}_{a,m}(\tau)$ is obtained by expanding the state $|Z^{\rm Nek}(\tau)\rangle$ in the states $|a,m\rangle$
(see also Sec.~\ref{sec.Liouville_rel} on this point).
Finally, from the correspondence for \eqref{Z4d3d} we have
\beq
Z_{\rm 3d}(a,a';m):=\langle a ,m| \hat{\mathcal{T}} | a',m \rangle
\longleftrightarrow \varphi_{\alpha, \alpha';E}:=\langle \alpha,E | \hat{\varphi} | \alpha' , E\rangle \ .
\label{Z=varphi}
\eeq
Here the identification of the parameters is, as before,
$a=\alpha, \quad a'=\alpha'$. Moreover, $m$ is, from the viewpoint of the 3d $\mathcal{N}=2$ theory, the parameter of the real mass deformation from $\mathcal{N}=4$ supersymmetry.

The operator $\hat{\varphi}$ represents the change of the conformal blocks
under $\varphi$. Assuming the natural relation
\beq
\hat{\varphi} |\scF(\tau)\rangle =|\scF(\tau') \rangle
\eeq
(this is the claim that $\varphi$ maps $\tau$ to $\tau'$, and its intuitive meaning is clear),
and inserting complete sets of states as appropriate, we obtain
\begin{equation}
	\begin{split}
		\scF_{\alpha, E}(q') & =\langle \alpha,E | \scF(q') \rangle
		= \langle \alpha,E | \hat{\varphi} | \scF(\tau) \rangle                                                                                                 \\
		                     & = \int \!d\alpha' \, \nu(\alpha') \, \langle \alpha,E| \hat{\varphi} | \alpha' , E\rangle \langle \alpha', E| \scF(q) \rangle \\
		                     & = \int \! d\alpha' \, \nu(\alpha') \, \varphi_{\alpha,\alpha';E}\scF_{\alpha',E}(\tau) \ .
		\label{F=Fderiv}
	\end{split}
\end{equation}
Readers familiar with conformal field theory can regard this as analogous to the \keyword{modular transformations}{modular transformation}
of characters in ordinary conformal field theories.

The $\varphi_{\alpha, \alpha';E}$ defined in this way can be
computed in Liouville theory and in the
quantum \Teichmuller theory discussed in Chap.~\ref{chap.Teichmuller}. In particular,
when $\Sigma=\Sigma_{1,1}$ and $\varphi$ is the S-transformation,\footnote{
	For a proposal of the 3d $\mathcal{N}=2$ theory corresponding to the integral kernel of the conformal blocks of the \keyword{four-punctured sphere}{four-punctured sphere} $\Sigma_{0,4}$,
	see Ref.~\cite{Vartanov:2013ima}.
} one can verify that the expression for
$\varphi_{\alpha, \alpha';E}$ computed in this way
reproduces the partition function \eqref{Skernelgauge} of $T[SU(2)]$ computed in the previous chapter
\cite{Hosomichi:2010vh,Terashima:2011qi}.\footnote{
	The parameters are identified as $\mu, \zeta \leftrightarrow \alpha, \alpha'$.}

\bigskip
Then, physically, elements of what kind of Hilbert space are states such as $|\alpha,E\rangle, |\scF(\tau)\rangle$?
As we will explain in Chap.~\ref{chap.Teichmuller}, in particular in Sec.~\ref{sec.Liouville_rel},
Liouville theory describes \Teichmuller space, which is (a certain connected component of)
the space of $SL(2, \bR)$ flat connections.
Its complexification is related to (a real slice of) the Hitchin system, which is the moduli of
$SL(2, \bC)$ flat connections,
and this is naturally the restriction to two dimensions of the space of
$SL(2, \bC)$ flat connections on 3-manifolds.
Therefore, we expect to obtain the desired Hilbert space by considering the
quantization of the space of $SL(2, \bC)$ flat connections on 3-manifolds, i.e.\ by canonically quantizing it.
As already stated in Sec.~\ref{subsec.twist},
this leads us to 3d $SL(2, \bC)$
Chern-Simons theory.\footnote{
	When we consider $SL(2, \mathbb{R})$ Chern-Simons theory on a manifold with boundary,
	one can show that, with appropriate boundary conditions,
	Liouville theory appears on the boundary \cite{Verlinde:1989ua}.
	It is well known that for Chern-Simons theories with compact groups,
	the WZNW (Wess-Zumino-Novikov-Witten) model appears on the boundary \cite{Witten:1988hf,Elitzur:1989nr};
	the claim above can be regarded as its extension to non-compact groups.
}

We have thus arrived at complex Chern-Simons theory.
The Hilbert space in Fig.~\ref{fig.Fubini}
was the Hilbert space of $\Sigma$ (see Ref.~\cite{Nekrasov:2010ka}).
To summarize this formally,
as elements of the Hilbert space $\scH_{S^3}\simeq \scH_{\Sigma}$ we have
\begin{equation}
	\begin{split}
		|Z^{\rm Nek}(\tau)\rangle & = |\scF(\tau) \rangle \ ,
	\end{split}
\end{equation}
from which it immediately follows that
\beq
\langle \overline{Z^{\rm Nek}(\tau)}| \hat{\varphi} |Z^{\rm Nek}(\tau) \rangle=
\langle \overline{\scF(\tau)}| \hat{\varphi} |\scF(\tau) \rangle \ .
\label{simple}
\eeq
As a slogan,
\textbf{the 3d-3d correspondence is the AGT correspondence made equivariant with respect to the duality group $=$ mapping class group},
or \textbf{the 3d-3d correspondence is a categorification of the AGT correspondence}.

The correspondences so far are summarized in Table~\ref{triality}.
The quantum \Teichmuller theory part of this table
will be explained in Sec.~\ref{sec.Liouville_rel}.

\begin{table}[t]
	\caption{Dictionary between supersymmetric gauge theories and Liouville theory/quantum \Teichmuller theory. See also Sec.~\ref{sec.Liouville_rel}.}
	\begin{center}
		\small
		\begin{tabular}{c|c}
			supersymmetric gauge theory                                 & Liouville theory (quantum \Teichmuller theory)                                     \\
			\hline
			\hline
			group of dualities                                          & mapping class group                                                                \\
			\hline
			real masses, FI parameters                                  & internal momenta (length parameters)                                               \\
			$a, a'$                                                     & $\alpha, \alpha'$                                                                  \\
			\hline
			supersymmetry-breaking masses                               & external momenta (puncture variables)                                              \\
			$m$                                                         & $E$                                                                                \\
			\hline
			Nekrasov partition function                                 & conformal block (change of basis)                                                  \\
			$Z^{\rm Nek}_{a,m}(\tau)=\langle a,m | Z^{\rm Nek}(\tau) \rangle$ & $\scF_{\alpha,E}(\tau)=\langle \alpha, E | \scF(\tau) \rangle$                           \\
			\hline
			3d partition function                                       & expectation value of the operator $\varphi$                                        \\
			$Z_{\rm 3d}(a,a';m)= \langle a,m | \hat{\mathcal{T}} | a' ,m\rangle $
			                                                            & $\varphi_{\alpha, \alpha';E}=\langle \alpha,E | \hat{\varphi} | \alpha', E\rangle$ \\
			\hline
			Hilbert space on $S^3$                                      & Hilbert space on $\Sigma$                                                          \\
			$\scH_{S^3}$                                                & $\scH_{\Sigma}$                                                                    \\
			\hline
		\end{tabular}
	\end{center}
	\label{triality}
\end{table}

\begin{practice}
	\item $[\bll]$ (S-duality of the 4d $\mathcal{N}=4$ theory) Compactifying the 6d $(2,0)$ theory on a circle of radius $R$ gave the 5d $\scN=2$ theory.
	Next, let us compactify this 5d theory on a circle of radius $R^{\prime}$.
	Express the gauge coupling constant of the resulting 4d $\scN=4$ theory in terms of $R$ and $R^{\prime}$ (see \eqref{g5R}). In particular, show that the exchange of the two radii corresponds to the 4d S-duality $g\to g^{-1}$.
	\label{ex.6d_Sdual}

	\item $[\bll]$ (Lagrangian for the 6d self-dual 2-form) \label{ex.6d_self_dual}
	As stated in the main text, it is believed to be difficult to write down
	a Lagrangian in the usual sense
	(e.g.\ one with manifest 6d Lorentz symmetry)
	for the 6d $A_N$-type $(2,0)$ theory. The 6d $(2,0)$ tensor multiplet contains
	the self-dual 2-form $B_{\mu\nu}$ satisfying $dB=*dB$. Let us then try to write down a Lagrangian (kinetic term) for this 2-form:
	if we take
	\begin{align}
		\mathcal{L}_{\textrm{self-dual 2-form}}=\frac{1}{g^2} \partial_{\mu} B_{\nu\rho} \partial^{\mu} B^{\nu\rho} \ ,
	\end{align}
	is there any problem with this Lagrangian?
	\item $[\bll]$ (Details of the topological twist) In this chapter we considered only scalar fields in the discussion of the topological twist.
	What fields, then, do the fermions become under the twist?
	Consider the cases of compactification on 2-manifolds and on 3-manifolds.
	In particular, show that as a result there exists a fermionic field transforming as a scalar
	under the Lorentz group of spacetime (a field which can be used as a BRST charge).

	\item $[\bll]$ (BPS equations)\label{ex.BPS_ASD}
	Show that the Hitchin equations \eqref{flat_F_2d_2} and \eqref{Hitchin_third} are the dimensional reduction of the 4d
	anti-self-dual (ASD) Yang-Mills equations
	\begin{align}
		F_{12}+F_{34}=F_{23}-F_{41}=F_{13}-F_{24}=0 \ ,
	\end{align}
	for fields independent of $x^3$ and $x^4$,
	where the gauge-field components $A_3, A_4$ along the two reduced directions
	become adjoint scalar fields on the 2d surface, and the Higgs field is
	the $1$-form on the 2d surface obtained from them as
	$\phi=\phi_1 dx^1 + \phi_2 dx^2$ with $(\phi_1, \phi_2)=(A_3, -A_4)$.
	Similarly, can \eqref{complex_flat} be regarded as the dimensional reduction of equations for gauge fields in higher dimensions (a higher-dimensional version of the (anti-)self-dual Yang-Mills equations)?

\end{practice}

\chapter{The World of 3-Manifolds}\label{chap.complexCS}

\begin{abstract}
	In this chapter, we discuss the moduli space of
	flat $G_{\bC}$ connections on a 3-manifold,
	as well as its quantization.
	The geometric moduli space of flat $G_{\bC}$-connections
	describes the geometry of 3-manifolds, in particular hyperbolic geometry,
	which is quantized by the complex Chern-Simons theory.
\end{abstract}

\section{Complex Chern-Simons Theory}\label{sec.complexCS}

In the previous chapter, we have seen that
\textbf{when we compactify the 6d theory on a 3-manifold $M$,
	complex Chern-Simons theory appears on $M$, and
	an $\mathcal{N}=2$ theory appears in the remaining three dimensions}.
We have thus arrived at the \keyword{3d-3d correspondence}{3d-3d correspondence}.\footnote{
	This name was given in Ref.~\cite{Terashima:2011xe}. Since in this correspondence both sides are 3d,
	and moreover in general theories with Chern-Simons terms appear on both sides, this can be confusing;
	we distinguish them by calling them ``3d $\mathcal{N}=2$ theory'', ``3d complex Chern-Simons theory'', and so on.
}
In the following we will gradually flesh out this correspondence.
One of the keys, in particular, is
the following claim \cite{Terashima:2011qi,Terashima:2011xe,Dimofte:2011jd,Dimofte:2011ju,Dimofte:2011py} (see also \eqref{5d3d_derive} in the previous chapter):
\begin{align}
	Z_{\textrm{3d $\scN=2$ theory $\mathcal{T}[M]$}}[S^3_b \textrm{ or } S^1\times S^2] = Z_{\textrm{3d $G_{\bC}$ CS theory}}[M] \ .
	\label{Z3d=Z3d}
\end{align}
Let us emphasize the non-triviality of this formula:
on the right-hand side there is only one theory under consideration
(Chern-Simons theory), whereas
on the left-hand side the theory $\scT[M]$ changes each time we change the choice of the manifold $M$ on the right-hand side.
Moreover,
the gauge group of the theory on the right-hand side is
the non-compact group $G_{\bC}$, whereas
the gauge group on the left-hand side is a compact group,
which in most cases is not $G$.
Furthermore, the left-hand side is a supersymmetric field theory, whereas
the right-hand side has only bosonic degrees of freedom.

In this chapter, we discuss \eqref{Z3d=Z3d} within the range where it can be discussed in general,
without relying on the details of the 3d $\scN=2$ theories.
For this purpose we first of all need to understand the right-hand side precisely.
What is complex Chern-Simons theory, and how is its partition function defined?
Interestingly, there are several subtle points in this seemingly trivial question.

\bparagraph{Non-Abelian Chern-Simons Term}\label{subsec.CS_term_2}

We have already touched upon \keyword{Chern-Simons theory}{Chern-Simons theory} in Chap.~\ref{chap.3dN2};
here let us look at its properties once again in a little more detail.

Let us consider a connection $A$ valued in the gauge group $G$, and Chern-Simons theory on a 3-manifold $M$.
Let us assume for the moment that $M$ is a closed manifold without boundary, and moreover that $G$ is a simple group (e.g.\
$SU(N)$). The Lagrangian is then given by
\begin{align}
	L[A] =\frac{k}{4\pi }
	L_{\rm CS}[A] \ ,
	\label{L_total}
\end{align}
with
\begin{align}
	\begin{split}
		L_{\rm CS}[A]
		 & :=
		\int_M \textrm{Tr}\left(A \wedge dA+\frac{2}{3}
		A\wedge A\wedge A\right)
		\\
		 & =\frac{1}{2}\int_M \epsilon^{ijk} \textrm{Tr}\left(A_i \left(\partial_j A_k-\partial_k A_j\right)+\frac{2}{3}
		A_i [A_j, A_k] \right) \ .
		\label{LCS_new}
	\end{split}
\end{align}
\nomenclature{$t$}{level of complex Chern-Simons theory}
\nomenclature{$\mathcal{A}$}{complexified gauge field}
Here the overall coefficient $k$ is called the \keyword{level}{level}.
The partition function of this theory is defined, assuming that the metric of $M$ is Euclidean, by the
path integral
\begin{align}
	Z_{\rm CS}=\int \mathcal{D}A \,\,\,  e^{iL[A]} \ .
\end{align}

Here we have chosen the normalizations
of $\textrm{Tr}$ and the Lie algebra generators
such that \eqref{Killing_form} holds.
This normalization, while a convention, is correlated with the
quantization condition of the parameter $k$ (as we will discuss shortly).

Care is needed since the gauge field itself appears in this Lagrangian.
In particular, when we consider
gauge fields (connections of principal bundles) valued in non-trivial bundles on the manifold,\footnote{When the gauge group is connected and simply connected (e.g.\ $SU(N)$), only trivial bundles exist; here, for the sake of understanding, let us consider a general gauge group $G$ which does not necessarily satisfy these conditions. For example, $U(1)$ is not simply connected.
} the form of the Lagrangian could change depending on the choice of their
trivialization.

To remove this ambiguity, let us choose a 4-manifold $V$ which has the 3-manifold $M$ as its boundary ($\partial V=M$),
and rewrite \eqref{L_total} by Stokes' theorem as
\begin{align}
	L & =\frac{k}{4\pi }\int_{\partial V} \textrm{Tr}\left(A
	\wedge dA+\frac{2}{3}
	A\wedge A\wedge A\right) \nonumber                       \\
	  & =2\pi k\int_V
	\textrm{Tr} \left(
	\frac{1}{8\pi^2} F\wedge F
	\right) \ .
	\label{LCS2}
\end{align}
Since only the field strength $F\, (=dA+A\wedge A)$ appears in this expression, it is manifestly
independent of the choice of the trivialization.

However, there is now an arbitrariness in the choice of the 4-manifold $V$.
The difference from the case where we choose another 4-manifold $V'$ is,
using the 4-manifold $X=V\cup (-V')$ without boundary,
\begin{align}
	\delta L =-2\pi k\int_X
	\textrm{Tr} \left(
	-\frac{1}{8\pi^2} F \wedge F
	\right)
	\in  2\pi k \mathbb{Z} \ .
\end{align}
Here, in the last step we used the fact that the integral of the \keyword{second Chern character}{second Chern character}
\begin{align}
	\textrm{ch}_2(F):= -\frac{1}{8\pi^2} \textrm{Tr}\left(F \wedge F\right)
	\label{ch2F_def}
\end{align}
over a 4-manifold without boundary ($\partial X=\emptyset$)
is an integer.\footnote{
	The precise quantization condition depends on the gauge group. Moreover,
	for example when the gauge group is $SO(n)$,
	it differs by a factor of $2$ depending on whether or not we require
	the 4-manifolds $V, V'$ to have \keyword{spin structures}{spin structure}.
}
Therefore, the action $e^{iL}$ is independent of the choice of $V$ if the level $k$ is an integer,
and we have derived that the level $k$ has to be quantized ($k\in \bZ$).
In exercise~\ref{ex.instanton} we will derive this quantization condition
by another method.

\bigskip
Since we did not need to use the metric of the manifold in the Lagrangian \eqref{LCS_new},
the theory is independent of the metric.
Such a field theory is called a \keyword{topological field theory}{topological field theory}.

\small
Topological theories are in a sense special theories; for example, unlike in usual theories,
correlation functions do not depend on distances. At the time of their discovery \cite{Witten:1988ze}, there seems to have been no small amount of skepticism regarding their physical importance.
However, through the many subsequent works, represented by the paper \cite{Witten:1988hf} by Witten,
relations to rich mathematics such as knot theory, topology, and representation theory have been discovered.
Moreover, in condensed matter physics as well, Chern-Simons theory has come to play a central role in the description of the quantum Hall effect and, more generally, of topological order,
and its usefulness can now be said to be established.
\normalsize

Since the level $k$ is a coefficient in front of the Lagrangian,
the expansion parameter when regarded as a quantum system
is $\hbar\sim \frac{1}{k}$.
In the semiclassical limit $\hbar\to 0, k \to \infty$,
the contributions to the path integral concentrate around the classical solutions of the equations of motion, which are given by \keyword[flat connection]{flat connections}{flat connection}:
\begin{align}
	F=dA+A\wedge A=0 \ .
\end{align}
In this sense, it can be said that Chern-Simons theory
gives a well-defined physical procedure to quantize the moduli space of flat connections
(mathematically as well, Chern-Simons theory was born from the study of
flat vector bundles).
Recall that we have already seen in Chap.~\ref{chap.6d}
that the compactification of the 6d $(2,0)$ theory gives rise to
a moduli space of flat connections.

\bparagraph{Complex Chern-Simons Term}

Since the flat connections in Chap.~\ref{chap.6d} were flat connections of the complexified gauge group
$G_{\bC}$, we also needed to make the gauge group $G_{\bC}$ on the Chern-Simons theory side.
Let us examine what happens in this case \cite{Witten:1989ip}.

In ordinary Yang-Mills theory we cannot complexify the gauge group.
This is because the non-compact group $G_{\bC}$
has no positive-definite quadratic form:
when we write
\begin{align}
	\textrm{Tr}(T^a T^b)=\epsilon_a \delta_{ab} \ ,
\end{align}
we cannot make all of the $\epsilon_a$ positive, and some of them become negative
(this can be easily seen by considering multiplying the generators $T^a$ by the imaginary unit $i$ in the usual Killing form \eqref{Killing_form}).
The kinetic term of the gauge field then becomes
\begin{align}
	\int \textrm{Tr} F_{\mu\nu} F^{\mu\nu}
	= \sum_a \int  \epsilon_a F^a_{\mu\nu} F^{a, \mu\nu} \ ,
\end{align}
so that the energy is no longer bounded from below, and
the theory becomes unstable. This is why we do not consider complex gauge groups for ordinary Yang-Mills fields.\footnote{For fermions, we can consider the Dirac sea, so that this problem does not arise; however, a problem with unitarity then arises instead.}

Fortunately, this problem does not exist in Chern-Simons theory,
since Chern-Simons theory is a topological theory and
its Hamiltonian is always zero.

Complex Chern-Simons theory is obtained by complexifying this gauge field $A$.
Here, the complexification $G_{\mathbb{C}}$ of the gauge group $G$ is,
concretely, $G_{\mathbb{C}}=SL(N, \mathbb{C})$ when $G=SU(N)$.
Denoting the connection of the $G_{\mathbb{C}}$ gauge group by $\mathcal{A}$, and
its complex conjugate by $\overline{\scA}$,
the Lagrangian of complex Chern-Simons theory is given by
\begin{align}
	\begin{split}
		L_{\textrm{complex CS}}[\scA, \overline{\scA}]=
		  & \frac{t}{8\pi } L_{\textrm{CS}}[\scA]+  \frac{\overline{t}}{8\pi } L_{\textrm{CS}}[\overline{\scA}]
		\\
		= & \frac{t}{8\pi }\int_M \textrm{Tr}\left(\mathcal{A} \wedge d\mathcal{A}+\frac{2}{3}
		\mathcal{A}\wedge \mathcal{A}\wedge \mathcal{A}\right)
		\\
		  & +
		\frac{\overline{t}}{8\pi }\int_M \textrm{Tr}\left(\overline{\mathcal{A}} \wedge d\overline{\mathcal{A}}+\frac{2}{3}
		\overline{\mathcal{A}}\wedge \overline{\mathcal{A}}\wedge
		\overline{\mathcal{A}}\right) \ .
		\label{LcCS}
	\end{split}
\end{align}
Here $t$ is the \keyword{complexified level}{complexified level}, and is written as\footnote{As stated later in footnote~\ref{foot.orientation},
	$s$ is either real or purely imaginary. Note in particular that when $s$ is purely imaginary, $t$ and
	$\overline{t}$ are not complex conjugates of each other.
	Note also that some references write $s$ for our $is$. This $s$ is unrelated to the vector multiplet scalar $\sigma$ of the 3d $\scN=2$ theory.
}
\nomenclature{$s$}{imaginary part of the complexified Chern-Simons level}
\begin{align}
	t:=k+i s \ , \quad
	\overline{t}:=k-is \ .
\end{align}
Moreover, the coefficient in \eqref{LcCS} was chosen to be
$\frac{t}{8\pi}$ for the convenience that the quantization condition discussed later becomes $k\in \bZ$.
The partition function of the theory is defined by the path integral
\begin{align}
	Z_{\textrm{complex CS}}:=\int \CD \CA \CD \bCA \,\,\,  e^{ i
			                                                       L_{\textrm{complex CS}}[\CA, \bCA]} \ .
	\label{CSint}
\end{align}

In the following we exclusively consider the case $G=SU(2), G_{\bC}=SL(2, \bC)$
(we will briefly comment on the case $G=SU(N), G_{\bC}=SL(N, \bC)$ in Sec.~\ref{subsec.AN}).

\small
Many readers may not be familiar with the non-compact group $SL(2, \bC)$;
however, one may feel more familiar with it upon noticing that $SL(2, \bC)\sim SO(3,1)$,
which agrees with the 4d Lorentz group $SO(3,1)$ up to the signature. When we consider unitary representations, however,
the difference in signature is crucial. For example, the compact group $SO(4)$ has
finite-dimensional unitary representations, whereas $SO(3,1)$ has
no finite-dimensional unitary representations.

\normalsize

\bparagraph{Integration Cycle}

We have explained complex Chern-Simons theory so far,
but the story is not yet over: there is an ambiguity in the definition \eqref{CSint} of the partition function.

In ordinary path integral problems, we consider
\begin{align}
	Z_{\textrm{path integral}}=\int \scD \Phi \,\,\, e^{i t  L[\Phi]} \ ,
\end{align}
where the coupling constant $t$ is a real parameter.
In this case the integrand is a phase factor (note the factor of $i$),
and oscillates rapidly as $\Phi$ varies;
its quantum-mechanical treatment is as taught in textbooks of quantum mechanics.

However, the situation is different once we start complexifying the coupling constants and the fields $\Phi$.
The integrand is not necessarily a phase factor, and its absolute value can also change.
In particular, as the values of the fields approach infinity, the value of the integrand
can also become large, so that the integral cannot be carried out.

In such situations, we can deform the integration contour
in the complex plane, and choose an integration contour which starts from a region at infinity where the integrand converges
and ends in another converging region, namely an
\keyword{integration cycle}{integration cycle} $\Gamma$:
\begin{align}
	Z_{\textrm{complex CS}}=\int_{\Gamma} \CD \CA \CD \bCA \,\,\,  e^{ i
			                                                               L_{\textrm{complex CS}}[\CA, \bCA]} \ .
	\label{CSintGamma}
\end{align}

Among such cycles there are several linearly independent ones, and
they can be constructed by Morse theory in infinite dimensions \cite{Witten:2010cx}.\footnote{
	More concretely, we start from a classical solution, and construct the integration cycle along the flow
	whose Morse function is the real part of the Lagrangian (this is an infinite-dimensional version of the relation between
	supersymmetric quantum mechanics and Morse theory \cite{Witten:1982im}). Such an integration cycle
	is called a \keyword{Lefschetz thimble}{Lefschetz thimble},
	and the integral over such an integration cycle has a clear semiclassical interpretation.
} We do not need the details here, but we will use below the fact that the independent integration cycles
are labeled by classical solutions (we do not directly explain this here;
it is sufficient to have the intuitive understanding that the WKB approximation around a classical solution,
made into an exact solution by including higher-order corrections,
gives the integral over an independent integration cycle).

In the present case, the Lagrangian \eqref{LcCS} was written as the sum of a holomorphic (i.e.\ depending on $t$) part and
an anti-holomorphic (i.e.\ depending on $\overline{t}$) part.
Therefore, the classical flat connections can be specified separately for the holomorphic part and the anti-holomorphic part.
More mathematically, this follows from
$(G_{\mathbb{C}})_{\mathbb{C}}=
	G_{\mathbb{C}}\times G_{\mathbb{C}}$.
Let us therefore label the classical solutions by
$\alpha, \bar\alpha$, and denote the corresponding integration cycles by
$\scC_{\alpha}, \scC_{\bar\alpha}$.
Then any integration cycle $\scC$ can be expanded in terms of
$\scC_{\alpha}, \scC_{\bar\alpha}$:
\begin{align}
	\scC=\sum_{\alpha, \bar\alpha} n_{\alpha, \bar\alpha} \left(\scC_{\alpha} \times \scC_{\bar\alpha}\right) \ ,
\end{align}
\nomenclature{$\scC$}{integration cycle}
\nomenclature{$\scC_{\alpha}, \scC_{\bar\alpha}$}{basis of integration cycles}
where $n_{\alpha, \bar\alpha}$ are integers,
given by the (signed) intersection numbers of
$\scC$ and $\scC_{\alpha} \times \scC_{\bar\alpha}$.
Therefore, if we define
\begin{align}
	Z_{\alpha}(t)=\int_{\mathcal{C}_{\alpha}} \mathcal{D}{\scA} \,\,\, e^{i
			                                                                   \frac{t}{8\pi} \int_{M_3} {\cal L}_{\rm CS}[\scA]}
	\label{CSblock}
\end{align}
and
\begin{align}
	\bar Z_{\bar\alpha}(\bar t)=\int_{\mathcal{C}_{\bar\alpha}} \mathcal{D}\overline\scA \,\,\, e^{i
			                                                                                            \frac{\overline{t}}{8\pi} \int_{M_3} {\cal L}_{\rm CS}[\overline{\scA}]} \ ,
	\label{CSantiblock}
\end{align}
the partition function is written as
\begin{align}
	Z_\textrm{complex CS}(t, \bar t)=\sum_{\alpha, \bar\alpha} n_{\alpha,\bar\alpha} Z_{\alpha}(t) \bar Z_{\bar \alpha}(\bar t) \ .
	\label{CSfactor}
\end{align}

We derived
\eqref{CSfactor}
in an abstract manner,
but it is not difficult to grasp its intuitive meaning.
First, the Lagrangian \eqref{LcCS} was written as the sum of a holomorphic part and
an anti-holomorphic part.
It therefore immediately follows that the expansion around a classical solution also
factorizes into the product of the expansion $Z_{\alpha}(t)$ around the holomorphic part
and the expansion $\bar Z_{\bar \alpha}(\bar t)$ around the anti-holomorphic part.
However, this is only perturbation theory around the classical solutions,
\begin{align}
	Z(t) \sim e^{t S_{-1} + S_0+t^{-1} S_1+\cdots } \ ,
	\label{t_asymp}
\end{align}
and once we take into account the classical solutions and the non-perturbative effects connecting them,
the simple factorization into holomorphic and anti-holomorphic parts no longer holds.
The integers $n_{\alpha,\bar\alpha}$ represent these non-perturbative effects.

What we call the partition function of $G_{\bC}$ Chern-Simons theory is
\eqref{CSfactor}.
On the other hand, $Z_{\alpha}(t)$ should precisely be called
\textbf{the partition function of analytically continued Chern-Simons theory}.
In the literature the latter is also often called
the partition function of complex Chern-Simons theory, so
care is needed as to which of the two is meant.
In this book, to avoid complications, we generally call both complex Chern-Simons theory,
but we would like to clearly distinguish which is meant:
for example, what appears from the localization of the 5d $\mathcal{N}=2$ theory in Sec.~\ref{subsec.twist} is
complex Chern-Simons theory (combining the holomorphic and anti-holomorphic parts),
whereas in Sec.~\ref{subsec.3d_as_classical} of this chapter, when we discuss
classical solutions, we consider analytically continued Chern-Simons theory (only the holomorphic part).

\section{6d Theory Revisited}\label{sec.6d_revisited}

With the discussion so far, we have precisely defined the right-hand side of \eqref{Z3d=Z3d}.
This is not merely for the sake of rigor; what is important is that in the process we obtained the decomposition \eqref{CSfactor}.
Since \eqref{Z3d=Z3d} is an equality, this at the same time means that the left-hand side of \eqref{Z3d=Z3d}, the
$S^3_b$ partition function of the 3d $\scN=2$ theory, also has a similar decomposition.
Here let us see that this can be understood from the compactification of the 6d theory.
This will clarify a point which we left ambiguous in the previous chapter.

In the discussion of the previous chapter we took $b\to 0$; here let us keep $b$
finite.
Following Chap.~\ref{chap.S3} (in particular Fig.~\ref{fig.S3Heegaard}), we decompose $S^3$ into two solid tori,
and consider $S^1\times D^2\times M$. $D^2$ can be regarded as $S^1$ fibered over the one-dimensional half-line $\bR_{\ge 0}$ with an end point.
We therefore obtain $S^1\times S^1\times \bR_{\ge 0} \times M$, and if we first compactify on
$T^2=S^1\times S^1$, we obtain
the 4d $\mathcal{N}=4$ theory on $\bR_{\ge 0}\times M$.
Of course, at the end point of $\bR_{\ge 0}$ one of the cycles of the torus $T^2$
shrinks, and correspondingly we have to assume that appropriate boundary conditions are imposed on the 4d $\mathcal{N}=4$ theory:
\begin{align}
	\begin{split}
		 & \textrm{6d $(2,0)$ theory: } S^1\times D^2 \times M \textrm{ (+ boundary conditions)}                                         \\
		 & \qquad \longrightarrow \textrm{4d $\mathcal{N}=4$ theory: } \mathbb{R}_{\ge 0}\times M \textrm{ (+ boundary conditions)}  \ .
	\end{split}
\end{align}

Since the 4d $\scN=4$ theory is placed on the curved manifold $M$,
it is topologically twisted along those directions. The twist identifies the
natural subgroup $SO(3)$ of $SO(6)\supset SO(2)\times SO(4)\simeq SO(2)\times SO(3) \times SO(3)$
with $SO(3)_M$ along the $M$-directions.\footnote{Under $SO(3)\times SO(3)\times SO(2)\simeq SU(2)\times SU(2)\times U(1)$, the $\bm{4}$-representation of $SO(6)\simeq SU(4)$ decomposes as $(\bm{2}, \bm{1})_{1} \oplus (\bm{1}, \bm{2})_{-1}$, and
	the $\bm{6}=\wedge^2 \bm{4}$-representation as $(\bm{2}, \bm{2})_0\oplus (\bm{1}, \bm{1})_{2}\oplus (\bm{1}, \bm{1})_{-2}$.} This is the topological twist first discussed in Refs.~\cite{Marcus:1995mq,Blau:1996bx}, and is the topological twist which appears in connection with the \keyword{geometric Langlands correspondence}{geometric Langlands correspondence} \cite{Kapustin:2006pk}.

When we choose $SO(3)$ from $SO(4)\simeq SO(3) \times SO(3)$, we may mix the original two $SO(3)$'s:
there are ways of twisting specified by points of $\mathbb{CP}^1$, which in the language of the moduli space
are nothing but the choices of a complex structure of the hyperK\"{a}hler manifold. The theory
constructed in this way is found to depend on a single parameter $\mathcal{K}$, combining the point
$t\in \bC\cup\{\infty\}$ of $\mathbb{CP}^1$ and the 4d complex gauge coupling constant $\tau$ \eqref{complex_gauge} \cite{Kapustin:2006pk}.\footnote{
	The precise relation is
	\begin{align}
		\mathcal{K}= \frac{\tau+\overline{\tau}}{2} +
		\frac{\tau-\overline{\tau}}{2} \left(
		\frac{t-t^{-1}}{t+t^{-1}}
		\right) \ .
	\end{align}
}

The Lagrangian of the theory thus obtained by the topological twist
is $\scQ$-exact with respect to a certain supercharge $\scQ$, and hence the theory is topological;
however, when there is a boundary, as in the present case,
we have to take the boundary terms there into account.
In fact, the boundary term is complex Chern-Simons theory with complexified level $t=\mathcal{K}/2$
\cite{Witten:2010cx,Witten:2010zr}.\footnote{
	Since the 4d $\mathcal{N}=4$ theory and its twist
	have kinetic terms for the gauge fields,
	the argument that a Chern-Simons term appears on the boundary from them is itself
	the same as in Sec.~\ref{subsec.4d_3d_SpZ}.}
This can be said to give a derivation of the fact that the theory on the 3-manifold is 3d Chern-Simons theory,
and is the computation of
Refs.~\cite{Lee:2013ida,Cordova:2013cea}, explained around \eqref{5d3d_derive} in the previous chapter,
with one dimension lowered.

Since $\bR_{\ge 0}$ has a point at infinity,
the partition function is not determined unless we also specify the boundary condition there.
For this purpose we can consider
flat $G_{\bC}$-connections labeled by $\alpha$
(as explained in \eqref{CSblock}). The claim is then
\begin{align}
	Z_{\alpha}^\textrm{4d $G$ $\scN\!=\!4$}[\mathbb{R}_{\ge 0}\times
		           M_3]=Z_{\alpha}^\textrm{3d $G_{\mathbb{C}}$ CS}[M_3](t)\ ,
	\label{4donRplus}\end{align}
where the right-hand side is the quantity introduced in \eqref{CSblock}.

Let us now return to the original $S^3_b$. $S^3_b$ was obtained by
gluing two solid tori by an $S$-transformation on their boundaries.
It is natural to identify the decomposition \eqref{CSfactor} of 3d Chern-Simons theory
with the decomposition into solid tori.

As a gluing of solid tori, let us first consider
gluing the two solid tori as they are.
What we obtain is then $S^1\times S^2$, and
from the result of the 5d localization in Ref.~\cite{Lee:2013ida}, $t$ is purely imaginary:
\begin{align}
	t=-i\frac{8\pi^2}{g_5^2}  \ , \quad \bar t=i \frac{8\pi^2}{g_5^2} \ .
	\label{t_S1S2}
\end{align}
$t$ and $\overline{t}$ have opposite signs; this means that, since
the orientations of the two boundaries are opposite, the Chern-Simons terms on the boundaries obtained by
integrating the bulk action by parts have opposite signs.
We do not explain the $S^1\times S^2$ partition function (mentioned at the end of Sec.~\ref{sec.S3_integral}) in this book,
but it may be worth knowing that this partition function has been computed by localization.

Let us next consider obtaining $S^3_b$.
The answer given by direct localization is\footnote{
	This follows Ref.~\cite{Cordova:2013cea}, but we have corrected a factor of $i$. See also Refs.~\cite{Dimofte:2014zga,Gang:2014ema}.
}
\begin{align}
	k=1 \ , \quad s=-i\frac{|b-b^{-1}|}{b+b^{-1}}
\end{align}
(in particular, the level $k$ is $1$), and hence (assuming $b\le 1$ without loss of generality)
\begin{align}
	t=1+\frac{|b-b^{-1}|}{b+b^{-1}} = \frac{2}{1+b^{2}} \ ,\quad
	\bar t=1- \frac{|b-b^{-1}|}{b+b^{-1}} = \frac{2}{1+b^{-2}} \ .
	\label{t_S3}
\end{align}
In particular, $t$ and $\bar t$ are exchanged under $b\to b^{-1}$;
this is consistent with the fact that the two cycles of the boundary tori of the two solid tori
are exchanged by the $S$-transformation, and that their sizes were $b$ and $b^{-1}$, respectively
(Fig.~\ref{fig.S3Heegaard}).

Returning to \eqref{Z3d=Z3d}, in both cases $S^1\times S^2$ and $S^3_b$,
there should exist the decomposition \eqref{CSfactor} of complex Chern-Simons theory:
\begin{align}
	Z_{S^1\times S^2, S^3_b}(t, \bar t)=\sum_{\alpha, \bar\alpha} n_{\alpha,\bar\alpha} Z_{\alpha}(t) \bar Z_{\bar \alpha}(\bar t) \ .
	\label{ZS3factor}
\end{align}
Indeed, this decomposition has been directly checked for concrete 3d $\scN=2$ theories using the explicit forms of the partition functions
\cite{Pasquetti:2011fj,Beem:2012mb}, and has also been derived by localization \cite{Benini:2013yva,Fujitsuka:2013fga} (the $Z_{\alpha}(t)$ appearing in this way are called \keyword{holomorphic blocks}{holomorphic block}, and the decomposition \eqref{ZS3factor}
is called \keyword{Higgs branch localization}{Higgs branch localization}\footnote{As in the case of ``Coulomb branch localization'', care is needed since the vacuum moduli space in flat spacetime is different from the vacuum moduli space in curved spacetime ($S^3_b$).}).

\bigskip

Let us now go back once again from the 5d theory to the 6d theory.
Since the 5d theory is expected to contain all the degrees of freedom of the 6d theory (at least the part preserving some of the supersymmetry),
the value of the partition function is expected not to change. However, the way of expansion should change.
The present expansion \eqref{t_asymp} in $t^{-1}$ is good when the value of $t$ is large,
which is when the value of $g_5$ is small, namely, referring to \eqref{g5d},
when the radius $R$ of the sixth direction is small.

When the radius $R$ of the sixth direction is large, $g_5$ is large, and hence the value of $t$ is small, so that the expansion in $t^{-1}$
becomes bad. If we regard $R\sim t^{-1}$ as the inverse temperature $\beta$, this is nothing but the low-temperature expansion.
The low-temperature expansion is a sum over excited states above the ground state, and each
excited state gives a contribution of the form $e^{-\beta (\textrm{constant})}$. Therefore,
Chern-Simons theory has an expansion in the parameters
\begin{align}
	q:=e^{4\pi i/ t}, \quad
	\overline{q}:=e^{-4\pi i/ \overline{t}} \ ,
	\label{qqbar}
\end{align}
and the decomposition \eqref{ZS3factor} is also rewritten as\footnote{Such a replacement of an infinite sum by another infinite sum valid in a different parameter region
	also appears, for example, when rewriting Gromov-Witten invariants in terms of Gopakumar-Vafa invariants.}
\begin{align}
	Z_{S^1\times S^2, S^3_b}(t, \bar t)=\sum_{\alpha, \bar\alpha} n_{\alpha,\bar\alpha} Z_{\alpha}(q) \bar Z_{\bar \alpha}(\overline{q}) \ .
	\label{ZS3factor_2}
\end{align}

Noticing the existence of a decomposition of the form \eqref{ZS3factor_2} resolves a contradiction.
Careful readers may have noticed that,
from \eqref{t_S3}, $t$ does not go to $\infty$ even when $b\to 0$,
which seems to contradict our storyline so far that the classical limit $t\to\infty$ of Chern-Simons theory
coincides with the dimensional reduction $b\to 0$ of 3d $\mathcal{N}=2$ theories.
In reality, however, the partition function is rewritten as a function of $q, \overline{q}$ in \eqref{qqbar},
and then
\begin{align}
	q=e^{4\pi i/ t}=e^{2\pi i(1+b^{2})} =e^{2\pi i b^{2}} \ ,
	\;  \overline{q}=e^{-4\pi i/\overline{t}}=e^{-2\pi i(1+b^{-2})}=e^{-2\pi i b^{-2}} \ .
	\label{qtb}
\end{align}
Namely, $b\to 0$ indeed corresponds to the classical limit $q\to 1$.\footnote{If we take $b\to 0$ on the real axis,
	$\overline{q}$ oscillates rapidly, so that we need to give $b$ a small imaginary part.}

\section{Semiclassical Limit}\label{subsec.CS_classical}

Let us now consider taking the limit $b\to 0$ on both sides of
\eqref{Z3d=Z3d}.
As already stated in Chap.~\ref{chap.S3},
the limit of the $S^3_b$ partition function takes the integral form \eqref{Zdiverge}, in which
the effective twisted superpotential appeared.

As already stated in the previous section,
what corresponds to this on the complex Chern-Simons theory side is
the classical limit $t\to \infty$.
The classical saddle points are
solutions of the equations of motion $\CF_{\CA}=\bar \CF_{\bCA}=0$,
and the moduli space formed by these solutions is given by the
moduli space $\mathcal{M}_{\rm flat}(M)$ of flat connections.
Since flat connections are always locally pure gauge,
the only non-trivial information is global, namely the Wilson loops.
Mathematically, this is written as
\begin{align}
	\mathcal{M}_{\rm flat}(M):=\textrm{Hom}(\pi_1(M), G_{\mathbb{C}})
	/ \textrm{(equivalence under $G_{\mathbb{C}}$-conjugation)} \ .
	\label{M_flat}
\end{align}
\nomenclature{$\mathcal{M}_{\textrm{flat}}$}{moduli space of flat connections}

To obtain a more concrete description,
it is useful to rewrite the Lagrangian \eqref{LcCS} as follows.
Let us write the real and imaginary parts of the $G_{\mathbb{C}}$ connection as
$\mathcal{A}=A+i\phi$.
We then obtain
\begin{align}
	\begin{split}
		L_{\textrm{complex CS}}= & \frac{k}{4\pi}
		\int_M
		\textrm{Tr}\left(
		A\wedge dA+\frac{2}{3} A\wedge A\wedge A-
		\phi \wedge D_A \phi
		\right)
		\\
		                         & -\frac{s}{2\pi}
		\int_M
		\textrm{Tr}\left(
		F_A \wedge \phi-\frac{1}{3} \phi\wedge \phi\wedge \phi
		\right) \ ,
	\end{split}
	\label{LcCS_re_im}
\end{align}
where $D_A$ is the covariant derivative
$D_A:=d+A$ with respect to $A$, and $F_A:=dA+A\wedge A$ is the field strength of $A$.
In this Lagrangian, the part which could change under gauge transformations is the Chern-Simons term for the real part $A$.
Since the gauge group $G_{\bC}=SL(N, \mathbb{C})$ is
contractible to its maximal compact subgroup $G=SU(N)$,
it follows from the same argument as in Sec.~\ref{subsec.CS_term_2} that the value of $k$ has to be
quantized to an integer.
On the other hand, the value of $s$ need not be quantized;
from the requirement of unitarity (its Euclidean version),
we have $s \in \mathbb{R}$ or
$s\in i\mathbb{R}$.\footnote{The requirement is that the action is preserved under complex conjugation combined with
	the orientation reversal of $M_3$.
	If $A, \phi$ do not change sign when we reverse the orientation of $M_3$, then $s\in \bR$,
	and if only $\phi$ changes sign while $A$ does not, then $s\in i \bR$
	\cite{Witten:1989ip}.\label{foot.orientation}}

Let us look at the second term, proportional to $s$, in \eqref{LcCS_re_im}.
When we rewrite $A=\omega, \phi=e$, this term becomes
\begin{align}
	-\frac{s}{2\pi }\,
	\textrm{Tr}
	\left(
	(d\omega+\omega\wedge \omega) \wedge e-\frac{1}{3}e\wedge e\wedge e
	\right) \ .
\end{align}
For $G_{\mathbb{C}}=SL(2, \mathbb{C})$, this is nothing but the Lagrangian of
\keyword{3d gravity}{3d gravity}, determined by the
\keyword{spin connection}{spin connection} $\omega$ and the
\keyword{dreibein}{dreibein} $e$,\footnote{
	For the meaning of the spin connection and the dreibein, see textbooks on general relativity.
	The relation between $\omega$ and $R_{\omega}$ is analogous to the relation between a gauge field and its field strength.
}
\begin{align}
	-\frac{1}{16\pi G_N}\textrm{Tr}
	\left(
	R_{\omega} \wedge e+\frac{\Lambda}{3}e\wedge e\wedge e
	\right)\ , \quad R_{\omega}:=d\omega+\omega \wedge \omega \ ,
	\label{L_3d_gravity}
\end{align}
with the Newton constant $G_N$ and the cosmological constant $\Lambda$ chosen as
\begin{align}
	G_{N}=\frac{1}{8s} \ , \quad \Lambda=-1 \ .
\end{align}
Moreover,
if we vary the Lagrangian \eqref{L_3d_gravity} of 3d gravity and examine the
classical solutions, we obtain
\begin{align}
	R_{\omega}=-\Lambda e\wedge e\ , \quad  D_{\omega}e=de+[\omega, e]=0 \ .
\end{align}
The second equation determines the spin connection from the dreibein,
and the first equation is nothing but the condition of constant negative curvature $R=\Lambda<0$.

In this way, complex $SL(2, \bC)$ Chern-Simons theory and
3d gravity agree at the level of classical solutions \cite{Witten:1988hc}.
However, not all classical solutions of complex Chern-Simons theory have an interpretation
in 3d gravity---in 3d gravity there was the condition that the metric is positive definite.
In fact, in $G_{\mathbb{C}}$ Chern-Simons theory the geometric interpretation is not clear;
for example, the solution $A_{\mu}=0$ also exists as a classical saddle point solution. In this sense,
3d gravity is sufficient as long as we consider classical solutions and the fluctuations around them;
however, as already explained in Chap.~\ref{chap.6d},
what is related to 3d $\mathcal{N}=2$ supersymmetric gauge theories in the discussion of this book is
complex Chern-Simons theory.
For reference, as a generalization of the discussion here,
it is known that $G=SU(N), G_{\mathbb{C}}=SL(N, \mathbb{C})$ Chern-Simons theory
is related to gravity theories allowing higher spins.

\bigskip
A classical solution of 3d gravity, namely a manifold admitting a metric
satisfying constant negative curvature $R=\Lambda<0$, is called
a hyperbolic manifold.
If we perform the semiclassical approximation around such a classical solution,
we find from \eqref{LcCS_re_im} and \eqref{L_3d_gravity} that its leading part is
\begin{align}
	Z_{\textrm{complex CS}} \longrightarrow \exp\left[
		                                            \frac{-is}{2\pi}\textrm{Vol}[M]+\frac{ik}{4\pi} \textrm{CS}[M]
		                                            \right] \ .
	\label{CSlimit}
\end{align}
Here $\textrm{CS}[M]$ is the quantity obtained by integrating the Chern-Simons form over the 3-manifold,
called the \keyword{Chern-Simons invariant}{Chern-Simons invariant}.\footnote{For closed manifolds,
	this quantity is well defined only modulo integers (in a suitable normalization), as we have already seen from its behavior under gauge transformations. When the manifold has a boundary,
	we need to specify boundary conditions, and the value of the invariant depends on the boundary conditions.}
In the situations considered in this book, the classical solutions corresponding to non-trivial hyperbolic metrics
give larger contributions to the partition function than the solutions corresponding to the trivial flat connection
(this is not a trivial statement, and can be said to be a rephrasing of the volume conjecture discussed later in Sec.~\ref{subsec.knot_complement}).
For these reasons, in the rest of this section and in Chap.~\ref{chap.3mfd}
we exclusively consider the complex flat connections corresponding to
non-trivial hyperbolic metrics.

\subsection{3d Hyperbolic Geometry as Classical Solution}\label{subsec.3d_as_classical}

In the previous section we have seen that
constant negative curvature metrics appear as
classical solutions of complex Chern-Simons theory.
Manifolds admitting such metrics are called
\keyword{hyperbolic manifolds}{hyperbolic manifold}.

Here we therefore briefly summarize
the basics of
3d hyperbolic geometry.
For readers who wish to learn more about 3d hyperbolic geometry,
the lecture notes \cite{ThurstonLecture} and the book \cite{ThurstonBook} by Thurston
are useful.
See also Refs.~\cite{KojimaBook,TaniguchiBook} for textbooks on 3d hyperbolic geometry.
However, the geometric facts needed in this book are very few,
and many of them have counterparts on the supersymmetric gauge theory side,
so that one can reason by analogy (if we reverse the argument and assume the correspondence between 3d $\mathcal{N}=2$ theories and
the geometry of 3-manifolds).

Among the geometric structures of 3-manifolds (see Sec.~\ref{sec.geometric_structure}),
the 3d hyperbolic structure is in a sense the most general one.\footnote{See footnote~\ref{foot.hyperbolic}.}
The 3d hyperbolic manifolds are locally identified with $\bH^3$.
The 3d hyperbolic space $\bH^3$ is defined, in the Poincar\'e upper half-space model, as
the upper half-space
\begin{align}
	\mathbb{R}^3_{> 0}:=\{(x_1, x_2, y) |\, x_1, x_2\in \mathbb{R}, y>0 \}
\end{align}
equipped with the metric
\begin{align}
	ds^2=\frac{(dx_1)^2+(dx_2)^2+dy^2}{y^2} \ .
	\label{H3metric}
\end{align}
\nomenclature{$\bH^3$}{3d hyperbolic space}
This manifold has the boundary $\partial \bH^3=\mathbb{C}\cup \{\infty\}$.
Moreover, $\mathbb{H}^3$ has the isometry group $PSL(2,\mathbb{C})$.
When we denote a matrix of $SL(2, \bC)$ by
$\left( \begin{array}{cc}
		a & b \\ c&d\end{array}\right)$, and represent a point $(x_1, x_2, y)$ of
$\bH^3$ as a quaternion
$h =x_1+ x_2 i +y j$,
the $SL(2, \bC)$ isometry acts on the metric \eqref{H3metric} as follows (this is called the \keyword{Poincar\'e extension}{Poincare extension}; see exercise~\ref{ex.Poincare}):
\begin{align}
	h\to (ah+b)(ch+d)^{-1}  \ .
	\label{eq.Poincare}
\end{align}
This action is transitive, and the stabilizer subgroup at each point is $SU(2)$ (exercise~\ref{ex.Poincare}). Namely, $\bH^3$ is a \keyword{homogeneous space}{homogeneous space} (coset):
\begin{align}
	\bH^3= SL(2, \bC)/SU(2) \ .
	\label{H3_coset}
\end{align}

\small
In ordinary 3d hyperbolic geometry, it seems common to include completeness (defined by the convergence of Cauchy sequences)
in the definition of a hyperbolic structure.
However, for our purposes it is important to consider
also incomplete hyperbolic structures, and hence
in this book we do not assume completeness unless otherwise stated.

\normalsize

A complete 3d hyperbolic manifold can be written, using a torsion-free discrete subgroup $\Gamma$ of $PSL(2, \mathbb{C})$, as
\begin{align}
	M=\mathbb{H}^3/\Gamma \ .
	\label{H3_Gamma}
\end{align}
\nomenclature{$\Gamma$}{discrete subgroup of $PSL(2, \mathbb{C})$ (Kleinian group)}
Since discrete subgroups of $PSL(2, \mathbb{C})$ are called \keyword{Kleinian groups}{Kleinian group}, we may also say that $\Gamma$ is a torsion-free Kleinian group. Being torsion-free means that the action of $\Gamma$ has no fixed points, which is a condition necessary for the quotient \eqref{H3_Gamma} to be a smooth manifold.

In this way, the study of 3d hyperbolic geometry is replaced by a more algebraic problem, the study of discrete groups.
In the actual applications treated in this book, 3d hyperbolic manifolds are rarely given from the beginning in the form $M=\bH^3/\Gamma$;
however, once $M$ is given,
$\Gamma$ can be extracted as $\Gamma=\pi_1(M)$.

In hyperbolic geometry, it is known that once we fix the topology,
the corresponding complete hyperbolic structure is uniquely determined:
complete hyperbolic structures are rigid geometries which admit no deformations.
This statement is known as \keyword{Mostow rigidity}{Mostow rigidity},
and an elegant proof by Gromov is known.

We have already seen the gauge theory counterpart of this statement.
The dimensional reduction limit $b\to 0$ of the 3d $S^3_b$ partition function discussed in Sec.~\ref{sec.3d_dim_red}
corresponded to the classical limit $t\to \infty$ of 3d Chern-Simons theory (see \eqref{t_S3} and \eqref{qtb}).
Moreover,
the complete hyperbolic structure is a saddle point in the classical limit, and was also a solution of the vacuum
equation \eqref{2dvac} of the 2d theory. The 2d theory under consideration is
a massive theory, and hence
has isolated vacua. Namely,
\textbf{the physical property that the vacua of 2d $\scN=(2,2)$ theories are isolated has been translated into the mathematical property of Mostow rigidity of complete hyperbolic manifolds in 3d classical hyperbolic geometry}.

\bparagraph{Wilson Lines and Knot Complements}\label{subsec.knot_complement}

So far we have considered the partition function of complex Chern-Simons theory; let us now consider the expectation values of more general physical observables.
Since there are no matter fields in the theory under consideration,
there are no spatially local operators,
and what we should consider are the Wilson lines in complex Chern-Simons theory (see \eqref{Wilson} in Chap.~\ref{chap.intro}):
\begin{align}
	W_{\gamma}:=\left\langle\textrm{Tr} \, P \,  \exp\left( \oint_{\gamma} \scA \right)\right\rangle  \ .
\end{align}
Here $\gamma$ is a one-dimensional
closed path in the 3-manifold $M$, namely an embedding of $S^1$ into $M$.
Since the theory is topological, the Wilson line
depends only on the isotopy class of the embedding (namely, it is invariant under continuous deformations of the knot).
Mathematically, this corresponds to considering
\keyword{knots}{knot} in $S^3$.\footnote{When we consider embeddings of several $S^1$'s, we call it a
	\keyword{link}{link}. In this book, unless otherwise stated, we collectively call knots (in the narrow sense) and
	links knots (in the broad sense).
}

\begin{figure}[t]
	\includegraphics[scale=0.18]{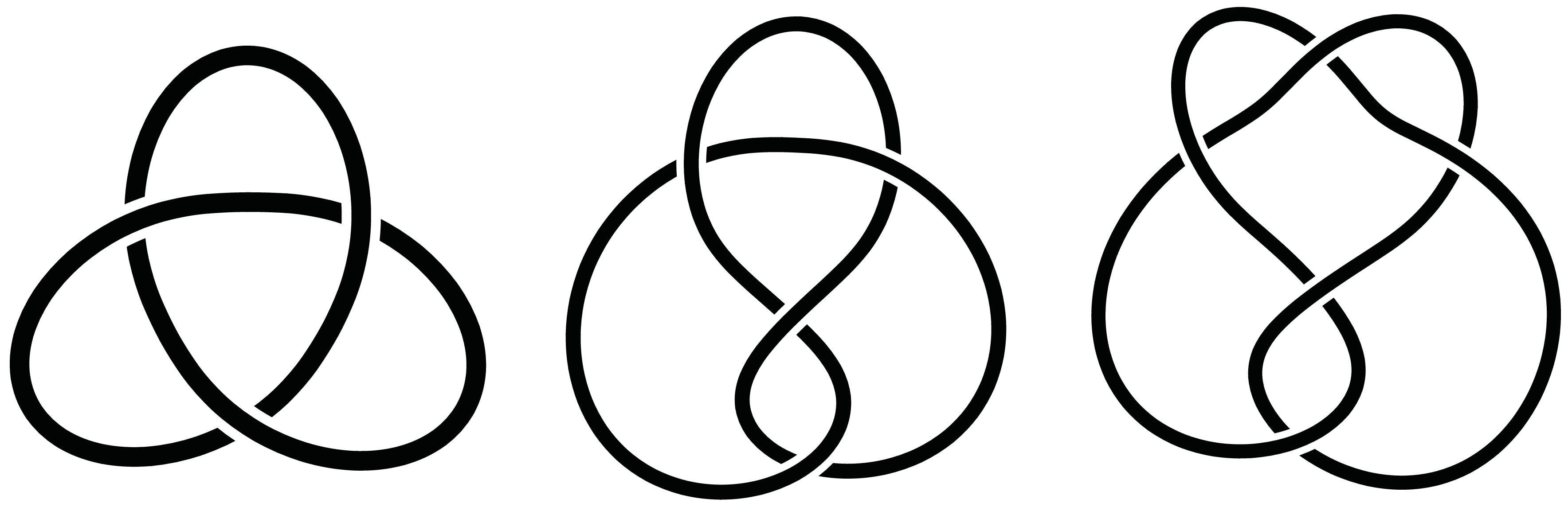}
	\caption{Examples of knots. From left to right, $3_1$ (the \keyword{trefoil knot}{trefoil knot}), $4_1$ (the \keyword{figure-eight knot}{figure-eight knot}), and $5_2$.
		The thickened knot is the tubular neighborhood,
		and its complement in $S^3$
		is called the knot complement. Tables of knots with small crossing numbers have been made,
		and the knots are given names.
		These were originally given in the textbook by Rolfsen \cite{RolfsenBook}, but
		errors were later found and corrected. Nowadays the web page
		\cite{KnotAtlas} is convenient.
	}
	\label{fig.knots}
\end{figure}

When there is a Wilson line, if we consider the equations of motion for the gauge field, we immediately see that
a delta-function-type singularity appears along the Wilson line. Namely, the effect of the Wilson line is equivalent to imposing a boundary condition on the gauge field with a singularity along it,
and performing the path integral under this condition. This means that Wilson loops in pure Chern-Simons theory
can also be understood as disorder operators (for disorder operators, recall also the monopole operators defined in Chap.~\ref{chap.3dN2}).

To handle the singularity nicely,
let us slightly thicken the one-dimensional knot
and consider a tubular region around the knot.
This is called the \keyword{tubular neighborhood}{tubular neighborhood},
and we denote it by $N(L)$.
This is in fact what we often have in mind when drawing pictures of knots (Fig.~\ref{fig.knots}); a mathematical knot is
the one-dimensional line obtained when the knot becomes thin. Specifying the values of the holonomies along the cycles of this tubular region
corresponds to the boundary condition.

The path integral is performed on the complement of the tubular neighborhood $N(L)$ in $M$,
\begin{align}
	M\backslash N(L) \ .
\end{align}
This is called the complement of the link $L$.
In this book we mainly consider
complements of links in $S^3$.\footnote{In the literature, one sometimes distinguishes more precisely between
	the knot complement $S^3\backslash K$ and the
	\keyword{exterior}{exterior} $M\backslash N(L)$ of the knot (they differ in whether the boundary is included).
}
A knot complement is itself a 3-manifold, and hence determines a corresponding 3d $\mathcal{N}=2$ theory.\footnote{In the late 19th century, Lord Kelvin tried to classify atoms using knots in the ether. His theory ended in failure, but his considerations stimulated the development of knot theory in mathematics, which, more than 100 years later, came to be used for the classification of 3d gauge theories.}
\nomenclature{$L$}{knot (link)}
\nomenclature{$N(L)$}{tubular neighborhood of $L$}

The
\keyword{knot complements}{knot complement} constructed in this way, and more generally \keyword{link complements}{link complement},
are among the most important examples of hyperbolic manifolds.
This is because, for knots in $S^3$, it is known that
the complement always admits a hyperbolic structure except when the knot is a \keyword{torus knot}{torus knot}
or a \keyword{satellite knot}{satellite knot} (a result due to Thurston).\footnote{
	In fact, it seems to be known that if we generate knots randomly in a certain way,
	they are satellite knots with probability close to $1$ \cite{Jungreis_random}.
	Even taking into account that it is not clear what measure we should put on the set of all knots to discuss probabilities, hyperbolic knots themselves are not necessarily generic.
	However, it has been proven that in any closed 3-manifold we can choose an appropriate hyperbolic knot such that the original manifold is
	constructed by Dehn filling the complement of the hyperbolic knot \cite{Myers_Simple},
	and in this sense hyperbolic knots can be said to be sufficiently general.
	\label{foot.hyperbolic}
}
Here, a torus knot is a knot which can be drawn on a torus,
and a satellite knot is a knot obtained by placing the torus carrying a torus knot
so that it becomes the tubular neighborhood of some knot in $S^3$.
Moreover, when the complement of a knot admits a hyperbolic structure, the knot is called a \keyword{hyperbolic knot}{hyperbolic knot}.

\bigskip

As already stated, the choice of the hyperbolic manifold was determined by the choice of the discrete group $\Gamma$
\eqref{H3_Gamma}.
This group $\Gamma$ is the fundamental group $\pi_1(M)$ of $M=S^3\backslash L$.
It is known that a presentation of the fundamental group $\pi_1(M)$ of $M$ by generators and relations
can be constructed mechanically from a projection diagram of the knot
(the Wirtinger presentation); however, since
the fundamental group is non-Abelian, it is not easy to handle directly.

We therefore focus on the boundary of $M=S^3\backslash L$.
The boundary of $S^3\backslash L$ coincides (up to orientation) with the boundary of the tubular region of $L$,
which by definition consists of
$n$ 2d tori $\overbrace{T^2\cup \ldots \cup T^2}^{n}$ for an $n$-component link;
changing the choice of knot or link only changes the value of $n$, and
what appears is always tori.
As for its fundamental group, for each torus there are two independent cycles,
the so-called $\alpha$-cycle and $\beta$-cycle, and
all the elements commute with each other.

Of course, $\pi_1(\partial M)$ is a different group from $\pi_1(M)$.
However, since there is the natural embedding $\partial M\rightarrowtail M$,
there is correspondingly a map
\begin{align}
	\pi_1(\partial M) \rightarrowtail \pi_1(M) \ .
\end{align}
Therefore, for the moduli \eqref{M_flat} of flat connections,
we obtain the embedding
\begin{align}
	\begin{split}
		 & \mathcal{M}_{\rm flat}(M)=\textrm{Hom}(\pi_1(M) , G_{\bC})/ \textrm{($G_{\mathbb{C}}$-conjugation)} \\
		 & \qquad  \rightarrowtail
		\mathcal{M}_{\rm flat}(\partial M) =\textrm{Hom}\left(\pi_1 (T^2)^n , G_{\bC}\right)/ \textrm{($G_{\mathbb{C}}$-conjugation)} \ .
	\end{split}
	\label{A_embed}
\end{align}

For simplicity, let us consider the case of a knot rather than a link ($n=1$),
and take $G_{\bC}$ to be $SL(2, \bC)$.
The holonomies corresponding to the $\alpha$-cycle and $\beta$-cycle of the boundary torus then
commute, and by conjugation with an element of $SL(2, \bC)$ they can be written as
\begin{align}
	\rho(\alpha)=
	\left(
	\begin{array}{cc}
		\mathfrak{m}^{\frac{1}{2}} & *                           \\
		0                          & \mathfrak{m}^{-\frac{1}{2}}
	\end{array}
	\right) \ ,
	\quad
	\quad \rho(\beta)=
	\left(
	\begin{array}{cc}
		\mathfrak{l} & *                 \\
		0            & \mathfrak{l}^{-1}
	\end{array}
	\right) \ .
	\label{m_l_def}
\end{align}
We then call $\mathfrak{m}$ ($\mathfrak{l}$) the
\keyword{meridian}{meridian}
(\keyword{longitude}{longitude}).
\nomenclature{$\mathfrak{m}$}{meridian}
\nomenclature{$\mathfrak{l}$}{longitude}
Note that even after gauge fixing as in \eqref{m_l_def}, the action of the discrete Weyl group
$(\mathfrak{m}, \mathfrak{l})\leftrightarrow (\mathfrak{m}^{-1}, \mathfrak{l}^{-1})$
remains unfixed. To fix this freedom as well,
we can consider the square of $\mathfrak{m}^{1/2}$, namely
$\mathfrak{m}$, and
$\mathfrak{l}$. In \eqref{m_l_def}, anticipating this point, we have taken the square root $\mathfrak{m}^{1/2}$ in advance in the definition of the meridian.

Now, returning to the embedding
\eqref{A_embed} in this case, we have
\begin{align}
	\scM_{\rm flat}(M) \to \scM_{\rm flat}(\partial M)=(\bC^{\times}\times \bC^{\times})/\bZ_2 \ ;
\end{align}
in fact, it is known that the image of the embedding is given by the zero locus of a Laurent polynomial $A( \mathfrak{l},\mathfrak{m})$ in $\mathfrak{m}\in \bC^{\times}$ and
$\mathfrak{l} \in \bC^{\times}$:
\begin{align}
	\{A (\mathfrak{l}, \mathfrak{m}) =0  \}
\end{align}
\cite{CooperApolynomial}.
This polynomial is called the \keyword{$A$-polynomial}{A-polynomial}.

\bigskip

When we consider complex Chern-Simons theory on a knot complement,
its classical limit \eqref{CSlimit} is reinterpreted as the mathematical statement called the \keyword{volume conjecture}{volume conjecture}
(original papers \cite{KashaevHyperbolic,MurakamiMurakami}; for a review see e.g.\ Ref.~\cite{MurakamiIntroduction}; for the relation to complex Chern-Simons theory see Ref.~\cite{Gukov:2003na}). What the volume conjecture claims is that,
for the knot invariant called the
\keyword{$N$-colored Jones polynomial}{N-colored Jones polynomial} $J_{L, N}(q)$ (where $L$ is a knot and $N > 1$ is an integer),
\begin{align}
	\lim_{N\to \infty} \frac{2\pi \log \left(J_{L,N}(q=e^{\frac{2\pi i}{N}}) \right) }{N}
	=\textrm{Vol}(L)+i \textrm{CS}(L) \ .
	\label{vol_conj}
\end{align}
From the fact that the right-hand side agrees with the classical limit \eqref{CSlimit} of complex Chern-Simons theory,
and from the fact that Chern-Simons theory with a compact group is related to the Jones invariants of knots \cite{Witten:1988hf},
one can imagine, even if only vaguely, the claim above.

So far we have exclusively discussed classical solutions;
complex Chern-Simons theory has both
classical solutions and the fluctuations around them.
For example, if we consider the one-loop correction, we obtain the
\keyword{Reidemeister torsion}{Reidemeister torsion}.
Moreover, if we consider higher loops,
we obtain infinitely many invariants of knots, which are interesting mathematical objects.\footnote{
	Similar quantities have been mathematically formulated, for compact groups $G$, as the so-called Chern-Simons perturbation theory.
	The story is similar when the gauge group is the non-compact $G_{\bC}$, but
	there are also essential differences; for example, the invariants defined are not of finite type.
}


\bparagraph{Column: Geometric Structures of 3-Manifolds}\label{sec.geometric_structure}

\small
Although this is somewhat outside the main subject of this book, let us take this opportunity to comment on geometric structures of manifolds.

Suppose that a topological manifold $M$ is given. A
\keyword{geometric structure}{geometric structure} on $M$
given by a manifold (model geometry) $H$ is a pair of an open cover
$\{U_i\}_{i\in I}$ of $M$ and a set of maps $\{\phi_i: U_i\to H\}_{i\in I}$
satisfying the following two conditions:\footnote{
	For our actual applications,
	we need to slightly extend the definition to include the case with boundaries,
	but we do not describe this here.
}
\begin{Kenumerate}
	\item For all $i\in I$, $\phi_i$ is a homeomorphism onto its image.
	\item When $U_i\cap U_j\ne \emptyset$, the transition function
	\begin{align}
		\phi_i \circ \phi_j^{-1}: \phi_j(U_i\cap U_j) \to \phi_i(U_i\cap U_j) \ ,
	\end{align}
	restricted to each connected component of $\phi_j(U_i\cap U_j)$, is an orientation-preserving
	\textbf{isometry} of $H$.
\end{Kenumerate}

Recall that the properties of transition functions determined the definitions of manifolds:
in the definition of a topological manifold the transition functions are
homeomorphisms, and in a differentiable manifold the transition functions are
diffeomorphisms.
In the present definition of a geometric structure, the transition functions have to be isometries of $H$,
so that the condition is stricter than in the usual definition of manifolds.
By changing the definition of the manifold $H$, we can define different geometric structures.
Classifying the geometric structures defined in this way is the idea of the
Erlangen program due to Klein.

Closed 2-manifolds are classified by their genus, but for 3-manifolds
the classification is more complicated.
It is known that every 3-manifold, when decomposed (along spheres and tori),
has at least one of the following eight geometric structures on each piece:
the eight manifolds $H$ are
\begin{align}
	H=\bS^3\ , \quad \bE^3 \ , \quad \bH^3\ ,  \quad \bS^2\times \bR\ ,\quad \bH^2\times \bR \ ,
	\quad \widetilde{SL(2, \bR)}\ , \quad \textrm{Nil}\ , \quad \textrm{Sol} \ .
\end{align}
Here $\widetilde{SL(2, \bR)}$ is the universal cover of $SL(2, \bR)$;
we do not explain Nil (nilmanifold) and Sol (solvmanifold) here, since
they are not used in this book.
Note that different pieces may carry different geometries, so that a 3-manifold as a whole need not admit a single geometric structure.

This result was originally proposed in the 1980s by W.~P.~Thurston as the \keyword{geometrization conjecture}{geometrization conjecture},
and was proven by G.~Perelman.
As a corollary, the Poincar\'e conjecture, which was one of the Millennium Problems (a simply connected compact 3-manifold is
homeomorphic to $S^3$), follows, so many people may have heard the news.

For Seifert manifolds (which include manifolds with $\bS^3$ geometry, as well as those with five other geometries), the discussion in, e.g.,
Ref.~\cite{Chung:2014qpa} is useful. The precise correspondence between the classification of 3-manifolds and that of 3d $\scN=2$ theories
is not known, but, for example, the known gravity solutions in the AdS/CFT correspondence are
constructed using hyperbolic manifolds only \cite{Gauntlett:2000ng}.

\normalsize

\begin{practice}

	\item $[\bll]$ (Quantization of the level)
	\label{ex.instanton}
	Let us derive the quantization of the Chern-Simons level by a method different from that in the main text.

	\begin{enumerate}

		\item $[\bll]$
		      Show by direct computation that under the gauge transformation
		      \begin{align}
			      A \to g^{-1} A g + g^{-1}dg \ ,
		      \end{align}
		      the field strength $F=dA + A\wedge A$ transforms as
		      \begin{align}
			      F \to g^{-1} F g \ ,
		      \end{align}
		      and that the action \eqref{L_total} then transforms as
		      \begin{align}
			      \begin{split}
				      iL[A]\to i L[A] & + \frac{i k}{4\pi}\int_M d\,  \textrm{Tr}(dg \wedge g^{-1} A)       \\
				                      & -\frac{i k}{12\pi} \int_M \textrm{Tr}\left((g^{-1} dg)^3\right) \ .
			      \end{split}
			      \label{CS_transf}
		      \end{align}
		      For the latter part, it may be convenient to write $L[A]$ as
		      \begin{align}
			      \frac{k}{4\pi} \int_M \textrm{Tr}\left(A\wedge F-\frac{1}{3} A\wedge A \wedge A\right) \ .
		      \end{align}

		      The second term of \eqref{CS_transf} is a total derivative and
		      vanishes when integrated over a closed manifold; moreover,
		      the third term $-\frac{i k}{12 \pi} \int_M \textrm{Tr}((g^{-1} dg)^3)$, or in components\footnote{This term is called the \keyword{Wess-Zumino term}{Wess-Zumino term}.}
		      \begin{align}
			      -2\pi i \frac{k}{24\pi^2} \int_M d^3x\, \epsilon^{\mu\nu\rho}
			      \textrm{Tr}\left( g^{-1} \partial_{\mu}g  g^{-1} \partial_{\nu}g g^{-1} \partial_{\rho}g\right) \ ,
			      \label{pi_3}
		      \end{align}
		      is known to be an element of $2 \pi i \mathbb{Z}$; hence
		      we find that the value of the action is uniquely determined when $k\in\mathbb{Z}$.

		\item $[\bll]$
		      Mathematically, \eqref{pi_3} represents an element of the
		      homotopy group $\pi_3(G)$ of the gauge group $G$.
		      Let us check with a simple example that its value (the \keyword{winding number}{winding number}) is an integer.
		      Take $G=SU(2)$, and take the 3-manifold to be $S^3$.
		      Since $S^3$ is also $SU(2)$ as a group manifold,
		      we can take the identity map as $g: S^3\to SU(2)$.
		      Convince yourself that, representing a point of
		      $S^3$ by $(x_1, \ldots, x_4)$ ($x_1^2+\cdots+x_4^2=r^2$), it can be written as
		      \begin{align}
			      g(x):=\frac{x^4+i x^{\mu}\sigma^{\mu}}{r} \ , \quad \mu=1,2,3 \ .
		      \end{align}
		      Moreover, compute the corresponding value of \eqref{pi_3}, and show that
		      \begin{align}
			      \begin{split}
				       & \frac{1}{24\pi^2} \epsilon^{\mu\nu\rho} \int_{S^3}
				      \textrm{Tr}\left( g^{-1} \partial_{\mu}g  g^{-1} \partial_{\nu}g g^{-1} \partial_{\rho}g\right)
				      \\
				       & \qquad =i^3 \frac{1}{24\pi^2} 3! \frac{1}{r^3} (2\pi^2 r^3)
				      \textrm{Tr}\left(\sigma_1 \sigma_2 \sigma_3 \right)
				      =1 \ .
			      \end{split}
		      \end{align}
		      Hint: $g(x)$ is the identity map on $S^3$, and the integrand does not depend on the coordinates of $S^3$; it is therefore useful to simplify the computation by
		      using the fact that we may evaluate it at, e.g., $x_4=1, x_1=x_2=x_3=0$.
		      Similarly, show that by considering
		      \begin{align}
			      g(x)=\left(\frac{x^4+ i x^{\mu} \sigma^{\mu}}{r}\right)^n\ , \quad n\in \bZ \ ,
		      \end{align}
		      we obtain the winding number $n$.

		\item $[\bll\bll]$ Careful readers may be bothered that the explanation here
		      seems different from the explanation in the main text.
		      Mathematically, \eqref{pi_3} represents an element of the
		      cohomology group $H^3(G, \mathbb{Z})$ of the gauge group $G$ (the fact that it is cubic in $g^{-1} \partial_{\mu}g$ is the reason for $H^3$),
		      whereas $\textrm{ch}_2(F)$ \eqref{ch2F_def} used in the main text
		      is a characteristic class represented as $H^4(BG, \mathbb{Z})$,
		      using the cohomology of the classifying space $BG$ of the gauge group $G$.
		      Show, using the exact sequence following from the universal bundle $G\to EG\to  BG$, that there exists a map $H^4(BG, \mathbb{Z})\to H^3(G; \mathbb{Z})$. (This map is a mathematical expression of the correspondence between 3d Chern-Simons theory and the \keyword{Wess-Zumino-Witten model}{Wess-Zumino-Witten model}. For $G=SU(N)$ this map is an isomorphism, but
		      for general groups which are not simply connected it is no longer surjective.
		      For these matters, as well as for the definition of more general\footnote{
			      In the definition in the main text we assumed that the bundle on the 3-manifold $M$ can be extended to the 4-manifold $V$;
			      more generally, it can happen that the vector bundle cannot be extended to a 4-manifold, so that
			      the definition in the main text is not the most general one. However, these subtleties are
			      not important for the purposes of this book.} Chern-Simons terms, see
		      Ref.~\cite{Dijkgraaf:1989pz}.)
	\end{enumerate}

	\item $[\bll\bll]$ (One-loop correction to the level)

	The value of the level $k$ of the Chern-Simons term for a compact group
	becomes, by the one-loop correction,\footnote{Care is needed since in some references $c_2(G)$ is defined to be twice the one here. In that case, for example, $c_2(SU(N))=2N$.}
	\begin{align}
		k\longrightarrow k+c_2(G) \ .
		\label{k_shift}
	\end{align} Here
	$c_2(G)$ is the \keyword{dual Coxeter number}{dual Coxeter number}
	(or the \keyword{quadratic Casimir}{quadratic Casimir})
	of the gauge group, defined, using the totally anti-symmetric structure constants $f_{abc}$ of the Lie algebra (with
	$[T^a, T^b]=i f^{abc} T^c$ for the generators $\{T^a\}$ of the Lie algebra), by
	\begin{align}
		f^{abc} f^{bc d}=c_2(G)\,  \delta^{ad} \ .
	\end{align}
	This represents the fact that, in the adjoint representation $(\rho(T^a))_{bc}=-i f^{abc}$ of the Lie algebra,
	the element $T^b T^b$ is proportional to the identity:
	\begin{align}
		\left(\rho(T^b)\rho(T^b)\right)_{ad} = c_2(G)\,\delta_{ad}
	\end{align}
	(since $T^b T^b$ commutes with all the elements of the Lie algebra, if the representation of the Lie algebra is irreducible, Schur's lemma implies that it has to coincide with the identity up to a constant multiple). For example, $c_2(SU(N))=N$ for $SU(N)$.

	For the one-loop correction \eqref{k_shift}, there is an explanation using the \keyword{index theorem}{index theorem} and the $\eta$-invariant in \cite{Witten:1988hf}; here let us check it by a direct loop computation.
	To derive only the shift $k\to k+c_2(G)$, it is sufficient to consider the gauge field propagator; to check the one-loop correction to the three-point vertex of the gauge field, however,
	\begin{align}
		f^{afd} f^{bde} f^{cef}=\frac{c_2(G)}{2} f^{abc}
	\end{align}
	is needed. To understand the last formula, it is useful to consider, e.g., $\rho(T^d T^c T^d)_{ab}$ in the adjoint representation. For details of these computations, see e.g.\ Refs.~\cite{AlvarezGaume:1989wk,Chen:1992ee}.

	\item $[\bll]$ (Integration cycles for the Airy function)
	To deepen our understanding of the choice of integration cycles in complex Chern-Simons theory,
	it is useful to consider the finite-dimensional case \cite{Witten:2010cx}.
	As an example, let us consider the Airy integral:
	\begin{align}
		Z:=\int_{-\infty}^{\infty} dx \,\, \exp\left(i\lambda\left( \frac{x^3}{3}-x\right)  \right)  \ .
	\end{align}
	Here let us take $\lambda$ to be a complex number.
	In which directions at infinity in the $x$-plane does the integrand converge? How many choices of linearly independent integration cycles are there? (For reference, the asymptotic expansion of this Airy integral takes
	different forms (related by linear transformations) in different regions of the complex plane.
	This is called the \keyword{Stokes phenomenon}{Stokes phenomenon}, and the Airy integral is a typical example.)

	\item $[\bll]$ (3d gravity)
	In 3d gravity with Euclidean and Lorentzian signatures,
	consider the cosmological constant $\Lambda=\pm 1$.
	What is the gauge group of the corresponding Chern-Simons theory in each case?
	How does the conclusion change depending on the signature of the metric and the sign of the cosmological constant?

	\item $[\bll]$ (Poincar\'e extension)\label{ex.Poincare}
	Verify by direct computation that the transformation \eqref{eq.Poincare}
	is an isometry of $\bH^3$.
	Moreover, confirm that the transformation \eqref{eq.Poincare} is
	the natural extension to $\bH^3$ of the linear fractional transformation
	\begin{align}
		z\to \frac{az+b}{cz+d} \ , \quad z\in \mathbb{C}
	\end{align}
	on the boundary $\partial \bH^3=\mathbb{C}\cup \{ \infty \}$.
	Also verify that the stabilizer subgroup of the action at each point is $SU(2)$.

	\item $[\bll]$ ($\bH^3$ as a homogeneous space) Find the metric on the homogeneous space \eqref{H3_coset}
	induced from the standard metric on $SL(2, \bC)$, and show that it is given by \eqref{H3metric}.

\end{practice}

\chapter{Classical and Quantum \Teichmuller Theory}
\label{chap.Teichmuller}

\begin{abstract}

	The deformation of the hyperbolic structure on a 2d Riemann surface
	is described by the \Teichmuller space.
	In this chapter, we summarize this \Teichmuller space and its quantization.
	Here the so-called Fock coordinates play a crucial role,
	which correspond to the VEVs of IR loop operators
	in 4d $\mathcal{N}=2$ supersymmetric theories.
\end{abstract}

\section{Classical \Teichmuller Theory}

In the previous chapter, we studied 3d complex Chern-Simons theory,
i.e.\ in more geometric terms, the
associated 3d hyperbolic geometry and the moduli space of
flat $G_{\bC}$-connections on a 3-manifold $M$.
In order to understand the connection with
3d $\mathcal{N}=2$ theories, however,
we need more detailed data than we have learned so far---for example,
explicit coordinates on the moduli space of
flat $G_{\bC}$-connections.

In this chapter, we study the moduli space of
flat $G_{\bC}$-connections on a 2d Riemann surface.
As we discussed already in
Chap.~\ref{chap.6d}, what appears there
is the Hitchin system, i.e.\ (in one complex structure)
the moduli space of
flat $SL(2, \bC)$-connections.
As we will explain momentarily, a connected component of its real slice
(the moduli space of flat $SL(2, \bR)$-connections) is the \keyword{\Teichmuller space}{Teichmuller space}.
While the real slice of the moduli space of flat $SL(2, \bC)$-connections gives the moduli space of
flat $SL(2, \bR)$-connections, the precise relation between them is a bit subtle.
For the purposes of this section, however,
we do not need to be overly sensitive to the distinction.
In this chapter we mainly use the language of the latter.


The bulk of the analysis of the 2d case in this chapter
can be translated into the language of 3-manifolds in the next chapter.
The 2d hyperbolic geometry is tightly connected with
3d hyperbolic geometry (discussed in the previous chapter),
and you will find some parallels between the contents of this chapter and
those of the previous chapter.\footnote{One of the reasons for this is that the
	orientation-preserving isometries of the hyperbolic spaces $\bH^2, \bH^3$
	are given by $PSL(2, \bR)$ and $PSL(2, \bC)$, respectively, and one is the real slice of the other.
	Note that such a relation does not hold for $\bH^{n}$ and $\bH^{n+1}$ with $n\ge 3$.
	\label{foot.SLRC}}

\subsection{\Teichmuller Space}\label{subsec.Teich}

Let us first quickly summarize the classical (i.e.\ before quantization)
\Teichmuller space.

In this chapter we specialize to the case of $G_{\bR}=SL(2, \bR)$, $G_{\bC}=SL(2, \bC)$.
See Sec.~\ref{subsec.AN} for comments concerning the more general case of $G_{\bR}=SL(N, \bR), G_{\bC}=SL(N, \bC)$.

The moduli space of flat $PSL(2, \bR)$ connections,
which is the real slice of the moduli space of
$PSL(2, \bC)$ flat connections,
is known to have several different connected components.
One of them, with the largest dimension, is the
\Teichmuller space,\footnote{
	Given a $PSL(2, \bR)$ flat connection, an $\bR \bP^1$-bundle on the 2d surface
	is determined accordingly.
	It is known that the connected components of $PSL(2, \bR)$ flat connections are
	labeled by the Euler number $e$ of this $\bR \bP^1$-bundle, and moreover that $e$
	satisfies $|e|\le 2g-2$ on a closed Riemann surface of genus $g$ \cite{MilnorFlat}.
	The \Teichmuller space is the component labeled by $e=2g-2$.
	In the \Teichmuller component
	the Euler number is positive, and $PSL(2, \bR)$ flat connections are
	at the same time $SL(2, \bR)$ flat connections.
	For this reason, in this chapter we do not strictly distinguish between $PSL(2, \bR)$ and
	$SL(2, \bR)$.
	However, similar statements do not hold for the other components.
	For example, $SL(2, \bR)$ flat connections are not classified
	by their Euler numbers; for example, for genus $2$ there seem to be
	33 components.
} which is the subject of study in \keyword{\Teichmuller theory}{Teichmuller theory}.
You can think of the \Teichmuller space as the component of the space of flat connections where the
metric constructed from the connection is invertible.
The \Teichmuller space can also be thought of as the space of \keyword{hyperbolic structures}{hyperbolic structure} on the surface, and also as the space of
equivalence classes of metrics under conformal transformations (see the comment at the end of Sec.~\ref{subsec.compact_twist}).

\small
\Teichmuller theory has a long history in mathematics, and has been discussed extensively
in the literature (see e.g.\ Refs.~\cite{ImayoshiTaniguchi,KohnoBook} for textbooks).
In physics it is often mentioned in the literature on
string perturbation theory,
but few references discuss the explicit coordinates to be discussed below.

\normalsize

\bigskip

It is known that 2d manifolds admit only three types of
geometric structures (see Sec.~\ref{sec.geometric_structure} of the previous chapter):
the spherical structure, corresponding to the 2d sphere $X=\mathbb{S}^2$,
the Euclidean structure, corresponding to the 2d Euclidean plane $X=\mathbb{E}^2$,
and the hyperbolic structure.
This is the statement of the \keyword{uniformization theorem}{uniformization theorem}.

In this book we mostly concentrate on hyperbolic structures.
For 2d surfaces, the hyperbolic structure, whose model geometry
is $H=\bH^2$, is the most general one, just as for 3-manifolds
the hyperbolic structure (with model geometry $H=\bH^3$) was the most general one.
To see this, let us consider a 2d surface $\Sigma$
with genus $g$ and $h$ punctures;
we sometimes denote the surface by $\Sigma_{g,h}$, to emphasize the
genus and the number of punctures.
This surface admits a hyperbolic structure if and only if its
Euler characteristic is negative:
\begin{align}
	\chi(\Sigma)=2-2g-h < 0 \ .
	\label{chiformula}
\end{align}
This condition is satisfied except for a small number of exceptions,
$g=0, h=0,1,2$ or $g=1, h=0$. In particular, we can always satisfy this condition
by adding an appropriate number of punctures if necessary.

When \eqref{chiformula} is satisfied, the 2d manifold
can locally be identified with the 2d
hyperbolic space (hyperbolic plane) $X=\bH^2$, i.e.\ it
admits a hyperbolic structure. Here $\bH^2$
is
\begin{align}
	\bH^2:=\{ (x, y)|\, x\in \bR, y\in \bR_{>0}\}
\end{align}
with the metric
\begin{align}
	ds^2=\frac{dx^2+dy^2}{y^2} \ .
	\label{H2metric}
\end{align}
This is known as the \keyword{Poincar\'{e} upper half-plane model}{Poincare upper half-plane model} of the hyperbolic plane.

\begin{figure}[t]
	\centering\includegraphics[scale=0.3]{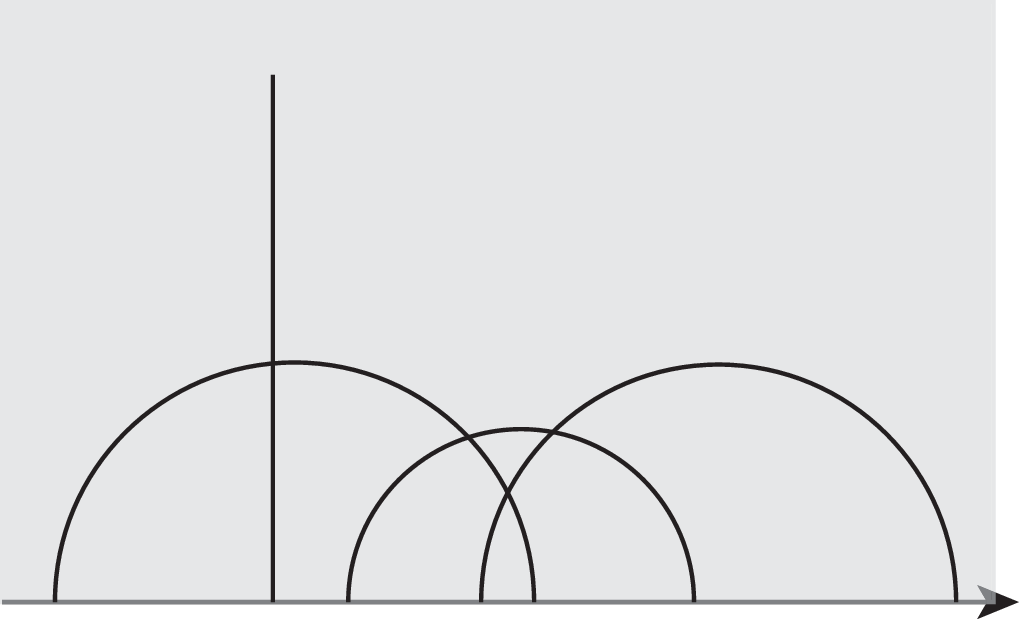}
	\caption{A geodesic of $\bH^2$ is either a vertical straight line or a semicircle intersecting the axis $y=0$ orthogonally.}
	\label{H2geodesic}
\end{figure}

The group of orientation-preserving isometries of $\bH^2$ is $\textrm{Isom}^+(\bH^2)=PSL(2, \bR)$,\footnote{
	When we map $\bH^2$ to the Riemann sphere $\bC\cup \{ \infty\}$ by the stereographic projection, the isometries are given by linear fractional transformations of the Riemann sphere.} and $\bH^2$ is the homogeneous space $PSL(2, \bR)/SO(2)$, where $SO(2)$ is the isotropy subgroup.
A geodesic in $\bH^2$ is either a semicircle orthogonal to the axis $y=0$, or a vertical straight line (see Fig.~\ref{H2geodesic} and exercise~\ref{ex_geodesic}).

\small
It is known that every complete (2d) hyperbolic manifold can be written in the form
$\bH^2/\Gamma$ (a theorem due to Cartan and Hadamard). Here $\Gamma$ is a discrete subgroup of $PSL(2, \bR)$ called a \keyword{Fuchsian group}{Fuchsian group}, which acts without fixed points (freely).
\normalsize

\nomenclature{$\mathcal{T}_{g,h}$}{\Teichmuller space}

\bigskip

The \Teichmuller space is a complex manifold, whose complex dimension is
\begin{align}
	\dim_{\bC} \mathcal{T}_{g,h} = 3g-3+h \ ,
	\label{undecorated_dim}
\end{align}
and it is given by a region inside
$\bR^{6g-6+2h}$.

\small

For example, for the sphere $g=0$, the dimension is $h-3$.
Since the sphere can be regarded as the complex plane with the point at infinity added, $\mathbb{C}\cup \{\infty\}$, it is natural to have one complex parameter for each puncture indicating its position. However, since the sphere has the automorphism group $SL(2, \mathbb{C})$
given by linear fractional transformations, subtracting its dimension we obtain $h-3$.

Similarly, for the torus $g=1$ the complex dimension is $h$,
which can be interpreted as the parameters representing the positions of the punctures on the torus $\mathbb{C}/\mathbb{Z}^2$.

\normalsize

For our purposes it is sometimes useful to include also the holonomies around the punctures, and to
consider the larger \keyword{decorated \Teichmuller space}{decorated Teichmuller space} $\tilde{\scT}_{g,h}$.
The real dimension of this space is
\begin{align}
	\dim_{\bR} \tilde{\mathcal{T}}_{g,h} = 2(3g-3+h)+h=3(2g-2+h) \ .
	\label{decorated_dim}
\end{align}
Note that this can be an odd number (and thus the decorated \Teichmuller space is in general
not a complex manifold).

\bparagraph{Fock Coordinates}

Let us next consider coordinates on the \Teichmuller space.
We first introduce the coordinates in a top-down manner, and we will
come back to the physical meaning of the coordinates later.

\begin{figure}[t]
	\centering\includegraphics[scale=0.3]{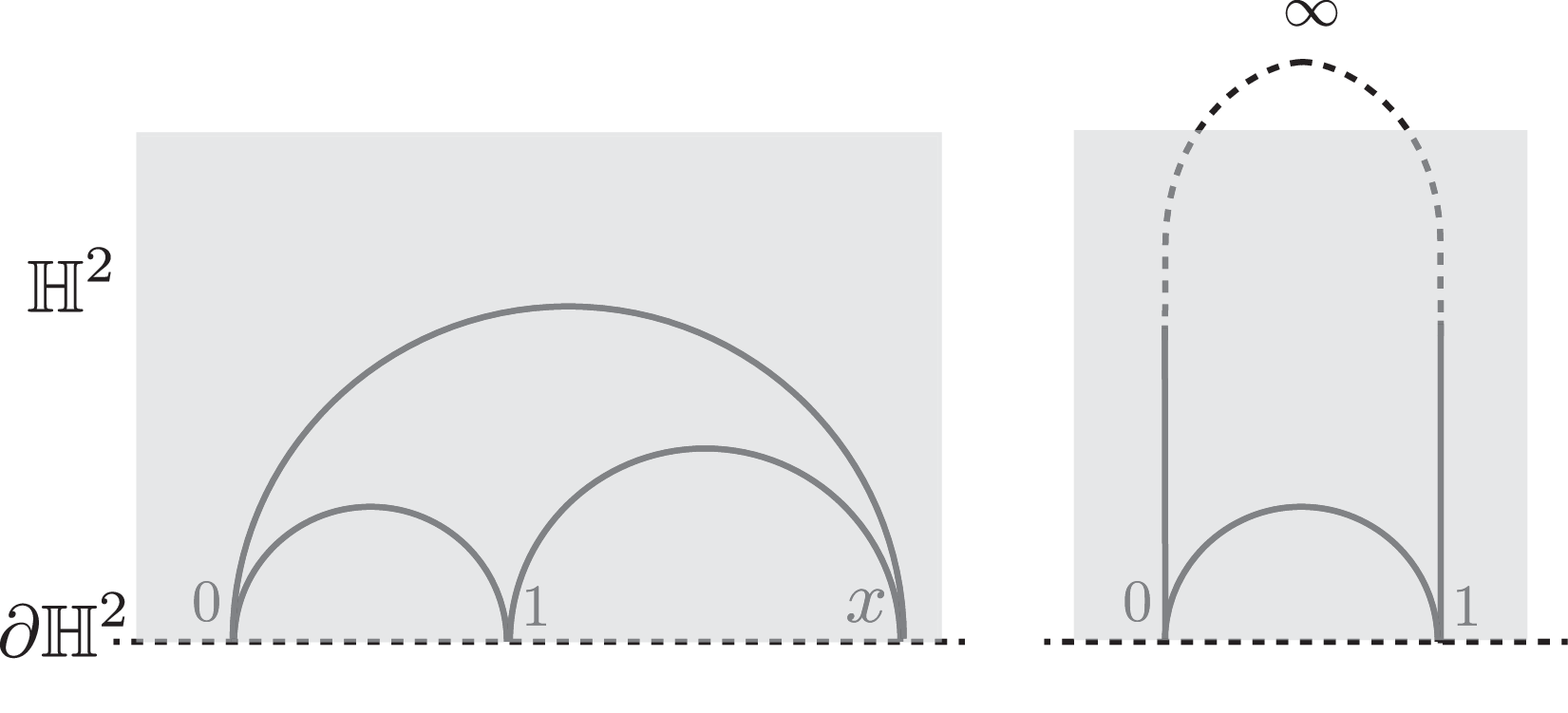}
	\caption{An ideal triangle in $\bH^2$ has its three vertices on the boundary $\partial \bH^2$,
		which can be written as $0, 1, x$. As on the right,
		we may also have $x=\infty$.}
	\label{fig.idealtriangle}
\end{figure}

Let us consider a triangulation of a 2-manifold
into ideal triangles. Such a triangulation is
called an \keyword{ideal triangulation}{ideal triangulation}:
\begin{align}
	\Sigma=\displaystyle\bigcup_i T_i  \ .
\end{align}
Here an ideal triangle (denoted here by $T_i$)
is a triangle in the hyperbolic plane $\bH^2$ all of whose vertices
are located on $\mathbb{R}\cup \{\infty\}=\partial \bH^2$ (Fig.~\ref{fig.idealtriangle}).

Using the isometries of $\bH^2$, we can, without loss of generality,
take the three vertices to be $0, 1, x$, respectively ($x\in \bR\backslash
	\{0,1 \}$).
By gluing ideal triangles we can obtain a 2d surface;
since the vertices of ideal triangles are on the boundary of $\bH^2$ and are not contained in $\bH^2$,
they become punctures of the 2d surface. Namely, an ideal triangulation is a triangulation of a 2d surface
all of whose vertices are punctures.

\begin{figure}[t]
	\centering{\includegraphics[scale=0.23]{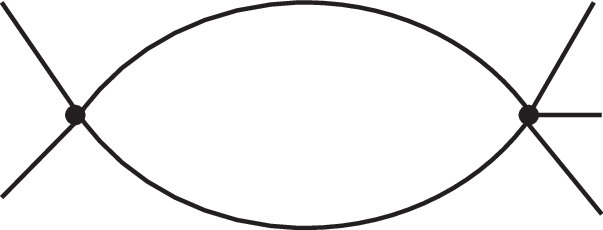}}
	\caption{In ideal triangulations,
		digons as in this figure are forbidden.}
	\label{fig.digon}
\end{figure}

Such a triangulation is possible if the 2d surface contains sufficiently many punctures.
In the following we assume the existence of an ideal triangulation (when we speak of triangulations, we do not allow digons as in Fig.~\ref{fig.digon}).

The numbers of faces ($F$), edges ($E$), and vertices ($V$) of an ideal triangulation are
determined by the genus $g$ and the number $h$ of punctures:
\begin{align}
	F=2(2g-2+h) \ , \quad E= 3(2g-2+h) \ , \quad V= h \ .
	\label{VEF}
\end{align}
To show this, we can use
\begin{align}
	3F=2E\ , \quad \chi(\Sigma_{g,0})=V-E+F=2-2g \ :
\end{align}
the first equation follows from the fact that every triangle has three edges,
and each edge is contained in two faces.

\bigskip
Now, the coordinates of the (decorated) \Teichmuller space which we wish to introduce here are called
\keyword{Fock coordinates}{Fock coordinate} (or shear coordinates; in particular, for $G_{\bC}$ connections, \keyword{Fock-Goncharov coordinates}{Fock-Goncharov coordinate}),\footnote{In the literature Fock coordinates and Fock-Goncharov coordinates are
	sometimes distinguished, and there seem to be variations in their precise definitions;
	for simplicity, in this book we exclusively use the term Fock coordinates.}
and the Fock coordinates assign a variable $Z_i$ (and $z_i:=e^{Z_i}$) to each edge $i$.
From \eqref{VEF} the number of edges is $3(2g-2+h)$; let us confirm that this
agrees with the real dimension \eqref{decorated_dim} of the decorated \Teichmuller space.

One big advantage of Fock coordinates is that
the \textbf{Weil-Petersson \keyword{Poisson bracket}{Poisson bracket} on the \Teichmuller space
	takes a constant (log-canonical) form}.

First, the \Teichmuller space is a
complex \textbf{K\"{a}hler manifold}, and in particular has a
\keyword{symplectic form}{symplectic form} (a non-degenerate closed 2-form) $\omega$. In the language of flat connections, it can be defined, for infinitesimal variations $\delta\mathcal{A}$, by
\begin{align}
	\omega:=\int_{\Sigma} \textrm{Tr}\left(
	\delta \mathcal{A}
	\wedge
	\delta \mathcal{A}
	\right) \ .
	\label{symplectic_flat}
\end{align}
This symplectic form is called the
\keyword{Weil-Petersson form}{Weil-Petersson form}.

\begin{figure}[t]
	\centering{\includegraphics[scale=0.3]{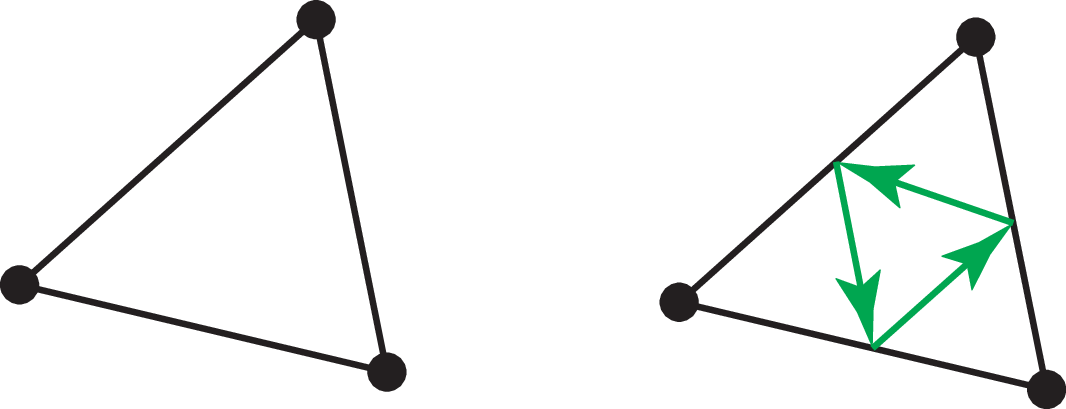}}
	\caption{
		Given an ideal triangulation of a 2d surface,
		we obtain the corresponding quiver by drawing a quiver for each triangle. The vertices of the quiver are associated with the edges of the triangles.
	}
	\label{nee}
\end{figure}

\begin{figure}[t]
	\centering\includegraphics[scale=0.28]{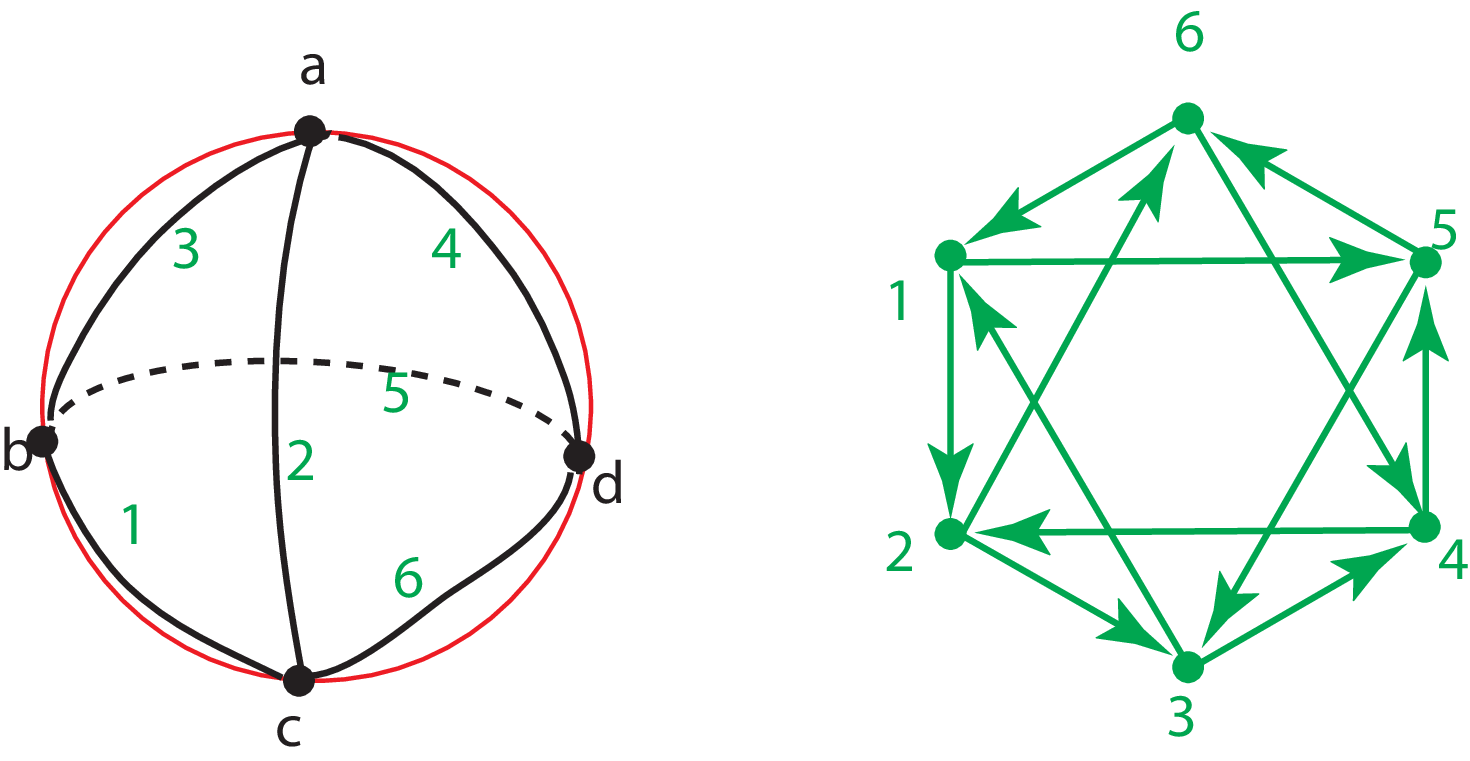}
	\caption{An example of a quiver given by a triangulation of the four-punctured sphere. The numbers of faces, edges, and vertices of the triangulation are $4, 6, 4$, respectively (see \eqref{VEF}).}
	\label{fig.quiver_4sphere}
\end{figure}

Given an ideal triangulation of a 2d surface,
we obtain a quiver (a directed graph) by drawing, for each triangle, the quiver given in Fig.~\ref{nee}.
Here, denoting by $\mathsf{Q}_{i,j}$ the signed number of arrows from edge $i$ to edge $j$,
we have, as is clear from the definition, $\mathsf{Q}_{j,i}=-\mathsf{Q}_{i,j}$, and always
\begin{align}
	\mathsf{Q}_{i,j}\in \{-2, -1,0,1,2\}
\end{align}
(see the examples in Fig.~\ref{fig.quivertrig} and Fig.~\ref{fig.quiver_4sphere}).
The mathematical claim is that, using the $\mathsf{Q}_{i, j}$ thus obtained,
the Poisson structure for the Weil-Petersson form \eqref{symplectic_flat} in terms of Fock variables becomes
\begin{align}
	\omega=\sum_{i<j} (\mathsf{Q}^{-1})_{i, j} dZ_i \wedge dZ_j \ ,
	\label{WPform}
\end{align}
giving rise to the Poisson brackets
\begin{align}
	\{Z_i, Z_j\}=\mathsf{Q}_{i,j} \ .
	\label{WPpoisson}
\end{align}

\bparagraph{Fock Coordinates as Loop Operators}\label{sec.length_twist}

Let us now comment, albeit very roughly, on the physical meaning of the constructions so far.
The comments in this section are useful for knowing the physical meaning of the mathematical constructions,
but it is possible to proceed without understanding them.

First, the \Teichmuller space was (a connected component of)
the space of $PSL(2, \bR)$ flat connections.
Flat connections are locally trivial but
not globally, and, by an argument similar to that in Chap.~\ref{chap.complexCS}, this information is
represented by the holonomies obtained by integrating the flat connection along closed paths.
These give the lengths of the closed paths, and by adding the twist coordinates paired with the lengths,
we obtain coordinates called the
\keyword{length-twist coordinates}{length-twist coordinate}.\footnote{These are coordinates associated with the decomposition of a Riemann surface called the \keyword{pants decomposition}{pants decomposition}. In a pants decomposition, three-punctured spheres are glued together by cylinders;
	the length coordinates determine the sizes of these cylinders, and the twist coordinates twist the cylinders.}

As an example, for the 4d $\scN=4$ theory,
the corresponding 2d surface $\Sigma$ was $T^2$, and
its independent closed paths are the two cycles,
the $\alpha$-cycle and the $\beta$-cycle of the 2d torus.
As explained in Chap.~\ref{chap.intro}, these corresponded to the electric and magnetic charges in the 4d $\scN=4$ theory,
so that the holonomies along them should also be quantities related to the electric and magnetic charges.

Natural physical quantities in gauge theories associated with charges are the expectation values of
Wilson loops and their S-duals, 't Hooft loops.
We are therefore led to the conjecture that the expectation values of these \keyword{loop operators}{loop operator} give the coordinates discussed so far.

This conjecture can be extended to general punctured Riemann surfaces $\Sigma$, not necessarily $T^2$.
Namely, it is conjectured that the classification of holonomies along closed paths of $\Sigma$
corresponds to the charges existing in the 4d $\mathcal{N}=2$ theory $\mathcal{T}[\Sigma]$.

\small
In the case of the $A_1$-type 6d $(2,0)$ theory discussed here,
$\mathcal{T}[\Sigma]$ has a known Lagrangian with gauge group $SU(2)^{3g-3+h}$,
and in this case the correspondence between closed paths and charges can be checked directly \cite{Drukker:2009tz}.
\normalsize

In particular, the dimension of the Coulomb branch of $\mathcal{T}[\Sigma]$ is
the same as the dimension of the \Teichmuller space,
\begin{align}
	\textrm{dim}_{\bC} \scM_{\rm Coulomb}=3g-3+h \ ,
\end{align}
and there are the same number of kinds of electric and magnetic charges.
Moreover, the decorated \Teichmuller space has $h$ more real dimensions,
which can be interpreted as corresponding to the existence of $h$ flavor symmetries in the theory $\mathcal{T}[\Sigma]$.

That the expectation values of loop operators become coordinates can be understood as follows.
As we saw in Chap.~\ref{chap.6d}, we are now considering
the $S^1$-compactification of 4d $\mathcal{N}=2$ theories,
and the loop operators wrap the $S^1$. Therefore, when we dimensionally reduce on $S^1$ \eqref{4d3d},
the loop operators become particles of the 3d theory.
The Coulomb branch in 4d was parametrized by the expectation values of the field $\phi$ taking values in the adjoint representation. When we dimensionally reduce to 3d, the integral of the gauge field along the $S^1$-direction (the expectation value of the Wilson line) gives a new complex scalar field $\phi'$ in 3d,\footnote{The Wilson line is complexified by combining with the 3d dual photon. See \eqref{monopoleop} in Chap.~\ref{chap.3dN2}.} and $\phi, \phi'$ form a pair parametrizing the 3d Coulomb branch.
We can therefore be convinced that it is natural that the complexified expectation values of loop operators
give coordinates specifying the Coulomb branch.

\bigskip

The explanation so far has been incomplete in one respect: the coordinates obtained in this way are the length-twist coordinates, and not the Fock coordinates which we wanted to use in this chapter.
The length-twist coordinates were, as discussed so far, the expectation values of loop operators of the UV theory (Lagrangian).
What about the Fock coordinates, then? The claim in the literature is that \textbf{the Fock coordinates are identified with the expectation values of loop operators of the IR theory on the Coulomb branch} \cite{Gaiotto:2010be}.
The ideal triangulations, which so far have been given top-down and combinatorially, are then constructed more physically:
BPS states of 4d $\mathcal{N}=2$ theories are given by geodesics on the Riemann surface $\Sigma$, and by considering families of such geodesics we can construct ideal triangulations (we do not need the details in this book, but see Ref.~\cite{Gaiotto:2010be}).
From this understanding, the Poisson bracket \eqref{WPpoisson}
is nothing but the Dirac quantization condition of the charges (including the charges of flavor symmetries).

\bparagraph{Transformation under Flip}\label{subsec.flip_transf}

In the definition of Fock variables so far, we have fixed the ideal triangulation.
However, ideal triangulations of a 2d surface are
not unique. What happens, then, if we change the ideal triangulation?

Such a change is not a matter of taste, but is
indispensable for a global description of the \Teichmuller space.
Since the \Teichmuller space is a manifold,
a single coordinate system does not cover the whole moduli space.
Fixing a triangulation determines one coordinate chart of the \Teichmuller space,
and in order to discuss the relations between different coordinate charts we have to change the triangulation.

\begin{figure}[t]
	\centering{\includegraphics[scale=0.3]{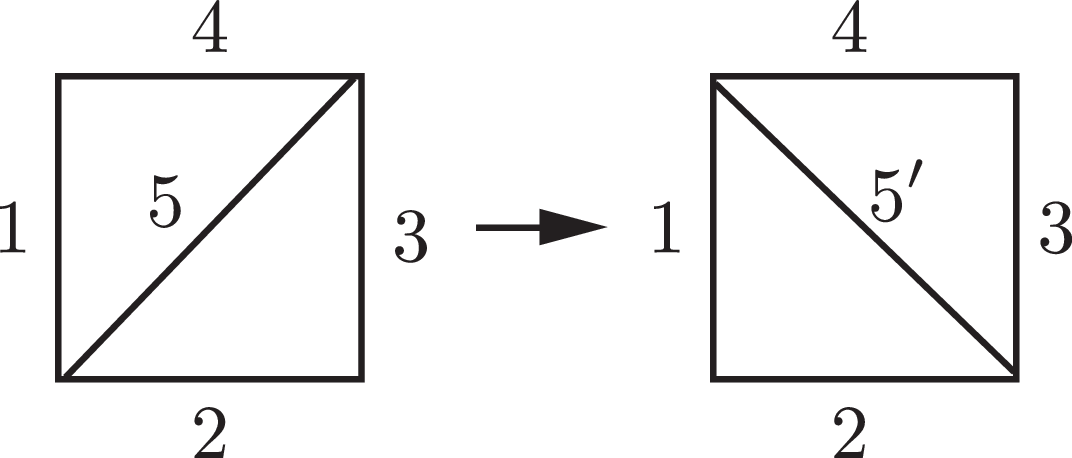}}
	\caption{The flip at the edge $5$ exchanges the diagonal $5$ of the ideal quadrilateral with the other diagonal $5'$.}
	\label{fig.flip}
\end{figure}

Here there is one convenient fact: it is known that any two ideal triangulations of a 2d surface are related by
a finite number of repetitions of the operation in Fig.~\ref{fig.flip} (exchanging the diagonals of an ideal quadrilateral),
called a \keyword{flip}{flip}\footnote{Since this maps two triangles to two other triangles, it is also called the
	\keyword{2--2 move}{2-2 move}. It is also called the \keyword{Whitehead move}{Whitehead move}.}
\cite[Prop.~4.5]{PennerUniversal}.

A flip can be performed once we specify an edge.
There are two triangles containing this edge, which together form
a quadrilateral. The flip is the operation of removing the diagonal and
adding the other diagonal (Fig.~\ref{fig.flip}).

\begin{figure}[t]
	\centering{\includegraphics[scale=0.3]{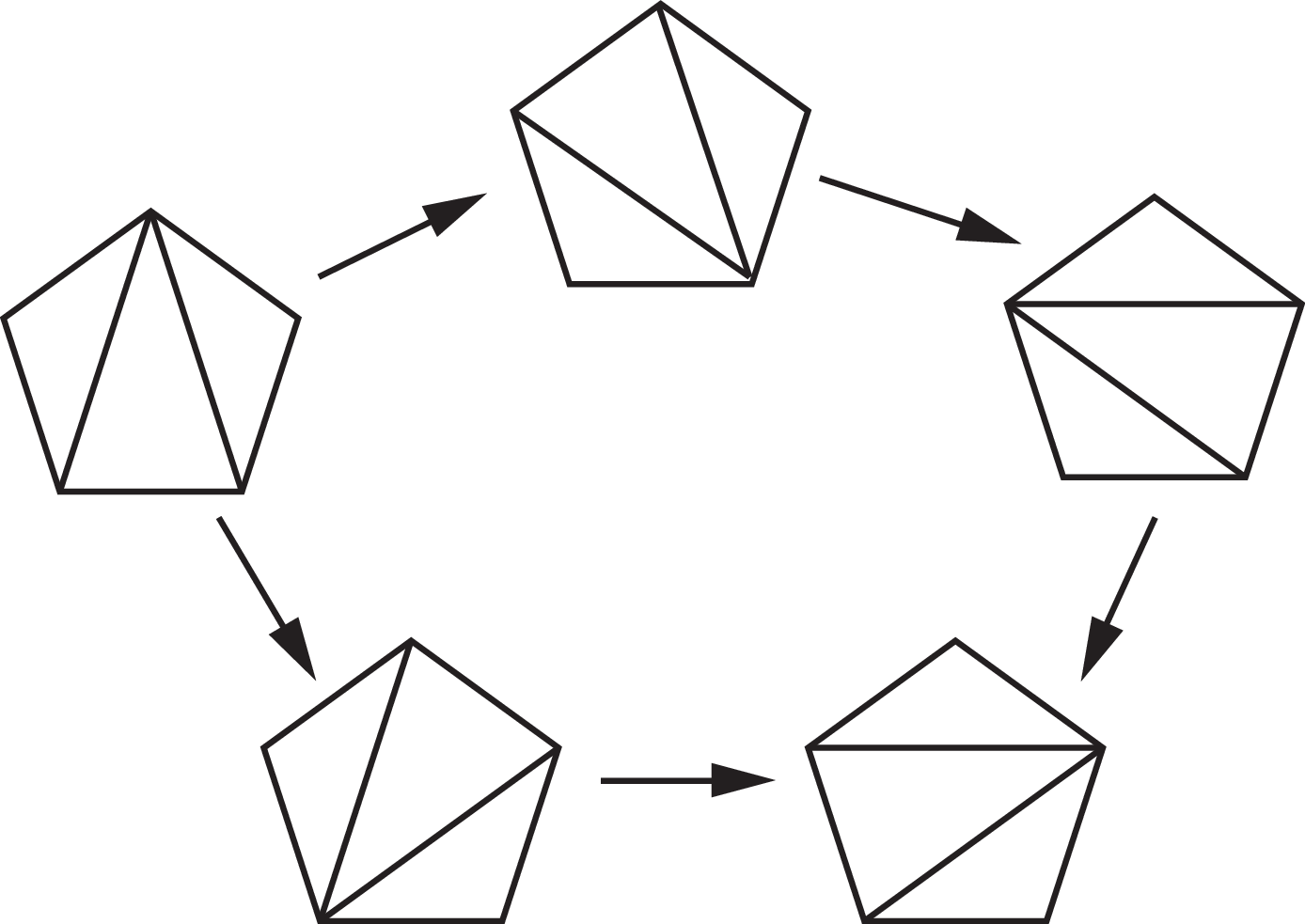}}
	\caption{The pentagon relation. When the edges are labeled,
		a relabeling is needed after the flips (see also Fig.~\ref{pentagon_id_SGC} in Appendix~\ref{app.clusterapp}).}
	\label{fig.pentagon}
\end{figure}

Flips satisfy the pentagon relation shown in Fig.~\ref{fig.pentagon}:
denoting by $T_{ij}$ the flip with respect to adjacent faces $i$ and $j$, we have\footnote{Invertible operations $T_{ij}$ satisfying \eqref{PennerPentagon} are called the \keyword{Ptolemy groupoid}{Ptolemy groupoid}.
	Incidentally, a groupoid is a category in which all morphisms are invertible.}
\begin{align}
	T_{12} T_{13} T_{23}=T_{23} T_{12} \ .
	\label{PennerPentagon}
\end{align}

Returning now to the coordinates,
the transformation rule of the Fock coordinates ($z_i=e^{Z_i}$) under a flip is known to be given by
\begin{align}
	\begin{split}
		z_1' & =z_1 (1+z_5) \ ,\quad
		z_2'=z_2 (1+z_5^{-1})^{-1}\ , \\
		z_3' & =z_3 (1+z_5) \ ,\quad
		z_4'=z_4 (1+z_5^{-1})^{-1}\ , \\
		z_5' & =z_5^{-1} \ .
	\end{split}
	\label{Focktransform}
\end{align}

One can immediately check that this coordinate transformation is compatible with the symplectic structure \eqref{WPform}. Moreover,
one can directly check that (1) two flips give back the original coordinates, and (2) the pentagon gives back the original coordinates
(exercise~\ref{classical_consistency}).

To derive the transformation rule \eqref{Focktransform},
the definition so far, as abstract coordinates, is not sufficient.
In Appendix~\ref{app.clusterapp} we explain that this can be interpreted, more generally, as a concrete example of the
mutation of BPS quivers.

More generally, the transformation rule \eqref{Focktransform} is
an example of the \keyword{cluster transformation}{cluster transformation} of
\keyword{cluster $y$-variables}{cluster y-variable}, discussed in
Appendix~\ref{app.clusterapp}, and quantum \Teichmuller theory can be
understood nicely within the framework of the general theory of cluster algebras.


\small

Let us here derive the transformation rule
\eqref{Focktransform} by geometric considerations.
For this purpose, let us first introduce the
\keyword{Penner coordinates}{Penner coordinate}
(also called the \keyword{$\lambda$-lengths}{lambda-length})
on the \Teichmuller space.

An edge $i$ of an ideal triangulation is a geodesic connecting two punctures,
a geodesic with respect to the metric determined by a point of the \Teichmuller space.
Let us denote its length by $L_i$.
The claim is that the set of these lengths
$\{L_i\}_{i\in E}$ can be taken as coordinates of the \Teichmuller space.

However, there is one problem with the description so far.
Since all the vertices of ideal triangles are on the boundary of $\bH^2$,
with this naive definition all the $L_i$ diverge: the metric \eqref{H2metric} diverges at $y=0$,
and the lengths have logarithmic divergences.

\begin{figure}[t]
	\centering\includegraphics[scale=0.29]{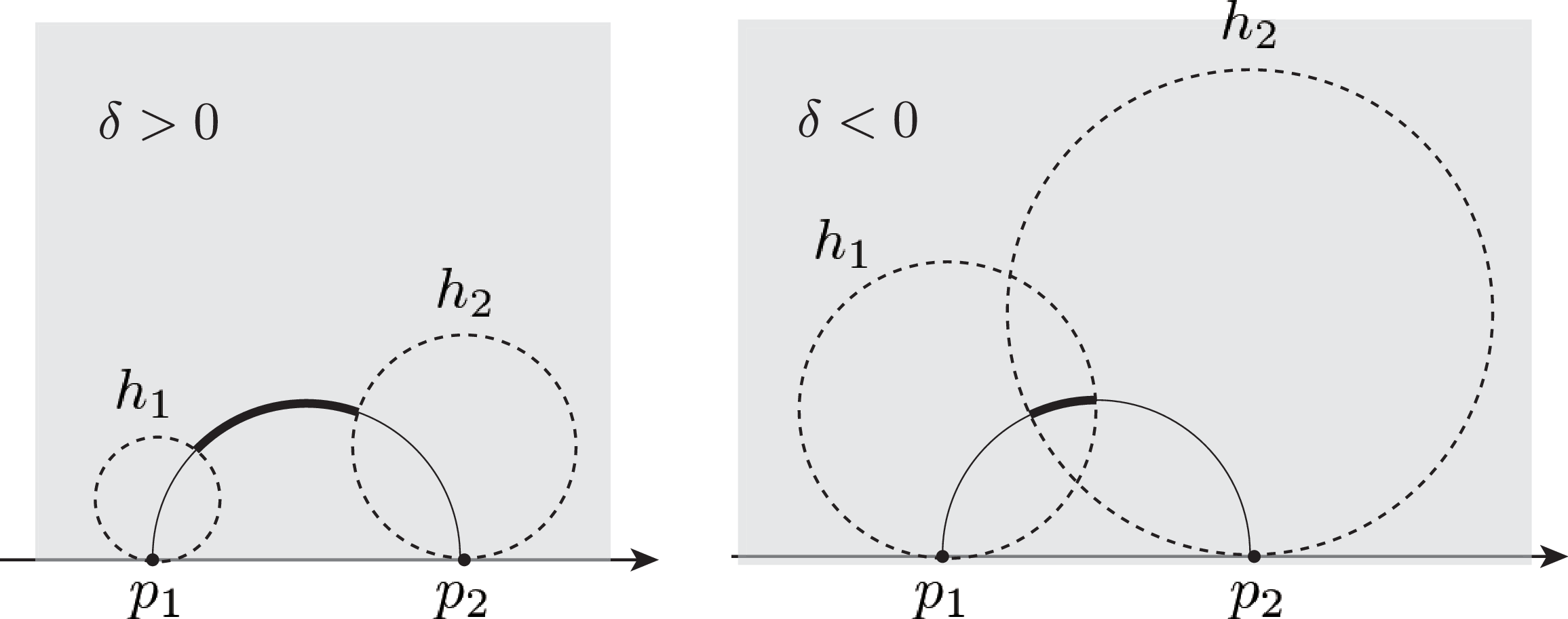}
	\caption{The length of the geodesic connecting two points $p_1, p_2$ of $\partial \bH^2$ diverges.
		To regularize this, we consider circles $h_1, h_2$ called horocycles (dotted lines in the figure),
		and measure the length of the part outside the two circles (left).
		When there is no part outside the circles, we attach a minus sign to the length of the overlapping part
		of the two circles (right).
		The Penner coordinate is defined from the signed length $\delta$ thus obtained by
		$L=e^{\delta/2}$.
	}
	\label{horocycle}
\end{figure}

Let us therefore consider the following regularization (Fig.~\ref{horocycle}).
Consider the end points $p_1, p_2$ of an edge $i$ of the ideal triangulation, and
take horocycles $h_1, h_2$ for each of them. Here a
\keyword{horocycle}{horocycle} for a point on the boundary of $\bH^2$
is a circle passing through that point and tangent to $\partial \bH^2$
(at the point at infinity, it is defined to be a horizontal straight line).

Consider the geodesic connecting $p_1$ and $p_2$, and
let the length of its part between the intersections with the two horocycles $h_1, h_2$ be the
absolute value $|\delta_i|$ of a quantity $\delta_i$. Moreover,
we determine the sign of $\delta_i$ by defining $\delta_i>0$ when $h_1, h_2$ do not intersect, and
$\delta_i<0$ when $h_1, h_2$ intersect.
We then define the Penner coordinates by $L_i:=e^{\delta_i/2}$.
In practice we can make $\delta_i>0$ by taking $h_1, h_2$ sufficiently small, but
it is often convenient to allow $\delta_i<0$ as well.

As already stated, any two ideal triangulations are related by
flips. When we choose the labels of the edges as in Fig.~\ref{fig.flip}, the Penner coordinates change under the flip as
\begin{align}
	L_5'=\frac{L_1 L_3+L_2 L_4}{L_5}
	\label{Ltransf}
\end{align}
(exercise~\ref{ex.Ptolemy}). This can be interpreted as (a regularized version of) the hyperbolic-geometry version of
Ptolemy's theorem, familiar from elementary geometry.

As in the case of Fock coordinates, we can directly verify that both
(1) two flips repeated on the same edge and
(2) the pentagon
keep the values of the coordinates (i.e.\ give the identity operation)
(exercise~\ref{classical_consistency}).

\bigskip

We have obtained finite quantities by regularization so far;
a remaining problem is that this regularization has an ambiguity.
Indeed,
if we change the length of the horocycle around a puncture $v\in V$, then $L_i$ is multiplied by a constant when the edge $i$
contains the vertex $v$, and
is unchanged when it does not contain $v$:
\begin{align}
	L_i \to c_vL_i \quad (v\in i) \ , \qquad
	L_i \to L_i  \quad (v\notin  i) \ .
	\label{Lambi}
\end{align}
This is the drawback of Penner coordinates.

The Fock coordinates $Z_i$ are coordinates with this ambiguity removed.
Any edge of an ideal triangulation (here the edge $5$) is, as in Fig.~\ref{fig.flip},
the diagonal of a certain ideal quadrilateral (here consisting of the edges $1, 2, 3, 4$).
The Fock variable $z_5$ of the edge $5$ is then defined as the
\keyword{cross ratio}{cross ratio} of
the Penner coordinates $L_1, \ldots, L_4$ of the edges $1, 2, 3, 4$:
\begin{align}
	z_5:=\frac{L_1 L_3}{L_2 L_4} \ .
	\label{ZasRatio}
\end{align}
With this definition, we can derive \eqref{Focktransform} from \eqref{Ltransf} and
\eqref{ZasRatio}. We have thus derived the transformation rule of the Fock variables.

\normalsize

\section{Quantum \Teichmuller Space}\label{subsec.qTeich}

\subsection{Quantization}\label{subsec.procedure}

Let us next quantize the classical \Teichmuller theory discussed above.
The quantization we discuss below is actually rather simple
if you are familiar with the standard quantization procedure in quantum mechanics:
the Poisson bracket is already of the simple form \eqref{WPpoisson},
and all we need to do is to replace it by
canonical commutation relations of operators.
Namely, the Fock coordinates $Z_i$ are promoted to operators $\hat{Z}_i$
satisfying the commutation relations
\begin{align}
	\left[ \hat{Z}_{i}, \hat{Z}_{j} \right]= -2i \hbar \, \mathsf{Q}_{i,j}  \ ,
	\label{omega_q}
\end{align}
where the ``Planck constant'' is denoted by $\hbar$, and we have included a factor of
$-2$ for later convenience.
In terms of the variables $\hat{z}_{i}:=e^{\hat{Z}_{i}}$ this reads
\begin{align}
	\hat{z}_{j} \hat{z}_{i} = q^{2 \mathsf{Q}_{i,j} }\hat{z}_{i}\hat{z}_{j} \ ,
	\label{qTorus}
\end{align}
where $q:=e^{i \hbar}$ (the classical limit is now $q\to 1$).
Note that, with $\hbar=\pi b^2$ as we will set below, this $q$ is the square root of the parameter $q=e^{2\pi i b^2}$ in \eqref{qtb} of the previous chapter.
For our purposes, the information on the arguments of $\hat{z}_i$ beyond modulo $2\pi$ is also important,
so that what is more directly important is \eqref{omega_q}.

By taking suitable linear combinations of the $\hat{Z}_i$ in \eqref{omega_q},
we can obtain a set $\{\hat{X}_a, \hat{P}_a, \hat{C}_{\alpha} \}$ satisfying
the canonical commutation relations
\begin{align}
	[\hat{X}_a, \hat{P}_b]=i \hbar \delta_{a,b} \ , \quad
	                                                [\hat{X}_a, \hat{C}_{\alpha}]=[\hat{P}_a, \hat{C}_{\alpha}]=0
	\label{XPdelta}
\end{align}
(cf.\ exercise~\ref{ex.4sphere_canonical}).
This is a rather standard finite-dimensional
quantum-mechanical system, and it is straightforward to quantize it.
Concretely, we can consider the Hilbert space of $L^2$-normalizable functions, on which $X_a$ are represented as coordinates and $P_b$ as their conjugate momenta
(derivatives with respect to $X_a$).
Here the $\hat{C}_{\alpha}$ are constants, and
represent conserved quantities of the system.
What we have obtained in this way is \keyword{quantum \Teichmuller theory}{quantum Teichmuller theory} (the original paper is Ref.~\cite{Chekhov:1999tn}).\footnote{
	As a general remark, care is needed since it is impossible to quantize general polynomials of $X_a, P_b$ consistently (the Groenewold-van Hove theorem).}\footnote{One might be interested to know that there is another quantization of \Teichmuller theory due to Kashaev \cite{KashaevQuantization}. In his formulation
	we use coordinates associated with the faces (not the edges) of the ideal triangulation.}

To get a better perspective, it is convenient to extend $\hat{Z}_i$ and $\hat{z}_i$
linearly. Namely, we consider elements (charges) $\gamma=\sum_i n_i \gamma_i$ of the lattice with basis $\gamma_i$, and set $\hat{Z}_{n_1 \gamma_1+n_2 \gamma_2}:=n_1 \hat{Z}_{\gamma_1}+n_2 \hat{Z}_{\gamma_2}, \hat{Z}_{\gamma_i}:=\hat{Z}_i$.
If we then define the
pairing (Dirac pairing) of two elements $\gamma=\sum_i m_i \gamma_i, \gamma'=\sum_i n_i \gamma_i$ by
\begin{align}
	\langle \gamma, \gamma' \rangle :=\sum_{i,j} m_i n_j \mathsf{Q}_{i,j} \ ,
	\label{Dirac_pairing}
\end{align}
\eqref{omega_q} and \eqref{qTorus} become
\begin{align}
	\left[ \hat{Z}_{\gamma}, \hat{Z}_{\gamma'} \right]=-2 i \hbar \, \langle \gamma, \gamma' \rangle
	\label{omega_q_linear}
\end{align}
and
\begin{align}
	\mathcal{A}_\mathsf{Q}: \hat{z}_{\gamma'} \hat{z}_{\gamma} = q^{2\langle \gamma, \gamma' \rangle} \hat{z}_{\gamma} \hat{z}_{\gamma'} \ .
	\label{qTorus_linear}
\end{align}
The algebra defined (from $\mathsf{Q}$) by the latter (suitably completed if necessary) is called the \keyword{quantum torus}{quantum torus}, and is denoted by $\mathcal{A}_\mathsf{Q}$ in the following.
\nomenclature{$\mathcal{A}_\mathsf{Q}$}{quantum torus}


\subsection{Physical Intuition for Quantization}

Let us now consider more physically the meaning of the ``quantization'' so far.
As discussed in Chap.~\ref{chap.6d}, we have been considering the setup where
4d $\mathcal{N}=2$ theories are compactified on $S^1$ (see \eqref{4d3d}).
What happens, then, if we do not consider strictly three dimensions, but keep
the radius $R$ of this $S^1$ finite?

Recalling the discussion of Chap.~\ref{chap.6d},
this radius $R$ is the radius of the $S^1$ in the sixth direction when lifting the 5d $\mathcal{N}=2$ theory to the 6d $(2,0)$ theory;
in the language of string theory, it was the radius of the M-theory direction when the five-dimensional branes of type IIA string theory (D4-branes)
are lifted to the six-dimensional branes of M-theory (M5-branes).

The particles in 3d are then lifted to loop operators in 4d.
We have been considering the 4d spacetime $\bR^3\times S^1$;
we can preserve supersymmetry by considering the spacetime with a twist by a subgroup $U(1)\subset SO(3)_{\rm rot}$ of the rotation group of $\bR^3$
as we go around the $S^1$-direction: if this $U(1)$ is the rotation of the plane $\bR^2\subset \bR^3$, the
spacetime becomes $\bR\times (\bR^2\times S^1)_q$. The parameter of this twist is
the parameter $q$ of the quantization, and it counts the value of the third component $J_3$ of the $SO(3)$ spin.

\small
In the explanation so far, the Fock coordinates were understood as the expectation values of IR loop operators.
To be compatible with the rotation symmetry by $J_3$, the loop operators have to be placed at the origin of $\bR^2$,
but they can be anywhere along the remaining $\bR$-direction.
\normalsize

Now, suppose that there are two loop operators.
The charges of the two loop operators then have to satisfy the Dirac quantization condition,
if electric and magnetic charges exist simultaneously. In the construction above, $\gamma$ represents the charges of the loop operators, and the fact that their pairing \eqref{Dirac_pairing} is an integer is
nothing but this quantization condition.
When they have a non-zero Dirac pairing (in which case they are called mutually non-local),
the two loop operators generate a Poynting vector $\mathbf{J}\sim  \mathbf{E}\times \mathbf{B}$ along $\bR$,
whose magnitude is given by $|\langle \gamma, \gamma' \rangle|$.
When we exchange the positions of the two loop operators, the value of $J_3$ changes sign, so that
the two loop operators do not commute, by a factor of $q$.
We have thus understood the commutation relation \eqref{qTorus_linear} of the quantized Fock operators $\hat{z}_{\gamma}$
\cite{Gaiotto:2010be}.\footnote{To be precise, however,
	whether the quantization of the Hitchin moduli in the geometric sense and the ``quantization''
	in the sense of counting angular momenta agree precisely should ultimately be checked by direct computation in field theory (at least for field theories whose Lagrangians are known). For this direction, see e.g.\ Ref.~\cite{Ito:2011ea}.}

\subsection{Example}

Before concluding this section, let us give a concrete example. Consider the four-punctured sphere of Fig.~\ref{fig.quiver_4sphere}. For each of
\begin{align*}
	\{a,b,c\}=\{1,2,3\} \ , \quad \{1,5,6\} \ , \quad \{2,6,4\} \ , \quad \{5,3,4\} \ ,
\end{align*}
we have
\begin{align}
	[\sfZ_{a}, \sfZ_{b}]=
	[\sfZ_{b}, \sfZ_{c}]=
	[\sfZ_{c}, \sfZ_{a}]=-2i\hbar \ .
\end{align}
The following four combinations commute with all the others:
\begin{align}
	\begin{split}
		\sfZ_{1}+\sfZ_{2}+\sfZ_{6} \ , \quad
		\sfZ_{1}+\sfZ_{3}+\sfZ_{5} \ , \quad
		\sfZ_{4}+\sfZ_{2}+\sfZ_{3} \ , \quad
		\sfZ_{4}+\sfZ_{5}+\sfZ_{6} \ .
	\end{split}
\end{align}
The remaining part is 2d; choosing, for example, $\sfZ_{1}$ and $\sfZ_{2}$ as a basis,
we have
\begin{align}
	[\sfZ_{1}, \sfZ_{2}]=-2i \hbar \ ,
\end{align}
which is nothing but a quantum-mechanical system with one degree of freedom.


We also need to quantize the change \eqref{Focktransform} of the Fock variables under a flip (Fig.~\ref{fig.flip}).
Here again we can replace the Fock coordinates by operators;
since the operators do not commute, there remains an ambiguity in the ordering of the operators.
However, what is important is that the operation thus defined is compatible with
the commutation relations \eqref{omega_q}.
The transformation satisfying this condition is as follows (exercise~\ref{comm_consistency}):
\begin{equation}
	\begin{split}
		 & \hat{z}_{1}\to  \hat{z}_1 (1+q\, \hat{z}_5)  \ ,\quad
		\hat{z}_{2}\to \hat{z}_2 (1+q^{-1}\hat{z}_5^{-1})^{-1} \ , \\
		 & \hat{z}_{3}\to \hat{z}_3 (1+q\, \hat{z}_5) \ , \quad
		\hat{z}_{4}\to \hat{z}_4 (1+q^{-1}\hat{z}_5^{-1})^{-1}  \ , \quad
		\quad \hat{z}_5\to \hat{z}_5^{-1} \ .
		\label{qxtransf}
	\end{split}
\end{equation}
It is easy to check that this reduces to \eqref{Focktransform} in the classical limit $q\to 1$.

As already stated, since the flip changes the commutation relations \eqref{omega_q},
\eqref{qxtransf} is a map between different quantum torus algebras $\mathcal{A}_{\rm before}, \mathcal{A}_{\rm after}$.
Its meaning becomes clearer if we decompose this transformation into the composition of the following two transformations.
First, one transformation is the map from $\mathcal{A}_{\rm before}$ to $\mathcal{A}_{\rm after}$
given by
\begin{equation}
	\begin{split}
		 & \hat{z}_{1}\to   \hat{z}_1   \ ,\quad
		\hat{z}_{2}\to q \hat{z}_2 \hat{z}_5 \ , \\
		 & \hat{z}_{3}\to  \hat{z}_3 \ , \quad
		\hat{z}_{4}\to q \hat{z}_4 \hat{z}_5  \ , \quad
		\quad \hat{z}_5\to \hat{z}_5^{-1} \ .
		\label{qxtransf_dilog}
	\end{split}
\end{equation}
Using the notation $\hat{Z}_{\gamma}$ introduced earlier in Sec.~\ref{subsec.procedure}, \eqref{qxtransf_dilog} becomes
\begin{equation}
	\begin{split}
		 & \hat{Z}_{\gamma_1 }\to   \hat{Z}_{\gamma_1}   \ ,\quad
		\hat{Z}_{\gamma_2}\to  \hat{Z}_{\gamma_2 + \gamma_5} \ ,  \\
		 & \hat{Z}_{\gamma_3}\to  \hat{Z}_{\gamma_3} \ , \quad
		\hat{Z}_{\gamma_4}\to \hat{Z}_{\gamma_4 +\gamma_5}  \ , \quad
		\quad \hat{Z}_{\gamma_5} \to \hat{Z}_{-\gamma_5} \ ,
		\label{qxtransf_dilog_c}
	\end{split}
\end{equation}
which is nothing but a certain linear transformation (change of basis) of $\gamma$.
As discussed in Sec.~\ref{subsec.BPS_quiver} of the appendix, this is
a linear transformation of the basis of charges in 4d $\mathcal{N}=2$ theories,
and is an operation changing the way the 3d field theory and the 4d field theory are coupled.

The next transformation is a map from $\mathcal{A}_{\rm after}$ to itself,
namely the conjugation by the
quantum dilogarithm function $e_b(x)$ (a function we have already encountered in Chap.~\ref{chap.S3}):
\begin{equation}
	\begin{split}
		 & \hat{z}_{i}\to e_b\left(\frac{\hat{Z}_5}{2\pi b}\right)^{-1}  \hat{z}_i \, e_b\left(\frac{\hat{Z}_5}{2\pi b}\right) \ ,
		\quad i=1,\cdots , 5 \ ,
		\label{dilog_conj}
	\end{split}
\end{equation}
where we defined the parameter $b$ by
\begin{align}
	\hbar=\pi b^2 \ .
\end{align}
As we will see in the next chapter, this can be interpreted as an operation changing the
3d $\mathcal{N}=2$ theory on the boundary.
In Appendix~\ref{app.clusterapp} we comment on this decomposition in a more general situation.
It is important that, in this way, one quantum dilogarithm function appears for each flip.

When we consider an element $\varphi$ of the mapping class group of a 2d surface, its effect on the ideal triangulation was, as already stated, represented by a sequence of flips. Since each flip is represented by an operator of the quantum dilogarithm function, the operator $\hat{\varphi}$ corresponding to the mapping class group element is in the end represented by a product of quantum dilogarithm functions.

In this chapter we have understood quantum \Teichmuller theory as a problem of
finite-dimensional quantum mechanics.
In the next chapter we will reinterpret it in the language of 3d geometry.

\section{Supplement: Relation with Liouville Theory}\label{sec.Liouville_rel}

\small

In this section let us comment on the relation between
quantum Liouville theory and quantum \Teichmuller theory.
This is not needed for understanding this chapter and the next;
however, it is needed for the discussion in Sec.~\ref{sec.3d_as_AGT}
(see also Table~\ref{triality} in Chap.~\ref{chap.6d}).

Let us first start with the classical theory.
The equation of motion derived from the Liouville Lagrangian \eqref{LiouvilleAction} reads
\beq
\partial \bar{\partial} \phi=2\pi \mu b e^{2b \phi} \ .
\label{metric_Ansatz}
\eeq
This is known as the \keyword{Liouville equation}{Liouville equation},
and it coincides with the condition that the metric
\beq
ds^2=e^{2 b \phi} dz d\bar{z}
\label{2dmetric}
\eeq
has constant negative curvature.

Now the celebrated
\keyword{uniformization theorem}{uniformization theorem} claims
that there is a unique solution to \eqref{metric_Ansatz}
at any point of the \Teichmuller space, giving rise to a relation between the
classical \Teichmuller space and the classical solutions of Liouville theory
(historically this is actually a roundabout explanation, since the Liouville equation arose in the study of the uniformization theorem).

Even if they agree as classical systems, it is not clear whether their quantizations agree.
That the quantization of \Teichmuller theory and
the quantization as Liouville theory agree was
conjectured by H.~Verlinde in the 1980s \cite{Verlinde:1989ua},
and more recently almost proven by Refs.~\cite{Teschner:2003at,Teschner:2005bz,Teschner:2010je}
(see also Ref.~\cite{FKV2} in the context of discrete Liouville theory).
The two theories, which share the same origin, have thus met again in the quantum world
after many years.

This theory is at present not mathematically rigorous, and we cannot explain its details
here; the basic idea is to construct
the conformal blocks of quantum Liouville theory
inside quantum \Teichmuller theory.
More concretely, the key claim is that the conformal blocks are given by the overlap of
the holomorphic basis $|\scF(\tau)\rangle$ of the quantum \Teichmuller space and
the length basis $|l\rangle=|\alpha\rangle=|\alpha, E\rangle$:
\beq
\scF_{\alpha,E}(\tau)=\langle \alpha, E | \scF(\tau) \rangle \ .
\label{FinTeichmuller}
\eeq

We have already assumed this formula in Sec.~\ref{sec.3d_as_AGT};
let us add an explanation here.
First, one basis (the holomorphic basis) $|\scF(\tau) \rangle$ is obtained
by representing the \Teichmuller space in terms of the holomorphic coordinates $\tau$
parametrizing the complex structure of $\Sigma$, and quantizing it.

The other basis is the length basis $|\alpha\rangle=|l \rangle$.
This is the basis naturally obtained when we write the \Teichmuller space in terms of the length-twist coordinates (Sec.~\ref{sec.length_twist})
and quantize it, where $l$ represents the length coordinates among the length-twist coordinates.

The Verlinde (E.~Verlinde) loop operators in Liouville theory
are the loop operators $\scL$ in \Teichmuller theory \cite{Drukker:2009id}.
The basis $|l\rangle$ is then, once a pants decomposition is fixed,
the simultaneous eigenstates of a maximal set of commuting Verlinde loop operators,
and forms a complete basis of the Hilbert space $\scH_{\Sigma}$ \cite{KashaevQuantum}:
\beq
\langle l | l' \rangle =\nu(l)^{-1} \delta(l-l') \ ,  \quad \int_{l>0} dl \,\nu(l)\, |l \rangle \langle l|=1 \ .
\label{nucomplete}
\eeq
Here $\nu(l)$ is the same as the integration measure $\nu(\alpha)$ in \eqref{Liouvillecorrelator}. The Liouville momentum $\alpha$ is identified with the length parameter $l$, and the external momentum $E$ with the holonomy $m$ at the puncture
\cite{Verlinde:1989ua,Witten:1990wn}.

The rough strategy for showing \eqref{FinTeichmuller} is to check that both sides have
the same transformation properties under the action of the mapping class group, and
the same asymptotic behavior.
A proof of AGT using a similar method has also been attempted \cite{Teschner:2010je}.

\normalsize


\begin{practice}

	\item $[\bll]$ ($\bH^2$ as a symmetric space)
	For a Lie group $G$ and its Lie subgroup $H$,
	the quotient (coset) $G/H$ is a homogeneous space, and
	has a canonical metric induced from that of $G$.
	Verify that for $G=PSL(2, \bR), H=SO(2)$
	this canonical metric coincides with the metric of $\bH^2$ given in \eqref{H2metric}.

	\item $[\bll]$ (Geodesics in $\bH^2$)
	Show that geodesics in $\bH^2$
	are either (1) semicircles which intersect the axis $y=0$ at right angles,
	or (2) vertical straight lines (Fig.~\ref{H2geodesic}).
	Hint: compute the length of the graph $y=y(x)$,
	and solve the Euler-Lagrange equation obtained by varying the length with respect to
	$y(x)$.
	\label{ex_geodesic}

	\item $[\bll]$ (Hyperbolic distance in $\bH^2$)\label{ex.distance}

	Let us represent two points in
	$\bH^2$ by two complex numbers $z, w$
	whose imaginary parts are positive.
	Show that the hyperbolic distance
	between $z,w$ under the metric \eqref{H2metric}
	is given by
	\begin{align}
		d(z,w)=\log \frac{|z-\overline{w}|+|z-w|}
		            {|z-\overline{w}|-|z-w|} \ .
		\label{eq.distance}
	\end{align}
	It is sometimes useful to know that this formula can also be written
	as
	\begin{align}
		 & \sinh\left(\frac{d(z,w)}{2}\right) =
		\frac{|z-w|}{2 \sqrt{\textrm{Im}(z) \textrm{Im}(w)}} \ , \nonumber            \\
		 & \cosh\left(\frac{d(z,w)}{2}\right) =
		\frac{|z-\overline{w}|}{2 \sqrt{\textrm{Im}(z) \textrm{Im}(w)}} \ , \nonumber \\
		 & \lambda(z,w):=\exp\left(\frac{d(z,w)}{2}\right)=
		\frac{|z-w|+|z-\overline{w}|}{2 \sqrt{\textrm{Im}(z) \textrm{Im}(w)}} \ .
	\end{align}

	Hint: first check that \eqref{eq.distance}
	is invariant under M\"obius transformations.
	Then show that $z,w$ can be mapped to
	purely imaginary numbers by an appropriate
	M\"obius transformation.
	Once this is done, the geodesic is simply
	the part of the imaginary axis between $z$ and $w$,
	and by writing $z=ip, w=iq$ the distance between the two is
	easily computed to be
	\begin{align}
		d(ip, iq)=\left| \log\left(\frac{p}{q}\right)\right| \ ,
		\label{eq.distance2}
	\end{align}
	matching \eqref{eq.distance}.

	\item $[\bll]$ (\keyword[generalized Ptolemy's theorem]{Generalized Ptolemy's theorem}{generalized Ptolemy's theorem})\label{ex.Ptolemy}

	Show that the Penner coordinates $L_i$,
	defined by regularizing the lengths with respect to the horocycles as in the main text,
	satisfy the generalized Ptolemy's theorem \eqref{Ltransf}.

	This is basically just a computation;
	see e.g.\ Penner's original paper \cite[Proposition 2.6]{PennerDecorated}.
	However, Penner uses a different coordinate system,
	and it is a bit cumbersome to rewrite his argument in our
	coordinates on the upper half-plane.

	In this exercise, for pedagogical purposes, let us try to be down to earth
	and verify \eqref{Ltransf} by rather direct computations.

	\begin{enumerate}

		\item To make the computation more efficient,
		      let us use M\"obius transformations to
		      move the four points to
		      $P_1=t_1, P_2=t_2, P_3=t_3$ and the point at infinity $P_4=\infty$
		      (see Fig.~\ref{fig.Ptolemy_ex}). Here we can choose $t_{1,2,3}$
		      to be real numbers satisfying $t_1<t_2<t_3$.

		\item
		      It is easy to write down the geodesics connecting two points.
		      For example, for $P_1$ and $P_4$ we have the straight line $\{z|\, \textrm{Re}(z)=t_1\}$;
		      for $P_1$ and $P_2$ we have the semicircle
		      \begin{align}
			      \left( x-\frac{t_1+t_2}{2}\right)^2+y^2=\left(
			      \frac{t_1-t_2}{2}\right)^2 \ .
		      \end{align}

		      Let us choose the horocycles around the points
		      $P_{1,2,3}$ to be
		      \begin{align}
			      (x-t_i)^2+\left(y-\frac{a_i^2}{2}\right)^2=\left(\frac{a_i^2}{2}\right)^2 \ , \quad (i=1,2,3) \ ,
		      \end{align}
		      and the one around $P_4$ to be $\{z|\, \textrm{Im}(z)=a_4^{-2} \}$, with $a_{1,2,3,4}>0$.

		\item
		      Compute the Penner coordinates from the formula \eqref{eq.distance2}.
		      Denoting by $L_{ij}$ the Penner coordinate
		      for the edge connecting $P_i$ and $P_j$, show that
		      \begin{align}
			      \begin{split}
				       & L_{14}=\frac{1}{a_1 a_4}\ , \quad
				      L_{24}=\frac{1}{a_2 a_4}\ , \quad
				      L_{34}=\frac{1}{a_3 a_4}\ , \quad
				      \\
				       & L_{12} =\frac{t_2-t_1}{a_1 a_2} \ , \quad
				      L_{23} =\frac{t_3-t_2}{a_2 a_3} \ , \quad
				      L_{13} =\frac{t_3-t_1}{a_1 a_3} \ .
			      \end{split}
			      \label{PtolemyEx1}
		      \end{align}
		      Hint: recall that $L$ is defined from a signed distance.

		\item
		      Eliminate the $t_i$'s from \eqref{PtolemyEx1} to obtain the generalized
		      Ptolemy's theorem
		      \begin{align}
			      L_{12} L_{34}+L_{23} L_{14}=L_{13} L_{24} \ .
		      \end{align}
	\end{enumerate}

	\begin{figure}[t]
		\centering\includegraphics[scale=0.35]{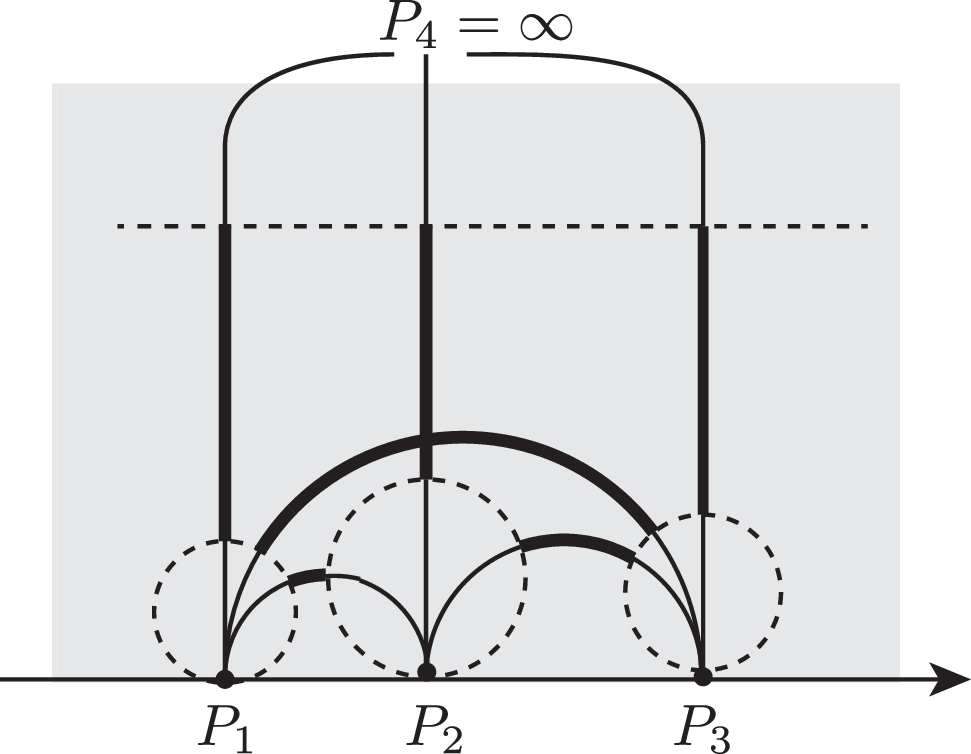}
		\caption{Setup for the generalized Ptolemy's theorem (exercise~\ref{ex.Ptolemy}).}
		\label{fig.Ptolemy_ex}
	\end{figure}

	\item  $[\bll]$ (Analogy between hyperbolic and plane geometry)
	Choose your favorite theorem in plane geometry,
	and ask if there is a counterpart of it in hyperbolic geometry.
	For example, what about Menelaus's theorem and Ceva's theorem?

	\item $[\bll]$ (Transformation rules of Fock/Penner coordinates)
	In classical \Teichmuller theory,
	verify that the transformation formulas for the
	Fock and Penner coordinates (\eqref{Focktransform} and \eqref{Ltransf})
	are consistent with the symplectic form \eqref{WPform} (before and after the flip).
	Also verify that the coordinates are preserved
	by (1) two flips on the same edge and (2) a pentagon move.
	Hint: after a pentagon move we need to exchange the labels of the edges to come back to the original coordinates.
	\label{classical_consistency}

	\item $[\bll]$  (Transformation of quantum Fock coordinates)\label{comm_consistency}
	For the quantum Fock coordinates,
	verify the consistency between the transformation property under a flip \eqref{qxtransf}
	and the commutation relation \eqref{omega_q}.
	Also verify that the quantum Fock coordinates are preserved
	by (1) two flips on the same edge and (2) a pentagon move.

	\item $[\bll]$   (Example: four-punctured sphere)\label{ex.4sphere_canonical}

	Consider the ideal triangulation given in Fig.~\ref{fig.quivertrig}
	(note the difference from Fig.~\ref{fig.quiver_4sphere}).
	Write down the Fock coordinates and their commutation relations
	in quantum \Teichmuller theory, and show that the resulting
	commutation relations can be written in the canonical form \eqref{XPdelta}
	after a suitable change of variables.

	If you are still not satisfied, repeat a similar exercise for your favorite ideal triangulation of the five-punctured sphere.

	\begin{figure}[t]
		\centering\includegraphics[scale=0.32]{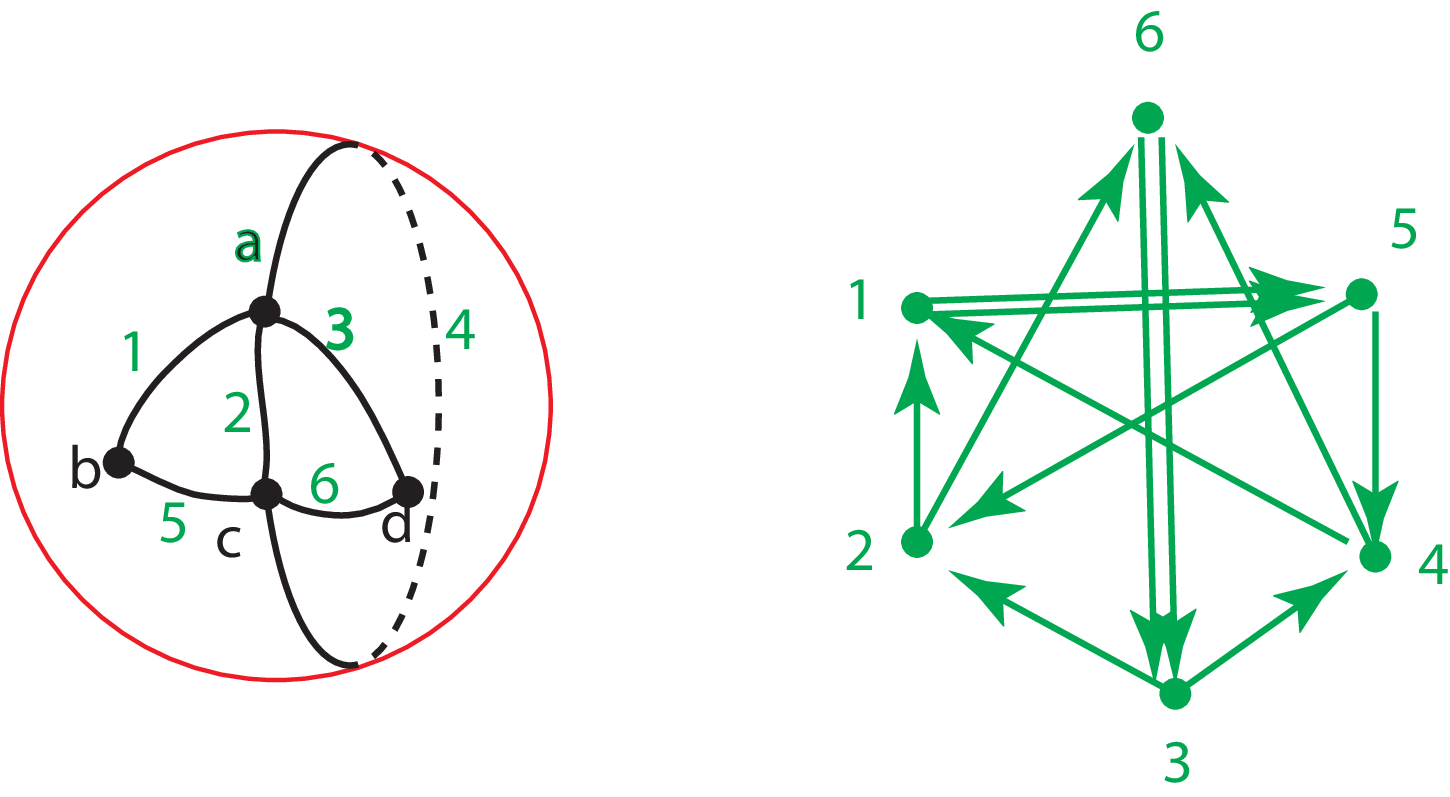}
		\caption{An example of a quiver associated with an ideal triangulation of the four-punctured sphere.
			This ideal triangulation is related to that in Fig.~\ref{fig.quiver_4sphere} by a flip;
			after a relabeling of the edges, their quantized Fock variables $\hat{z}_i$ are related by the transformation rule \eqref{qxtransf}.}
		\label{fig.quivertrig}
	\end{figure}

\end{practice}

\nextchaptermark{Tetrahedron Decomposition and 3d Theories}
\chapter{Tetrahedron Decomposition and 3d $\scN=2$ Theories}\label{chap.3mfd}

\begin{abstract}
	In this chapter, we synthesize the
	discussions of the previous chapters,
	and summarize the correspondence
	between 3d supersymmetric field theories and
	3d topological theories.
	We discuss the moduli space of
	flat $G_{\bC}$-connections on a 3-manifold
	and its quantization.
	Geometric flat $G_{\bC}$-connections describe the geometry of 3-manifolds,
	in particular of hyperbolic 3-manifolds (for the case $G_{\bC}=SL(2, \bC)$),
	and this geometry is quantized by complex Chern-Simons theory.
\end{abstract}

\section{3d as ``Movie'' of 2d}\label{sec.3d_as_2d}

In the previous chapter
we discussed flat connections on 2d manifolds.
In this chapter, we apply the knowledge from the previous chapter
to the geometry of 3-manifolds,
and find their counterparts in
3d $\mathcal{N}=2$ theories.

The basic idea is rather simple.
In the Hamiltonian formalism,
a 3d theory
is specified first by a choice of a state
in the Hilbert space associated with a spatial 2d slice,
and then by the time evolution of the state.

For 3d Chern-Simons theory,
this is exactly what we discussed in Sec.~\ref{subsec.TQFT}
and Chap.~\ref{chap.Teichmuller} (the previous chapter),
where the ``time evolution'' in this case is the
action of the mapping class group---a mapping class group element maps
one ideal triangulation to another, which is equivalently represented by a
sequence of flips of the ideal triangulation.

\begin{figure}[t]
	\centering{\includegraphics[scale=0.25]{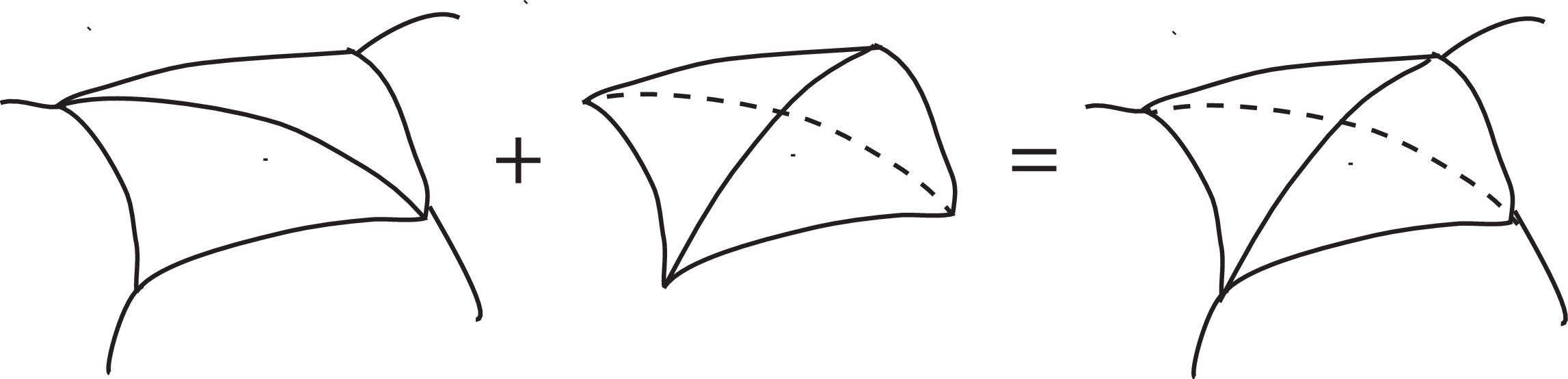}}
	\caption{Attaching a (3d) tetrahedron to a (2d) triangulation. The diagonal of a certain quadrilateral in two dimensions is then
		replaced by the other diagonal (flip), and as a result the triangulation changes into
		another triangulation. From a 3d viewpoint, this corresponds to
		adding one ideal tetrahedron.}
	\label{fig.pillow}
\end{figure}

Let us translate this into 3d language.
This is intuitively not too difficult: as in Fig.~\ref{fig.pillow},
a flip of the 2d surface
is equivalent, in more 3d language, to attaching a tetrahedron.
Note that a tetrahedron has four faces, and when squeezed enough,
we have two faces on the front and two on the back; the faces on the back
are glued to the 2d surface we started with, leaving
two new faces open (we can regard a tetrahedron as a cushion, with diagonals
drawn on its front and back).

Since the vertices of ideal triangles on the 2d surface were on the boundary of $\bH^2$, the same should be true for the
vertices of the tetrahedra. We can therefore expect that the 3d tetrahedra are also tetrahedra with vertices on the boundary,
namely ideal tetrahedra.\footnote{We have not yet taken into account
	the metric and hyperbolic structure of ideal tetrahedra here;
	this will be discussed shortly.} To summarize,
\begin{align}
	\textrm{flip of 2d surface} \longleftrightarrow \textrm{3d ideal tetrahedron} \ .
	\label{flip_2d_3d_dict}
\end{align}

In the description above we decomposed the action of the mapping class group into flips;
correspondingly, in three dimensions we glue together several ideal tetrahedra.
Namely, we consider an ideal tetrahedron decomposition of the 3-manifold $M$:\footnote{The idea of decomposing 3-manifolds into tetrahedra has been considered in the context of 3d quantum gravity.
	In that case, however, one often considers finer and finer tetrahedron decompositions,
	whereas the situation here is different. Relatedly, since we
	consider ideal tetrahedra, whose vertices are always on the boundary of $\bH^3$,
	we do not need to consider the operation of adding a new vertex inside a tetrahedron (also called the 1--4 move).
}
\begin{align}
	M=\displaystyle \bigcup_i \Delta_i  \ ,
\end{align}
where each $\Delta_i$ is an \keyword{ideal tetrahedron}{ideal tetrahedron}.
\nomenclature{$\Delta, \Delta_i$}{ideal tetrahedron}

\begin{figure}[t]
	\centering{\includegraphics[scale=0.21]{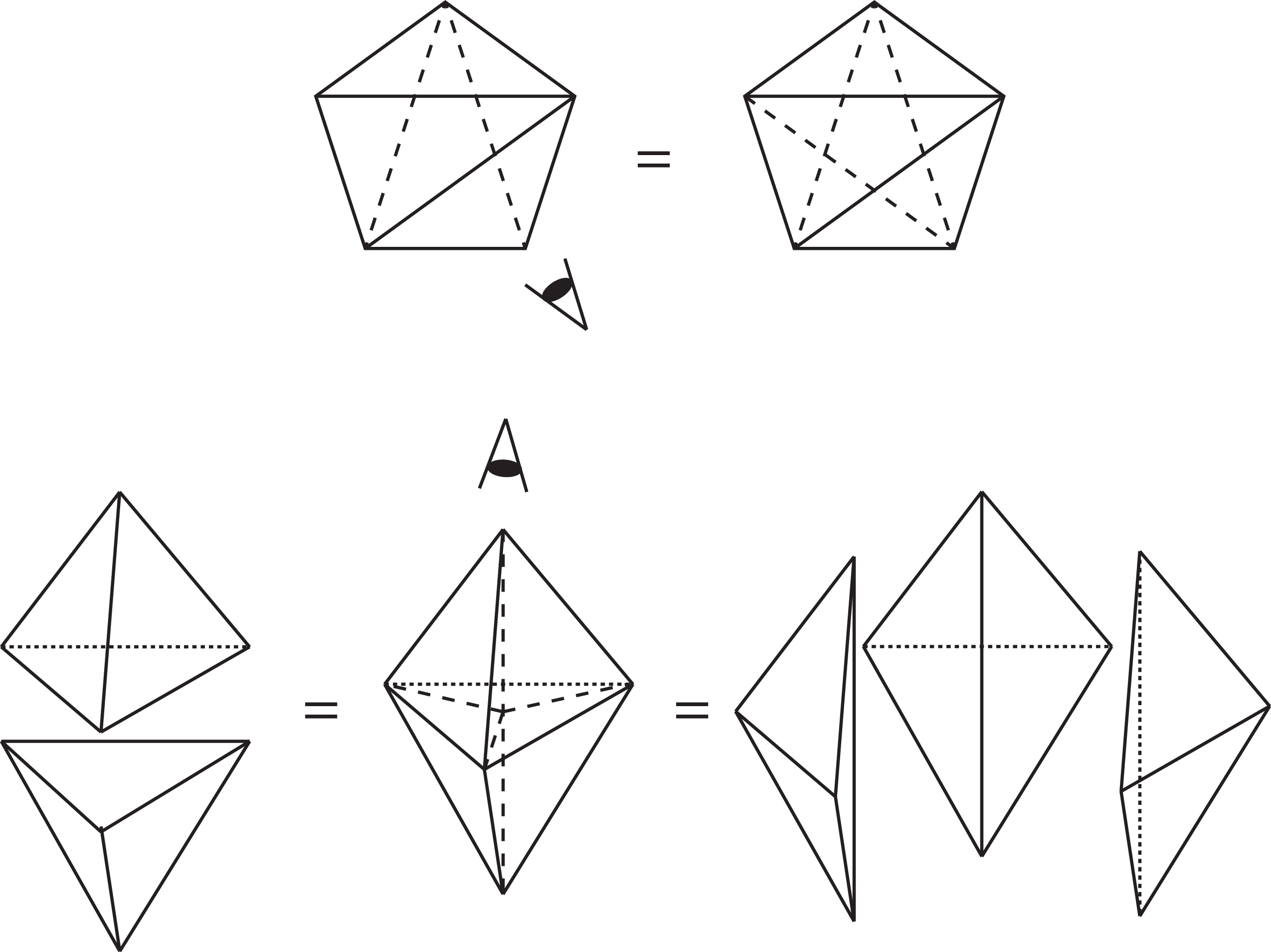}}
	\caption{Viewing the pentagon on a 2d Riemann surface (Fig.~\ref{fig.pentagon}) stereoscopically,
		we obtain the 2--3 Pachner move for 3d ideal tetrahedra.}
	\label{fig.Pachner}
\end{figure}

As we have seen already in Chap.~\ref{chap.Teichmuller},
the decomposition of a mapping class group action into a sequence of flips is
not unique, and the ambiguity is essentially exhausted by pentagons.
We can also translate this ambiguity into more 3d language:
as in Fig.~\ref{fig.Pachner}, a pentagon now
replaces two tetrahedra by three tetrahedra.
This operation is known in the literature
as the \keyword{Pachner move}{Pachner move} (or the 2--3 move, or the 2--3 Pachner move). Summarizing,
\begin{align}
	\begin{split}
		 & \textrm{pentagon on 2d surface}   \\
		 & \qquad \qquad \longleftrightarrow
		\textrm{2--3 Pachner move of 3d ideal tetrahedron decomposition} \ .
	\end{split}
\end{align}

Pushing the correspondence so far further,
we expect that the coordinates themselves of the \Teichmuller space on the 2d surface (and its quantization)
also have 3d counterparts. Since in two dimensions the coordinates were defined for the
edges of the ideal triangulation, in three dimensions coordinates should also be defined for each edge of the ideal tetrahedron decomposition,
and they are expected to satisfy properties similar to those in two dimensions.
Moreover, once we arrive at this understanding,
we should be able to carry out a similar construction also for more
general 3-manifolds, which cannot be regarded as time evolutions of 2d surfaces.
This expectation turns out to be correct, as we will explain below.

\small
Among the discussions so far, the parts concerning the gluing of topological triangles and polyhedra
have counterparts for more general $d$-dimensional and $(d+1)$-dimensional hyperbolic geometries.
However, once we start discussing their metrics and hyperbolic structures, the relation becomes more complicated.
That the correspondence between two and three dimensions works nicely is partly due to the
special circumstance that, for the isometry group $PSL(2, \bR)$ of the 2d hyperbolic space $\bH^2$ and
the isometry group $PSL(2, \bC)$ of the 3d hyperbolic space $\bH^3$, the former is the real slice of the latter
(see also footnote~\ref{foot.SLRC} in Chap.~\ref{chap.Teichmuller}).

\normalsize

\section{Ideal Tetrahedra and 3d $\scN=2$ Theories}
\bparagraph{Ideal Tetrahedra}

\begin{figure}[t]
	\centering{\includegraphics[scale=0.28]{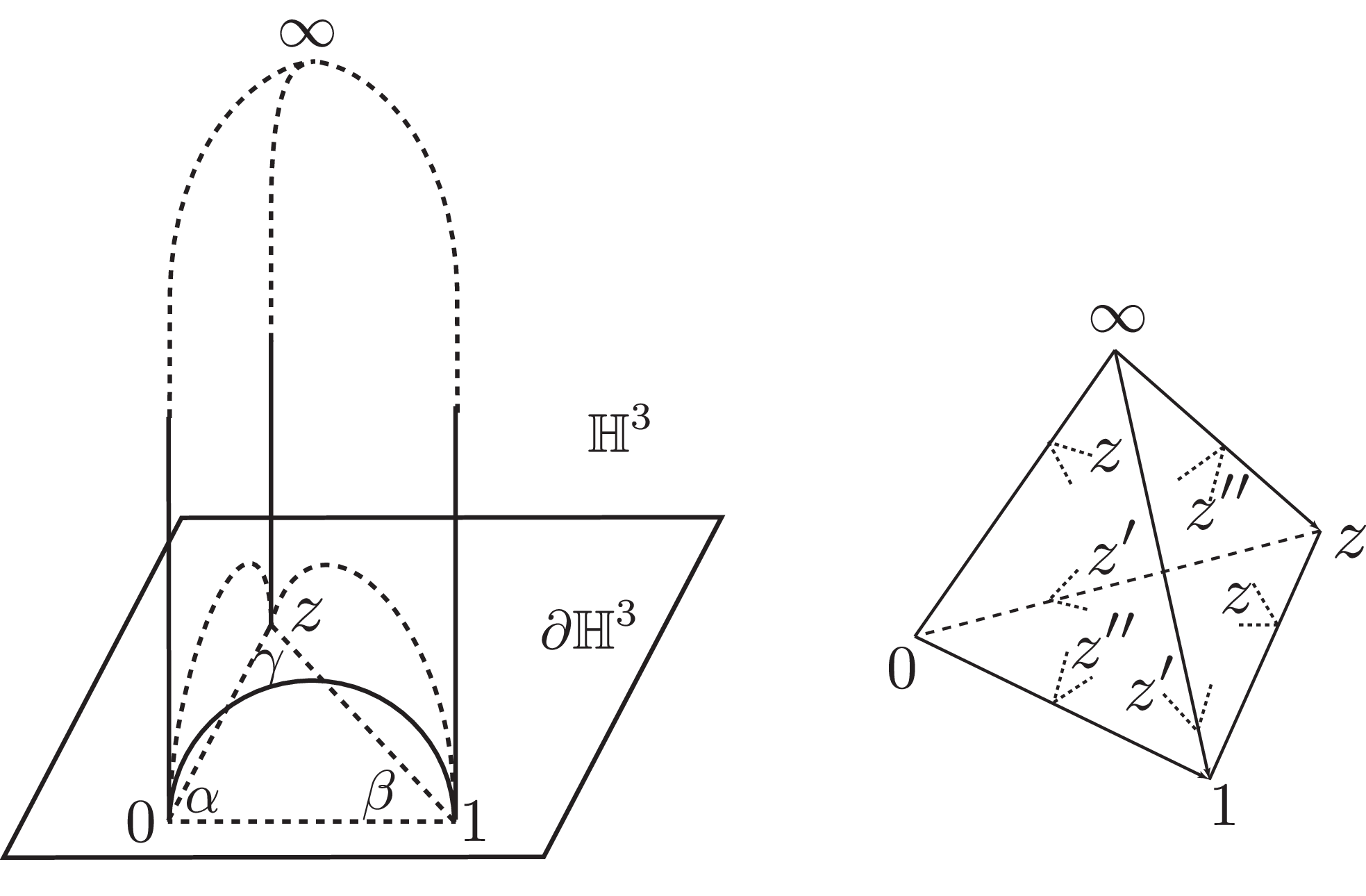}}
	\caption{(Left) An ideal tetrahedron in the 3d hyperbolic space $\mathbb{H}^3$
		has all its vertices on
		the boundary of $\mathbb{H}^3$,
		and the vertices can be taken to be
		$\{0,1,z,\infty\} \in
			\mathbb{C}\cup \{\infty\}$. (Right) Illustration of the three moduli $z, z', z''$ of the ideal tetrahedron.
		The three dihedral angles are given by $\alpha=\textrm{Arg}(z), \beta= \textrm{Arg}(z'), \gamma=\textrm{Arg}(z'')$.}
	\label{idealtetrahedron}
\end{figure}

Let us first consider ideal tetrahedra.
These are the natural higher-dimensional generalization of the ideal triangles explained in Chap.~\ref{chap.Teichmuller},
and are tetrahedra in $\mathbb{H}^3$
all of whose vertices are on the
boundary $\partial \bH^3$ of $\mathbb{H}^3$ (Fig.~\ref{idealtetrahedron}).
As already stated in Chap.~\ref{chap.complexCS},
$\bH^3$ has the isometries of M\"obius transformations, and
by using them we can bring the vertices of the tetrahedron to
$0,1,z$ and infinity ($\infty$).
This complex parameter $z$ is called the
\keyword{modulus}{modulus} of the ideal tetrahedron.

If we fix three of the four vertices of the tetrahedron to
$0, 1, \infty$, and the remaining one to $z$, the
freedom of the M\"obius transformations $PSL(2, \bC)$ is completely fixed.
However, as to which of the three vertices are identified with $0,1,\infty$,
respectively, there is the freedom of the $3!=6$ permutations of the three.
Half of them preserve the orientation of the tetrahedron, and
send the modulus $z$ to
\begin{align}
	z \ , \quad z':=\frac{1}{1-z} \ , \quad z'':=1-z^{-1} \ ,
	\label{zzz}
\end{align}
\nomenclature{$z, z', z''$}{moduli of an ideal tetrahedron}
respectively (please check each of these). Replacing $z$ by $z'$ or $z''$ merely relabels the vertices,
so that the actual tetrahedron does not change. The remaining three permutations
reverse the orientation of the tetrahedron, and send the modulus to
\begin{align}
	z^{-1} \ , \quad z'^{-1}=1-z \ , \quad z''^{-1}=\frac{1}{1-z^{-1}} \ .
	\label{modulus_triple}
\end{align}
Note that which of the three vertices we regard as $z$ is
a matter of taste, and the freedom of cyclically permuting $z, z', z''$ remains.

The variables $z, z', z''$ satisfy the relation
\begin{align}
	z z' z''=-1 \ ,
\end{align}
which follows from \eqref{zzz}. This means that the sum of the arguments of $z, z', z''$ is
$\pi$, which is nothing but the statement that the sum of the three interior angles
\begin{align}
	{\rm Arg}(z) \ , \quad {\rm Arg}(z') \ , \quad
	                                         {\rm Arg}(z'')
	\label{three_angles}
\end{align}
of the triangle in $\bC$ determined by $0,1, z$ is $\pi$.
More precisely, in order to make explicit that the sum of the interior angles is $\pi$ and not, for example, $3\pi$,
it is better to write
\begin{align}
	z=:e^Z\ , \quad z'=:e^{Z'}\ , \quad z''=:e^{Z''}
	\label{Z_triple}
\end{align}
\nomenclature{$Z, Z', Z''$}{moduli of an ideal tetrahedron ($z=e^Z\ , \quad z'=e^{Z'}\ , \quad z''=e^{Z''}$)}
and
\begin{align}
	Z + Z'+ Z'' = i \pi \ .
	\label{sumZ=pi}
\end{align}

The three angles \eqref{three_angles} are also the \keyword{dihedral angles}{dihedral angle} of the ideal tetrahedron at the three edges ending at the vertex $\infty$ (a horospherical cross-section near $\infty$ is a Euclidean triangle similar to that with vertices $0,1,z$).
Namely, a tetrahedron has six edges, each of which has
a dihedral angle; for an ideal tetrahedron with modulus $z$,
the two dihedral angles at opposite edges of the tetrahedron coincide (see Fig.~\ref{idealtetrahedron}; it is instructive to confirm
this by direct computation).

\subsection{3d $\mathcal{N}=2$ Counterpart of Ideal Tetrahedra}

Let us now relate the geometric explanations so far to
quantum \Teichmuller theory in the previous chapter.

First, let us consider the boundary
$\Sigma_{0,4}$ of an ideal tetrahedron. As we have already seen,
in this case quantum \Teichmuller theory has
$\hat{Z}, \hat{Z}', \hat{Z}''$ satisfying
\begin{align}
	[\hat{Z}, \hat{Z}']=[\hat{Z}', \hat{Z}'']=[\hat{Z}'', \hat{Z}]=2i \hbar= 2i \pi b^2 \ ,
	\label{Z_comm_here}
\end{align}
and their sum $\hat{Z}+\hat{Z}'+\hat{Z}''$ commutes with everything and
could be regarded as a constant.\footnote{
	Reversing the orientation of the surface reverses the sign of $\mathsf{Q}_{i,j}$ in \eqref{WPpoisson},
	and hence the overall sign of the commutation relations.
	Here we choose the orientation of $\Sigma_{0,4}$ induced from the ideal tetrahedron with $\textrm{Im}(Z)>0$,
	which is opposite to the convention of the previous chapter; this is the choice for which the dictionary
	\eqref{Z_sigma_id_2} below matches the geometry (positive dihedral angles correspond to positive R-charges).}
We can therefore quantize by regarding, for example, $\hat{Z}$ as the coordinate and
$\hat{Z}''$ as the momentum.

It is natural to identify these $\hat{Z}, \hat{Z}', \hat{Z}''$ with the quantization of the
moduli of the tetrahedron introduced in \eqref{Z_triple} and \eqref{sumZ=pi}.

Following \eqref{flip_2d_3d_dict},
let us associate a flip with an ideal tetrahedron.
As we saw in \eqref{dilog_conj} in the previous chapter, a flip corresponded to
the quantum dilogarithm operator $e_b\left(\frac{\hat{Z}}{2\pi b}\right)$.
Let us interpret this as a wave function on the ideal tetrahedron (whose argument is $z$, or its logarithm $Z$).
Since \eqref{dilog_conj} took the Heisenberg form for operators,
in the Schr\"odinger picture for wave functions we expect
\begin{align}
	\psi_{\textrm{ideal tetrahedron}}(Z)=e_b\left(\frac{iQ}{2}-\frac{Z}{2\pi b}\right) \ ,
	\label{psi_Delta}
\end{align}
where, for later convenience, we have shifted $Z$ by a constant and chosen the sign of $Z$
(correlated with the orientation convention in \eqref{Z_comm_here}).

Let us examine whether this identification is natural.
Recall that, from \eqref{recursion_eb} in the appendix,
\begin{align}
	e_b\left(\frac{iQ}{2}-\frac{x+2\pi i b^2}{2\pi b}\right)=(1- e^{-x})\, e_b\left(\frac{iQ}{2}-\frac{x}{2\pi b}\right)
\end{align}
holds.
Using $\hat{z}=e^{\hat{Z}}, \hat{z}''=e^{\hat{Z}''}$, this can be written as (note that, due to the commutation relations \eqref{Z_comm_here}, $\hat{z}''$ is a difference operator shifting the argument $Z$ of the wave function $\psi(Z)$ by $2\pi i b^2$)
\begin{align}
	\left( 1-\hat{z}^{-1} - \hat{z}'' \right) \psi_{\textrm{ideal tetrahedron}}(Z)=0 \ .
\end{align}
Namely, we find that this is the quantization, as an operator acting on the wave function, of the relation
$1-z^{-1}-z''=0$ between $z$ and $z''$ in \eqref{zzz}.

Once we have come this far, it is easy to identify the
3d $\mathcal{N}=2$ theory corresponding to the wave function \eqref{psi_Delta}.
First, the theory $T_{\Delta}$ corresponding to the ideal tetrahedron should have the following
partition function:
\begin{align}
	\begin{split}
		Z_{S^3_b}(T_{\Delta}) & = e_b\left(\frac{iQ}{2}-\frac{Z}{2\pi b}\right)                                                                                               \\
		                      & =s_b\left(\frac{iQ}{2}-\frac{Z}{2\pi b}\right) \, e^{\frac{i\pi}{2} \left(\frac{iQ}{2}-\frac{Z}{2\pi b}\right)^2-\frac{i\pi}{24}(2-Q^2)} \ .
	\end{split}
\end{align}
Here we used the formula \eqref{sbeb} replacing $e_b(x)$ by $s_b(x)$.
The identification of the parameters is, using $\overline{\sigma}$ in \eqref{bar_sigma},
\begin{align}
	\frac{iQ}{2}-\frac{Z}{2\pi b} \leftrightarrow \overline{\sigma} \ ,
	\quad
	Z_{S^3_b}(T_{\Delta}) = s_b\left(\overline{\sigma} \right) \, e^{\frac{i\pi}{2} \overline{\sigma}^2-\frac{i\pi}{24}(2-Q^2)} \ .
	\label{Z_sigma_id}
\end{align}
In particular, using $\overline{\sigma}=\frac{iQ}{2}(1-r)-\sigma$ (see \eqref{bar_sigma}), we have
\begin{align}
	\textrm{Im}(Z) \leftrightarrow \pi bQr   \ ,
	\quad
	\textrm{Re}(Z) \leftrightarrow 2\pi b\sigma \ .
	\label{Z_sigma_id_2}
\end{align}
That $\sigma$ is multiplied by $2\pi b$ here is natural from the viewpoint of the dimensional reduction from 3d $\mathcal{N}=2$ theories to 2d $\mathcal{N}=(2,2)$ theories ($2\pi b\sigma=\sigma_{\rm 2d}$ in \eqref{sigma_2d}; see Sec.~\ref{sec.3d_dim_red}).

Using the rules for $S^3$ partition functions in Chap.~\ref{chap.S3}, we find that $T_{\Delta}$ is
a non-interacting $\mathcal{N}=2$ chiral multiplet $\Phi$ \cite{Dimofte:2011ju}. Since $\overline{\sigma}=-\sigma_R-\sigma$
included the effects of both the R-symmetry and the flavor symmetry (see \eqref{bar_sigma}),
this chiral multiplet has a $U(1)_F$ flavor symmetry and a $U(1)$ R-symmetry,
and the charges of the field $\Phi$ are
\begin{align}
	\begin{array}{c|cc}
		     & F  & R \\
		\hline
		\Phi & +1 & 0
	\end{array}  \ .
\end{align}
From the partition function, following the rule \eqref{ZCS_offdiagonal}, we can read off the following levels of the Chern-Simons terms:
\begin{align}
	k_{\rm bare}=
	\left(
	\begin{array}{c|cc}
		  & F            & R            \\
		\hline
		F & -\frac{1}{2} & -\frac{1}{2} \\
		R & -\frac{1}{2} & -\frac{1}{2}
	\end{array}
	\right) \ .
	\label{Delta_level}
\end{align}
Note that this choice of Chern-Simons levels satisfies the constraints from the parity anomaly
(see Sec.~\ref{subsec.parity_anomaly} and \eqref{keff_2}).
Indeed, $\Delta k_{FF}=\frac{1}{2}$ is generated quantum mechanically by the parity anomaly,
which cancels against $k_{\textrm{bare}, FF}=-\frac{1}{2}$, so that the total Chern-Simons level is an integer. Similarly, for the Chern-Simons terms involving the R-symmetry,
the total levels are integers once quantum corrections are included,
and the flavor symmetry and the R-symmetry of the theory are preserved.
Even if the flavor symmetry were broken, the theory itself would not be inconsistent;
in practice, however, as we will see shortly, it is indispensable to gauge this flavor symmetry,
so that the analysis here is important.


\small
In the discussion so far we have fixed one dihedral angle as the modulus $z$ of the tetrahedron.
When we replace $z$ by $z'$ or $z''$,
the wave function is again written in terms of the quantum dilogarithm function
$e_b(x)$, by the first formula of \eqref{ebFourier} in Appendix~\ref{app.dilog}. On the Chern-Simons theory side, this is
the action of the \keyword{affine symplectic group}{affine symplectic group} $ISp(2, \bZ)$, combining the change of
polarization by an $SL(2, \bZ)=Sp(2, \bZ)$ matrix and a shift of variables. On the 3d $\mathcal{N}=2$ theory side, this can be interpreted as
a kind of 3d mirror symmetry \cite{Dimofte:2011ju}.

\normalsize


\section{Gluing Tetrahedra}

So far we have considered a single ideal tetrahedron; what happens, then, when we glue
several tetrahedra together?

\begin{figure}[t]
	\centering{\includegraphics[scale=0.25]{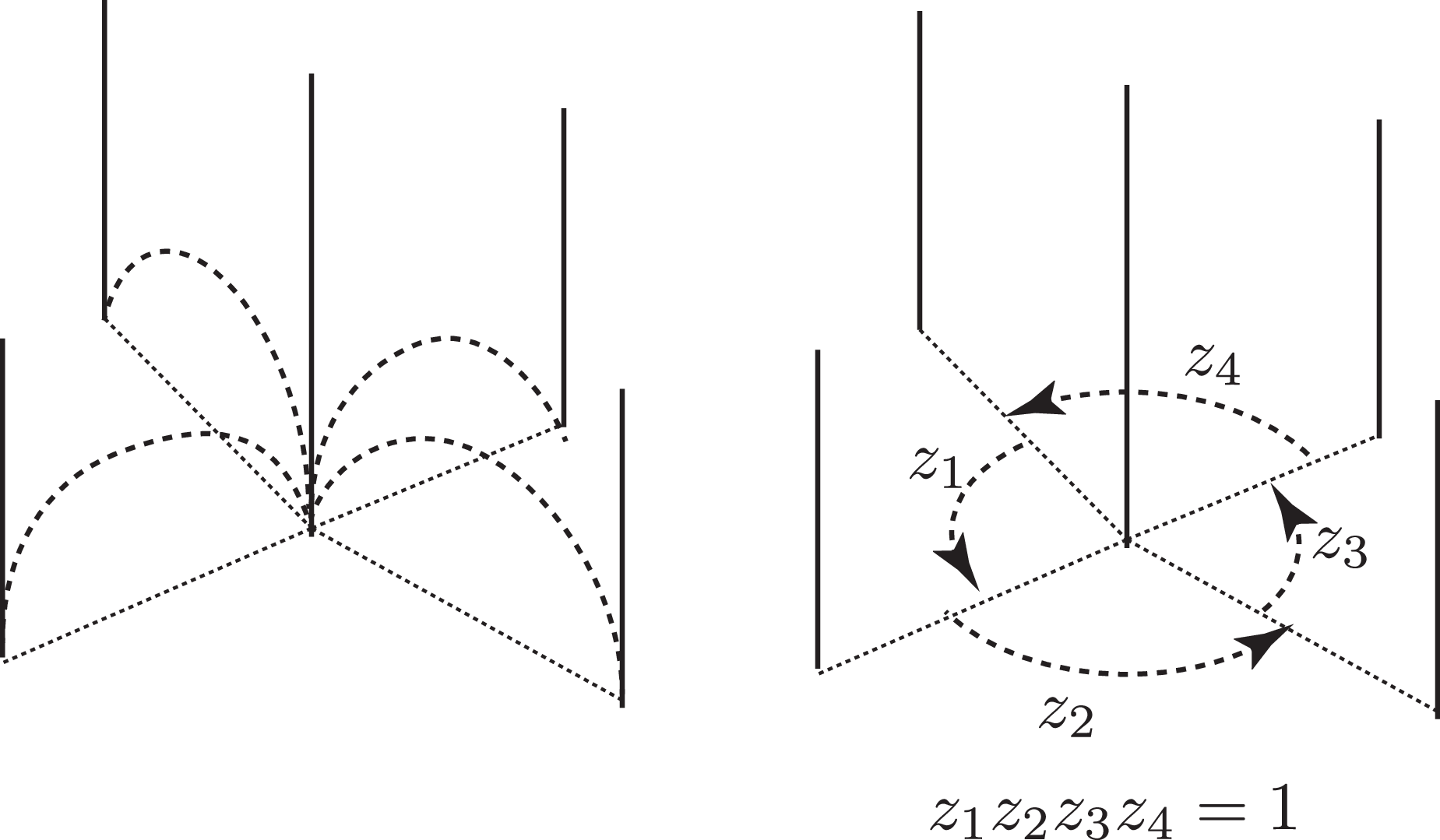}}
	\caption{Gluing of ideal tetrahedra. The total sum of the angles around an edge has to be $2 \pi$.
		For a complete hyperbolic structure,
		the sum of the torsions (the real parts of the moduli) also has to vanish.}
	\label{fig.gluing}
\end{figure}

Let us first consider this using classical geometry.
Since the modulus $z$ has an interpretation as a dihedral angle,
the gluing conditions have to be satisfied when we glue ideal tetrahedra.
First, the tetrahedra are glued around each edge of the tetrahedra,
and the sum of the angles has to be $2\pi$ (Fig.~\ref{fig.gluing}).
This can be naturally expressed as
\begin{align}
	\sum_\textrm{$i$: edge} Z_i=2\pi i \ , \quad \prod_\textrm{$i$: edge} z_i=1 \ .
	\label{bulkgluing}
\end{align}
We call this the \keyword{gluing condition}{gluing condition}\index{gluing equations@gluing equations|see{gluing condition}}\index{edge equations@edge equations|see{gluing condition}}.
Eq.~\eqref{bulkgluing} contains not only the condition on the sum of the angles (i.e.\ the imaginary parts of $Z$),
but also the condition on the real parts of $Z$ (the \keyword{torsion}{torsion}). When the latter condition is not satisfied, going once around the edge we come back with the same angle
but to a different point. Repeating this,
we can consider a sequence of points which is not complete. Therefore, if we wish to consider
complete hyperbolic structures, the condition on the real parts also has to be satisfied \cite[Proposition 3.4.19]{ThurstonLecture}.

Now, what does this formula mean in the language of 3d $\mathcal{N}=2$ theories?
The imaginary parts of $Z_i$ were identified with the R-symmetry charges of the fields, and the real parts
with the charges of the other global symmetries.
If, quantum mechanically, we have
\begin{align}
	\sum_\textrm{$i$: edge} \hat{Z}_i=2\pi i b Q=2\pi i (b^2+1)
	\label{bulkgluing_quantum}
\end{align}
(note that the combination $b^2+1$ also appeared in \eqref{qtb}), then using \eqref{Z_sigma_id_2} we obtain
\begin{align}
	\sum_\textrm{$i$: edge} r_i =2 \ , \quad \sum_\textrm{$i$: edge} \sigma_i =0 \ .
\end{align}
This is the condition for the superpotential $W=\prod_i \Phi_i$ to preserve the symmetries,
where $r_i$ is the R-charge of $\Phi_i$ and
$F_i$ is the charge under the $U(1)$ global symmetry (with $\sigma_i=F_i \sigma$ for an $i$-independent $\sigma$).
We have thus found that the gluing of ideal tetrahedra can be identified with the operation of
gluing the corresponding field theories by (Yukawa-type) superpotential interactions:\footnote{Even if the fields used in the superpotential are fields appearing directly in the Lagrangian, they can be transformed into monopole operators when we move to a different duality frame; hence, in general,
	the superpotential used for the gluing needs to involve monopole operators.
	In many cases, as long as we focus on a single term of the superpotential, we can use an appropriate duality frame so that all the fields appearing in that term are fields appearing in the Lagrangian (but we cannot write all the terms simultaneously in terms of the fields in the Lagrangian). However, for ideal tetrahedron decompositions in which an edge of an ideal tetrahedron is glued to another edge of the same tetrahedron,
	this is not the case, and monopole operators are needed in every duality frame even if we focus on a particular term.}
\begin{align}
	\begin{split}
		 & \textrm{gluing of field theories by superpotential} \\
		 & \qquad \qquad \longleftrightarrow
		\textrm{gluing of 3d ideal tetrahedra} \ .
	\end{split}
\end{align}
Here the terms of the superpotential in general involve
monopole operators (see Sec.~\ref{subsec.monopole}), namely fields which do not exist in the original Lagrangian.
As an exception, if the dihedral angles at the glued edge of the ideal tetrahedra
are chosen as the moduli of all the ideal tetrahedra around it,
then all the fields appearing in the superpotential appear in the Lagrangian.
This depends on the choice of polarization, and there are also cases where such a choice of polarization
is not possible in the first place.

\subsection{2--3 Mirror Symmetry}\label{subsec.23mirror}

We have identified the theory corresponding to an ideal tetrahedron, and learned about its
changes under changes of the polarization of quantization. We have also understood the gluing of ideal tetrahedra.
Therefore, by repeatedly gluing ideal tetrahedra, we have identified, for complicated 3-manifolds,
the corresponding 3d $\mathcal{N}=2$ supersymmetric field theories (in the sense of gluing from basic theories).\footnote{To be precise, what can be understood in this way using the moduli $z$ of ideal tetrahedra is limited to flat connections with geometric interpretations; for example, the so-called Abelian flat connections, which are conjugate to elements of the Cartan of the gauge group, are not included. On this point, see also Sec.~\ref{subsec.CS_classical}, where we discussed the relation with 3d gravity, and Ref.~\cite{Chung:2014qpa}.}

What, then, is the field theory corresponding to the pentagon?
The pentagon replaces two tetrahedra by three tetrahedra; since
a 3d $\scN=2$ theory corresponds to each of them, the pentagon should mean
the equivalence (duality) of two theories.

The identity of quantum dilogarithm functions arising from the pentagon of ideal tetrahedra (which in two dimensions was also the pentagon of ideal triangulations) is, as explained in Appendix~\ref{app.clusterapp},
the pentagon of quantum dilogarithm functions. This formula is given in \eqref{ebpentagon} in Appendix~\ref{app.dilog};
since it is an important formula, let us write it here again, at the cost of repetition:
for operators $\sfP, \sfQ$ with $[\sfP, \sfQ]=\frac{1}{2\pi i}$,
\begin{align}
	e_b(\sfP) e_b(\sfQ) =e_b(\sfQ) e_b(\sfP+\sfQ) e_b(\sfP) \ .
	\label{ebpentagonbody}
\end{align}
This is an operator identity, but by taking expectation values it can be converted into an identity of c-numbers,
and as a result we obtain the relation \eqref{ZQED_ZXYZ} satisfied by quantum dilogarithm functions (see Appendix~\ref{app.dilog}). Since this represented the 2--3 mirror symmetry,
we have in the end obtained the following conclusion \cite{Dimofte:2011ju}:
\begin{align}
	\begin{split}
		 & \textrm{2--3 mirror symmetry in 3d $\mathcal{N}=2$ theory} \\
		 & \qquad \qquad \longleftrightarrow
		\textrm{pentagon of ideal tetrahedra}   \ .
	\end{split}
\end{align}

In this section we have used the quantum dilogarithm function and the
$S^3_b$ partition function as tools in the identification of 3d $\mathcal{N}=2$ theories;
note, however, that we have already established a direct correspondence between the defining data of the field theories and
the geometric operations on 3-manifolds. Therefore, the correspondence also holds if we compute other physical quantities,
for example the $S^1\times S^2$ partition function \cite{Dimofte:2011py}.

\bparagraph{Semiclassical Limit}

Let us next consider the semiclassical limit $q\to 1, b\to 0$,
which we also considered in Sec.~\ref{subsec.CS_classical}.
The wave function of the ideal tetrahedron
was written, in the semiclassical limit, in terms of the classical dilogarithm function:
\begin{align}
	\psi_{\textrm{ideal tetrahedron}} \sim \exp\left[\frac{1}{2\pi i b^2} \textrm{Li}_2(e^{-Z})
		                                           \right] \ .
\end{align}
Moreover, depending on the choice of polarization of the quantization,
this needs to be supplemented by a quadratic function, which is the generating function of a canonical transformation (recall Sec.~\ref{subsec.wave_function}).
Therefore, the semiclassical limit for 3-manifolds is written
as a sum of classical dilogarithm functions and a quadratic function.
We have given an interpretation of this
in the context of 2d $\mathcal{N}=(2,2)$ theories in Chap.~\ref{chap.S3},
around \eqref{scW_sum}.

Let us here clarify the relation between 3d $\mathcal{N}=2$ theories and 3d Chern-Simons theory.
As explained around Sec.~\ref{sec.6d_revisited},
the classical limit of Chern-Simons theory is the $b\to 0$ limit of 3d $\mathcal{N}=2$ theories on $S^3_b$,
namely the dimensional reduction (Sec.~\ref{sec.3d_dim_red}). Therefore, first,
\begin{align}
	\begin{split}
		 & \textrm{vacua of 3d $\mathcal{N}=2$ theory $\mathcal{T}[M]$ on $S^1\times \bR^2$} \\
		 & \qquad \qquad\longleftrightarrow
		\textrm{flat connections on 3-manifold $M$} \ .
	\end{split}
\end{align}
Next, \textbf{we identify the equations appearing in this limit, namely the vacuum conditions \eqref{2dvac} of the 2d theory, with the gluing conditions \eqref{bulkgluing} of the 3-manifold}. Corresponding to the fact that the vacua of the 2d theory were obtained from the twisted superpotential \eqref{scW_sum} via \eqref{2dvac},
when we consider the classical limit of the wave function of the 3-manifold we obtain a certain potential,
and the gluing conditions of ideal tetrahedra are obtained as its saddle point equations. Such a potential is known in hyperbolic geometry as the \keyword{Neumann-Zagier potential}{Neumann-Zagier potential} \cite{NeumannZagier}:
\begin{align}
	\begin{split}
		 & \textrm{2d twisted superpotential} \\
		 & \qquad \qquad\longleftrightarrow
		\textrm{Neumann-Zagier potential} \ .
	\end{split}
\end{align}


\section{Knots and Braids}\label{subsec.braid}

We have arrived at the foundations of the relation between the theory of 3-manifolds and 3d $\mathcal{N}=2$ theories,
but we are already running out of pages. In the rest of this chapter, therefore,
let us quickly touch upon several (mainly mathematical-physics) topics (see also Ref.~\cite{Dimofte:2014ija} for a review).

\subsection{Knot Complement and AJ Conjecture}

An important example of hyperbolic manifolds is the knot complement
discussed in Sec.~\ref{subsec.knot_complement}.
How can we represent this in the language of ideal tetrahedron decompositions?
Where is the knot?

\begin{figure}[t]
	\centering\includegraphics[scale=0.32]{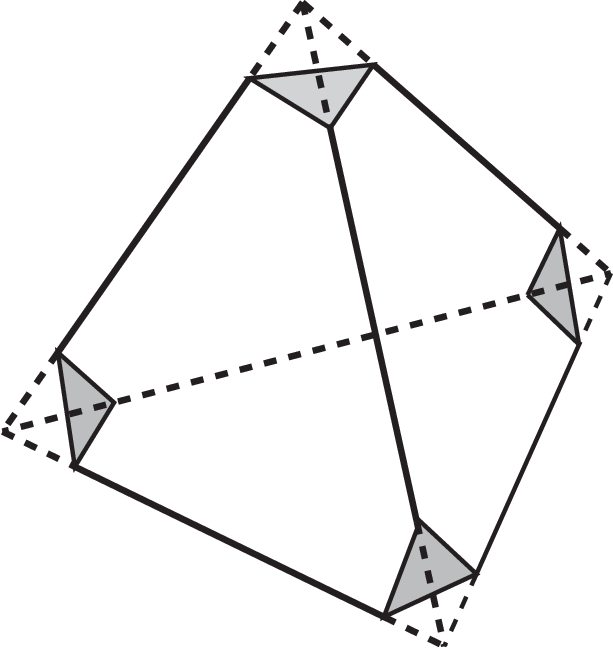}
	\caption{The regions around the vertices of ideal tetrahedra are cut off, and the knot passes through there.
		Collecting the cut surfaces, we obtain the boundary torus of the knot complement.}
	\label{fig.tetrahedron_cut}
\end{figure}

In fact, in the discussion so far we have been vague
about the treatment of ideal tetrahedra around their vertices.
Recall that the vertices of an ideal tetrahedron
are on the boundary $\partial \bH^3$ of $\bH^3$.
In particular, for example, the distance between two vertices (measured with the metric canonically determined by the hyperbolic structure)
diverges as it stands.
Just as we considered horocycles of finite size around punctures in 2d hyperbolic geometry,
we now consider ideal tetrahedra with the regions around their vertices
cut off (Fig.~\ref{fig.tetrahedron_cut}).
These are glued together to form the torus boundary of the knot complement (see Sec.~\ref{subsec.knot_complement})
(for an example of this, see exercise~\ref{ex.fig8}).

The boundary torus has two independent non-trivial cycles, namely
the $\alpha$-cycle and the $\beta$-cycle.
The holonomies of the gauge field in complex Chern-Simons theory
are the meridian $\mathfrak{m}$ and the longitude $\mathfrak{l}$ defined previously in
Chap.~\ref{chap.complexCS}, which can be computed as products of the
moduli of the tetrahedra around the cycles on the boundary torus:
\begin{align}
	\prod_{\rm meridian} z_i=\mathfrak{m} \ , \quad \prod_{\rm longitude} z_i=\mathfrak{l}^2 \ .
	\label{boundarygluing}
\end{align}
When we impose the completeness of the hyperbolic structure,
we further impose $\mathfrak{m}=1$.
We then find that there are only isolated solutions to
\eqref{bulkgluing} and
\eqref{boundarygluing}, as suggested by Mostow rigidity (see Sec.~\ref{subsec.3d_as_classical}).
The parameter $\mathfrak{m}$ gives a one-parameter family of
deformations of the hyperbolic structure \cite{NeumannZagier}.

After quantization, the longitude $\mathfrak{l}$
and the meridian $\mathfrak{m}$ are promoted to
non-commuting operators $\hat{\mathfrak{l}}$ and $\hat{\mathfrak{m}}$.
Similarly, the $A$-polynomial $A(\mathfrak{l}, \mathfrak{m})$ is promoted to a
difference operator
(the \keyword{quantum $A$-polynomial}{quantum A-polynomial}) $\hat{A}(\hat{\mathfrak{l}}, \hat{\mathfrak{m}})$
acting on the wave function $\psi$.
Moreover, this wave function $\psi$ satisfies the following difference equation:\footnote{This is the
	Wheeler-DeWitt equation in 3d gravity.}
\begin{align}
	\hat{A}(\hat{\mathfrak{l}}, \hat{\mathfrak{m}}) \psi=0 \ ,
	\label{AJ_conj}
\end{align}
where $\psi$ is the partition function constructed by gluing ideal tetrahedra,
and can be related to the Jones polynomial,
as discussed in Sec.~\ref{subsec.3d_as_classical}.
This means that \eqref{AJ_conj} can be understood as the statement that the Jones polynomial
satisfies the difference equation determined by the $A$-polynomial
(the \keyword{AJ conjecture}{AJ conjecture}).

In the language of 3d $\mathcal{N}=2$ theories, the wave function $\psi$ is the $S^3_b$ partition function,
and hence $\hat{\mathfrak{l}}, \hat{\mathfrak{m}}$ should be operators acting on it.
In fact, these are the Wilson loop operators encountered in \eqref{Wilson_line_S3},
and their duals, the vortex loop operators.
Namely, \eqref{AJ_conj} can be interpreted as representing the algebraic relations formed by
Wilson loops and vortex loops.

\subsection{Mapping Torus and Braids}\label{subsec.map_torus}

As another class of 3-manifolds,
let us consider the mapping torus $M=(\Sigma\times S^1)_{\varphi}$
defined previously in \eqref{mapping_torus}
(see Refs.~\cite{Terashima:2011xe,Terashima:2013fg,HikamiInoue} for detailed discussions of the 3d-3d correspondence in this case).
In this case, since the definition itself uses a 2d surface, as discussed in Sec.~\ref{sec.3d_as_2d}
we can represent the mapping class $\varphi$ of the 2d surface by flips,
and associate an ideal tetrahedron $\Delta_m$ with each flip $m$.
By repeating this, we obtain a tetrahedron decomposition of the mapping torus:
\begin{align}
	M=\bigcup_{m: \textrm{flip}} \Delta_m \ .
\end{align}
This is known in the literature
as the canonical ideal tetrahedron decomposition \cite{FloydHatcher,Lackenby,Gueritaud}.
When $\varphi$ satisfies the condition called \keyword{pseudo-Anosov}{pseudo-Anosov},\footnote{
	For example, when we take the torus $T^2$ as $\Sigma$,
	$\varphi$ is pseudo-Anosov when it satisfies $|\textrm{Tr}(\varphi)| >2$ as a $PSL(2, \bZ)$ matrix. In this sense, we may think that a generic $\varphi$ is pseudo-Anosov.}
it is known that the resulting mapping torus
admits a hyperbolic structure (a result due to Thurston;
see e.g.\ Ref.~\cite{Otal} for a proof),
and moreover we can choose the tetrahedron decomposition above
as an ideal tetrahedron decomposition compatible with the hyperbolic
structure.

\begin{figure}[t]
	\centering\includegraphics[scale=0.31]{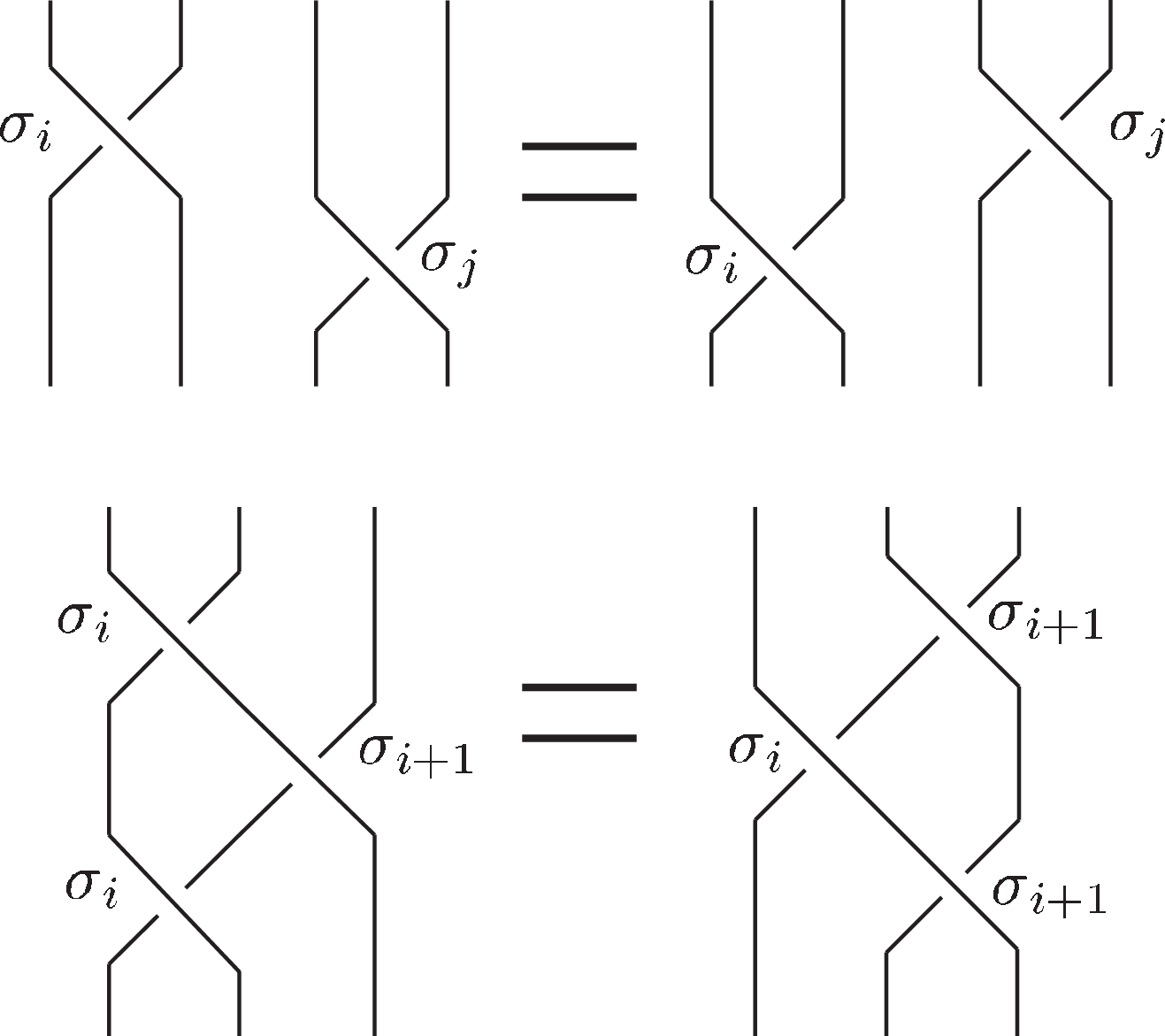}
	\caption{Pictorial representation of the relations of the braid group. In our context, a braid represents the time evolution of the punctures of the $(n+1)$-punctured sphere,
		and $\sigma_i$ represents the Dehn twist around a closed curve surrounding the punctures $i, i+1$.
		When the braid is closed in an appropriate way into a knot, these relations become \keyword{Reidemeister moves}{Reidemeister move} (more precisely, two of the three).}
	\label{fig.braid_relation}
\end{figure}

Let us now consider, in particular, the $(n+1)$-punctured sphere ($n+1\ge 4$) as the Riemann surface. It is known that, in this case, the subset of the mapping class group
fixing one point (out of the $n+1$ points) is
the \keyword{braid group}{braid group} (divided by its center),
which can be written, using the generators $\sigma_i$, as
\begin{align}
	\begin{split}
		 & \sigma_i \sigma_{i+1}\sigma_i=\sigma_{i+1}  \sigma_i \sigma_{i+1}  \ , \\
		 & \sigma_i \sigma_j=\sigma_j \sigma_i \ , \quad |i-j|> 1
	\end{split}
	\label{braid_relation}
\end{align}
(see Fig.~\ref{fig.braid_relation}).

\begin{figure}[t]
	\centering\includegraphics[scale=0.25]{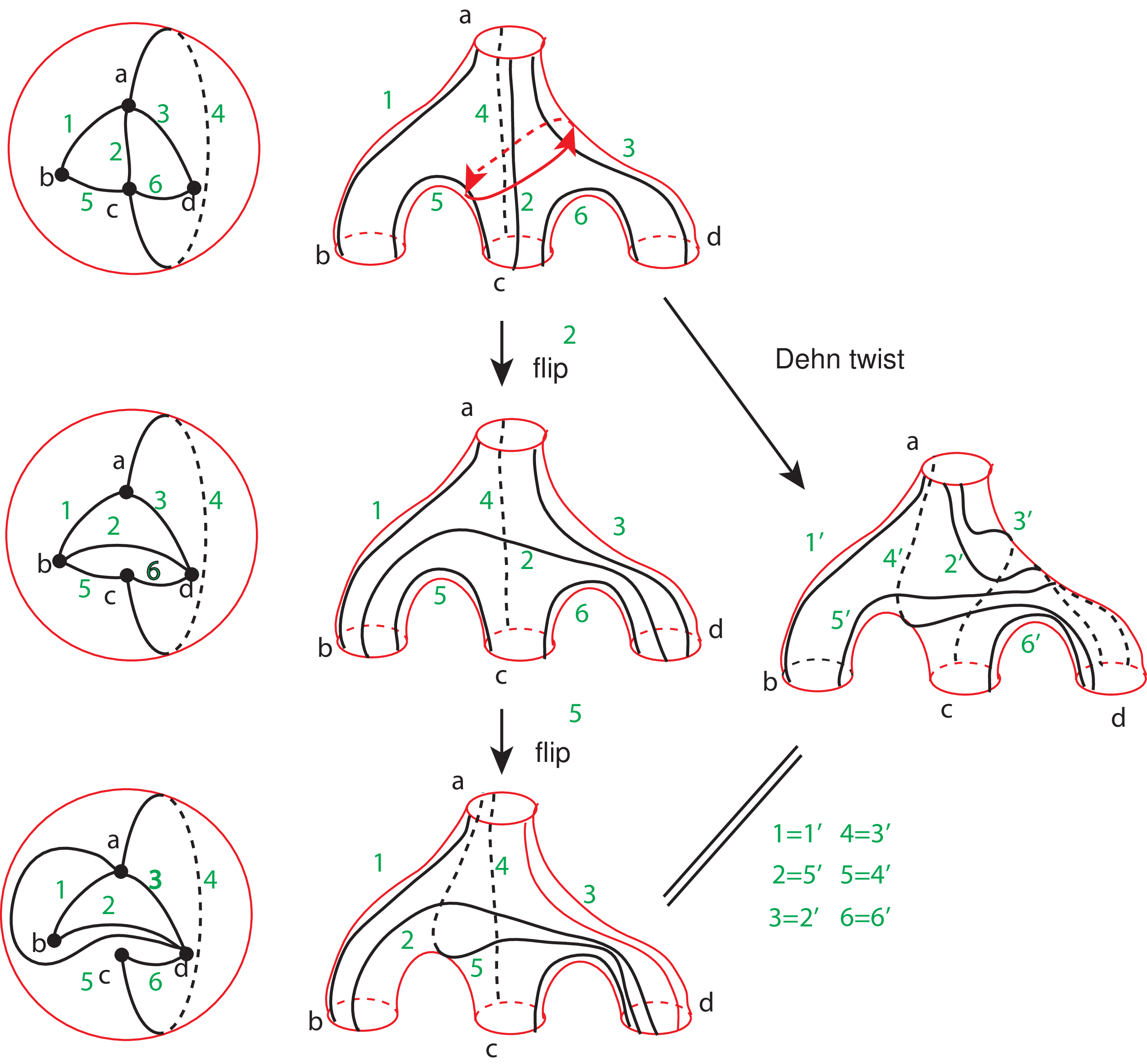}
	\caption{The action of a Dehn twist on an ideal triangulation of the four-punctured sphere. We performed the Dehn twist around the closed curve surrounding two, $c, d$, of the four punctures of the four-punctured sphere. This Dehn twist is
		equivalent to the composition of two flips of the triangulation and a relabeling of the edges; in more general cases, four flips are needed (see Fig.~C4 of Ref.~\cite{Terashima:2013fg}).}
	\label{fig.Dehn_twist}
\end{figure}

Let us label by $1, 2, \ldots, n$ the $n$ vertices other than the fixed one among the $n+1$ vertices. The action exchanging the adjacent vertices $i$ and $i+1$ is then
an element of the mapping class group. Let us denote it by $\sigma_i$. More precisely, this element of the mapping class group is
called the \keyword{Dehn twist}{Dehn twist} around a curve surrounding the two vertices (see the example in Fig.~\ref{fig.Dehn_twist}).

A Dehn twist, which is an element of the mapping class group, maps an ideal
triangulation into another, and can be represented as a sequence of flips.
Correspondingly, an operator $\hat{\mathcal{R}}_i$ in quantum \Teichmuller theory is determined (see Fig.~\ref{fig.Dehn_twist} and Chap.~\ref{chap.Teichmuller}). Since the mapping class group elements $\sigma_i$
satisfy the relations \eqref{braid_relation},
the corresponding operators also
satisfy the same relations:
\begin{align}
	\hat{\scR}_i \hat{\scR}_{i+1} \hat{\scR}_i =
	\hat{\scR}_{i+1} \hat{\scR}_{i} \hat{\scR}_{i+1} \ .
\end{align}
The operators $\hat{\scR}_i$ constructed in this way give an infinite-dimensional (projective) representation of the braid group;
in the limit where the value of $q$ becomes a root of unity,
these operators reduce to operators acting on a finite-dimensional space (i.e.\ matrices),
and they can be shown to coincide with the
$R$-matrix constructed by Kashaev \cite{HikamiInoue_Braid}.
From this Kashaev $R$-matrix one can
construct invariants of knots and links \cite{Kashaev6jsymbol}, which
coincide with the $N$-colored Jones polynomial evaluated at $q=e^{2\pi i/N}$.
These are the quantities appearing in the
volume conjecture \eqref{vol_conj} in Chap.~\ref{chap.complexCS}.

Now, given a representation of an element of the braid group (an element $\varphi$ of the mapping class group of the $(n+1)$-punctured sphere),
by regarding the direction in which $\varphi$ acts as the time direction,
we obtain a picture of $n$ entangled lines.
This is a knot with both ends cut open, and is called a \keyword{braid}{braid}.

By closing both ends appropriately,
we can construct any knot in $S^3$, and moreover,
by performing the corresponding operation on our partition function,
we can also construct knot invariants.
The way to do this is not unique;\footnote{
	One way is to decompose a closed 3-manifold by a
	\keyword{Heegaard decomposition}{Heegaard decomposition} as follows:
	\begin{align}
		M=H_1 \cup_{\varphi} H_2\ , \quad
		\partial H_1=\partial H_2=\Sigma_{g,0} \ ,
		\label{Heegaard}
	\end{align}
	where $H_1$ and $H_2$ are the basic building blocks called \keyword{handlebodies}{handlebody};
	roughly speaking, they are the ``simplest''
	manifolds with the 2d surface $\Sigma_g$ as their boundary. For example, when $g=0$, i.e.\ when $\Sigma$ is
	a sphere, it is the 3d ball $B^3$.
	Moreover, $\Sigma$ is called the \keyword{Heegaard surface}{Heegaard surface}, and $g$ the \keyword{Heegaard genus}{Heegaard genus}.

	$\varphi$ is a map from $\partial H_1$ to $\partial H_2$, and
	by gluing the two handlebodies using this map
	we can obtain a closed 3-manifold (for the topological type of the Heegaard decomposition,
	only the mapping class of $\Sigma_{g,0}$ matters).

	It is known that any closed orientable 3-manifold has a (non-unique)
	Heegaard decomposition.
	As the simplest example, $S^3$ can be divided into the northern hemisphere and
	the southern hemisphere, which are glued along the boundary $S^2$. An example with a Heegaard decomposition of genus $1$ is
	the lens space $S^3/\bZ_p$.

	In our setup we need to consider not closed 3-manifolds but
	knots (Wilson lines), and
	we need to allow handlebodies with braids removed. By extending the notion of Heegaard decompositions
	appropriately in this way, we can also discuss any knot in $S^3$ by this decomposition.}
for example, one way is as follows.
For simplicity, consider the $4$-punctured sphere as an example:
it is sufficient to identify the four faces of the tetrahedron in pairs,
and then the four braids passing through the four points are identified in pairs \cite{SakumaWeeks,FuterGueritaud,Terashima:2013fg}. In this way we can construct knots called
\keyword{2-bridge knots}{2-bridge knot}.

\small

A similar construction can be generalized to the case of $2n$ points.
In this case, we obtain a knot by identifying the $2n$ braids at their ends; this is called the
\keyword{plat representation}{plat representation} of the knot. It is known that any knot in $S^3$ has a plat representation.
Moreover, the plat representation is ambiguous even if we fix the knot, but this can be resolved by
considering moves called \keyword{Birman moves}{Birman move}. This is analogous to the
\keyword{braid representation}{braid representation} of knots and the \keyword{Reidemeister moves}{Reidemeister move} on it.

\normalsize

\subsection{Generalization to $A_N$-Type}\label{subsec.AN}

So far we have discussed the case where the 6d theory is of $A_1$-type (the case of two M5-branes).
What happens, then, more generally for $A_{N}$-type with $N>1$ (the case of $N+1$ M5-branes)? It follows from Sec.~\ref{subsec.twist} that we should consider $SL(N+1, \bC)$ Chern-Simons theory.

\begin{figure}[t]
	\centering\includegraphics[scale=0.35]{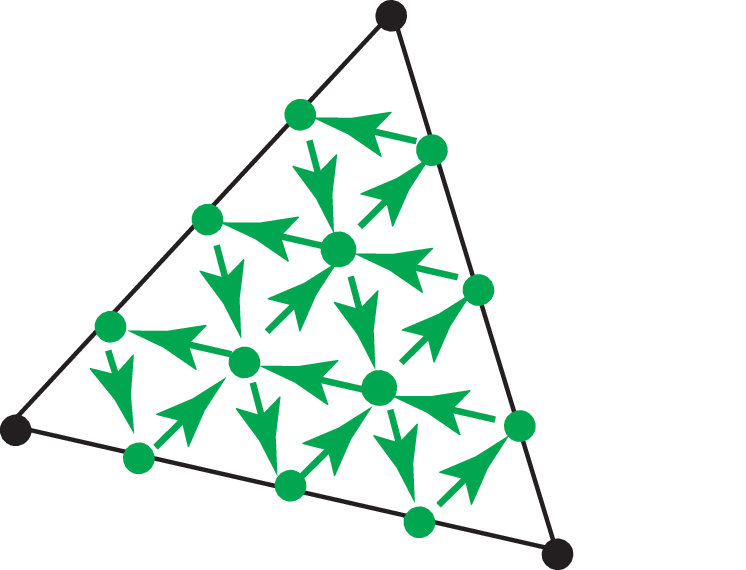}
	\caption{When we consider compactifications of the $A_N$-type $(2,0)$ theory, or $PSL(N+1, \bC)$ flat connections, we consider the quivers obtained by modifying the rule of Fig.~\ref{nee} as in this figure. In this example we took $N=3$. For general $N$, we place $N$ vertices on each edge of the triangle,
		and the total number of vertices is $(N+5)N/2$.}
	\label{fig.higher}
\end{figure}

The counterpart of the (quantum) \Teichmuller theory of Chap.~\ref{chap.Teichmuller} is,
in this case, what is called \keyword{higher \Teichmuller theory}{higher Teichmuller theory} \cite{FockGoncharovHigher}.
Higher \Teichmuller theory has rich contents, but a quick way to understand one aspect of it is to
look at Fig.~\ref{fig.higher}. Namely, we replace the rule of Fig.~\ref{nee} for constructing quivers
by the rule of Fig.~\ref{fig.higher}.
This can also be translated into the language of 3-manifolds and reinterpreted in terms of ideal tetrahedron decompositions \cite{Dimofte:2013iv}.

In the $A_{N}$-type case, there is a discrete freedom in how the holonomy at the knot is specified,
given by the shape of the Jordan blocks of the holonomy (a partition of $N+1$), and taking this into account there is an even richer structure (see e.g.\ Ref.~\cite{Xie:2012dw}).
To discuss these more general situations, the $T[SU(N+1)]$ theory discussed in Sec.~\ref{subsec.TSUN} and
its extensions play important roles.

\subsection{Generalization to Cluster Algebras}

In higher \Teichmuller theory of $A_N$-type,
we modified the rule for determining quivers from ideal triangulations as in Fig.~\ref{fig.higher};
what if, more generally, we consider quivers which do not come from ideal triangulations?

In Sec.~\ref{subsec.BPS_quiver} of the appendix, we interpreted a quiver $\mathsf{Q}$ as the BPS quiver of a 4d $\mathcal{N}=2$ theory;
if we consider 4d $\mathcal{N}=2$ theories which do not come from compactifications of M5-branes,
we can obtain quivers as their BPS quivers.
Even for a general quiver $\mathsf{Q}$, we can define from it the algebra $\mathcal{A}_\mathsf{Q}$ and
the Hilbert space on which it acts
in exactly the same way (see Appendix~\ref{app.clusterapp}).
Mutations of the quiver define operators, and their expectation values, the partition functions,
are again written in terms of quantum dilogarithm functions,
and can be interpreted as $S^3_b$ partition functions of 3d $\mathcal{N}=2$ theories \cite{Terashima:2011xe}.
Namely, a quiver and its mutations determine a 3d $\scN=2$ theory (up to its $Sp(2n, \bZ)$-equivalence class).

This 3d $\mathcal{N}=2$ theory can be interpreted as the theory appearing on a boundary or a domain wall of the
4d $\mathcal{N}=2$ theory we started with. The quiver, one of the defining data, is the BPS quiver of the 4d $\mathcal{N}=2$ theory,
and specifies the gauge symmetries of the 4d theory, namely the charges of the global symmetries of the 3d theory.
Next, each mutation corresponds to a quantum dilogarithm function,
and hence to a 3d $\scN=2$ chiral multiplet. Moreover, the quantum $y$-variables were placed at the
vertices of the quiver; these can be interpreted as the
loop operators (Wilson loops and vortex loops) for the corresponding global symmetries.
The part where changes of the polarization of the quantum-mechanical system determined by the quiver
are interpreted as $Sp(2n, \bZ)$-transformations of the 3d theory is the same as before.


\begin{practice}

	\item $[\bll]$  (Hyperbolic volume of an ideal tetrahedron)\label{ex.tetra_volume}

	By integrating the canonical volume form associated with the hyperbolic metric,
	show that the hyperbolic volume of an ideal tetrahedron
	specified by a modulus $z$ is given by the \keyword{Bloch-Wigner dilogarithm}{Bloch-Wigner dilogarithm}\footnote{The function $\textrm{Li}_2(z)$ has a branch cut on the complex plane, but
		$D(z)$ has no cut.}
	\begin{align}
		D(z):=\textrm{Im}\, \textrm{Li}_2(z)+\arg(1-z) \log|z| \ .
	\end{align} Moreover, examine when the volume becomes maximal as we vary the value of $z$,
	and what the value of the hyperbolic volume is then.
	Show also that the function $D(z)$ satisfies
	\begin{align}
		D(z)= D\left(\frac{1}{1-z}\right)=D\left(\frac{z-1}{z}\right)
		=-D\left(\frac{1}{z}\right)=-D(1-z)=-D\left(\frac{z}{z-1}\right)
	\end{align}
	and
	\begin{align}
		D(x)+D(y)=D\left(\frac{x(1-y)}{1-xy}\right) +D(xy)+D\left( \frac{y(1-x)}{1-xy}\right) \ .
	\end{align}
	What is the geometric meaning of these identities in the language of ideal tetrahedra?

	\item $[\bll]$ (Figure-eight knot complement)\label{ex.fig8}

	One of the most famous examples of decompositions of hyperbolic manifolds into ideal tetrahedra
	is the figure-eight knot in the middle of Fig.~\ref{fig.knots}.\footnote{
		The complement of this knot can also be realized as a mapping torus of the once-punctured torus.
	} The complement of this knot is a famous example
	discussed in Thurston's lecture notes \cite{ThurstonLecture}.

	\begin{enumerate}
		\item The complement of this knot is decomposed into the following two ideal tetrahedra (Fig.~\ref{fig.fig8trig}),
		      and the picture of the boundary torus in this case is
		      given in Fig.~\ref{fig.fig8_torus}.
		      Convince yourself of this. Hint: to help convince yourself of Fig.~\ref{fig.fig8trig},
		      see e.g.\ \cite{FrancisBook}.

		      \begin{figure}[t]
			      \centering\includegraphics[scale=0.33]{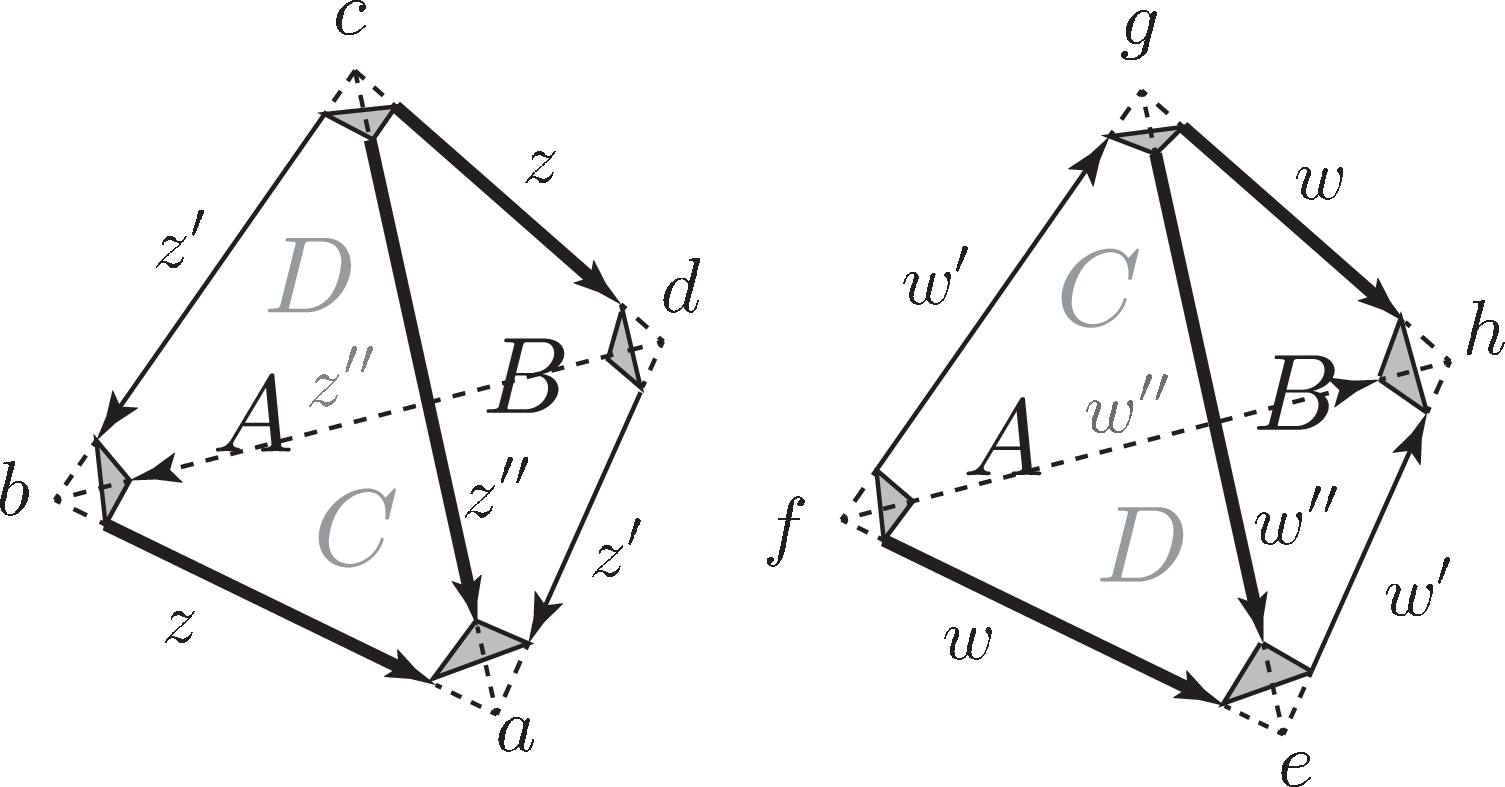}
			      \caption{The complement of the figure-eight knot is decomposed into
				      two ideal tetrahedra. Here faces with the same labels are identified, and so are
				      thin lines with thin lines and thick lines with thick lines.}
			      \label{fig.fig8trig}
		      \end{figure}

		      \begin{figure}[t]
			      \centering\includegraphics[scale=0.34]{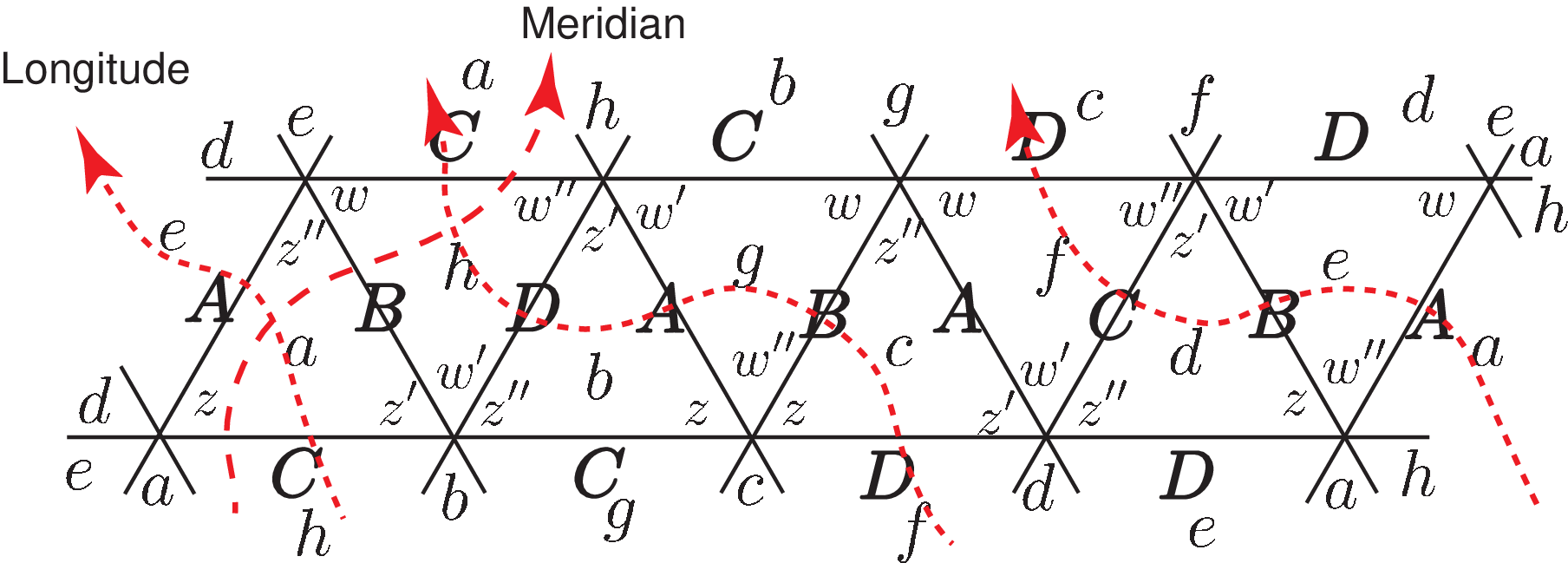}
			      \caption{The triangulation of the boundary torus of the figure-eight knot complement.
				      The longitude and meridian can be computed from products of the moduli.
				      Note that the longitude is not a horizontal straight line in this figure.}
			      \label{fig.fig8_torus}
		      \end{figure}

		\item
		      Show from Fig.~\ref{fig.fig8_torus} that, for the figure-eight knot, the equations determining the hyperbolic structure are
		      \begin{align}
			      z^2 z'' w^2 w''=1 \ , \quad
			      z'^2 z'' w'^2 w''=1 \ ,
		      \end{align}
		      or, after simplification,
		      \begin{align}
			      z(1-z) w(1-w)=1
			      \label{fig8_str_1}
		      \end{align}
		      (the two equations give the same condition).

		\item
		      Compute the meridian and the longitude to be
		      \begin{align}
			      \mathfrak{m}=z'^{-1} w = (1-z) w \ , \quad
			      \mathfrak{l}^2=(z w'' z'^{-1}  w''^{-1})^2 =\left(\frac{z}{z'}\right)^2 \ .
			      \label{fig8_str_2}
		      \end{align}
		      Show that the elimination of $z$ and $w$ from
		      \eqref{fig8_str_1} and \eqref{fig8_str_2} gives\footnote{We choose the sign of
			      $\mathfrak{l}$ to be $\mathfrak{l}=-{z}/{z'}$.}
		      \begin{align}
			      A_{4_1}(\mathfrak{m}, \mathfrak{l})
			      =\mathfrak{l}+\mathfrak{l}^{-1}+(-\mathfrak{m}^{-2}+\mathfrak{m}^{-1}+2+\mathfrak{m}-\mathfrak{m}^{2})=0 \ .
			      \label{A_41}
		      \end{align}
		      Eq.~\eqref{A_41} is known as the $A$-polynomial of the figure-eight knot.
		      Impose the completeness condition $\mathfrak{m}=1$ of the hyperbolic structure to determine the
		      values of the tetrahedron moduli $z$ and $w$.\footnote{That there are no moduli in the solution space is a manifestation of the Mostow rigidity of complete hyperbolic structures.}
		      Use the values of $z, w$ and exercise~\ref{ex.tetra_volume} to obtain the
		      hyperbolic volume of the figure-eight knot complement.
	\end{enumerate}


	\item $[\bll]$  (3d hyperbolic geometry by SnapPea/SnapPy)\label{SnapPea}

	In the area of 3-manifolds
	there has been active research assisted by computer programs,
	and fortunately many tools have been developed for this purpose.
	One such program is \keyword[SnapPy]{SnapPea/SnapPy}{SnapPy}\footnote{\url{http://snappy.computop.org}} (see Fig.~\ref{fig.SnapPea} for a screenshot).

	Let us download SnapPea/SnapPy and play around for a while.
	For example, input the knots in Fig.~\ref{fig.knots} (e.g.\ with a mouse),
	and numerically compute the hyperbolic volume and the Chern-Simons invariant
	of the knot complement; for the figure-eight knot the hyperbolic volume
	should of course match the exact value obtained in exercise~\ref{ex.fig8}.
	SnapPea/SnapPy also computes the fundamental group $\pi_1$ of the knot complement.

	\begin{figure}[t]
		\includegraphics[scale=0.22]{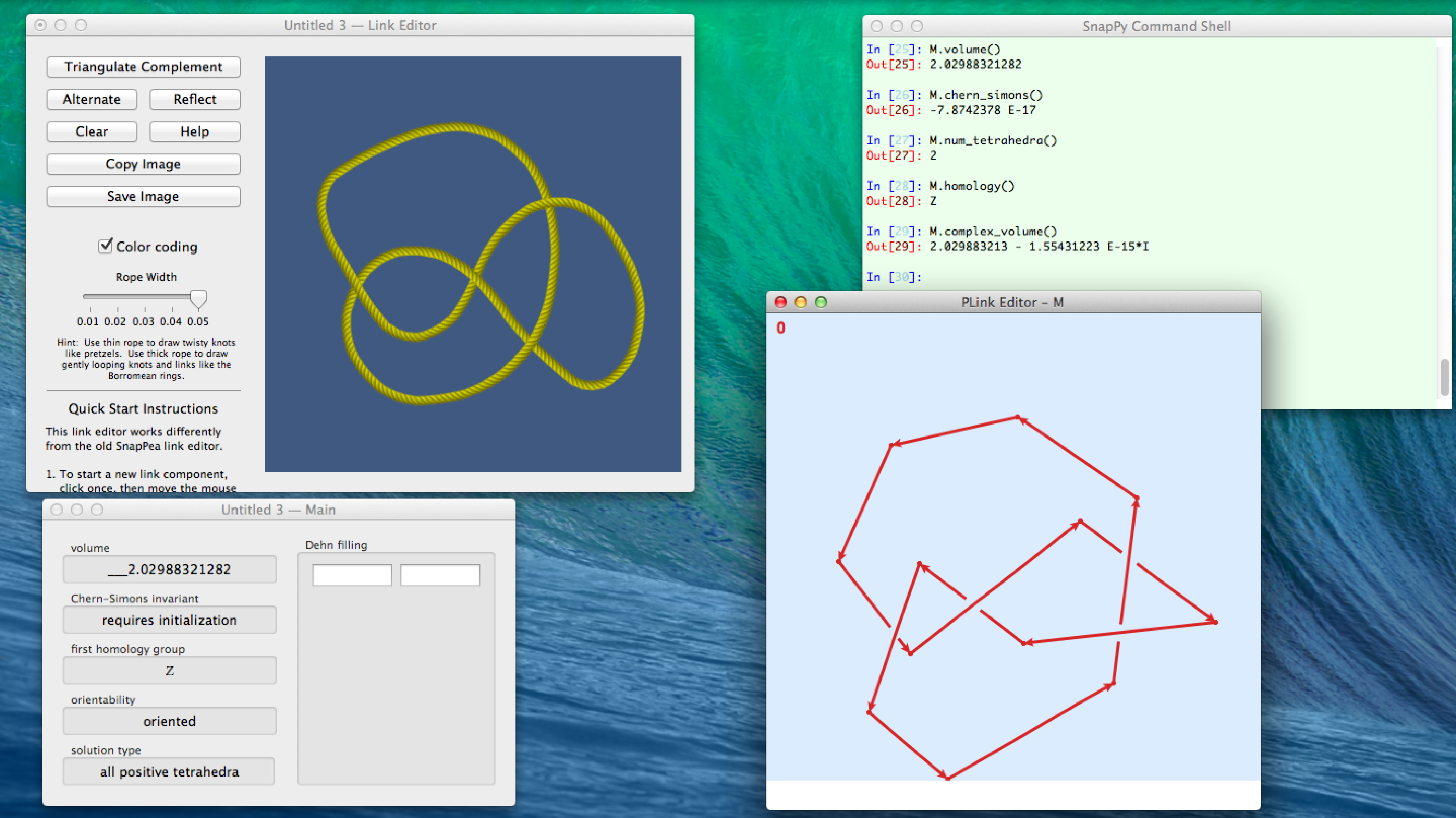}
		\caption{Screenshots of SnapPea (left) and SnapPy (right). SnapPea is currently no longer maintained and was integrated into SnapPy.}
		\label{fig.SnapPea}
	\end{figure}

	\item $[\bll]$ (Limit of 2--3 mirror symmetry)

	In Appendix~\ref{app.dilog} we derived \eqref{ebFourier}
	as a limit of the pentagon identity \eqref{Ramanujan},
	where the latter represents the 2--3 mirror symmetry.
	Can we understand each step of the limiting procedure
	in the language of 3d $\scN=2$ theories?

\end{practice}
 
\chapter{Finale: Beyond 3-Manifolds}\label{chap.conclusion}

This book started with the following question: \textbf{what is quantum field theory}?

Quantum field theory is a huge system with a long history, and
it may be difficult to give a definitive answer to this question.
However, the essence of physics is to understand the complicated world from a very small number of principles,
and history teaches us that, however complicated a theoretical system may be,
its essence is extracted through repeated cycles of scrapping and building.

Of course, at present the author does not have a satisfactory answer to the question above.
In this book, however, the author has tried to face quantum field theory anew
by shedding light on one aspect of field theory (a limited one, but one which the author believes to be essential)---starting
from its conceptual issues, the author has tried to discuss them in a physically and mathematically precise manner
in concrete setups of supersymmetric field theories.

In this book we have emphasized several aspects of field theory.
One is the concept of \keyword{renormalization}{renormalization},
and the recognition that field theories are \keyword[low-energy effective field theories]{low-energy effective theories}{low-energy effective field theory}. One manifestation of the \keyword{universality}{universality} of the renormalization group is the \keyword{duality}{duality} of field theories.
Moreover, by combining and splitting field theories through \keyword{gauging}{gauging},
we have been led to the geometric and algebraic structures existing in the \keyword{theory space}{theory space} of field theories. Namely,
there are structures which appear only when we consider the \textbf{theory space} of field theories.

To explore these structures, we started from the compactification of the 6d $(2,0)$ theory (M5-branes)
on 3-manifolds (Chap.~\ref{chap.6d}).
The resulting
3d $\mathcal{N}=2$ theories (Chap.~\ref{chap.3dN2}) can be
analyzed freely with the tool of the $S^3_b$ partition function (Chap.~\ref{chap.S3}), and
the structure of the quantum-mechanical system appearing in the resulting theory space (Chap.~\ref{chap.wall})
corresponded to the quantum-mechanical structure arising from the quantization of 3-manifolds (Chaps.~\ref{chap.Teichmuller} and \ref{chap.3mfd}).

In the compactification of the 6d theory, since we started from the beginning with
the geometry of the manifold on which we compactify,
it can be said that it is guaranteed (at least in a rough sense) that
the resulting theory space carries a geometric structure.
However, the existence of rich structures on the theory space of field theories
is not limited to direct compactifications of M5-branes.

As such an example, let us consider the supersymmetrization of the quiver gauge theories discussed in Sec.~\ref{subsec.repeat},
to 4d $\mathcal{N}=1$.\footnote{We also need to specify the superpotential, which can be naturally determined from the realization of the quiver on a plane. For details, see e.g.\ Refs.~\cite{Yamazaki:2008bt,Xie:2012dw,Franco:2012mm}.} In this case, since the only data needed to define the field theory is a graph,
it is not at all clear whether there is any structure beyond the simple graph.

Interestingly, however, there exists an integrable structure on the theory space of field theories defined in this way
\cite{Yamazaki:2013nra}.\footnote{
	See also Refs.~\cite{Yamazaki:2012cp,Terashima:2012cx}.
	Realizations of these field theories from branes are known \cite{Heckman:2012jh}.}
The key to seeing the integrable structure is Seiberg duality (see Sec.~\ref{subsec.Seiberg}). Translated into the language of quivers, this duality becomes an operation on quivers;
in fact, exactly the same transformation of graphs had been known in integrable systems (where it is called the star-star relation). Focusing on this fact,
the \keyword{Yang-Baxter equation}{Yang-Baxter equation}, which determines integrability,
is translated into a duality between 4d quiver gauge theories (\keyword{Yang-Baxter duality}{Yang-Baxter duality}).
Once integrability is translated into dualities of field theories,
we can in turn systematically construct new integrable models (related to elliptic quantum groups)
by considering partition functions of the field theories, for example the $S^1\times S^3$ partition function (the \keyword{superconformal index}{superconformal index}) and its extension to $S^1\times S^3/\bZ_r$ \cite{Benini:2011nc}.

Here again, as in the case of 3-manifolds,
the integrable structure does not exist in a single field theory,
but becomes apparent only when we consider the \textbf{theory space} of theories.

At present there are not many examples where such structures have been found in theory spaces of field theories;
however, the author believes that the fact that the universality of the renormalization group is
a general phenomenon suggests that such structures are more general.\footnote{
	Here quantum effects are in general essential (for example, most of the examples discussed in this book are strongly coupled around their IR fixed points). More generally, it may be essential whether we can directly understand
	quantum systems themselves, not limited to field theories.

	In the usual approach of quantum mechanics,
	we first start from a classical system,
	which exists by itself without any problem.
	Next, by ``quantizing'' it,
	we obtain a quantum system. Of course, given a classical system,
	its quantization is not necessarily unique. However, it has been an implicit understanding in field theory that,
	if we start from a quantum system, there correspondingly exists a unique classical limit.
	Moreover, as seen in the discussion of anomalies, it is well known that
	even if a classical system is consistent, inconsistencies can arise in the quantum system.

	However, the situation is more complicated than that:
	as we have emphasized in this book,
	one quantum theory may have several (in general infinitely many)
	classical limits, and
	conversely there also exist quantum theories which have no classical limit (Lagrangian) in the first place.
	The phenomenon of duality can also be regarded as a redundancy which appears essentially
	when we, who cannot directly understand quantum theories,
	try to understand them using classical tools.

	To be sure, note that dualities themselves exist also in classical theories.
	One of the points of this book is that there are also dualities for which quantum effects are
	essential.

	Incidentally, the relation between the classical and the quantum is more subtle in string theory.
	For example, in the example by Strominger \cite{Strominger:1995cz},
	the classical geometry has a singularity, but the singularity is resolved
	only by adding non-perturbative effects,
	and the quantum theory exists without any problem.
} The ``structures'' here may be geometric structures such as manifolds,
integrable structures, or algebraic structures such as cluster algebras \cite{Terashima:2011xe,Benini:2014mia}.

The study of such structures of the theory spaces of field theories is an area which has remained almost untouched,
despite the long history of field theory. Of course,
there are some exceptions; the space of marginal deformations of conformal field theories and
its metric (the Zamolodchikov metric) are examples of this kind.
However, in order to fully understand the structure of combining and decomposing theories by gauging,
a framework beyond traditional field theory may be needed.

The author thinks that one clue for this lies in string theory.
String theory is a theory containing both field theories and gravity;
by taking appropriate limits, we can decouple the gravitational degrees of freedom and
extract pure field theories. In this process,
the information of field theories is replaced by the information of geometry (for example, the geometry of Calabi-Yau manifolds).
The correspondence between 3-manifolds and 3d $\mathcal{N}=2$ theories treated in this book is
one variation of this.

In this way, string theory is useful as a tool for systematically constructing field theories by changing the choice of geometry,
and this idea has been applied in various contexts since the 1990s. However,
if we take the ideas of string theory more seriously,
we may also regard them as \textbf{suggesting the very structure of the theory space of field theories, which is usually neglected}.
Indeed, in string theory field theories are given as effective theories on branes. However,
branes are themselves dynamical degrees of freedom of string theory; for example,
the number of branes changes according to the dynamics,
and in string theory different field theories seem to appear as different states of a single theory.

Admittedly, these ideas are still vague, and it is not clear whether they will actually help in understanding the theory space of field theories.
What is hidden in the theory space of field theories?
What does string theory teach us about the theory space of field theories? And
how will it rewrite our understanding of field theory?
The story spun in this book is only
a tiny step toward this.
The author sincerely hopes that the readers of this book will become interested in these questions, and one day
bring surprises to the world.
 
%
\appendix
\chapter{Roadmap of the Supersymmetric World}\label{chap.SUSYbasic}

\begin{abstract}
	In this appendix, we will provide an overview of supersymmetries in various dimensions discussed in the main text.
\end{abstract}

In this book, we will encounter quantum field theories
in various dimensions (from 1d to 6d) and with
various amounts of supersymmetry.
That many different supersymmetric field theories appear
makes the story exciting,
but this might also scare away beginners interested in the subject.
To alleviate this problem,
we include this brief summary
of supersymmetric field theories discussed in this book.
This appendix refers to only those theories directly relevant for this book (Table~\ref{fig.SUSYbasic});
see e.g.\ Ref.~\cite[Appendix B]{Polchinski:1998rr} for a more complete discussion.
In the following we will concentrate on the on-shell matter content
and do not mention auxiliary fields in the supersymmetric multiplets
needed for off-shell supersymmetry.
Also, $\phi$ always denotes a real scalar, and
$\lambda$ a fermion with the minimal number of components allowed in the spacetime dimension.

\begin{table}[t]
	\caption{Supersymmetric quantum field theories discussed in this book.
		We only list those theories directly relevant for this book.}
	\centering
	\begin{tabular}{c|c|c|c}
		\backslashbox{spacetime\\dimension}{number of\\supersymmetries} & 16           & 8        & 4            \\
		\hline
		6                                                               & $\scN=(2,0)$ & -        & -            \\
		\hline
		5                                                               & $\scN=2$     & -        & -            \\
		\hline
		4                                                               & $\scN=4$     & $\scN=2$ & $\scN=1$     \\
		\hline
		3                                                               & -            & $\scN=4$ & $\scN=2$     \\
		\hline
		2                                                               & -            & -        & $\scN=(2,2)$ \\
		\hline
		1                                                               & -            & -        & $\scN=4$     \\
	\end{tabular}
	\label{fig.SUSYbasic}
\end{table}

\bigskip
\paragraph{6d}

In 6d, we have a Weyl fermion,
with $2^{\left[\frac{6}{2}\right]}/2\times 2=8$ real components (we divided by $2$
for the Weyl condition, and multiplied by $2$ for counting complex components
in terms of real components).

In this book, we consider the 6d $(2,0)$ theory of $A_N$ type; for the most part we discuss the $A_1$ type.
The $(2,0)$ tensor multiplet is given by $(B_{\mu\nu}, \phi_{1, \cdots, 5}, \lambda_{1, 2})$,
containing an anti-symmetric self-dual 2-form satisfying $dB=*dB$
and five real scalars. The R-symmetry is $Sp(4)_R\simeq SO(5)_R$,
and the five scalars $\phi_{1, \cdots, 5}$ transform in the $\bm{5}$ representation of $SO(5)_R$.

The $(2,0)$ tensor multiplet decomposes into
a $(1,0)$ tensor multiplet and a $(1,0)$ hypermultiplet.
A $(1,0)$ tensor multiplet is given by $(B_{\mu\nu}, \phi_1, \lambda_1)$.
A $(1,0)$ hypermultiplet $(\phi_{2, \cdots, 5}, \lambda_2)$ contains two
complex scalars $\phi_2+i\phi_3, \phi_4+i\phi_5$.
The R-symmetry of the $(1,0)$ theory is $Sp(2)_R\simeq SO(3)_R$.


\bigskip
\paragraph{5d}

In 5d there are neither Weyl nor Majorana fermions.
The Dirac fermion has $2^{\left[\frac{5}{2}\right]}\times 2=8$ real components.

In this book we discuss 5d $\scN=2$ theories (Chap.~\ref{chap.6d}),
obtained by dimensional reduction of the 6d $(2,0)$ theory.
As we discussed in Chap.~\ref{chap.6d},
the self-dual 2-form $B_{\mu\nu}$ reduces to a
gauge field $A_{\mu}$ after dimensional reduction.
This implies that the $\mathcal{N}=2$ vector multiplet, which is obtained from the dimensional
reduction of the 6d $(2,0)$ tensor multiplet,
is given by $(A_{\mu}, \phi_{1, \cdots, 5}, \lambda_{1,2})$.
The R-symmetry of this theory is the same as that of the
6d $(2,0)$ theory, namely $Sp(4)_R\simeq SO(5)_R$.

The decomposition of this multiplet under
$\mathcal{N}=1$ supersymmetry is similar to the case of 6d:
the $\mathcal{N}=2$ vector multiplet decomposes into an
$\mathcal{N}=1$ vector multiplet
$(A_{\mu}, \phi_1, \lambda_1)$
and an $\mathcal{N}=1$ hypermultiplet $(\phi_{2, \cdots, 5}, \lambda_2)$.

\bigskip
\paragraph{4d}

In 4d there exist both Weyl and Majorana fermions,
and both have $2^{\left[\frac{4}{2}\right]}\times 2 /2 =4$ real components.

In this book, we encountered 4d $\mathcal{N}=4$ theories in Chaps.~\ref{chap.intro} and \ref{chap.wall}.
The 4d $\mathcal{N}=4$ vector multiplet consists of $(A_{\mu}, \phi_{1, \cdots, 6}, \lambda_{1, \cdots, 4})$,
as expected from the dimensional reduction of a 5d $\mathcal{N}=2$ vector multiplet.
Note that a 5d fermion decomposes into
two 4d Majorana fermions, and hence now we have
$4$ Majorana fermions. The R-symmetry is given by $SU(4)\simeq SO(6)$,
and the $6$ scalars transform in the $\bm{6}$ representation of the $SO(6)$ R-symmetry.

As in 5d and 6d, the 4d $\mathcal{N}=4$ multiplet decomposes under 4d $\mathcal{N}=2$ supersymmetry
into a 4d $\mathcal{N}=2$ vector multiplet $(A_{\mu}, \phi_{1,2}, \lambda_{1,2})$ and an $\mathcal{N}=2$
hypermultiplet $(\phi_{3, \cdots, 6}, \lambda_{3,4})$.

In this book, we consider a general 4d $\mathcal{N}=2$ theory
$\scT[\Sigma]$ associated with the compactification of the
6d $(2,0)$ theory on a 2d Riemann surface.
These theories have $U(1)\times SU(2)\simeq SO(2)\times SO(3)$
R-symmetry.

If we lower the supersymmetry further, we arrive at $\mathcal{N}=1$ supersymmetry.
This is the most typical supersymmetry in textbooks,
and in this book it is mentioned only briefly in the discussion of the Seiberg duality.
An $\mathcal{N}=1$ vector multiplet contains $(A_{\mu}, \lambda)$,
and an $\mathcal{N}=1$ chiral multiplet contains $(\phi_{1,2}, \lambda)$.
When we decompose $\mathcal{N}=2$ multiplets into
$\mathcal{N}=1$ multiplets,
an $\mathcal{N}=2$ vector multiplet decomposes into an
$\mathcal{N}=1$ vector multiplet and an adjoint-valued $\mathcal{N}=1$
chiral multiplet. An $\mathcal{N}=2$ hypermultiplet decomposes into
two $\mathcal{N}=1$ chiral multiplets
transforming in the conjugate representations.

\bigskip
\paragraph{3d}

In three dimensions we have Majorana fermions, which have
$2^{\left[\frac{3}{2}\right]}\times 2 /2 =2$ real components.
This means that a 4d Majorana fermion, when dimensionally reduced to 3d,
decomposes into two Majorana fermions.
It then follows that the dimensional reduction of the
4d $\mathcal{N}=1$ theory is the 3d $\mathcal{N}=2$ theory (see Chap.~\ref{chap.3dN2}).

The multiplet structure can again be obtained by dimensional reduction:
3d $\mathcal{N}=4$ vector multiplet contains $(A_{\mu}, \phi_{1,2,3}, \lambda_{1, \cdots, 4})$,
3d $\mathcal{N}=4$ hypermultiplet $(\phi_{1, \cdots, 4}, \lambda_{1, \cdots, 4})$,
3d $\mathcal{N}=2$ vector multiplet $(A_{\mu}, \phi, \lambda_{1, 2})$,
and 3d $\mathcal{N}=2$ chiral multiplet $(\phi_{1,2},  \lambda_{1, 2})$.
The vector multiplet scalar $\phi$ inside the $\mathcal{N}=2$ vector multiplet
is denoted by $\sigma$ in the rest of this book.
The 3d $\mathcal{N}=2, 4$ theories have $SO(2), SO(4)$ R-symmetry, respectively.

The decomposition pattern is the same as in four dimensions.
Namely, an $\mathcal{N}=4$ vector multiplet
decomposes into an $\mathcal{N}=2$ vector multiplet
as well as an adjoint-valued $\mathcal{N}=2$ chiral multiplet.
An $\mathcal{N}=4$ hypermultiplet decomposes into two
$\mathcal{N}=2$ chiral multiplets transforming in conjugate representations.

The 3d $\scN=2$ theories are the major players in this book.
They appear as a theory $\scT[M]$ associated with the compactification on a 3-manifold.
The $S^3_b$ partition function of Chap.~\ref{chap.S3}
is a tool applicable to general 3d $\mathcal{N}=2$ theories.

\bigskip
\paragraph{2d}

In 2d we have a Majorana-Weyl fermion (with both Majorana and Weyl conditions imposed simultaneously),
and it has $2^{\left[\frac{2}{2}\right]}\times 2 /2/2 =1$ real component.
Thus right-movers and left-movers can have different numbers of fermions,
and for example
2d $\scN=(2,2)$ supersymmetry means that the right-movers and the left-movers
have two supercharges each, four in total.
In the following we denote the right-handed (left-handed) fermions by $\lambda$ ($\overline{\lambda}$).

In this book, 2d $\scN=(2,2)$ theory appears in the dimensional reduction of
3d $\scN=2$ theories. This dimensional reduction can be understood as the classical limit of the
complex Chern-Simons theory.

The 2d $\mathcal{N}=(2,2)$ vector multiplet contains the fields $(A_{\mu}, \phi_{1,2}, \lambda_{1,2}, \overline{\lambda}_{1,2})$,
while an $\mathcal{N}=(2,2)$ chiral multiplet contains $(\phi_{1,2}, \lambda_{1,2}, \overline{\lambda}_{1,2})$.

\chapter{Classical and Quantum Dilogarithm}\label{app.dilog}

\begin{abstract}
	In this appendix, we briefly summarize definitions and properties of the classical and quantum dilogarithm functions.
	As discussed in the main text,
	many of the properties of these special dilogarithm functions
	have direct gauge-theoretic/geometric interpretations,
	and hence can be derived from purely physical considerations.
\end{abstract}

\section{Classical Dilogarithm Function}

\keyword[classical dilogarithm function]{The classical dilogarithm function}{classical dilogarithm function} can mean several different functions.
First, we have the
\keyword{Euler dilogarithm function}{Euler dilogarithm function} $\textrm{Li}_2(x)$\footnote{
	This function is built into many mathematical software packages, e.g.\
	\texttt{PolyLog[2,x]} in Mathematica.}
\begin{align}
	\Li(x):=\sum_{n=1}^{\infty}\frac{x^n}{n^2}
	=-\int_0^x\! dt \,\, \frac{\log(1-t)}{t} \ .
	\label{LEuler}
\end{align}
\nomenclature{$\textrm{Li}_2(x)$}{classical dilogarithm function (Euler dilogarithm)}
The sum in \eqref{LEuler} is convergent only when $|x|<1$;
however, we can extend the definition of $\Li(x)$ to the whole complex
plane using the integral expression of \eqref{LEuler}.
Note that $\Li(x)$ so defined has a cut along the real axis $x>1$.

Another classical dilogarithm function is the \keyword{Rogers dilogarithm function}{Rogers dilogarithm function} $L(x)$,
which is defined by
\begin{align}
	\begin{split}
		L(x) & := \Li(x)+\frac{1}{2} \log x \log (1-x)
		\\
		     & =-\frac{1}{2}\int_0^x \! dt\,
		\left(
		\frac{\log(1-t)}{t}
		+ \frac{\log t}{1-t}
		\right) \ .
	\end{split}
	\label{LRogers}
\end{align}
\nomenclature{$L(x)$}{classical dilogarithm function (Rogers dilogarithm)}
Note that this function does not have a cut in the complex plane,
which fact is useful for remembering the definition of the logarithmic term in \eqref{LRogers}.

The classical dilogarithm functions satisfy dozens of highly non-trivial formulas~\cite{Kirillov:1994en}.
The most important among them
is the following \keyword{pentagon identity}{pentagon identity}, originally due to Abel:
\begin{align}
	L(x)+L(y)=L\left(
	\frac{x(1-y)}{1-xy}
	\right)
	+
	L\left(
	xy
	\right)
	+
	L\left(
	\frac{y(1-x)}{1-xy}
	\right) \ .
	\label{Lpentagon}
\end{align}
This formula, when expressed in terms of $\Li$, reads
\begin{align}
	\begin{split}
		\Li(x) & +\Li(y)=\Li\left(
		\frac{x(1-y)}{1-xy}
		\right)
		+
		\Li\left(
		xy
		\right)                    \\
		       &
		+
		\Li\left(
		\frac{y(1-x)}{1-xy}
		\right)
		+
		\log\left(
		\frac{1-x}{1-xy}
		\right)
		\log\left(
		\frac{1-y}{1-xy}
		\right) \ .
		\label{L2pentagon}
	\end{split}
\end{align}

We also have another formula:
\begin{align}
	L(x)+L(1-x)=L(1)=\frac{\pi^2}{6} \ ,
	\label{Ldilong}
\end{align}
or rewritten in terms of $\Li$:
\begin{align}
	\Li(x)+\Li(1-x)+\log x \log (1-x)=\frac{\pi^2}{6} \ .
\end{align}
\small
Conversely, we can prove that a one-variable function satisfying two relations \eqref{Lpentagon}, \eqref{Ldilong} and differentiable three times or more
is only $L(x)$. To show this, we assume \eqref{Lpentagon},
take its derivative with respect to $x$ and $y$, and solve the
resulting differential equation (see Ref.~\cite[section 4]{RogersOld}).
\normalsize

\section{Quantum Dilogarithm Function}

Let us next consider the \keyword{quantum dilogarithm function}{quantum dilogarithm function}. As the name suggests, this is a one-parameter
deformation of the classical dilogarithm function.

Somewhat confusingly,
the name quantum dilogarithm refers to
several different functions in the literature,
which are respectively denoted by $(x;q)_{\infty}, s_b(x), e_b(x)$
in the conventions of this book.\footnote{
	In Mathematica, $(x;q)_{\infty}$ is defined as \texttt{qPochhammer[x,q]}.
	The functions $e_b(x), s_b(x)$ are not defined as built-in functions, but can be
	built out of $(x;q)_{\infty}$ (we need to be careful with convergence, however; see below).
	See Ref.~\cite{IpPlot} for a visualization of the quantum dilogarithm function.
}
In this book, we mainly use $s_b(x), e_b(x)$.
In the literature, $(x;q)_{\infty}$ is sometimes called the \keyword{compact quantum dilogarithm}{compact quantum dilogarithm},
	while
$e_b(x), s_b(x)$ are called the non-compact quantum dilogarithm, or the
modular quantum dilogarithm.
The latter two functions are also called the
\keyword{double sine function}{double sine function}, hyperbolic gamma function, quantum exponential function, etc., and
are denoted by several different symbols.
The notations in this book are one of the most standard in the
literature on Liouville theory and supersymmetric gauge theories.

After a pioneering work by Shintani \cite{Shintani},
the quantum dilogarithm function was defined by
Kurokawa \cite{Kurokawa_multiple} in the early nineties.
The definition here is due to Faddeev and his collaborators,
who also discovered quantum pentagon identities \cite{FaddeevVolkovAbelian,FaddeevKashaevQuantum,Faddeev95}.
See e.g.\ Refs.~\cite{VolkovNoncommutative,Ponsot:2000mt,SpiridonovEssays}, Ref.~\cite[section
	III]{Ruijsenaars} and Ref.~\cite[Appendix]{Kharchev:2001rs} for properties of quantum dilogarithm.

\subsection{$(x;q)_{\infty}$}

\paragraph{Definition}

The function $(x;q)_{\infty}$ is the $n\to\infty$ limit of
the \keyword{$q$-Pochhammer symbol}{q-Pochhammer symbol}
\begin{align}(x;q)_n:=\prod_{k=0}^{n-1} (1-x q^k) \ ,
\end{align}
namely
\begin{align}
	(x;q)_{\infty}:=\prod_{n=0}^{\infty}
	(1-x q^n) \ .
\end{align}
\nomenclature{$(x;q)_{n}, (x;q)_{\infty}$}{$q$-Pochhammer symbol}
Here $q$ is taken to be $|q|<1$, and $x$ is an arbitrary complex number.
\small
This function is also called the $q$-shifted factorial.
This name originates from the fact that
when we define the $q$-integer as
\begin{align}
	[n]_q:=\frac{1-q^n}{1-q}
\end{align}
($\lim_{q\to 1}[n]_q\to n$ as $q\to 1$),
we have $[n]_q!\, (1-q)^n=(q;q)_n$.
\normalsize

\bigskip
\paragraph{Functional Relation}

This function satisfies a functional relation
\begin{align}
	(x;q)_{\infty}=(1-x) \, (qx;q)_{\infty} \ , \quad
	                                            (0;q)_{\infty}=1 \ ,
	\label{PochRecursion}
\end{align}
from which we can show the series ($q$-exponential) expansions
\begin{align}
	\begin{split}
		(x;q)_{\infty}           & =\sum_{n=0}^{\infty} \frac{(-1)^n
			                                                q^{n(n-1)/2}}{(q;q)_n}x^n \ , \quad
		\\
		\frac{1}{(x;q)_{\infty}} & =\sum_{n=0}^{\infty} \frac{1}{(q;q)_n}x^n \ .
	\end{split}
\end{align}

\bigskip
\paragraph{Classical Limit}

The limit $q\to 1^-$ ($q$ approaching $1$ from inside $|q|<1$)
is important for the purpose of this book: (exercise~\ref{ex.PochAsymp})
\begin{align}
	(x;q)_{\infty}
	\xrightarrow[]{q\to 1^{-}} \exp\left(
	\frac{1}{\log q}
	\sum_{n=0}^{\infty} \frac{B_n (\log q)^n}{n!} \mathrm{Li}_{2-n}(x)
	\right) \ .
	\label{PochAsymp}
\end{align}
Here $B_n$ is the $n$-th order \keyword{Bernoulli(-Seki) number}{Bernoulli number}
\begin{align}
	\begin{split}
		 & B_0=1 \ , \quad B_1=-\frac{1}{2} \ , \quad B_2=\frac{1}{6}\ ,
		\quad B_4=-\frac{1}{30} \ , \cdots                               \\
		 & B_{2k+1}=0 \quad (k\ge 1) \ .
	\end{split}
\end{align}
The $k$-th \keyword{polylogarithm}{polylogarithm} $\mathrm{Li}_k(x)$ is defined to be
\begin{align}
	\mathrm{Li}_k(x):=\sum_{n=1}^{\infty} \frac{x^n}{n^k} \ .
	\label{polylog}
\end{align}
This reduces to the previously-defined
Euler classical dilogarithm \eqref{LEuler} when we have $k=2$.
In particular, the leading contribution to the expansion \eqref{PochAsymp}
is given by the classical dilogarithm function,
and $(x;q)_{\infty}$ can be regarded as its one-parameter deformation.

\bigskip
\paragraph{Quantum Pentagon Identity}

One of the most important properties of the
$q$-Pochhammer symbol is the
(quantum) pentagon identity: for operators $\sfX, \sfY$ satisfying $\sfX \sfY=q \sfY \sfX$,
the identity
\begin{align}
	(\sfY;q)_{\infty} (\sfX;q)_{\infty}=(\sfX;q)_{\infty}(-\sfY
	                                    \sfX;q)_{\infty} (\sfY;q)_{\infty}
	\label{PochPentagon}
\end{align}
holds.

\small

Let us comment on the proof of \eqref{PochPentagon}.
First,  using the commutation relations of $\sfX, \sfY$
we can easily show the following relation (e.g.\ by induction):
\begin{align}
	(\sfX+\sfY)^n=\sum_{k=0}^n \qbinom{n}{k}_q
	\sfY^k \sfX^{n-k} \ , \quad \qbinom{n}{k}_q:=\frac{[n]_q!}{[k]_q!\, [n-k]_q!}
	\label{qbinom}
\end{align}
This is a $q$-deformation of the familiar expansion in terms of binomial coefficients.
From this equation \eqref{qbinom} we first obtain
\begin{align}
	(\sfX;q)_{\infty} (\sfY;q)_{\infty}=(\sfX+\sfY;q)_{\infty}
	\label{tmpPoch1}
\end{align}
Next,
\begin{align}
	\begin{split}
		 & (\sfY;q)_{\infty} (\sfX;q)_{\infty}
		   (\sfY;q)_{\infty}^{-1}
		=\left( (\sfY;q)_{\infty} \sfX (\sfY;q)_{\infty}^{-1}
		; q \right)_{\infty}                                               \\
		 & \quad=\left( \sfX (q^{-1}\sfY;q)_{\infty}(\sfY;q)_{\infty}^{-1}
		; q \right)_{\infty}
		=\left( \sfX (1-q^{-1}\sfY) ; q \right)_{\infty}                   \\
		 & \quad =\left( \sfX -\sfY \sfX ; q \right)_{\infty} \ .
		\label{tmpPoch2}
	\end{split}
\end{align}
Finally, by using \eqref{tmpPoch2} and then \eqref{tmpPoch1} we show
\begin{align}
	\begin{split}
		(\sfY;q)_{\infty} (\sfX;q)_{\infty}
		 & =
		(\sfX-\sfY \sfX;q)_{\infty} (\sfY;q)_{\infty}                       \\
		 & =(\sfX;q)_{\infty} (-\sfY \sfX;q)_{\infty} (\sfY;q)_{\infty} \ ,
	\end{split}
\end{align}
where in the last line we used $\sfX (\sfY \sfX)=q (\sfY \sfX) \sfX$.

\normalsize

We can show that the semiclassical limit  of the
pentagon identity \eqref{PochPentagon} for the
$q$-Pochhammer symbol gives the pentagon identity for the classical dilogarithm \eqref{Lpentagon}
(exercise~\ref{KNex}). This means that $(x;q)_{\infty}$
is a good ``quantization'' keeping fundamental properties of the
classical dilogarithm function.
\subsection{$e_b(x), s_b(x)$}
\paragraph{Definition}

Let us next define functions $e_b(x)$ and $s_b(x)$.
These are built out of $(x;q)_{\infty}$.

First, given a complex parameter $q$, let us define related complex parameters $b, Q$ and
$\overline{q}$ by
\begin{align}
	q=e^{i\pi b^2} \ , \quad \overline{q}:=e^{-i\pi b^{-2}}\ , \quad
	Q:=b+b^{-1}
\end{align}
Note that the classical limit $q\to 1$ is then given by $b\to 0$,
and that $Q$ is preserved under the exchange of $b$ and $b^{-1}$.
The quantum dilogarithm function $e_b(x)$ is defined as a ratio of
two $q$-Pochhammer symbols:
\begin{align}
	e_b(x):=
	\frac{(e^{2\pi \left(x+\frac{iQ}{2}\right)b}; q^2)_{\infty}}{(e^{2\pi \left(x-\frac{iQ}{2}\right)b^{-1}
			}; \overline{q}^2)_{\infty}}
	=
	\frac{(-q e^{2\pi b x} ; q^2)_{\infty}}{(-\overline{q} e^{2\pi b^{-1} x}; \overline{q}^2)_{\infty}}  \ .
	\label{ebratio}
\end{align}
\nomenclature{$e_b(x)$}{quantum dilogarithm function}
Here, the convergence requires us to choose $|q|, |\overline{q}|<1$,
which is satisfied e.g.\ when $b^2\in i \bR_{>0}$.
Another function $s_b(x)$ is the same as this $e_b(x)$,
up to an exponential of a quadratic expression:
\begin{align}
	\begin{split}
		s_b(x): & =e^{-\frac{i\pi x^2}{2}} e^{\frac{i\pi}{24}(2-Q^2)}e_b(x)                                                                                                  \\
		        & =e^{-\frac{i\pi x^2}{2}} e^{\frac{i\pi}{24}(2-Q^2)} \frac{(e^{2\pi \left(x+\frac{iQ}{2}\right)b}; q^2)_{\infty}}{(e^{2\pi \left(x-\frac{iQ}{2}\right)b^{-1}
				                                                              }; \overline{q}^2)_{\infty}} \\
		        & =e^{-\frac{i\pi x^2}{2}} e^{\frac{i\pi}{24}(2-Q^2)}
		\frac{(-q e^{2\pi b x} ; q^2)_{\infty}}{(-\overline{q} e^{2\pi b^{-1} x}; \overline{q}^2)_{\infty}}  \ .
		\label{sbeb}
	\end{split}
\end{align}
\nomenclature{$s_b(x)$}{quantum dilogarithm function}

In the following we use both $e_b(x)$ and $s_b(x)$, whichever suits best for the
consideration.
For example, $s_b(z)$ is most natural for the consideration of the $S^3$
partition function, while the connection with the classical dilogarithm and the pentagon
identity is simpler in terms of $e_b(z)$.

\small
As a side remark, the combination $-\frac{i\pi x^2}{2}+
	\frac{i\pi}{24}(2-Q^2)$ appearing in the definition of
$s_b(x)$ is given by the second Bernoulli polynomial $B_{2,2}(x;\omega_1, \omega_2)$:
\begin{align}
	 & B_{2,2}(x;\omega_1, \omega_2)
	:=\frac{x^2}{\omega_1 \omega_2}-\frac{x}{\omega_1}
	-\frac{x}{\omega_2}+
	\frac{\omega_1}{6\omega_2}
	+\frac{\omega_2}{6\omega_1}
	+\frac{1}{2}  \ ,                \\
	 &
	\frac{i\pi }{2}B_{2,2}\left(ix+\frac{Q}{2};b,b^{-1}\right)=-\frac{i\pi}{2}x^2+\frac{i\pi (2-Q^2)}{24} \ .
	\label{B_22}
\end{align}
In general, the Bernoulli polynomial $B_{m,n}(x;\omega_1, \ldots, \omega_m)$
is defined by
\begin{align}
	\frac{y^m e^{x y}}{\prod_{k=1}^m (e^{\omega_k y}-1)} =
	\sum_{n=0}^{\infty} B_{m,n}(x; \omega_1, \ldots , \omega_m) \frac{y^n }{n!} \ .
	\label{foot.Bernoulli}
\end{align}
What is usually called the $n$-th Bernoulli polynomial in the literature is $B_n(x):=B_{1,n}(x,1)$.

\normalsize
\bigskip
\paragraph{Classical Limit}

Let us consider the limit $b\to 0$
of the function $e_b(x)$.
For this purpose, we can choose e.g.\ $\epsilon\to 0^+$ limit of
$b^2=i \epsilon$. We then have $\overline{q}\to 0$,
which trivializes the denominator of \eqref{ebratio}.
The leading classical contribution, as
read off from \eqref{PochAsymp},
reads
\begin{align}
	e_b(x) \xrightarrow[]{b\to 0}
	\exp\left(
	\frac{1}{2\pi i b^2} \mathrm{Li}_2(-e^{2\pi b x})
	+\mathcal{O}(b^{0})
	\right) \ .
\end{align}
This, in combination with \eqref{sbeb}, immediately gives
\begin{align}
	s_b(x) \xrightarrow[]{b\to 0}
	\exp\left(
	\frac{1}{2\pi i b^2}
	\left(
	\mathrm{Li}_2(-e^{2\pi b x}) +\frac{1}{4} (2\pi b x)^2 +\frac{\pi^2}{12}
	\right)
	+\mathcal{O}(b^{0})
	\right) \ .
	\label{sblimit}
\end{align}

\bigskip
\paragraph{Functional Relations}
We can immediately show from the definitions \eqref{ebratio} and \eqref{sbeb}
that
\begin{align}
	e_b\left(x-\frac{i b^{\pm 1}}{2}\right) & =(1+e^{2\pi b^{\pm 1} x})\,
	e_b\left(x+\frac{ib^{\pm 1}}{2}\right) \ ,
	\label{recursion_eb}                                                   \\
	s_b\left(x-\frac{i b^{\pm 1}}{2}\right) & =(2\cosh \pi b^{\pm 1} x) \,
	s_b\left(x+\frac{ib^{\pm 1}}{2}\right)  \ .
	\label{recursion}
\end{align}

\bigskip
\paragraph{Analytic Continuation}

We can choose the deformation parameter $b$ and the argument $z$
to be arbitrary complex numbers.
This requires analytic continuation in the complex plane, since
so far there have been restrictions on $b$ and $z$ (for convergence).

This is rather important for applications to
supersymmetric gauge theories, since
there we need to choose $b$
to be real, where the infinite product \eqref{ebratio}
does not converge.

\begin{figure}[t]
	\centering{\includegraphics[scale=0.8]{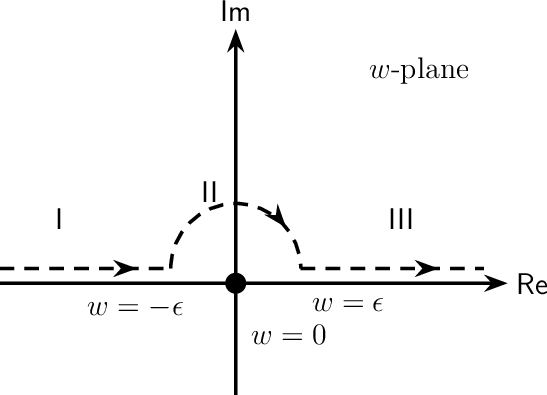}}
	\caption{The integration contour for $e_b(x)$ in the integral expression \eqref{ebdef}.
		We can divide the integration contour into three parts:
		I ($w<0$), II (the semi-circle of radius $\epsilon$ around $w=0$), and III ($w>0$).}
	\label{fig.ebpath}
\end{figure}

After analytic continuation, the function
$e_b(x)$ has the following integral expression
in the region $|\mathrm{Im}(x)|<\frac{|\textrm{Re} Q|}{2}$
of the complex $x$-plane:
\begin{align}
	e_b(x) & =\exp\left(\frac{1}{4} \int_{-\infty+i0}^{\infty+i0}
	\frac{dw}{w} \frac{e^{-i 2xw} }{\sinh(bw) \sinh(b^{-1}w)}\right)
	\label{ebdef}                                                 \\
	       & = \exp\left(\int_{-\infty+i0}^{\infty+i0}
	\frac{dw}{w} \frac{e^{-(ix+\frac{Q}{2})w} }{(1-e^{-bw})
		             (1-e^{-b^{-1}w })}\right) \nonumber        \\
	       & = \exp\left(\int_{-\infty+i0}^{\infty+i0}
	\frac{dw}{w} \frac{e^{-(ix-\frac{Q}{2})w} }{(e^{bw}-1)
		             (e^{b^{-1}w }-1)}\right) \ , \nonumber
\end{align}
where the integration contour avoids the pole $w=0$ by passing above
(Fig.~\ref{fig.ebpath}).

\small

In the equation above, $b$ is taken to be a positive real number.
In general, as long as $b$ is not purely imaginary
we can deform the integration contour as we continuously change the value of $b$,
to define this function.

\normalsize

We also have a similar relation for the function $s_b(x)$:
\begin{align}
	s_b(x)=\exp\left[ \frac{1}{i} \int_0^{\infty} \frac{dw}{w}
		           \left( \frac{\sin 2xw}{2\sinh (bw) \sinh (b^{-1}w)}-\frac{x}{w} \right) \right].
	\label{sbdef}
\end{align}

\bigskip
\small

Let us discuss these two equations in detail.\footnote{
	Readers in a hurry can skip to \eqref{ebnew1}.}

First, in \eqref{ebdef} thanks to the condition $|\mathrm{Im}(x)|<\frac{|\textrm{Re} Q|}{2}$
the integrand approaches zero exponentially in the limit $w\to \pm \infty +i 0$,
and the integral is well-defined.

Next, the definition of the integral \eqref{sbdef}
is potentially dangerous near $w\sim 0$ and $w\sim \infty$.
First, near $w\sim 0$ we have
\begin{align}
	\frac{\sin 2 x w}{2\sinh b w \sinh b^{-1} w}
	\sim \frac{2 x w}{2 (b w)(b^{-1} w)}
	\sim \frac{x}{w} \ ,
\end{align}
and the divergent $\frac{1}{w}$  term cancels out inside the integrand.
Also, near $w\sim \infty$ the integrand is
\begin{align}
	\frac{1}{w} \frac{\sin 2xw}{2\sinh (bw) \sinh (b^{-1}w)}\sim \frac{1}{w} \frac{e^{-i 2 x w}}{\frac{e^{\pm bw}}{2} \frac{e^{\pm b^{-1}w}}{2}}
	\sim \frac{1}{w} e^{(-i 2 x\mp(b+b^{-1})) w } \ ,
\end{align}
where the sign depends on the sign of $\textrm{Re}(b)$.
Since $\int^{\infty} \frac{dw}{w}e^{-\alpha w}\sim \log \alpha$,
there is again no divergence under the condition
$|\mathrm{Im}(x)|<\frac{|\textrm{Re} Q|}{2}$.

\bigskip
Let us check the consistency between \eqref{ebdef}, \eqref{sbdef} and
\eqref{sbeb}.
Let us divide the integration contour of \eqref{ebdef}
into three parts (Fig.~\ref{fig.ebpath}).
In the half-infinite line we have
\begin{align}
	\textrm{I}=\frac{1}{4}
	\int_{-\infty}^{-\epsilon} \frac{dw}{w}\, \frac{e^{-2 i xw}}{\sinh b w \sinh b^{-1}
		                                          w}
	=-\frac{1}{4}\int_{\epsilon}^{\infty} \frac{dw}{w}\, \frac{e^{2 i xw}}{\sinh b w \sinh b^{-1}
		                                                     w} \ ,
\end{align}
and when combined with the similar contribution from the
other half-line $w>0$, we have
\begin{align}
	\begin{split}
		\textrm{I}+\textrm{III}
		 & =-\frac{1}{4}\int_{\epsilon}^{\infty} \frac{dw}{w}\, \frac{(e^{2i xw}-e^{-2 i xw})}{\sinh b w \sinh b^{-1}
			                                                        w} \\
		 & =\frac{1}{2i}\int_{\epsilon}^{\infty} \frac{dw}{w}\, \frac{\sin 2 x w}{\sinh b w \sinh b^{-1}
			                                                        w} \ .
	\end{split}
\end{align}
The remaining contribution from the small contour around
$w=0$ is, by changing the integration variable to
$w=\epsilon e^{i \theta }, \theta\in [0, \pi]$,
\begin{align}
	\textrm{II}=-\frac{1}{4}\int_{0}^{\pi}  (i d\theta)\, \frac{e^{-2i x (\epsilon
				                                                      e^{i\theta})}}{\sinh (b \epsilon e^{i \theta}) \sinh (b^{-1} \epsilon e^{i\theta})} \ .
	\label{contour_II}
\end{align}
By collecting these, we obtain in the limit $\epsilon\to 0$
(exercise~\ref{residue_ex}),
\begin{align}
	\eqref{ebdef}=
	\frac{1}{i}\int_{\epsilon}^{\infty} \frac{dw}{w}\,
	\left(
	\frac{\sin 2 x w}{2\sinh b w \sinh b^{-1}
		w}
	-\frac{x}{w}
	\right)
	-\frac{i\pi}{24}(2-Q^2)+\frac{i\pi}{2}x^2 \ .
	\label{sum_of_three}
\end{align}
Note that as mentioned already the integrand is smooth in the limit $\epsilon\to 0$,
and hence we can take the limit $\epsilon\to 0$
without any problem, giving rise to \eqref{sbeb}
under \eqref{sbdef}.

\normalsize

\bigskip

Let us next evaluate the integral in \eqref{ebdef}, to obtain \eqref{ebratio}.
Let us choose the parameter $b$ to be a complex parameter satisfying $|q|, |\overline{q}|<1$,
and let us moreover assume $b\notin i \bR$.
Thanks to the residue theorem, we can evaluate the integral by closing the integration contour;
as mentioned already the integration contour in \eqref{ebdef}
does not cross a pole as long as $b$ is not purely imaginary.
This means that for computational purposes we can first choose
$b$ to be real, evaluate the integral along the real axis, and then
analytically continue the resulting answer to a more general complex value of $b$.

We should close the contour either above or below the real axis, depending on the value of $\textrm{Re}(x)$.
If $\textrm{Re}(x)<0$ we close the contour in the semi-circle in the upper half plane,
and by summing over residues at poles $w=i b^{-1} m \pi$ and $w=i b n
	\pi$ ($m, n>0$) we obtain
\begin{align*}
	\eqref{ebdef}=\exp\left[
		                  \frac{1}{2} \sum_{n=1}^{\infty} \frac{1}{n}
		                  \left(
		                  \frac{e^{2b^{-1} \pi x n}}{(-1)^n \sinh(i b^{-2} \pi n)}
		                  +
		                  \frac{e^{2b \pi x n}}{(-1)^n \sinh(i b^2 \pi n)}
		                  \right)
		                  \right] \ .
\end{align*}
Here
\begin{align}
	\frac{1}{2}\sum_{n=1}^{\infty}\frac{1}{n}
	\frac{e^{\alpha n}}{\sinh\beta n}
	=
	\begin{cases}
		\displaystyle-\log \prod_{n=1}^{\infty} \left(
		1-e^{\alpha-(2n-1)\beta}\right) & (\beta>0)  \ , \\
		\displaystyle\log \prod_{n=1}^{\infty} \left(
		1-e^{\alpha+(2n-1)\beta}\right) & (\beta<0)      \\
	\end{cases}
	\label{sumformula}
\end{align}
(to show this, expand the $\sinh$ as an exponential, and change the order of the sums),
to obtain (for $\textrm{Re}(x)<0$)
\begin{align}
	e_b(x)=
	\begin{cases}
		\displaystyle\frac{(-q e^{2\pi b x}; q^2)_{\infty}}{(-\overline{q}e^{2\pi b^{-1}x};
			             \overline{q}^2)_{\infty}}                         & (|q|, |\overline{q}|<1)  \ ,             \\
		\displaystyle\frac{(-\overline{q}^{-1} e^{2\pi b^{-1} x}; \overline{q}^{-2})_{\infty}}{(-q^{-1}e^{2\pi bx};
			             q^{-2})_{\infty}} & (|q|, |\overline{q}| > 1)  \ .
		\label{ebnew1}
	\end{cases}
\end{align}
Here, depending on the two cases in
\eqref{sumformula} we have two different equations, one for the
case $|q|, |\overline{q}|>1$.

We can apply the same argument to the case $\textrm{Re}(x)>0$.
In this case, we have an extra residue from pole $w=0$, to obtain
\begin{align}
	e_b(x)=
	\begin{cases}
		e^{i\pi x^2-\frac{i\pi}{12}(2-Q^2)}
		\displaystyle
		\frac
		{(-\overline{q}e^{-2\pi b^{-1}x}; \overline{q}^2)_{\infty}}
		{(-q e^{-2\pi b x}; q^2)_{\infty}}
		 & (|q|, |\overline{q}|<1) \ ,    \\
		e^{i\pi x^2-\frac{i\pi}{12}(2-Q^2)}
		\displaystyle\frac
		             {(-q^{-1}e^{-2\pi bx}; q^{-2})_{\infty}}
		             {(-\overline{q}^{-1} e^{-2\pi b^{-1} x}; \overline{q}^{-2})_{\infty}}
		 & (|q|, |\overline{q}| > 1)  \ .
		\label{ebnew2}
	\end{cases}
\end{align}
These equations are all analytically continued to the whole complex plane.

For completeness
let us rewrite these equations in terms of $s_b(x)$.
When $|q|, |\overline{q}|<1$
\begin{align}
	\begin{split}
		s_b(x) & =
		e^{-\frac{i\pi}{2} x^2+\frac{i\pi}{24}(2-Q^2)}
		\displaystyle\frac{(-q e^{2\pi b x}; q^2)_{\infty}}{(-\overline{q}e^{2\pi b^{-1}x};
			             \overline{q}^2)_{\infty}} \\
		       & =e^{\frac{i\pi}{2} x^2-\frac{i\pi}{24}(2-Q^2)}
		\displaystyle
		\frac
		{(-\overline{q}e^{-2\pi b^{-1}x}; \overline{q}^2)_{\infty}}
		{(-q e^{-2\pi b x}; q^2)_{\infty}} \ .
		\label{sbnew1}
	\end{split}
\end{align}
When $|q|, |\overline{q}|>1$,
\begin{align}
	\begin{split}
		s_b(x) & =
		e^{-\frac{i\pi}{2} x^2+\frac{i\pi}{24}(2-Q^2)}
		\displaystyle
		\frac
		{(-\overline{q}^{-1}e^{2\pi b^{-1}x}; \overline{q}^{-2})_{\infty}}
		{(-q^{-1} e^{2\pi b x}; q^{-2})_{\infty}} \\
		       & = e^{\frac{i\pi}{2} x^2-\frac{i\pi}{24}(2-Q^2)}
		\displaystyle\frac
		             {(-q^{-1}e^{-2\pi bx}; q^{-2})_{\infty}}
		             {(-\overline{q}^{-1} e^{-2\pi b^{-1} x}; \overline{q}^{-2})_{\infty}} \ .
		\label{sbnew2}
	\end{split}
\end{align}

\bigskip
\paragraph{Poles and Zeros}

It is straightforward to identify the positions of the poles
and zeros of $e_b(z)$ and $s_b(z)$
from the expression \eqref{ebratio}.
The poles are located at
\begin{align}
	z=\frac{iQ}{2}+i (n b+mb^{-1}) \ , \quad n, m \in \mathbb{Z}_{\ge 0} \ ,
\end{align}
and the zeros at
\begin{align}
	z=-\frac{iQ}{2}-i (n b+mb^{-1}) \ , \quad n, m \in \mathbb{Z}_{\ge 0} \ .
\end{align}
Note that approximately half of the zeros from the
numerator of \eqref{ebratio} cancel out poles
from the denominator.
The poles and the zeros above
suggest that the function
$s_b(z)$ can be thought of as a regularization of an infinite product
\begin{align}
	\prod_{m,n\in \mathbb{Z}_{\ge 0}}
	\frac{mb+nb^{-1}+\frac{Q}{2}-iz }{mb+nb^{-1}+\frac{Q}{2}+iz }
	\label{naive}
\end{align}
This infinite product appears in the computation of the
one-loop determinant of the $S^3$ partition function,
where the integers $m, n$ represent a product over the
modes (spherical harmonics) of $S^3$,
those uncanceled modes coming either from a boson or a fermion.

The equation \eqref{naive} implies that the function $s_b(x)$
can be written as a ratio of two \keyword{Barnes double gamma functions}{Barnes double gamma functions} $\Gamma_2(x|\epsilon_1, \epsilon_2)$~\cite{Barnes}:
\begin{align}
	 & s_b(x)=\frac
	          {
		          \Gamma_2\left(\frac{Q}{2}+ix| b, b^{-1}\right)
	          }
	          {
		          \Gamma_2\left(\frac{Q}{2}-ix| b, b^{-1}\right)
	          } \ , \\
	 & \Gamma_2(x|\epsilon_1, \epsilon_2) \sim
	\prod_{m,n\ge 0}(x+m\epsilon_1+n \epsilon_2)^{-1} \ .
	\label{sbdef_2}
\end{align}
Here a precise definition of $\Gamma_2(x|\epsilon_1, \epsilon_2)$
is given by
\begin{align}
	\Gamma_2(x|\epsilon_1, \epsilon_2)
	:=\exp\frac{d}{ds}\Big|_{s=0}
	\zeta_2(s,x|\epsilon_1, \epsilon_2)  \ ,
	\label{sbdef_1}
\end{align}
where $\zeta_2(s,x|\epsilon_1, \epsilon_2)$
is the Barnes double zeta function\begin{align}
	\begin{split}
		\zeta_2(s,x|\epsilon_1, \epsilon_2)
		 & :=\sum_{m,n\ge 0} (x+m\epsilon_1+n\epsilon _2)^{-s}           \\
		 & =\frac{1}{\Gamma(s)}\int_{0}^{\infty} \!\frac{dt}{t}\, t^s \,
		\frac{e^{-t x}}{(1-e^{-\epsilon_1 t})(1-e^{-\epsilon_2 t} )}
	\end{split}
\end{align}
The Barnes double functions appear in several contexts in
physics, including the conformal blocks of the Liouville theory (see Sec.~\ref{sec.3d_as_AGT})
and the Nekrasov partition function for
4d $\mathcal{N}=2$ theory (Sec.~\ref{subsec.S3inS4}).

\small

We can show that the definitions \eqref{sbdef_2} and \eqref{sbdef_1}
coincide with the previous definition \eqref{sbdef}.
Let us demonstrate this explicitly (cf.\ Ref.~\cite{Ruijsenaars_Barnes}).

First, from the definition of the multiple Bernoulli polynomial \eqref{foot.Bernoulli}
we obtain
\begin{align}
	\frac{e^{-xt}}{(1-e^{-\epsilon_1 t}) (1-e^{-\epsilon_2 t})}
	=\sum_{n=0}^{\infty} \frac{t^{n-2}}{n!} B_{2,n}(\epsilon_1+\epsilon_2-x+1; \epsilon_1, \epsilon_2) \, e^{-t}  \ .
	\label{Bernoullli_expand}
\end{align}
Note that this has a non-positive power of $t$ only when
$n=0, 1, 2$.
Let us plug this into the previous definition.
We then need to evaluate
\begin{align}
	\frac{d}{ds}\Big|_{s=0} \frac{1}{\Gamma(s)} \int_{0}^{\infty}\! \!\frac{dt}{t}\, t^s t^{n-2} e^{-t}
	=  \frac{d}{ds}\Big|_{s=0} \frac{\Gamma(s+n-2)}{\Gamma(s)}
	\label{tmp_Barnes}
\end{align}
for each $n$.
Using $\Gamma(x+1)=x\Gamma(x)$, we compute
\begin{align}
	\frac{d}{ds}\Big|_{s=0} \frac{\Gamma(s+n-2)}{\Gamma(s)}
	=\begin{cases}
		 \frac{3}{4}                                                                  & n=0 \ , \\
		 -1                                                                           & n=1 \ , \\
		 0                                                                            & n=2 \ , \\
		 \Gamma(n-2) =\displaystyle\int_0^{\infty}\!\! \frac{dt}{t}\,  t^{n-2} e^{-t} & n\ge 3
	 \end{cases}
	\ .
	\label{case_by_case}
\end{align}
For $n\ge 3$, we can after all forget about $\frac{d}{ds}\frac{1}{\Gamma(s)}$ in \eqref{tmp_Barnes},
and simply set $s=0$ in the rest of the expression.
This means that in the expansion \eqref{Bernoullli_expand}
we can evaluate only the
$n=0,1,2$ parts by \eqref{case_by_case},
while for all other terms
we can simply drop the factor $\frac{d}{ds}\frac{1}{\Gamma(s)}$
and set $s=0$.
This procedure gives
\begin{align}
	\begin{split}
		 & \log \Gamma_2(x|b,b^{-1}) = \frac{3}{4}-\frac{2+Q-2 x}{2}                                                                      \\
		 & \qquad +\int_0^{\infty} \frac{dt}{t} \, e^{-t} \left[
			                                                  \frac{e^{-(x-1)t}}{(1-e^{-bt})(1-e^{-b^{-1}t})} -\frac{1}{t^2} -\frac{2+Q-2x}{2t} \right.\\
			                                                  &\qquad\qquad\qquad\left. - \frac{7+6Q+Q^2-12 x-6Q x+6 x^2}{12}
			                                                  \right]  \ .
	\end{split}
\end{align}
and thus
\begin{align}
	\begin{split}
		 & \log \frac{\Gamma_2\left(\frac{Q}{2}+ix |b,b^{-1}\right)}{\Gamma_2\left(\frac{Q}{2}-i x |b,b^{-1}\right)}      \\
		 & \qquad=2xi -2i \int_0^{\infty} \frac{dt}{t} \, \left[
			                                                  \frac{\sin(xt)}{ 4\sinh(\frac{bt}{2}) \sinh(\frac{b^{-1}t}{2}) }
			                                                  -e^{-t}  x \left(\frac{1}{t}+1\right)
			                                                  \right]  \ .
	\end{split}
\end{align}
Using
\begin{align}
	\int_0^{\infty} \frac{dt}{t} \left(\frac{1}{t}-e^{-t}\left(\frac{1}{t}+1\right)\right) =1 \ ,
\end{align}
we obtain
\begin{align}
	\begin{split}
		 & \log \frac{\Gamma_2\left(\frac{Q}{2}+ix |b,b^{-1}\right)}{\Gamma_2\left(\frac{Q}{2}-i x |b,b^{-1}\right)}       \\
		 & \qquad =-2i \int_0^{\infty} \frac{dt}{t} \, \left[
			                                                   \frac{\sin(xt)}{ 4\sinh(\frac{bt}{2}) \sinh(\frac{b^{-1}t}{2}) }
			                                                   -\frac{x}{t} \right]  \ .
	\end{split}
\end{align}
We obtain the previous definition when we change the integration variable to
$t=2w$.

\normalsize

\bigskip
\paragraph{Inversion Formula}

The functions
$e_b(x)$ and $s_b(x)$
are kept invariant under the exchange of $b$ and $b^{-1}$:
\begin{align}
	e_b(x)=e_{b^{-1}}(x) \ , \quad
	s_b(x)=s_{b^{-1}}(x) \ .
	\label{sbsym}
\end{align}
This symmetry reflects the
geometry of the $S^3_b$, and also of the
algebraic structure known as the \keyword{modular double}{modular double} of the quantum group \cite{FaddeevModular}.

Note that the symmetry \eqref{sbsym} is manifest in \eqref{ebdef}, \eqref{sbdef},
but not in the infinite-product expressions \eqref{ebratio}, \eqref{sbeb}.

The three functions
$(x;q)_{\infty}, e_b(x), s_b(x)$
satisfy essentially the same functional relations
\eqref{PochRecursion} and \eqref{recursion}.
The function $(x;q)_{\infty}$, however, does not have the inversion property \eqref{sbsym}.
In this sense, the functions $s_b(z), e_b(z)$
can be thought of as an extension of $(x;q)_{\infty}$
to have the symmetry
$b\to b^{-1}$.

\bigskip
\paragraph{Asymptotic Form}

It is straightforward to prove the following from \eqref{sbdef} and \eqref{sbeb}:
\begin{align}
	e_b(x) e_b(-x)= e^{i\pi x^2-\frac{i\pi}{12}(2-Q^2)} \ , \quad
	s_b(-x)s_b(x)=1 \ .
	\label{sbinv}
\end{align}

The asymptotic form of the function $s_b(x)$, as its
argument $x$ approaches infinity along the real axis,
can be obtained from \eqref{sbnew1} and \eqref{sbnew2}:
\begin{align}
	s_b(x) \to
	\begin{cases}
		e^{\frac{i\pi x^2}{2}}  e^{-\frac{i\pi}{24}(2-Q^2)} & (x\to \infty) \ ,  \\
		e^{-\frac{i \pi x^2}{2}} e^{\frac{i\pi}{24}(2-Q^2)} & (x\to -\infty) \ .
	\end{cases}
	\label{sb_asymp}
\end{align}

\bigskip
\paragraph{$b=1$ Limit}

Recall that the quantum dilogarithm function reduces to
the classical dilogarithm function in the semiclassical limit
$b\to 0$ (or equivalently $b\to \infty$).
There is another special value of $b$:
for $b=1$ the quantum dilogarithm function specializes to the
classical dilogarithm function (see exercise~\ref{ex.b1} for derivation):
\begin{align}
	\log e_{b=1}(x) & = 
	\frac{i}{2 \pi }\mathrm{Li}_2(e^{2\pi x})
	+ix \log(1-e^{2 \pi x}) \ .
	\\
	\log s_{b=1}(x) & = 
	\frac{i}{2 \pi }\mathrm{Li}_2(e^{2\pi x})
	+ix \log(1-e^{2 \pi x})-\frac{i\pi }{2}x^2-\frac{i\pi}{12} \ .
	\label{b1}
\end{align}
\small
The specialization
$b=1$ is important for many practical applications, including the
applications to the 3d $F$-theorem and the entanglement entropies.\footnote{
	The function $\log s_{b=1}(ix)$ is denoted by $l(x)$ in the paper \cite{Jafferis:2010un} on the $S^3$ partition functions.}
\normalsize

\bigskip
\paragraph{Pentagon Identity}

One of the most important (and probably the most important)
properties of the quantum dilogarithm function is the
\keyword{pentagon identity}{pentagon identity}.
Suppose that we have two operators $\sfP, \sfQ$
satisfying the canonical commutation relation $[\sfP, \sfQ]=\frac{1}{2\pi i}$.
The quantum dilogarithm function
then satisfies the following identity:
\begin{align}
	e_b(\sfP) e_b(\sfQ) =
	e_b(\sfQ) e_b(\sfP+\sfQ) e_b(\sfP) \ .
	\label{ebpentagon}
\end{align}
This equation can be proven as the pentagon identity for the
$q$-Pochhammer symbol.
(In the language of Appendix~\ref{app.clusterapp},
this identity is the quantum dilogarithm identity for the
$A_2$-type quiver.)

While this identity is an operator identity, we can convert this into a
c-number identity by taking an expectation value under a
suitable state (exercise~\ref{ex.pentagon_relate}).
The resulting identity reads
\begin{align}
	\begin{split}
		\int\! d\sigma\, & e^{2\pi i \zeta \sigma} s_b\left(\sigma+r\right)
		s_b\left(-\sigma-s\right)                                                                                                          \\
		                 & =
		e^{-\frac{i\pi}{12}(1+Q^2)} e^{-\pi i \zeta (r+s)}
		s_b\left(\zeta-\frac{r}{2}+\frac{s}{2} +\frac{iQ}{2} \right)                                                                       \\
		                 & \qquad \times s_b\left(-\zeta-\frac{r}{2}+\frac{s}{2}+\frac{iQ}{2}\right) s_b\left( r-s-\frac{iQ}{2}\right) \ .
	\end{split}
	\label{Ramanujan}
\end{align}
In this book, we also refer to this identity as the
\keyword{pentagon identity}{pentagon identity}.
Eq.~\eqref{Ramanujan} is sometimes called the \keyword{integral Ramanujan identity}{integral Ramanujan identity}.

\bigskip
\paragraph{Fourier Transform}

By taking an appropriate limit of the pentagon identity
\eqref{Ramanujan},
we can derive the following Fourier-transformation formula
for the function $e_b(z)^{\pm 1}$~\cite{Faddeev:2000if} (exercise~\ref{ex.ebFourier}):
\begin{align}
	\begin{split}
		 & \int dx\, e_b(x)\, e^{2\pi i  w x}=e^{-i\pi w^2+\frac{i\pi}{12} (1+Q^2)}
		\, e_b\left(w+i \frac{Q}{2}\right)
		\ ,                                                                                         \\
		 & \int dx\, e_b(x)^{-1}\, e^{2\pi i  w x}=e^{i\pi w^2- \frac{i\pi }{12}(1+Q^2)}
		\, e_b\left(-w-i \frac{Q}{2}\right)^{-1}
		\ ,                                                                                         \\
		 & \int dx\, s_b(x)\, e^{\frac{i\pi}{2}x^2+ 2\pi i  w x}                                    \\
		 & \qquad\qquad =e^{-\frac{i\pi}{2} \left(w-\frac{iQ}{2}\right)^2-\frac{i\pi}{12} (2Q^2-1)}
		\, s_b\left(w+i \frac{Q}{2}\right)
		\ ,                                                                                         \\
		 & \int dx\, s_b(x)^{-1}\, e^{-\frac{i\pi}{2}x^2+ 2\pi i  w x}                              \\
		 & \qquad \qquad=e^{\frac{i\pi}{2} \left(w-\frac{iQ}{2}\right)^2+\frac{i\pi}{12} (2Q^2-1)}
		\, s_b\left(-w-i \frac{Q}{2}\right)^{-1}
		\ .                                                                                         \\
	\end{split}
	\label{ebFourier}
\end{align}


\begin{practice}
	\item $[\bll]$ (Classical Dilogarithm Identity)
	Show the classical dilogarithm identities \eqref{Lpentagon} and \eqref{Ldilong}.
	Hint: check the identity for the special values $x=y=0$,
	as well as the derivatives of the identities with respect to $x$ and $y$, respectively.
	Many classical dilogarithm identities can be proven by this method.\label{ex.L}

	\item $[\bll]$ ($q\to 1$ Limit of $(x;q)_{\infty}$)\label{ex.PochAsymp}
		Prove the $q\to 1$ expansion \eqref{PochAsymp} of the $q$-Pochhammer symbol $(x;q)_{\infty}$.
		Hint: take the logarithm of the infinite product expansion of $(qx;q)_{\infty}$,
		to obtain a Riemann sum.
		Apply the Euler-Maclaurin formula
		\begin{align}
			\begin{split}
				\sum_{m=M}^N f(m) & = \int_{M}^N f(t) dt +\frac{1}{2}(f(N)+f(M))         \\
				                  & +\sum_{k=1}^{p} \frac{B_{2k}}{(2k)!}
				\left(f^{(2k-1)}(N)-f^{(2k-1)}(M)\right)
				\\
				                  & -\int_M^N \frac{B_{2p}(t-[t])}{(2p)!} f^{(2p)}(t) dt
			\end{split}
		\end{align}
		where in the last term (remainder term)
	$B_{2p}(t)$ is a Bernoulli polynomial ($B_k(0)=B_k$, see \eqref{foot.Bernoulli})
		and $t-[t]$ is the non-integer part of $t$.
		In the computation, use the following identity for the
		polylogarithm function $\mathrm{Li}_s(x)$ in \eqref{polylog}:
		\begin{align}
			\mathrm{Li}_1(x)=-\log(1-x)\ , \quad
			\frac{d}{dx} \mathrm{Li}_{s}(e^x)=\mathrm{Li}_{s-1}(e^x) \ .
		\end{align}

		\item $[\bll]$ (Pentagon Identity for $(x;q)_{\infty}$)

		Take the semiclassical limit of the pentagon identity \eqref{PochPentagon}
		for the $q$-Pochhammer symbol, to
		obtain the pentagon identity \eqref{Lpentagon}
		for the classical dilogarithm function.
		\label{KNex}

		\item $[\bll]$ (Relation between $e_b(x)$ and $s_b(x)$)
		Evaluate the expression \eqref{contour_II} in the limit $\epsilon\to 0$,
		and derive the integral \eqref{sum_of_three}.
		\label{residue_ex}

		\item $[\bll]$ (Functional Relation for $e_b(x), s_b(x)$)
		In the body we derived the function relation \eqref{recursion}
		from the infinite-product expression
		\eqref{ebratio}, \eqref{sbeb}.
		Derive the same functional relation \eqref{recursion},
		this time from the integral expressions \eqref{ebdef}
		and \eqref{sbdef}.

		\item $[\bll]$ (Equivalence of Infinite-Product Expressions)
		Note that the two equations \eqref{sbnew1} and \eqref{sbnew2}
		imply the equivalence of the following
		two infinite-product expressions:
		\begin{align}
			\begin{split}
				 &
				(-q e^{2\pi b x};q^2)_{\infty}
				(-q e^{-2\pi b x}; q^2)_{\infty}
				\\
				 & \qquad
				=e^{i\pi x^2-\frac{i\pi }{12}(2-Q^2)}
					 (-\overline{q} e^{2\pi b^{-1} x}; \overline{q}^2)_{\infty}
					 (-\overline{q} e^{-2\pi b^{-1} x}; \overline{q}^2)_{\infty} \ .
			\end{split}
			\label{to_be_shown_dilog}
		\end{align}
		Demonstrate this identity directly.
		Hint: in terms of the \keyword{theta function}{theta function} $\theta_{00}(x|q)$ and the Dedekind eta function $\eta(q)$
		\begin{align}
			\theta_{00}(x|q):                      & =\prod_{m=1}^{\infty}(1-q^{2m})
			\left(1+2\cos (2\pi x) q^{2m-1}+q^{4m-2}
			\right)  \ ,                                                                                \\
			\eta\left(\frac{\log q}{2\pi i}\right) & :=e^{\frac{\log q}{24}}\prod_{n=1}^{\infty}(1-q^n)
		\end{align}
		we can rewrite \eqref{to_be_shown_dilog} into
		\begin{align}
			\begin{split}
				\frac{\eta(-b^{-2})}{\eta(b^2)}=
				e^{i\pi x^2}
				\frac{\theta_{00}(-i b^{-1}x|\overline{q})}{\theta_{00}(-ib x|q)}
			\end{split}
		\end{align}
		This follows from the modular transformations of
	$\eta(q)$ and $\theta_{00}(x|q)$:
		\begin{align}
			\frac{\eta(-b^{-2})}{\eta(b^2)}=\sqrt{-ib^2} \ ,  \quad
			\frac{\theta_{00}(\frac{x}{b^2}; \overline{q})}
			{\theta_{00}(x; q)}=\sqrt{-ib^2} e^{\frac{i \pi x^2}{b^2}} \ .
		\end{align}

		\item $[\bll]$ ($b=1$ Specialization of $e_b(x), s_b(x)$)
		Derive the expressions  \eqref{b1} for $e_{b=1}(x)$ and $s_{b=1}(x)$.
		Hint: Set $b=1$ in the integral expression \eqref{ebdef} for $e_b(x)$.
		Close the integration contour and use the residue theorem.
		This computation here is similar to the computation of
		deriving \eqref{ebratio} from \eqref{ebdef}; however note that
		for $b=1$ several poles collide and thus we have double poles,
		to modify the computation of the residue theorem.
		In general, whenever $b^2$ is a rational number,
	$s_b(x)$ and $e_b(x)$ can be written in terms of the classical dilogarithm function
	$\textrm{Li}_2(x)$ \cite{Garoufalidis:2014ifa,Ip:2014pva}.
		\label{ex.b1}

		\item $[\bll]$ (Pentagon Identity for $e_b(x)$)
		Prove the quantum dilogarithm identity  \eqref{ebpentagon} for $e_b(x)$.
		Hint: use \eqref{PochPentagon} and
		\eqref{ebratio}.
		\label{ex.ebpentagon}

		\item $[\bll]$ (Semiclassical Limit of the Quantum Pentagon Identity)

		Show that the quantum dilogarithm identity \eqref{ZQED_ZXYZ}
		in the semiclassical limit $b\to 0$
		gives the classical dilogarithm identity \eqref{Lpentagon}.
		We can do this as follows:

		\begin{enumerate}

		\item
		Let us consider the limit $b\to 0$,
		with combinations $\Sigma, Z, M$
		\begin{align}
			\Sigma:=e^{2\pi b \sigma} \ , \quad
			Z:=e^{-2\pi b \zeta} \ , \quad
			M:=e^{2\pi b \mu} \
		\end{align}
		held fixed; here we take $\mu, \zeta$ to be real, so that $M, Z>0$.
		Note that this is natural scaling in the discussion of the dimensional reduction of
		3d $\mathcal{N}=2$ theories in Sec.~\ref{sec.3d_dim_red}.
		By taking the semiclassical limit \eqref{sblimit} of the function $s_b(z)$,\footnote{
		There is a $b$-dependent contribution from the change of the integral variable from
	$\sigma$ to $\Sigma$; however, this is
		subleading in the expansion with respect to $b$.
		}
		\begin{align}
			\begin{split}
				\int \! d\Sigma \,
				 & \exp\left[\frac{1}{2\pi i b^2}
					       \mathcal{W}_1(\Sigma, Z, M)
					       + \mathcal{O}(b^0)
					       \right]     \\
				 & \qquad=\exp\left[\frac{1}{2\pi i b^2}
					              \mathcal{W}_2(Z, M)
					              +\mathcal{O}(b^0)
					              \right] \ .
			\end{split}
		\end{align}
		Here $\mathcal{W}_1$ and $\mathcal{W}_2$
		are twisted superpotentials in two dimensions,
		whose explicit forms are given by
		\begin{align}
			\begin{split}
				\mathcal{W}_1(\Sigma, Z, M)&=
				  -\log Z \log \Sigma+\Li\left(\frac{\Sigma}{M}\right)+\Li\left(\frac{1}{\Sigma M}\right)
				\\
				 & \quad + \frac{1}{2}\left(\log\Sigma \right)^2+ \frac{1}{2}\left(\log M \right)^2 - i\pi \log M-\frac{\pi^2}{3}
				\ ,                                                                                                         \\
				\mathcal{W}_2(Z, M)& =
				  \Li\left(\frac{1}{M^2}\right)+ \Li(-MZ)+\Li\left(-\frac{M}{Z}\right)
				\\
				 & \quad +\frac{3}{2} \left( \log M \right)^2 + \frac{1}{2}\left(\log Z \right)^2 - i \pi \log M
			\end{split}
		\end{align}

		\item
		Let us evaluate the integral in the saddle-point approximation.
		The saddle point with respect to $\Sigma$ is computed to be:
		\begin{align}
			\exp\left(\frac{\partial \mathcal{W}_1(\Sigma, Z, M)}{\partial \log \Sigma}
			\right)
			=1 \longrightarrow
			\Sigma=\frac{1+MZ}{M+Z} \ .
		\end{align}
		The value of $\scW_1$ at this saddle point is
		\begin{align}
			\begin{split}
				\scW_1\big|_{\textrm{saddle point}} & =\textrm{Li}_2\left(\frac{1+MZ}{M(M+Z)}\right)+\Li\left(\frac{M+Z}{M(1+MZ)}\right) \\
				                                    & \quad-\log Z \log \frac{1+MZ}{M+Z}
				+\frac{1}{2}\left( \log \frac{1+MZ}{M+Z} \right)^2                                                                       \\
				                                    & \quad +\frac{1}{2}\left(\log M\right)^2-i\pi \log(M) -\frac{\pi^2}{3} \ .
			\end{split}
		\end{align}

		\item
		Equating this
	$\scW_1\big|_{\rm saddle}$ with $\scW_2$,
		and rewriting the identity in terms of the
		Rogers dilogarithm function \eqref{LRogers},
		we obtain
		\begin{align}
			\begin{split}
				 & L\left(\frac{1}{M^2} \right)+L\left(-MZ \right)+L\left(-\frac{M}{Z} \right)
				-L\left(\frac{1}{M}\frac{1+MZ}{M+Z} \right)-L\left(\frac{1}{M}\frac{M+Z}{1+MZ} \right) \\
				 & \quad +\frac{\pi^2}{3}+\frac{i\pi}{2} \log\left( \frac{Z}{(1+MZ)(M+Z)}\right)=0 \ .
			\end{split}
		\end{align}

		Defining
		\begin{align}
			 & x=\frac{1}{M}\frac{1+MZ}{M+Z} \ , \quad
			y=\frac{1}{M}\frac{M+Z}{1+MZ} \ ,          \\
			 & u=\frac{x(1-y)}{1-xy}\ , \quad
			v=\frac{y(1-x)}{1-xy} \ ,
		\end{align}
		we obtain
		\begin{align}
			\begin{split}
				 & L\left(xy \right)+L\left(1-u^{-1} \right)+L\left(1-v^{-1} \right)
				-L\left(x \right)-L\left(y \right)                                                  \\
				 & \quad+\frac{\pi^2}{3}+\frac{i\pi }{2} \log\left(\frac{Z}{(1+MZ)(M+Z)} \right) =0
				\ .
			\end{split}
			\label{dilog_tmp}
		\end{align}
		Finally, since $M, Z>0$ implies $0<u, v<1$, we can use the identities (valid for $0<x<1$)
		\begin{align}
			 & L(1-x)+L(x)=\frac{\pi^2}{6}  \ ,                                          \\
			 & L\left(\frac{1}{x}\right)+L(x)=\frac{\pi^2}{3}+\frac{i\pi}{2} \log(x) \ ,
		\end{align}
		and hence
		\begin{align}
			L(1-x^{-1})=L(x)-\frac{\pi^2}{6} -\frac{i\pi}{2} \log(x) \ ,
		\end{align}
		for $x=u, v$; then
		\eqref{dilog_tmp} reduces to
		\begin{align}
			\begin{split}
				 & L\left(xy \right)+L\left(u \right)+L\left(v \right)
				-L\left(x \right)-L\left(y \right) = 0  \ .
			\end{split}
			\label{dilog_tmp_2}
		\end{align}
		This coincides with the pentagon identity \eqref{Lpentagon}
		for the classical dilogarithm.
		\end{enumerate}

		\item $[\bll]$  (Relation between Two Quantum Pentagon Identities)
		Evaluate the expectation value of the operator-version of the
		quantum pentagon identity \eqref{ebpentagon},
		to obtain the $c$-number version of the pentagon identity
		\eqref{Ramanujan}.

		\label{ex.pentagon_relate}

		\item $[\bll]$ (Fourier Transform of $e_b(x)$)
		Start with \eqref{Ramanujan}, shift $\zeta\to \zeta-\frac{s}{2}$ and
		set $r=0, s\to -\infty$, to obtain the third identity (and thus also the first identity) of \eqref{ebFourier}.
	Show also the second and the fourth identities of \eqref{ebFourier}
	in a similar manner.
	\label{ex.ebFourier}

\end{practice}

\chapter{Crash Course on Cluster Algebras}\label{app.clusterapp}

\begin{abstract}
  In this appendix we summarize the mathematical facts on cluster algebras
  minimally needed for the understanding of this book.
  Many of the structures discussed in the main text, such as Fock variables, their quantization,
  and their transformation properties under flips in \Teichmuller theory,
  can be regarded as special instances of the general theory of cluster algebras.
\end{abstract}

\section{Quiver and Mutation}

In this appendix we briefly summarize
the mathematical concept of the
\keyword{cluster algebra}{cluster algebra} \cite{FominZelevinsky1}.
The cluster algebra is a mathematical structure
which has increasing significance in
a number of different areas of mathematical physics
(see e.g.\ Ref.~\cite{KellerSurvey} for a review,
and Ref.~\cite{JPA_special}
for a sample of recent activities in mathematical physics),
and is the underlying structure of many of the ingredients of this book,
in particular 2d hyperbolic geometry,
2d \Teichmuller theory, and 3d hyperbolic geometry
(see Refs.~\cite{GSV,FST1} for the connection with 2d hyperbolic geometry,
and Ref.~\cite{Nagao:2011aa} for 3d hyperbolic geometry).
Cluster algebras can be thought of as defining data
for 3d $\scN=2$ theories, as we also commented briefly in Chap.~\ref{chap.conclusion}.

Let us begin with a
\keyword{quiver}{quiver},
namely an oriented graph (Fig.~\ref{fig.quiver}).
In the main text we encountered this as the defining data of
quiver gauge theories.

\begin{figure}[t]
  \centering
  \includegraphics[scale=0.27]{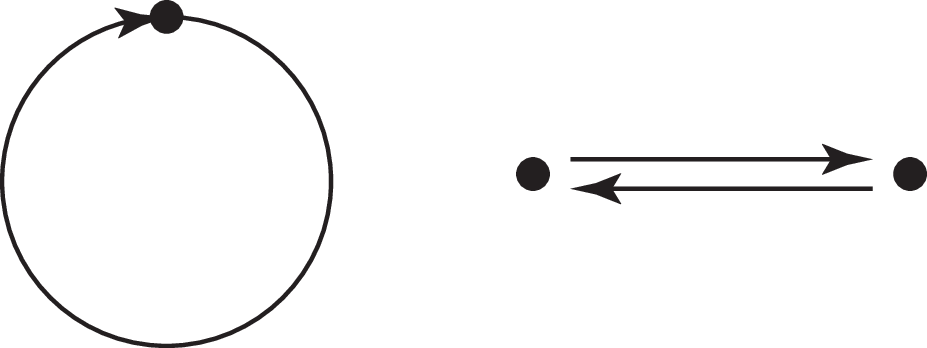}
  \caption{We assume that there are no loops (left) or oriented $2$-cycles (right).}
  \label{fig_loops}
\end{figure}

Let us denote the set of vertices by $I$,
with its elements denoted by
$i,j, \ldots \in I$.
For technical reasons let us assume that the
quiver has no loops or oriented $2$-cycles (Fig.~\ref{fig_loops}).\footnote{Note that this condition is not always satisfied in practical applications.}
Note that this condition is satisfied for the
quantum \Teichmuller theory discussed in Chap.~\ref{chap.Teichmuller}.

Given a quiver $\mathsf{Q}$,
let us define an anti-symmetric matrix $\mathsf{Q}=(\mathsf{Q}_{i,j})_{i,j\in I}$
by\footnote{In this book, we use the same symbol $\mathsf{Q}$ both for a quiver and for a matrix.}
\begin{align}
  \mathsf{Q}_{i,j}:=\sharp\{\text{arrows from $i$ to $j$}\}-
  \sharp\{\text{arrows from $j$ to $i$}\}  \ .
\end{align}
\nomenclature{$\mathsf{Q}=(\mathsf{Q}_{i,j})$}{quiver}
Namely, $\mathsf{Q}_{i,j}$ is the number of arrows from vertex $i$ to vertex $j$, counted with signs.
Notice that the quiver $\mathsf{Q}$ is determined completely by the anti-symmetric matrix $\mathsf{Q}_{i,j}$
under the assumptions of Fig.~\ref{fig_loops}.

\begin{figure}[t]
  \centering
  \includegraphics[scale=0.6]{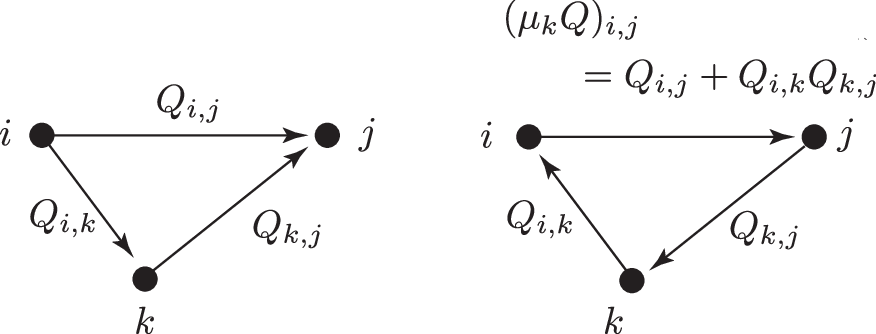}
  \caption{Quiver mutation \eqref{Qmutate} at vertex $k$ changes the quiver on the left into the one on the right.
    Here we have taken $\mathsf{Q}_{i,k}, \mathsf{Q}_{k,j}, \mathsf{Q}_{i,j}>0$.}
  \label{fig.quiver_mutation}
\end{figure}

For a vertex $k$ of the quiver,
we define the \keyword{mutation}{mutation}\index{quiver mutation@quiver mutation} $\mu_k$
at the vertex $k$ to be the operation of
replacing the quiver $\mathsf{Q}$ by another quiver $\mu_k\mathsf{Q}$,
where the new quiver $\mu_k\mathsf{Q}$ is defined by (Fig.~\ref{fig.quiver_mutation})
\begin{align}
  (\mu_k\mathsf{Q})_{i,j}:=\!
  \begin{cases}
    -\mathsf{Q}_{i,j} & \!\!\!\!\!\!\!\!\!\!\!\!\!\!\!\!\!\!\!\text{($i=k$ or $j=k$)} \ , \\
    \mathsf{Q}_{i,j}\! +\!  [\mathsf{Q}{}_{i,k}]_{+}[\mathsf{Q}_{k,j}]_{+} \!-\! [\mathsf{Q}_{j,k}]_+[\mathsf{Q}_{k,i}]_+
    \,       & \!\! \text{$(i,j\neq k$)}\ ,
  \end{cases}
  \label{Qmutate}
\end{align}
where we defined $[x]_+:=\textrm{max}(x,0)$.
\nomenclature{$\mu_k$}{quiver mutation at vertex $k$}

The most direct gauge-theoretic interpretation of the quiver mutation
\eqref{Qmutate} is found in the context of
2d $\mathcal{N}=(2,2)$ theories:\footnote{The Seiberg duality \cite{Seiberg:1994pq} of 4d $\mathcal{N}=1$ theories (see Chap.~\ref{chap.intro})
  also induces a quiver mutation; see exercise~\ref{ex.Seiberg}.
  However, these mutations are not the most general quiver mutations, since the
  ranks of the gauge groups at the quiver vertices are constrained by
  anomalies.}
when we regard the quiver as the defining data of a
2d $\mathcal{N}=(2,2)$ quiver gauge theory
(the definition here is similar to the case of 4d quiver gauge theories in
Chap.~\ref{chap.intro}),
the 2d $\mathcal{N}=(2,2)$ theories defined from the quiver and from its mutation
are dual, and flow to the same IR fixed point under the renormalization group flow \cite{Benini:2012ui,Benini:2014mia}.

\small
\subsection{Supplement: Physical Meaning of BPS Quivers}\label{subsec.BPS_quiver}

While this appendix basically summarizes mathematical formulations and results,
it is often useful to understand their physical significance when we apply the mathematics to physics.
For this purpose, we explain the physical interpretation of quivers and their mutations
in 4d $\mathcal{N}=2$ theories.

Let us consider 4d $\mathcal{N}=2$ theories, as also considered in this book, and
their Coulomb branches (see Sec.~\ref{subsec.Hitchin}).
Here let us study the spectrum of the theory;
namely, we consider how many stable particles with which charges exist.
In order to make the most of the constraints from $\mathcal{N}=2$ supersymmetry, let us consider in particular only particles preserving the maximal part (in the present case, half) of the $\mathcal{N}=2$ supersymmetry.
Such particles are called
\keyword{BPS particles}{BPS particle} (in the present case, more precisely, $1/2$-BPS particles).

A characteristic feature of BPS particles is that the mass of a BPS particle with charge $\gamma$ is determined by
the absolute value $\big| Z_{\gamma}  \big|$ of a complex number $Z_{\gamma}$ (the \keyword{central charge}{central charge}) determined from its charge.\footnote{
  The central charge also appeared in Sec.~\ref{subsec.SUSY_algebra}, where we considered 3d $\mathcal{N}=2$ theories. In general, central charges appear in the supersymmetry algebra for $\mathcal{N}>1$ supersymmetry. However,
  whether they are complex or real depends on the dimension and on the number of supersymmetries.}
On the other hand, the phase of $Z_{\gamma}$ specifies which $\mathcal{N}=1$ supersymmetry among the $\mathcal{N}=2$ supersymmetry is preserved. Moreover, $Z_{\gamma}$ is linear in $\gamma$: $Z_{\gamma_{1}+\gamma_2}=Z_{\gamma_1}+Z_{\gamma_2}, Z_{n \gamma}=n Z_{\gamma}$.

Since on the Coulomb branch there is the unbroken Abelian gauge group $U(1)^r$,
particles have charges $\gamma\in \Gamma$ under this gauge group.
Here $\Gamma$ is the set of all allowed charges, which has an anti-symmetric pairing, under which
the Dirac quantization condition is satisfied.

\begin{figure}[t]
  \centering\includegraphics[scale=0.24]{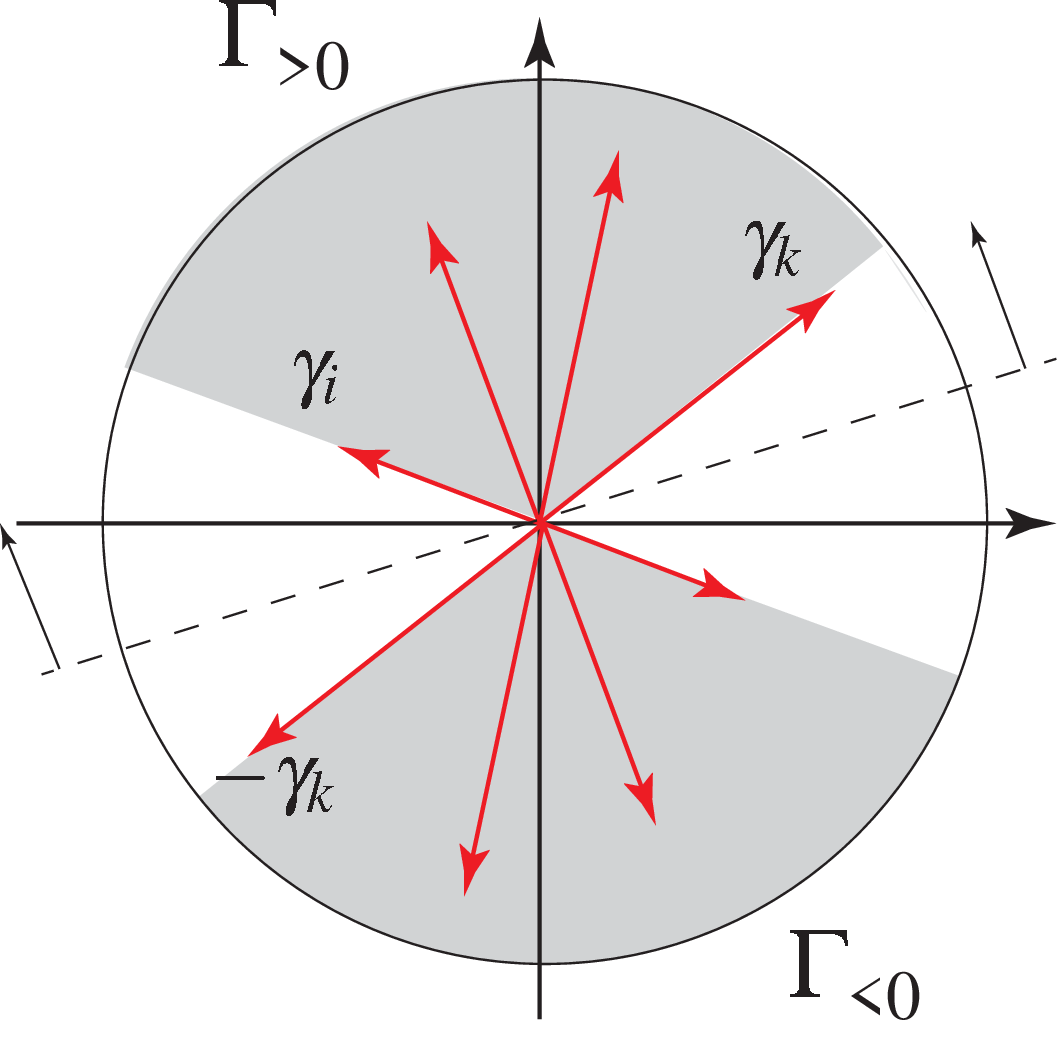}
  \caption{The set $\Gamma$ of charges is divided into positive and negative parts according to the argument of $Z_{\gamma}$.}
  \label{half_plane}
\end{figure}

Since we are starting from relativistic local field theories,
the CPT theorem holds; in particular,
if there is a particle with charge $\gamma$,
there must also be an anti-particle with the opposite charge $-\gamma$.
Therefore, when counting particles, it is sufficient to consider only one of the two,
and $\Gamma$ decomposes into a disjoint union of two parts:
\begin{align}
  \Gamma=\Gamma_{\rm >0} \cup \Gamma_{\rm <0} \ .
\end{align}
More practically,
we can collect those charges for which the complex number $Z_{\gamma}$ introduced above has
argument in $(\zeta, \zeta+\pi)$ on the complex plane, and call the set $\Gamma_{\rm >0}$ (Fig.~\ref{half_plane}):
\begin{align}
  \gamma\in \Gamma_{\rm >0} \longleftrightarrow \textrm{Im} (e^{- i\zeta} Z(\gamma))>0 \ .
  \label{positive}
\end{align}
Here we have fixed a certain argument $\zeta$. As we will comment further later,
note that the definition of $\Gamma_{\rm >0}$ explicitly depends on the value of the argument $\zeta$.

Let us now define the quiver.
Let $\{ \gamma_i \}$ be a basis generating $\Gamma_{\rm >0}$:\footnote{
  It is not clear whether such a basis exists. For example, the $\mathcal{N}=4$ theory does not have such a finite basis.}
\begin{align}
  \Gamma_{\rm >0} = \displaystyle\bigoplus_i \mathbb{Z}_{\ge 0} \gamma_i \ . \label{positive_span}
\end{align}
We then define the anti-symmetric matrix $\mathsf{Q}=\left(\mathsf{Q}_{i,j} \right)$, and hence the quiver, by
\begin{align}
  \mathsf{Q}_{i,j}:=\langle \gamma_i, \gamma_j \rangle \ .
  \label{bdef}
\end{align}
Let us call this quiver the \keyword{BPS quiver}{BPS quiver}.
Note that, since the pairing was anti-symmetric as already stated, $\mathsf{Q}_{i,j}$ also gives
an anti-symmetric matrix. Moreover, that $\mathsf{Q}_{i,j}$ are
integers is a consequence of the Dirac quantization condition.

The definition \eqref{bdef} is somewhat top-down, but there is a proper physical meaning behind it.
A BPS particle is a particle existing in a 4d $\mathcal{N}=2$ supersymmetric gauge theory (with eight supercharges), but it itself preserves half of the supersymmetry. Therefore, from the standpoint of someone sitting on the particle, it looks as if a supersymmetric quantum mechanics with $\mathcal{N}=4$ appears (the field theory for a particle is quantum mechanics).
The quiver defined above gives the defining data of this supersymmetric quantum mechanics (\textbf{quiver supersymmetric quantum mechanics}).

Concretely, this goes as follows: each vertex of the quiver corresponds to one element $\gamma_i$ of the basis defined above for the charges.
When we consider the charge $\gamma=\sum_i n_i \gamma_i$,
we consider the gauge group $U(n_i)$ at the $i$-th vertex, and associate with an edge from vertex $i$ to $j$
a bifundamental field transforming as $(n_i, \bar{n}_j)$ under $U(n_i) \times U(n_j)$ (see Sec.~\ref{subsec.repeat}).\footnote{To be precise, we also need to specify interactions by a superpotential.}\footnote{The vacuum moduli space of the supersymmetric quantum mechanics is mathematically the moduli space of stable representations of the quiver with potential,
  and is closely related to quiver varieties.
} The number of fields in the bifundamental representation is given by the quiver:
\begin{align}
  \mathsf{Q}_{i,j}:=\langle \gamma_i, \gamma_j  \rangle  \ .
  \label{Q_from_gamma}
\end{align}
This shows which particles with which charges exist in the 4d theory,
and gives the directed graph called the BPS quiver.

For the 4d $\scN=2$ theories associated with a Riemann surface $\Sigma$,
the BPS quiver is believed to coincide with the quiver constructed from an ideal triangulation of $\Sigma$
as in Fig.~\ref{nee}.\footnote{The quivers constructed in this way give only finitely many quivers even under repeated mutations. Quivers satisfying this condition are called quivers of finite mutation type, and
  their mathematical classification is known. For a gauge-theoretic interpretation of this condition, see
  Ref.~\cite{Cecotti:2011rv}.} Moreover, the Fock variables were associated with the vertices of the quiver;
from the present discussion they carry charges of the 4d $\mathcal{N}=2$ theory,
which matches nicely with the discussion in Sec.~\ref{sec.length_twist} (identifying Fock variables with expectation values of loop operators) \cite{Gaiotto:2009hg}.

\bigskip

\begin{figure}[t]
  \centering\includegraphics[scale=0.24]{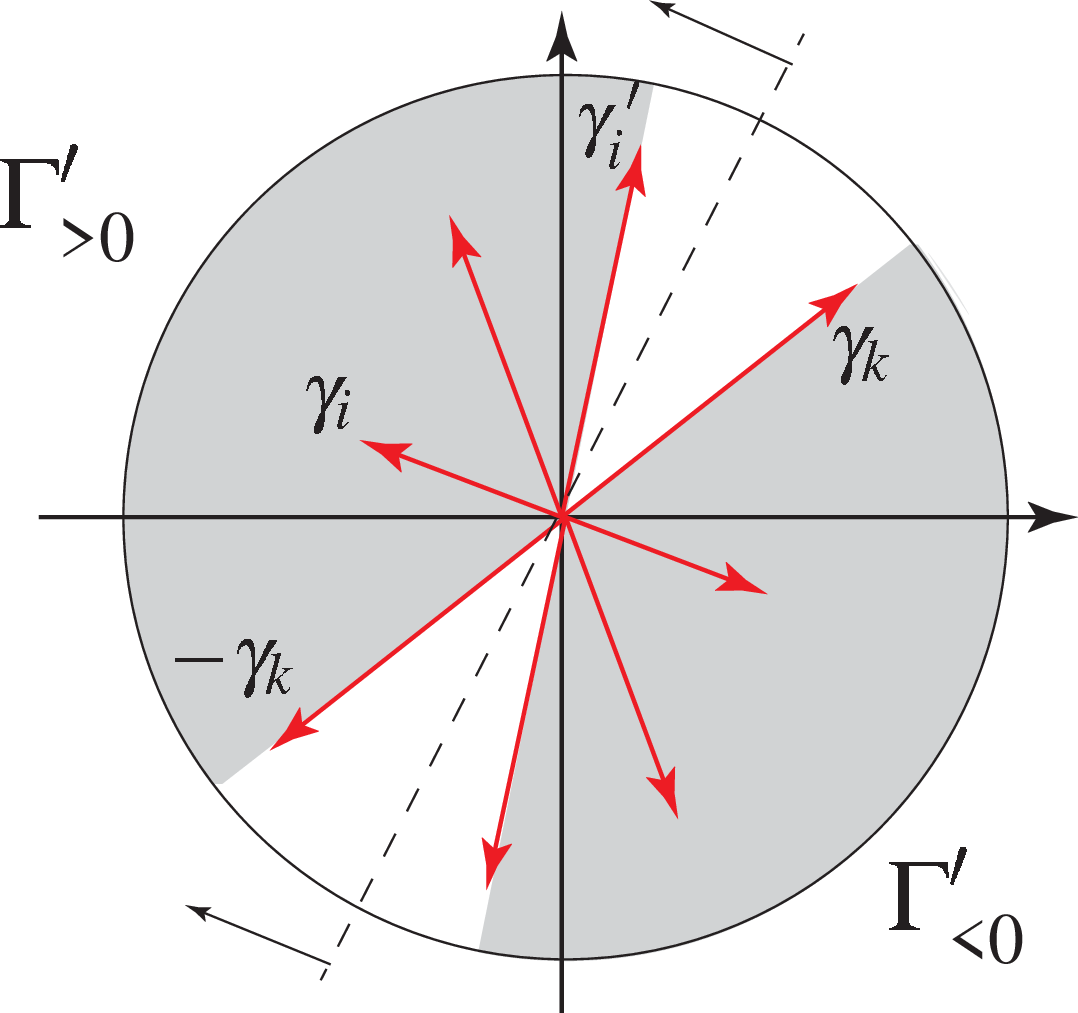}
  \caption{By changing the value of $\zeta$, the division of charges into positive and negative ones changes.
    In this case, we also need to change the basis of $\Gamma_{\rm >0}$.}
  \label{half_plane_change}
\end{figure}

We have explained quivers and their physical meaning, but
there is an unsatisfactory point in the explanation so far:
the division into positive and negative charges is not unique.
This is also clear from the fact that \eqref{positive} depends on the argument $\zeta$.
In particular, a change occurs when $\zeta$ varies across the argument of a certain element $\gamma_k$ of the basis:
$\Gamma_{\rm >0}$, and hence the quiver, then changes (Fig.~\ref{half_plane_change}).

Suppose that $\gamma_k$, which had positive charge until then,
now has negative charge.
Then, clearly, we have to add $-\gamma_k$ to the new basis:
\begin{align}
  \gamma_k\to -\gamma_k \ .
  \label{gchange_1}
\end{align}
However, since we need to preserve \eqref{positive_span},
we also need to change the other elements of the basis. The result is known to be
\begin{align}
  \gamma_i \to \gamma_i + \left[\mathsf{Q}_{k,i} \right]_{+}  \gamma_k \ .
  \label{gchange_2}
\end{align}
The transformation rules \eqref{gchange_1} and \eqref{gchange_2}
define the so-called \keyword{$c$-vector}{c-vector}, where physically the $c$-vectors are charge vectors.\footnote{I myself do not know why it is called a $c$-vector; it is probably a coincidence
  that the first letter $c$ of the word ``charge'' coincides with the
  $c$ in $c$-vector.}
From this transformation rule and \eqref{Q_from_gamma},
we can immediately derive the transformation rule \eqref{Qmutate} of quivers stated earlier (exercise~\ref{ex.cvector_mutation}).

We do not prove \eqref{gchange_2} here; see, for example, section 3.1.2 of Ref.~\cite{Alim:2011kw} from the viewpoint of the representation theory of quivers.

The discussion in this section has been hasty, and we had no room to explain all the details;
however, we hope the reader could at least feel that the various pieces fit together nicely.

\normalsize

\section{Cluster $x$ and $y$ Variables and Their Quantizations}
\subsection{Classical Theory}

In quantum \Teichmuller theory, in addition to the ideal triangulation,
another essential ingredient was the action of the mapping class group.
The substitute for this action in a
general cluster algebra is a sequence of mutations.

Let us consider a mutation sequence
$\mu_{\bm{k}}=(\mu_{k_1}, \ldots, \mu_{k_L})$
defined from a sequence of vertices
$\bm{k}=(k_1, \ldots, k_L)$.
This represents a ``time evolution'' of the quiver,
where the quiver at ``time'' $t$ is given by
\begin{align}
  \mathsf{Q}(t):=\mu_{k_t} \mu_{k_{t-1}}\ldots \mu_{k_1} \mathsf{Q}\ ,
  \quad
  \mathsf{Q}(0):=\mathsf{Q} \ .
\end{align}

Let us define cluster variables $x(t)$ and $y(t)$ recursively,
starting with the initial values $x_i(t=0):=x_i$ and $y_i(t=0):=y_i$,
and then defining (with $k:=k_{t+1}$)
\begin{align}
  x_i(t+1):=\begin{cases}
              \dfrac{\prod_{j\in I} x_j(t)^{\left[\mathsf{Q}(t)_{k,j}\right]_{+} } +\prod_{j\in I} x_j(t)^{\left[\mathsf{Q}(t)_{j,k}\right]_+ }}{x_k(t)} & (i=k) \ , \\
              x_i(t)                                                                                                                   & (i\ne k)
            \end{cases}
  \label{cluster_x_def}
\end{align}
\nomenclature{$x_i(t)$}{(classical) $x$-variable}
and\footnote{For the case $i\ne k$ in \eqref{cluster_y_def},
  we can also write
  \begin{align}
    y_i(t+1):=\begin{cases}
                y_i(t) \,  (1+y_k(t)^{-1})^{-|\mathsf{Q}(t)_{k,i}|} & (\mathsf{Q}(t)_{k,i}\ge 0) \ , \\
                y_i(t) \,  (1+y_k(t))^{|\mathsf{Q}(t)_{k,i}|}       & (\mathsf{Q}(t)_{k,i} <0) \ .
              \end{cases}
    \label{cluster_y_def_explicit}
  \end{align}
}
\begin{align}
  y_i(t+1):=\begin{cases}
              y_k(t)^{-1}                                                             & (i=k) \ , \\
              y_i(t) \, y_k(t)^{\left[\mathsf{Q}(t)_{k,i}\right]_{+}} (1+y_k(t))^{-\mathsf{Q}(t)_{k,i}} & (i\ne k)  \\
            \end{cases} \ .
  \label{cluster_y_def}
\end{align}\nomenclature{$y_i(t)$}{(classical) $y$-variable}

For a given quiver, the sets of $x_i(t)$'s and $y_i(t)$'s
obtained by repeated applications of quiver mutations are called
\keyword{cluster $x$-variables}{cluster x-variable}
and \keyword{cluster $y$-variables}{cluster y-variable}, respectively.
The cluster algebra is defined as the algebra generated by the former.

There is also a convention in which the $x$-variables are called $\scA$-variables and
the $y$-variables $\scX$-variables \cite{FockGoncharovHigher};
note in particular that the $x$-variables are not the same as the $\scX$-variables.\footnote{We need extra care in spoken language,
  where it is hard to distinguish between $x$ and $\scX$.} Historically, the $y$-variables are also sometimes called the \keyword{coefficients}{coefficient} of cluster variables.

Note that in the \Teichmuller theory of Chap.~\ref{chap.Teichmuller}
the cluster $x$-variables are the Penner variables,
and the cluster $y$-variables the Fock variables.

\subsection{Quantum Theory}

Let us next consider the quantization of cluster variables \cite{FockGoncharovEnsembles,FockGoncharovQuantumCluster}.
There is more than one quantization known in the literature,
and they are not equivalent. In the following we use the version
where only the $y$-variables are quantized.

In order to quantize the cluster $y$-variables,
we replace the variables by operators
(\keyword{quantum cluster $y$-variables}{quantum cluster y-variable}) $\{\sfx_i\}_{i\in I}$, and
impose the commutation relations
\begin{align}
  \sfx_j \sfx_i=q^{2\mathsf{Q}_{i,j}} \sfx_i
  \sfx_j \ .
  \label{xCCR}
\end{align}
\nomenclature{$\sfx_i$}{quantum cluster $y$-variable}
The algebra generated by these $\{\sfx_i\}_{i\in I}$
(or, if necessary, its suitable completion)
will be denoted by $\scA_\mathsf{Q}$.

In terms of the variables defined by $\sfx_i=e^{\sfY_i}$,
the commutation relations \eqref{xCCR} read
\begin{equation}
  [\sfY_j, \sfY_i]= 2 i \hbar \,  \mathsf{Q}_{i, j} \ , \quad q=e^{i \hbar}   \ .
  \label{xCCR_2}
\end{equation}
If we regard the anti-symmetric matrix
$\mathsf{Q}_{i,j}$ as (in general degenerate) Poisson brackets in classical dynamics,
then \eqref{xCCR} is the standard quantization of the
phase space of a classical mechanical system with
finitely many degrees of freedom,
and hence the system can easily be quantized
once we choose a polarization, i.e.\ choose a
set of canonically conjugate pairs of coordinates and momenta.
This gives an explicit construction of the Hilbert space $\scH_\mathsf{Q}$,
and the algebra $\scA_\mathsf{Q}$ of operators on $\scH_\mathsf{Q}$.

In the commutation relation \eqref{xCCR}, $q$ is a quantization parameter, the exponential of the ``Planck constant''.
In applications to
3d $\scN=2$ theories, this is related to the deformation parameter
$b$ of $S^3_b$ \eqref{S3bmetric} by the following relation:
\beq
q=e^{ i \pi b^2 } \ .
\label{qb}
\eeq
The semiclassical limit, or the dimensional reduction to 2d,
is given by $q\to 1$ (i.e.\ $b\to 0$).

As in the discussion in Sec.~\ref{subsec.qTeich}, it is convenient to consider
the natural linear extensions of $\hat{y}_i$ and $\hat{Y}_i$. The discussion in Sec.~\ref{subsec.qTeich}
can be used as it is, by replacing $\hat{Z}_i$ and $\hat{z}_i$ with $\hat{Y}_i$ and $\hat{y}_i$.

Let us write this explicitly.
Let us choose a basis $\gamma_i$, which spans an element $\gamma$ of the charge lattice as $\gamma=\sum_i n_i \gamma_i$.
We define $\hat{Y}_{n_1 \gamma_1+n_2 \gamma_2}=n_1 \hat{Y}_{\gamma_1}+n_2 \hat{Y}_{\gamma_2}, \hat{Y}_{\gamma_i}:=\hat{Y}_i$.
Let us define the pairing (Dirac-Schwinger-Zwanziger (DSZ) pairing)
between two elements $\gamma=\sum_i m_i \gamma_i, \gamma'=\sum_i n_i \gamma_i$ by
\begin{align}
  \langle \gamma, \gamma' \rangle :=\sum_{i,j} m_i n_j \mathsf{Q}_{i,j} \ .
  \label{Dirac_pairing_cluster}
\end{align}
Then the commutation relations read
\begin{align}
  \left[ \hat{Y}_{\gamma}, \hat{Y}_{\gamma'} \right]= -2\pi b^2 i  \, \langle \gamma, \gamma' \rangle
  \label{omega_q_linear_cluster}
\end{align}
and, when exponentiated, they define the quantum torus algebra $\mathcal{A}_\mathsf{Q}$:
\begin{align}
  \mathcal{A}_\mathsf{Q}: \hat{y}_{\gamma} \hat{y}_{\gamma'} = q^{2\langle \gamma', \gamma \rangle} \hat{y}_{\gamma'} \hat{y}_{\gamma}  \ .
  \label{qTorus_linear_cluster}
\end{align}

\subsection{Quantized Mutations}

A mutation will also be replaced by an operator
acting on the algebra $\scA_\mathsf{Q}$.
The action of this operator gives the map
\begin{align}
  \hat{\mu}_k : \quad \scA_{\mathsf{Q}}\to \scA_{\mu_k \mathsf{Q}} \ ,
\end{align}
and, more concretely, the action of the operator $\hat{\mu}_k$ on $\sfx_i$ is given, for $i\ne k$, by\footnote{
  This can also be written as
  \begin{align}
     & \hat{\mu}_k (\sfx_i):=
    \begin{cases}
      \displaystyle
      \sfx_i
      \prod_{m=1}^{|\mathsf{Q}_{k,i}|}
      \big(1+ q^{2m-1}
      \sfx_k{}^{-1}\big)^{-1}
       & \mathsf{Q}_{k,i} \ge 0 \ , \\
      \displaystyle
      \sfx_i
      \prod_{m=1}^{|\mathsf{Q}_{k,i}|}
      \big(1+ q^{2m-1}
      \sfx_k\big)
       & \mathsf{Q}_{k,i} <0 \ .
    \end{cases}
  \end{align}
}
\begin{align}
  \begin{split}
     & \hat{\mu}_k (\sfx_i): \\
     & \quad = \displaystyle
    q^{
        \mathsf{Q}_{i,k}\left[\mathsf{Q}_{k,i}\right]_+
      }
    \sfx_i \sfx_k{}^{[\mathsf{Q}_{k,i}]_+}
    \prod_{m=1}^{|\mathsf{Q}_{k,i}|}
    \big(1+ q^{-\mathrm{sgn}(\mathsf{Q}_{k,i})(2m-1)}
    \sfx_k\big)^{-\mathrm{sgn}({\mathsf{Q}_{k,i}})}
    \\
     & \quad =\displaystyle
    q^{
        \mathsf{Q}_{i,k}\left[-\mathsf{Q}_{k,i}\right]_+
      }
    \sfx_i \sfx_k{}^{[-\mathsf{Q}_{k,i}]_+}
    \prod_{m=1}^{|\mathsf{Q}_{k,i}|}
    \big(1+ q^{\mathrm{sgn}(\mathsf{Q}_{k,i})(2m-1)}
    \sfx_k{}^{-1}\big)^{-\mathrm{sgn}({\mathsf{Q}_{k,i}})} \ ,
    \label{ymutation1}
  \end{split}
\end{align}
and, for $i=k$, by
\begin{align}
  \hat{\mu}_k (\sfx_k) :=\displaystyle
                         {\sfx_k}{}^{-1} \ ;
  \label{ymutation2}
\end{align}
this can be naturally extended to an action on the whole of $\scA_\mathsf{Q}$.\footnote{
  Care is needed since the convention here is different from, e.g., that of Ref.~\cite{Yamazaki_science_2015}.
  In some references, what we denote by $q^2$ here is defined to be $q$.}

We can explicitly verify that the
$\sfx'_i$ thus defined
satisfy the commutation relations
\eqref{xCCR}
for the new quiver $\mu_k \mathsf{Q}$ (exercise~\ref{q_mutation_consistency}).
This means that $\hat{\mu}_k$ indeed gives a map from
$\scA_\mathsf{Q}$ to $\scA_{\mu_k \mathsf{Q}}$.

\bigskip

The transformation rules
\eqref{ymutation1} and \eqref{ymutation2} look complicated,
and their meanings are easier to see when they are decomposed into two, as follows.

The first transformation is
\begin{align}
  \hat{c}_k:
  \begin{array}{l}
    \hat{y}_{\gamma_k} \to \hat{y}_{-\gamma_k}=\hat{y}_{\gamma_k}^{-1} \ , \\
    \hat{y}_{\gamma_i} \to \hat{y}_{\gamma_i+ [\mathsf{Q}_{k,i}]_{+} \gamma_k} \quad (i\ne k) \ .
  \end{array}
  \label{Xchange_1}
\end{align}
This transformation follows the transformation rules \eqref{gchange_1} and \eqref{gchange_2} of the charge lattice,
and coincides with the transformation rules of the
tropical $y$-variables defined in the next subsection.
Indeed, when $\hat{y}_{\gamma_i}$ satisfy \eqref{xCCR_2}, one can check that the $\hat{y}_{\gamma}$ after this transformation satisfy \eqref{xCCR_2} for the new quiver $\mathsf{Q}'=\mu_k(\mathsf{Q})$. For this, we use
\begin{align}
  \hat{y}_{\gamma_i+ [\mathsf{Q}_{k,i}]_{+} \gamma_k}
  =q^{[\mathsf{Q}_{k,i}]_{+} \mathsf{Q}_{i,k}}  \hat{y}_{i}   \hat{y}_{k}^{[\mathsf{Q}_{k,i}]_{+} } \ ,
\end{align}
which follows from \eqref{qTorus_linear_cluster}.

The next transformation is the conjugation by the quantum dilogarithm function:\footnote{
  Here we used, in the definition of $\hat{K}_{k}$, the quantum dilogarithm function satisfying $e_b(x)=e_{1/b}(x)$;
  this is not necessary, and, for example, using
  \begin{align}
    \begin{split}
       & \Psi_q(x;q):=(-q x; q^2)_{\infty} \ ,    \\
       & \Psi_q(q^2 x;q)=(1+q x)^{-1} \Psi_q(x;q)
    \end{split}
  \end{align}
  (see \eqref{PochRecursion}), the transformation
  \begin{align}
    \hat{K}_{k}: \hat{y}_{\gamma_i} \to \Psi_q(\hat{y}_{\gamma_k})^{-1} \hat{y}_{\gamma_i} \Psi_q(\hat{y}_{\gamma_k})
    \label{Xchange_3}
  \end{align}
  gives the same effect. Therefore, for example, the quantum dilogarithm identities discussed in the next subsection also
  hold with $e_b\left(\frac{\hat{Y}}{2\pi b}\right)$ replaced by $\Psi_q(\hat{y})$ ($\hat{y}=e^{\hat{Y}}$).
}
\begin{align}
  \hat{K}_{k}: \hat{y}_{\gamma_i} \to e_b\left(\frac{\hat{Y}_{\gamma_k}}{2\pi b}\right)^{-1} \hat{y}_{\gamma_i} \, e_b\left(\frac{\hat{Y}_{\gamma_k}}{2\pi b}\right) \ .
  \label{Xchange_2}
\end{align}
Here $e_b(x)$ is the quantum dilogarithm function defined in Appendix~\ref{app.dilog},
which in particular satisfies the difference equation \eqref{recursion_eb}
\begin{align}
  \begin{split}
    e_b(x+ib)=(1+q e^{2\pi b x})^{-1} e_b(x) \ .
  \end{split}
\end{align}

We can directly verify that the composition of the two transformations
$\hat{c}_k$ and $\hat{K}_{k}$ gives the quantized mutation defined in
\eqref{ymutation1} and \eqref{ymutation2}:
\begin{align}
  \hat{\mu}_k := \hat{K}_k \hat{c}_k \ .
  \label{mu_decompose}
\end{align}

So far we have considered only a single quiver mutation.
We can also consider a sequence of repeated quiver mutations.
These change the quiver
$\mathsf{Q}$ into another quiver $\mathsf{Q}'$,
\begin{align}
  \mathsf{Q}'=\mu_{k_L} \cdots \mu_{k_1}(\mathsf{Q}) \ ,
\end{align}
and the composition of the operators
$\hat{\mu}_k$ defines a map from the
quantum torus of $\mathsf{Q}$ to that of $\mathsf{Q}'$:
\begin{align}
  \hat{\mu}_{k_L}\cdots \hat{\mu}_{k_1}: \scA_{\mathsf{Q}(0)} \to \scA_{\mathsf{Q}(L)} \ .
  \label{mutation_sequence_map}
\end{align}

\subsection{Quantum Dilogarithm Identities}

For a given quiver $\mathsf{Q}$,
it sometimes happens that two different mutation sequences $\bm{k}=(k_1, \cdots, k_L)$
and $\bm{l}=(l_1, \cdots, l_{L'})$ generate the same quiver $\mathsf{Q}'$ and the
same set of $y$-variables, up to a relabeling of the quiver vertices.
In this case, the maps \eqref{mutation_sequence_map} associated with
$\mu_{\bm{k}}$ and $\mu_{\bm{l}}$ should coincide.
What holds in this case is the \keyword{quantum dilogarithm identity}{quantum dilogarithm identity} \cite{KellerOn,Reineke_Poisson,Kashaev:2011se}.

\small
Quantum dilogarithm identities are interesting identities appearing in various contexts of
mathematical physics.
For example, they appear in the \keyword{Kontsevich-Soibelman formula}{Kontsevich-Soibelman formula} \cite{KontsevichSoibelman}, which governs the counting of BPS states of 4d $\scN=2$ theories, or, more mathematically, the
\keyword{wall crossing phenomenon}{wall crossing phenomenon} of (generalized) \keyword{Donaldson-Thomas invariants}{Donaldson-Thomas invariant}. That the same quantum dilogarithm function appears both in the counting of BPS states of 4d $\scN=2$ theories and in the
$S^3$ partition functions of 3d $\mathcal{N}=2$ theories is not a coincidence \cite{Cecotti:2010fi,Terashima:2013fg}.

\normalsize

The claim that $\bm{k}$ and $\bm{l}$ give the same quiver and $y$-variables is
equivalent to the claim that the new sequence $\bm{l}^{-1}\bm{k}$, obtained by first performing $\bm{k}$ and then the inverse $\bm{l}^{-1}$ of $\bm{l}$,
keeps the quiver and the $y$-variables trivially invariant (up to a permutation of the elements of $I$); in the following we therefore use the assumption in this form.

Let us here write down the example of the quantum dilogarithm identity associated with the
$A_2$ quiver. More general cases are notationally somewhat cumbersome, so we do not state them explicitly here;
we have tried to explain things so that the general rule becomes clear.

Let us consider the $A_2$ quiver
\begin{align}
  1\longleftarrow 2 \ .
\end{align}
There are two independent quantum $y$-variables $\sfx_1, \sfx_2$, corresponding to the two vertices,
satisfying $\sfx_1\sfx_2 = q^2 \sfx_2 \sfx_1$.

\begin{figure}[t]
  \centering\includegraphics[scale=0.45]{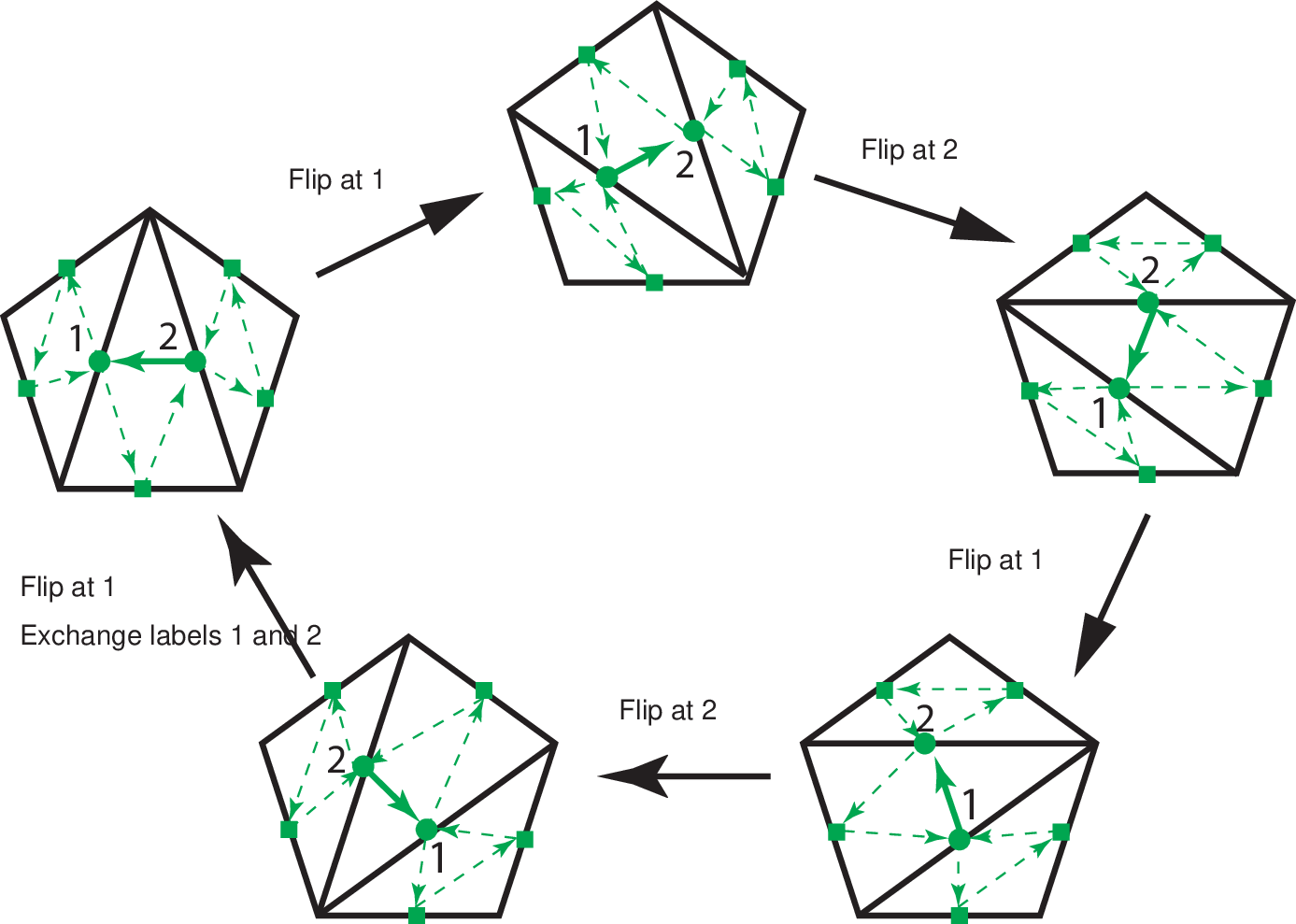}
  \caption{Change of the quiver under the pentagon.
    The two internal edges correspond to the two vertices of the $A_2$ quiver,
    and the pentagon corresponds to five mutations.
    The quantum dilogarithm identity associated with this mutation sequence is nothing but the
    pentagon identity for the quantum dilogarithm.}
  \label{pentagon_id_SGC}
\end{figure}

We consider the mutation sequence at the vertices $(1,2,1,2,1)$.
This in fact corresponds to the pentagon satisfied by flips. Fig.~\ref{pentagon_id_SGC} shows the quivers
determined, following the rule of Fig.~\ref{nee}, for Fig.~\ref{fig.pentagon} of Sec.~\ref{subsec.flip_transf}.
If we draw only the two vertices corresponding to the internal edges of the pentagon and
the arrows between them, we obtain the $A_2$ quiver and its
mutations at the vertices $(1,2,1,2,1)$.\footnote{The quiver vertices corresponding to the five edges of the pentagon, and the edges connected to them, are not used here. In general, such quiver vertices at which no mutation is performed are called \keyword{frozen vertices}{frozen vertex}.}

We can compute the $y$-variables from
\eqref{ymutation1} and \eqref{ymutation2}:
\begin{align}
  \begin{split}
    \hat{y}(1) & = (\sfx_1, \sfx_2) \ ,                                                    \\
    \hat{y}(2) & =(\sfx_1^{-1},\sfx_2(1+q\sfx_1) ) \ ,                                        \\
    \hat{y}(3) & =(\sfx_1\inv(1+q\sfx_2+\sfx_1\sfx_2), \sfx_2\inv(1+q\inv \sfx_1)\inv) \ ,             \\
    \hat{y}(4) & =((1+\! q\sfx_2+\! \sfx_1\sfx_2)\inv \sfx_1,q\inv \sfx_1\inv \sfx_2\inv(1+\!q\sfx_2)) \ , \\
    \hat{y}(5) & =(\sfx_2\inv,q\inv \sfx_1\sfx_2(1+q\inv \sfx_2)\inv) \ ,                        \\
    \hat{y}(6) & =(\sfx_2, \sfx_1)  \ .
  \end{split}
  \label{Y_A2}
\end{align}
Note that, although the $y$-variables undergo complicated transformations, in the end they come back to
the original $y$-variables, up to the exchange of $\sfx_1$ and $\sfx_2$. This is what we wanted to assume.

Now, for the quantum $y$-variables thus obtained and their mutations,
let us consider the operation
\begin{align}
  1+\textrm{(sum of monomials with positive powers of $y$)} \longrightarrow 1 \ .
\end{align}
For example, in this case $\sfx_1 \inv(1+q\sfx_2+\sfx_1\sfx_2)\to \sfx_1^{-1}$.
This operation is called \keyword{tropicalization}{tropicalization},
and what is obtained from (quantum) $y$-variables by tropicalization is called the \keyword{tropical $y$-variables}{tropical y-variable}, which we denote here by $\overline{y}$.

Such a tropicalization applied to \eqref{Y_A2} gives
\begin{align}
  \begin{split}
    \hat{y}(1) & \longrightarrow \overline{y}(1)=  (\underline{\sfx_1}, \, \sfx_2) \ ,                                           \\
    \hat{y}(2) & \longrightarrow  \overline{y}(2)=  (\sfx_1^{-1},\, \underline{\sfx_2}) \ ,                                      \\
    \hat{y}(3) & \longrightarrow  \overline{y}(3)= (\underline{\sfx_1^{-1}}, \, \sfx_2^{-1} )\ ,                                      \\
    \hat{y}(4) & \longrightarrow  \overline{y}(4)= (\sfx_1,\,  \underline{q\inv \sfx_1\inv \sfx_2^{-1}=\hat{y}_{-\gamma_1-\gamma_2}}) \ , \\
    \hat{y}(5) & \longrightarrow  \overline{y}(5)= (\underline{\sfx_2}\inv,\, q\inv \sfx_1\sfx_2=\hat{y}_{\gamma_1+\gamma_2}) \ ,         \\
    \hat{y}(6) & \longrightarrow  \overline{y}(6)= (\sfx_2, \, \sfx_1)  \ .
  \end{split}
\end{align}
Here the underlined ones are the
tropical $y$-variables corresponding to the vertices to be mutated. Reading off the charges of these variables
(namely, writing them as $\hat{y}_{\sum_i (c_t)_i \gamma_i}$ and reading off $(c_t)_i$), we obtain
\begin{align}
  \begin{array}{lll}
    c_1 & =(1,0) \ ,   & \epsilon_1=+ \ ,  \\
    c_2 & =(0,1) \ ,   & \epsilon_2=+  \ , \\
    c_3 & =(-1,0) \ ,  & \epsilon_3=-  \ , \\
    c_4 & =(-1,-1) \ , & \epsilon_4=-  \ , \\
    c_5 & =(0,-1) \ ,  & \epsilon_5=-  \ . \\
  \end{array}
\end{align}
The $c_t$ obtained in this way are what are called $c$-vectors, and their transformation rules agree with the transformation rules of the charges introduced in Sec.~\ref{subsec.BPS_quiver}.
Moreover, interestingly, all the non-zero components of $c_t$ have the same sign,
which we denoted by $\epsilon_t$ (this sign is called the \keyword{tropical sign}{tropical sign}).

The quantum dilogarithm identity claims that the product of the operators
$e_b\left(\frac{Y_{c_t \epsilon_t} }{2\pi b}\right)^{\epsilon_t}$,
ordered from right to left following the time ordering (i.e.\ the operator for $t=1$ is placed rightmost), is the identity.
In our case, we have
\begin{align}
  e_b\!\left(\frac{Y_2}{2\pi b}\right)^{-1}
  e_b\!\left(\frac{Y_1+Y_2}{2\pi b}\right)^{-1}
  e_b\!\left(\frac{Y_1}{2\pi b}\right)^{-1}
  e_b\!\left(\frac{Y_2}{2\pi b}\right)
  e_b\!\left(\frac{Y_1}{2\pi b}\right) =1 \ .
\end{align}
In terms of $\sfP:=\frac{Y_1}{2\pi b}, \sfQ:=\frac{Y_2}{2\pi b}$
we have $[\sfP,\sfQ]=-\frac{1}{2\pi i}$ (from \eqref{omega_q_linear_cluster} with $\mathsf{Q}_{1,2}=-1$), and the identity above becomes
\begin{align}
  e_b\left(\sfQ\right)
  e_b\left(\sfP\right)
  =
  e_b\left(\sfP\right)
  e_b\left(\sfP+\sfQ\right)
  e_b\left(\sfQ\right) \ .
\end{align}
Since $[\sfQ, \sfP]=\frac{1}{2\pi i}$, this is our pentagon identity \eqref{ebpentagon} for the quantum dilogarithm, with the roles of $\sfP$ and $\sfQ$ exchanged.

\small

It is actually not difficult to prove quantum dilogarithm identities once we have the formalism of cluster algebras. Let us here comment only on the strategy, leaving the details to interested readers.

First, since the two mutation sequences give the same quantum $y$-variables, and
each mutation was written in terms of $\hat{\mu}_k$,
an identity for the products of the $\hat{\mu}_{k}$'s holds (here we used the irreducibility of the representation (for generic values of $b$) and Schur's lemma).
Each map $\hat{\mu}_{\bm{k}}$ was,
as in \eqref{mu_decompose},
written in terms of the linear transformations $\hat{c}_k$ of the basis $\hat{Y}_i$ and
the actions $\hat{K}_k$ by quantum dilogarithm functions:
\begin{align}
  \cdots \hat{K}_{k_2} \hat{c}_{k_2} \hat{K}_{k_1} \hat{c}_{k_1} \ .
\end{align}
The operators $\hat{K}$ and $\hat{c}$ now alternate; however,
since we know that conjugation by $\hat{c}_k$ induces linear transformations,
we can collect all of them to the right (or to the left)\footnote{What we described in the main text is the form of the quantum dilogarithm identity called the \keyword{``tropical form''}{tropical form}; by collecting the $\hat{c}_k$ to the right or to the left, with the left and right
  reversed, we can obtain another (but equivalent) quantum dilogarithm identity called the \keyword{``universal form''}{universal form} \cite{Kashaev:2011se}.} (this is sometimes called the \keyword{shuffling argument}{shuffling argument}\footnote{This is a simple argument but has various applications;
  for example, when the quantum parameter $q$ is a root of unity, the same argument can be used to derive
  root-of-unity versions of quantum dilogarithm identities \cite{Ip:2014pva}.}).
Moreover, as can be seen from the discussion above,
$\hat{c}$ induces the transformations of tropicalized quantum $y$-variables;
since, if the quantum $y$-variables agree, the tropicalized quantum $y$-variables
trivially agree as well, the $\hat{c}$'s cancel among themselves.
Therefore, the remaining product of the $\hat{K}$'s must also cancel by itself.
This is the quantum dilogarithm identity.

\normalsize

\begin{practice}

  \item $[\bll]$ (Examples of mutations)
  The definition of a mutation of a quiver, while it
  looks complicated at first,
  is not really complicated, and after
  a bit of familiarity you will be able to mutate a quiver directly on pictures
  of the quiver. For complicated quivers it is useful to
  use the Java applet by B.~Keller,
  which can be downloaded from his web page
  \url{https://webusers.imj-prg.fr/~bernhard.keller/quivermutation/}.\footnote{I am tempted to add that this web page also lists unsolved problems,
    and if you become the first person to solve any of them you will be awarded a
    prize of 20 Euros. Ambitious readers should definitely look into some of these problems.}

  Draw a simple quiver of your choice, choose a vertex, and mutate the quiver at that vertex.
  Repeat the process several times, and you will find rather intricate patterns.
  For example, what happens in general when we change the order of two mutations?
  For a quiver associated with an ideal triangulation of the four-punctured sphere,
  how many quivers can you get if you mutate many times?\footnote{One rather non-trivial
    fact in the theory of cluster algebras is that we obtain only finitely many quivers in this example.
    Such a quiver is said to be of \keyword{finite mutation type}{finite mutation type}, and such quivers are classified (up to equivalence under mutations) \cite{FST_finite}.
  }

  \item $[\bll]$ (Mutation and flip)

  \begin{enumerate}
    \item A flip of an ideal triangulation of a 2d surface
          changes the quiver (determined as in Fig.~\ref{nee}) as in Fig.~\ref{trig_mutation}.
          Show that this coincides with a mutation of the quiver
          at the quiver vertex corresponding to the diagonal of the quadrilateral.

          \begin{figure}[t]
            \centering\includegraphics[scale=0.4]{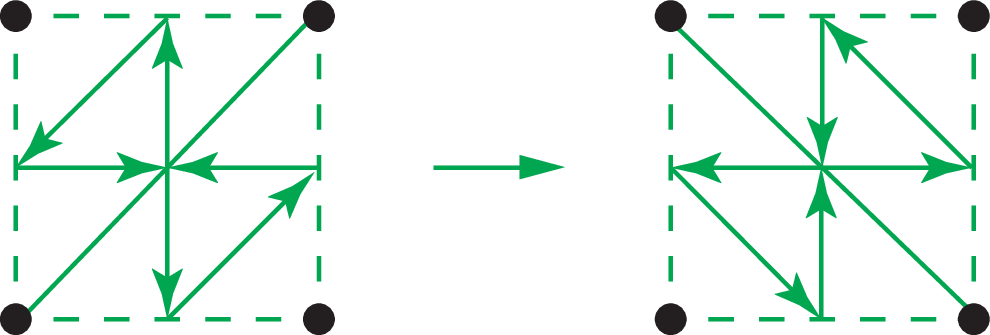}
            \caption{A flip of the ideal triangulation induces a mutation of the corresponding quiver.}
            \label{trig_mutation}
          \end{figure}

    \item Let us repeat the exercise for the $A_N$-type quivers associated with ideal triangulations of surfaces (Fig.~\ref{fig.higher}, generalizing those for $N=2$ in Fig.~\ref{nee}).
          Can we realize a flip of the ideal triangulation by a suitable mutation sequence? If so, how many mutations do we need at minimum?
  \end{enumerate}

  \item $[\bll]$ ($c$-vectors and mutation)

  Derive the quiver transformation rule \eqref{Qmutate}
  from the charge transformation rules \eqref{gchange_1}, \eqref{gchange_2}
  and the relation \eqref{bdef}.
  \label{ex.cvector_mutation}

  \item $[\bll]$ (Mutation and Seiberg duality)

  Convince yourself that the matter content of Seiberg duality
  (discussed in Sec.~\ref{subsec.Seiberg})
  can be represented in the language of quiver mutations, say for $N_f=2 N_c$.
  Hint: refer to the quiver in Fig.~\ref{trig_mutation}. Consider the case $N_f=2 N_c$, and keep manifest only the
  $SU(N_c)^4$ global symmetry among the $SU(N_f)\times SU(N_f)$ global symmetry acting on the quarks and anti-quarks, respectively. To check the agreement with the mutation rule,
  the information of the superpotential is also indispensable.
  \label{ex.Seiberg}

  \item $[\bll]$ (Flips in cluster algebras)
  Consider the quiver given by an ideal triangulation of the four-punctured sphere (Fig.~\ref{fig.quiver_4sphere}).
  Verify that, when we apply a flip to the ideal triangulation,
  the corresponding quiver changes by the mutation at the quiver vertex corresponding to the flipped edge.
  Moreover, verify that the corresponding transformation rules \eqref{cluster_x_def} and \eqref{cluster_y_def} of the cluster $x$- and $y$-variables
  agree with the transformation rules \eqref{Ltransf} and \eqref{Focktransform} of the Penner and Fock variables.

  \item $[\bll]$ (Transformation rule for quantum cluster $y$-variables)
  Verify that the variables $\sfx'_i$ given by
  \eqref{ymutation1} and \eqref{ymutation2}
  satisfy the commutation relations \eqref{xCCR}
  for the new quiver $\mu_k \mathsf{Q}$ defined in \eqref{Qmutate}.

  \label{q_mutation_consistency}

  \item $[\bll]$ (Quantum dilogarithm identities of type $A_3, A_4$)

  In Fig.~\ref{pentagon_id_SGC} we discussed the ideal triangulation of the pentagon.
  Replace the pentagon by a hexagon or a heptagon.
  Find mutation sequences generating non-trivial quantum dilogarithm identities in these cases,
  this time for the $A_3$ and $A_4$ quivers
  \begin{align}
    1\longleftarrow 2 \longrightarrow 3 \ ,  \quad
    1\longleftarrow 2 \longrightarrow 3 \longleftarrow 4 \ .
  \end{align}
  Show also that the resulting identities follow from repeated use of the pentagon identity (see \cite{Kim:2016eox} for more on this topic).

\end{practice}

%
\backmatter
\clearpage
\phantomsection\addcontentsline{toc}{chapter}{Bibliography}
\bibliographystyle{ytphys}
\bibliography{SGCbook_e}

\clearpage
\phantomsection\addcontentsline{toc}{chapter}{List of Symbols}
\printnomenclature

\clearpage
\phantomsection\addcontentsline{toc}{chapter}{Index}
\index{duality@duality|seealso{Seiberg duality, S-duality, 3d mirror symmetry, 2--3 mirror symmetry}}
\printindex 

\end{document}